\documentclass[twocolumn]{aastex62}

\newcommand\Ha{\ifmmode {\mathrm H}\alpha \else H$\alpha$\fi}
\newcommand\Hb{\ifmmode {\mathrm H}\beta \else H$\beta$\fi}
\newcommand{\halpha}{H{$\alpha$}}
\newcommand{\hbeta}{H{$\beta$}}

\def\FeII{Fe\,{\sc ii}}
\def \OIII {[O\,{\sc iii}]}

\def\ergs{${\rm erg\,s^{-1}}$}

\newcommand{\aox}{\ifmmode{\alpha_{\mathrm{ox}}} \else $\alpha_{\mathrm{ox}}$\fi} 
\newcommand{\atoms}{\ifmmode{\mathrm{\,atoms~cm^{-2}}} \else \,atoms cm$^{-2}$\fi}
\newcommand{\ax}{\ifmmode{\alpha_x} \else $\alpha_x$\fi} 
\newcommand{\bprp}{\ifmmode{B_P-R_P} \else $B_P-R_P$\fi} 
\newcommand{\cmsq}{\ifmmode{\mathrm{cm^{-2}}} \else cm$^{-2}$\fi}
\newcommand{\degsq}{\ifmmode {\mathrm{deg^2}} \else deg$^2$\fi}
\newcommand{\perdegsq}{\ifmmode {\mathrm{deg^{-2}}} \else deg$^{-2}$\fi}
\newcommand{\ew}{\ifmmode{W_{\lambda}} \else $W_{\lambda}$\fi}
\newcommand{\fbol}{\ifmmode f_{\mathrm{bol}} \else $f_{\mathrm{bol}}$\fi} 
\newcommand{\fcgs}{\ifmmode \mathrm{erg~cm^{-2}~s^{-1}}\else erg~cm$^{-2}$~s$^{-1}$\fi}
\newcommand{\flamcgs}{\ifmmode \mathrm{erg\,cm^{-2}\,s^{-1}\,\AA^{-1}}\else erg\,cm$^{-2}$\,s$^{-1}$\,\AA$^{-1}$)\fi}
\newcommand{\fnucgs}{\ifmmode {\mathrm{erg~cm^{-2}~s^{-1}~Hz^{-1}}}\else erg~cm$^{-2}$~s$^{-1}$~Hz$^{-1}$\fi}
\newcommand{\gax }{{\lower0.8ex\hbox{$\buildrel >\over\sim$}}}
\newcommand{\kms}{\ifmmode~{\mathrm{km~s}}^{-1}\else ~km~s$^{-1}~$\fi}
\newcommand{\lax }{{\lower0.8ex\hbox{$\buildrel <\over\sim$}}}
\newcommand{\lcgs}{\ifmmode \mathrm{erg~s^{-1}}\else erg~s$^{-1}$\fi}
\newcommand{\lnucgs}{\ifmmode erg~s^{-1}~Hz^{-1}\else erg~s$^{-1}$~Hz$^{-1}$\fi}

\newcommand{\logz}{\ifmmode{\mathrm{log}}~z \else log$~z$\fi}
\newcommand{\lo}{\ifmmode l_o \else $~l_o$\fi}
\newcommand{\Lo}{\ifmmode L_o \else $~L_o$\fi}
\newcommand{\lx}{\ifmmode l_x \else $~l_x$\fi}
\newcommand{\Lx}{\ifmmode L_x \else $~L_x$\fi}
\newcommand{\lbol}{\ifmmode L_{\mathrm{bol}} \else $L_{\mathrm{bol}}$\fi}
\newcommand{\Lbol}{\ifmmode L_{\mathrm{bol}} \else $L_{\mathrm{bol}}$\fi}
\newcommand{\LBol}{\ifmmode L_{\mathrm{bol}} \else $L_{\mathrm{bol}}$\fi}
\newcommand{\LEdd}{\ifmmode L_{\mathrm{Edd}} \else $L_{\mathrm{Edd}}$\fi}
\newcommand{\LxLbol}{\ifmmode L_x/L_{\mathrm{bol}} \else $L_x/L_{\mathrm{bol}}$\fi}
\newcommand{\rEdd}{\ifmmode L_{\mathrm{bol}}/L_{\mathrm{Edd}} \else $L_{\mathrm{bol}}/L_{\mathrm{Edd}}$\fi}
\newcommand{\REdd}{\ifmmode L_{\mathrm{bol}}/L_{\mathrm{Edd}} \else $L_{\mathrm{bol}}/L_{\mathrm{Edd}}$\fi}
\newcommand{\Rblr}{\ifmmode {R_{\mathrm BLR}} \else $R_{\mathrm BLR}$\fi}
\newcommand{\lamEdd}{\ifmmode \lambda_{\mathrm{Edd}} \else $\lambda_{\mathrm{Edd}}$\fi}
\newcommand{\mbh}{\ifmmode {M_{\rm BH}}\else${M_{\rm BH}}$\fi}
\newcommand{\Mbh}{\ifmmode {M_{\rm BH}}\else${M_{\rm BH}}$\fi}
\newcommand{\mdot}{\ifmmode \dot{m} \else $\dot{m}$\fi}
\newcommand{\mdote}{\ifmmode \dot{m}_{E} \else $\dot{m}_{E}$\fi}
\newcommand{\mone}{\ifmmode ^{-1}\else$^{-1}$\fi}
\newcommand{\msun}{\ifmmode {M_{\odot}}\else${M_{\odot}}$\fi}
\newcommand{\Msun}{\ifmmode {M_{\odot}}\else${M_{\odot}}$\fi}
\newcommand{\mtwo}{\ifmmode ^{-2}\else$^{-2}$\fi}
\newcommand{\Mvir}{\ifmmode {M_{\rm BH}^{\mathrm SE}}\else${M_{\rm BH}^{\mathrm SE}}$\fi}
\newcommand{\nhgal}{\ifmmode{ N_{H}^{Gal}} \else N$_{H}^{Gal}$\fi}
\newcommand{\nh}{\ifmmode{\mathrm N_{H}} \else N$_{H}$\fi}
\newcommand{\nhintr}{\ifmmode{ N_{H}^{intr}} \else N$_{H}^{intr}$\fi}
\newcommand{\nhtot}{\ifmmode{ N_{H}^{tot}} \else N$_{H}^{tot}$\fi}
\newcommand{\nhz}{\ifmmode{ N_{H}^z} \else N$_{H}^z$\fi}
\newcommand{\oi}{\ifmmode{\mathrm [O\,II]} \else [O\,II]\fi}
\newcommand{\oii}{\ifmmode{\mathrm [O\,II]} \else [O\,II]\fi}
\newcommand{\oiii}{\ifmmode{\mathrm [O\,III]} \else [O\,III]\fi}

\graphicspath{{./}{figures/}}

\usepackage{color,subfigure}

\usepackage{ulem}
\usepackage{amsmath,bm}
\usepackage{commath}
\usepackage[figuresright]{rotating}
\usepackage{tabularx}
\usepackage[vlines]{tabularht}
\usepackage{multirow}
\usepackage{longtable}
\usepackage{textcomp}
\usepackage{booktabs}
\DeclareUnicodeCharacter{2212}{-}
\usepackage{savesym}
\savesymbol{tablenum}
\usepackage{siunitx}
\restoresymbol{SIX}{tablenum}
\shorttitle{Turn-on CLQs}
\shortauthors{Yang et al.}
\begin{document}

\title{Galaxies Lighting Up: Discovery of Seventy New Turn-on Changing-look Quasars} 

\correspondingauthor{Qian Yang}
\email{qian.yang@cfa.harvard.edu} 

\author[0000-0002-6893-3742]{Qian Yang}
\affiliation{Center for Astrophysics $\vert$ Harvard \& Smithsonian, 60 Garden Street, Cambridge, MA 02138, USA}

\author[0000-0002-8179-9445]{Paul J. Green}
\affiliation{Center for Astrophysics $\vert$ Harvard \& Smithsonian, 60 Garden Street, Cambridge, MA 02138, USA}

\author[0000-0002-7350-6913]{Xue-Bing Wu}
\affil{Department of Astronomy, School of Physics, Peking University, Beijing 100871, P. R. China}
\affil{Kavli Institute for Astronomy and Astrophysics, Peking University, Beijing, 100871, P. R. China}

\author[0000-0002-3719-940X]{Michael Eracleous}
\affiliation{Department of Astronomy and Astrophysics and Institute for Gravitation
and the Cosmos, Penn State University, 525 Davey Lab, 251 Pollock Road, University Park, PA
16802}

\author[0000-0003-4176-6486]{Linhua Jiang}
\affil{Kavli Institute for Astronomy and Astrophysics, Peking University, Beijing, 100871, P. R. China}

\author[0000-0002-0759-0504]{Yuming Fu}
\affiliation{Leiden Observatory, Leiden University, P.O. Box 9513, NL-2300 RA Leiden, The Netherlands}

\begin{abstract}
``Changing-look quasars" (CLQs), discovered less than a decade ago, show dramatic, rapid changes in optical/UV continuum and broad line emission.  The majority of CLQs have been found dimming as ``turn-off" CLQs because most selection methods start from samples of spectroscopically-confirmed quasars.
We present here a sample of 82 spectroscopically confirmed ``turn-on" CLQs, 70 of which are newly identified. 
The turn-on CLQs are selected from spectroscopically classified galaxies with subsequent significant and dramatic variability in both the optical and mid-infrared bands, indicating a mechanism of changing accretion rate of the supermassive black holes rather than variable obscuration.  
Based on their bright state Eddington ratios, turn-on CLQs are associated with lower accretion rates compared to turn-off CLQs or typical SDSS quasars with similar redshift and magnitude distributions, even though turn-on CLQs have lower black hole masses.  
Most turn-on CLQs reside in host galaxies that follow local relations between the central black hole mass and host galaxy properties, such as stellar mass and velocity dispersion.  However, their host galaxies have higher mass than normal inactive galaxies, with star formation rates more similar to hosts of Type 2 AGN than to the overall galaxy population.
\end{abstract}
%** abstract length limit is 250 words

% Quasars (1319); Active galactic nuclei (16); Supermassive black holes (1663); X-ray binary stars (1811)

\section{Introduction} \label{sec:introduction}

Supermassive black holes (SMBHs) are ubiquitous in the centers of massive galaxies \citep[e.g.,][]{Kormendy1995, Magorrian1998}
and predominantly grow by accreting gas from surrounding materials. 
Actively accreting SMBHs may produce radiation that spans the entire electromagnetic spectrum, depending on their accretion rate and obscuring materials. 

Accretion onto black holes (BHs), from stellar mass BHs in Galactic X-ray binaries (XRBs) to supermassive black holes (SMBHs) in Seyferts and quasars, illuminates fascinating and important physics. 
Theoretical viscous timescales in accretion theory \citep{Krolik1999, Frank2002}, along with mass scaling from XRBs to SMBHs \citep{Sobolewska2011}, suggest that dramatic state changes in AGN should span $10^{4-7}$ years. However, just in the last few years, ``changing-look'' (CL) AGN were discovered shifting between bright and dim states, on timescales from only months to years \citep[e.g.,][]{Denney2014, LaMassa2015, Runnoe2016, MacLeod2016, Gezari2017, Yang2018, Green2022, Yang2023}. 
These phenomena challenge and invigorate debates about both accretion theory and the AGN unification model \citep{Antonucci1993, Urry1995}.

The term ``changing-look" originated from the study of X-ray variability, referring to AGN changing between Compton thick and Compton thin, due to changing obscuration \citep[e.g.,][]{Matt2003, Bianchi2003, Piconcelli2007, Ballo2008, Marchese2012, Ricci2016}. More recently, dramatic optical changes in continuum and broad emission line (BEL) strengths have been recognized in a small fraction of AGN, dubbed changing look AGN (CL AGN), where the optical spectra show changes in type e.g., as defined by \citet{Osterbrock1993}.  Seyfert Type 1s show both broad and narrow emission lines, while Seyfert 2s have only narrow emission lines (NELs).  Intermediate types include Seyfert 1.5 (where broad and narrow H$\beta$ are of comparable strength), 1.8 (very weak but detectable broad lines in both H$\beta$ and H$\alpha$) and 1.9 (broad emission detectable only in H$\alpha$).

With long-term observations, more and more AGN were reported to show changing-look phenomena, e.g., 
Mrk 1018 \citep{McElroy2016},
1ES 1927+654 \citep{Trakhtenbrot2019, Ricci2020, Ricci2021},
NGC 1566 \citep{Oknyansky2019, Oknyansky2020, Parker2019, Jana2021, Tripathi2022}, 
NGC 3516 \citep{Shapovalova2019, Mehdipour2022, Popovic2023}, 
HE 1136-2304 \citep{Parker2016, Zetzl2018}, 
NGC 4151 \citep{Mahmoud2020},
NGC 2992 \citep{Guolo2021},
NGC 4156 \citep{Tozzi2022},
and NGC 5273 \citep{Neustadt2023}.

These highly variable Seyferts typically have optical luminosities $10^{42-44.5}$\ergs\, in the bright state \citep{Yee1980}.  
Seyferts with lower luminosities than quasars ($\gtrsim 10^{45}$\ergs), often have fractionally stronger and more rapid variability than quasars.  However, \citet{LaMassa2015} reported the discovery of the first ``changing-look quasar" (CLQ) J015957.64+003310.5 (hereafter J0159+0033), which transitioned from a Type 1 quasar to a Type 1.9 AGN, demonstrating that dramatic spectral transitions are possible in luminous quasars. The Time Domain Spectroscopic Survey (TDSS) identified another CLQ, SDSS J101152.98+544206.4 (hereafter J1011+5442). Initially observed as a typical Type 1 broad-line quasar, a later spectrum revealed a galaxy spectrum with only a weak quasar continuum and residual broad H$\alpha$ \citep{Runnoe2016}.  A number of intensive studies of individual CLQs have been published since then e.g., \citealt{Gezari2017, Stern2018, Wang2018, Nagoshi2021, Saha2023}.  

Subsequent systematic searches for CLQs have been led to the discovery of dozens to hundreds of new CLQs, using multi-object multi-epoch spectroscopic surveys, including SDSS \citep{MacLeod2016, Ruan2016, MacLeod2019, Green2022}, the Large Sky Area Multi-Object Fiber Spectroscopic Telescope \citep[LAMOST; ][]{Yang2018}, the Dark Energy Spectroscopic Instrument \citep[DESI; ][]{Guo2024}, and the SDSS-V Black Hole Mapper \citep{Zeltyn2024}. However, the majority of them are turn-off CLQs. This is due to the selection bias, with parent samples selected from the SDSS quasar catalogs. 

The accretion inflow timescale is shorter in the low state than in the high state \citep[e.g.,][]{Dexter2019_mag, Feng2021b}. Therefore, the time-scales observed in turn-on CLQ transitions provide tighter constraints on the models.
The number of fading (`turn-off') quasars in a large sample can be used to measure or set a lower limit on the episodic lifetime of quasars \citep{Martini2003}. On human timescales, it must be balanced by the `turn-on' rate, to maintain the observed quasar sky density. Significant dimming on much longer timescales ($10^{3-5}$ years) is also known, discovered through the identification of strongly ionized clouds in extended emission line regions with no currently active nuclear emission source nearby \citep[e.g.,][]{Lintott2009,Keel2024}.

In this paper, we present the largest sample to date of spectroscopically confirmed turn-on CLQs, comprising 70 new discoveries. These turn-on CLQs were identified from a parent sample of spectroscopically classified galaxies in SDSS, selected based on subsequent dramatic variability at both the optical and mid-infrared wavelengths. The unique nature of turn-on CLQs offers opportunities to study the evolution of AGN over relatively short timescales, thereby contributing to a more dynamic picture of SMBH accretion processes.

This paper is organized as follows. Section \ref{sec:selection} describes our target selection process. Section \ref{sec:data} describes the spectroscopic follow-up observations, as well as our spectral fitting techniques. In Section \ref{sec:results}, we present the turn-on CLQ results. In Section \ref{sec:discussion}, we discuss the host properties and time scale of turn-on CLQs. We summarize the paper in Section \ref{sec:summary}. In this work we adopt a standard $\Lambda$CDM cosmology with $\Omega_{\Lambda}=0.7$, $\Omega_m=0.3$, and $H_0=70$ km s$^{-1}$ Mpc$^{-1}$.

\section{CLQ Target Selection} \label{sec:selection}

We here describe our selection procedures and sample size, which are summarized in Table\,\ref{tab:selection}. 

\begin{deluxetable*}{llr}[htbp]
\tablecaption{\ Turn-on CLQ Candidate Selection and Observation \label{tab:selection}}
\tablewidth{0pt}
\tablehead{
\colhead{Description} &
\colhead{Criteria} &
\colhead{Number}
}
\startdata
% \\
Spectroscopically observed in SDSS DR17 & All & 5,801,200 spectra\\
Spectroscopically classified as galaxy & Class == ``GALAXY" & 3,237,535 spectra \\
Spectra with good quality & $zWarning = 0$ \& SNR $>$ 1 & 2,702,761 spectra \\
% \\
Significantly variable in MIR & WISE SN$_{W1} \geq 5$ \& SN$_{W2} \geq 5$ & 12,025 spectra\\
WISE variability amplitude & $\Delta_{W1} \geq 0.2~$mag \& $\Delta_{W2} \geq 0.2~$mag &  5,236 spectra\\
% \\
Unique sources & 2$\arcsec$ coordinates cross-match & 4,750 galaxies \\
Not radio source & Not detected by FIRST & 3,929 galaxies \\
% \\
Significantly variable in optical & ZTF SN$_g \geq 4$ & 653 galaxies\\
% \\
\hline
Known CLQs in literature & & 36 confirmed CLQs\\
\hline
Spectroscopic Follow-up & & 115 + 2 objects \\
Spectroscopy confirmed turn-on CLQs & CLQ quantitive criteria in \textsection \ref{sec:criteria} & $82$ turn-on CLQs \\ % removed J1027
Newly discovered turn-on CLQs events & & \textbf{70 New turn-on CLQs} \\
\hline
% \\
\enddata
\end{deluxetable*}

\subsection{The SDSS Parent Galaxy Sample}
Since we seek turn-on CLQs, galaxies whose supermassive black holes have begun to accrete within recent decades, we start from SDSS spectroscopically observed objects classified as galaxies. 
There are 5,801,200 spectra in the SDSS Seventeenth Data Release \citep[DR17;][]{SDSS_DR17} taken by the Sloan Foundation 2.5m telescope \citep{Gunn2006} at the Apache Point Observatory. The spectra are from the SDSS-I/II with a wavelength coverage from 3800 to 9100\,\AA, and the Baryon Oscillation Spectroscopic Survey \citep[BOSS;][]{Dawson2013} spectrograph of the SDSS-III \citep{Eisenstein2011} with a wavelength coverage from 3600 to 10400\,\AA\ \citep{Smee2013}. The spectral resolution is 1500 at 3800\,\AA and 2500 at 9000\,\AA. 
SDSS-I/II has 640 fibers, with a fiber diameter of 3$\arcsec$, and BOSS has 1000 fibers, with a fiber diameter of 2\arcsec.

The SDSS spectroscopic pipelines classify the objects as galaxies (``GALAXY"), stars (``STAR"), or quasars (``QSO"), through the comparison of individual spectrum with galaxy, QSO, and stellar templates \citep{Bolton2012, Hutchinson2016}. 
We first require an SDSS spectroscopic pipeline classification of “GALAXY”, which encompasses  
56\% (3.2 million) objects within DR17.   We further impose spectral quality criteria, requiring ``zWarning" as zero for good data without identified problems \citep[e.g.,][]{Stoughton2002} and the median signal-to-noise (SNR) larger than 1, resulting in 2.7 million spectra. 

\subsection{Photometric Selection Criteria} \label{sec:photo}

Variability is a hallmark of accreting black holes, while normal galaxies are usually quiescent. To find galaxies whose supermassive black holes have begun to accrete within the last decade or so, we select galaxies showing strong intrinsic mid-IR variability, which has proven successful for finding CL-AGN  (e.g., \citealt{Yang2018, Stern2018}).

We use multi-epoch MIR observations from WISE to select CLQ candidates.
The {\it WISE} mission scanned the full sky from January to July in 2010 in four bands centered at wavelengths of 3.4, 4.6, 12, and 22 $\mu$m ($W1$, $W2$, $W3$, and $W4$). The secondary cryogen survey and Near-Earth Object Wide-field Infrared Survey Explorer \citep[{\it NEOWISE};][]{Mainzer2011} Post-Cryogenic Mission mapped the sky from August, 2010 to February, 2011. The {\it NEOWISE} Reactivation Mission \citep[{\it NEOWISE-R};][]{Mainzer2014} surveys the sky in $W1$ and $W2$ bands from 2013 twice a year. {\it WISE} obtains $\sim 10-20$ observations within a 36-hrs window in each visit. We calculate the median magnitude and magnitude error, specifically the semi-amplitude of the range enclosing the 16th and 84th percentiles of all flux measurements within a 36-hrs window. We limit to good quality single-epoch data points with the best frame image quality score ($qi\_fact=1$), observed far away from the South Atlantic Anomaly ($saa\_sep \geq 5$), with no contamination from the moon ($moon\_masked=0$), and excluding spurious detection ($cc\_flags=0$). The {\it WISE} magnitudes are profile-fitting magnitudes, and are converted from Vega to AB magnitude as $m_{\rm AB} = m_{\rm Vega} + \Delta m$, where $\Delta m$ is 2.699 and 3.339 in $W1$ and $W2$ bands, respectively \citep{Jarrett2011}.

Observed variability includes both intrinsic variability and photometric uncertainties.
To characterize the intrinsic variability of each light curve we use the method of \citet{Yang2020}. Briefly, we use the maximum-likelihood estimator detailed in \citet[][Equations 5--9]{Shen2019}\footnote{We correct a typo in their Eqn. (9): ${\rm Var}[\mu]=\sigma_0^2/\Sigma g_i$.}. The estimate of the intrinsic variability in one band, $\sigma_{\rm band}$, and its uncertainty $\Delta\sigma_{\rm band}$ are defined by Eqn.\ (8) of \citet{Shen2019}. We then define the SNR of the estimated intrinsic variability as ${\rm SN_{band}} = \sigma_{\rm band}/\Delta\sigma_{\rm band}$. We require significant variability detection in the MIR light curve as ${\rm SN_{W1}} \geq 5$ and ${\rm SN_{W2}} \geq 5$. Since galaxies normally show weak or undetectable variability, this criterion rules out the vast majority of galaxies, yielding of order 12k objects. \citet{Yang2018} found that varations of more than 0.2 mag in WISE were detected in most CL AGN. Therefore, we further require the objects must brighten by 0.2 mag both in WISE $W1$ and $W2$ bands. This criterion leaves 4750 objects. 

Since we are interested in accretion- rather than jet- related variability, we cross-match the FIRST survey \citep{Becker1995} to rule out radio-detected objects. This criterion results in 3929 objects. 

Some AGN were found with dramatic MIR variability but little optical variability. A systematic search for MIR flares in nearby galaxies shows only a small fraction (11\%) has corresponding optical flares \citep{Jiang2021, Wang2022_MIR}. One extreme case shows $>3$ mag WISE variability without changes in the optical over the same period ($\leq 0.2$ mag); \citealt{Yang2019}). These objects were interpreted as turn-on CLQs or tidal disruption events that are heavily obscured. 

To select CLQs with high purity (a low false identification rate), we require the objects be significantly variable in  both MIR and optical light curves. Since in this study we are interested in the effects of strong variability on the broad emission line region (BLR), our confirmation of CLQ status requires dramatic optical variability.  Below, we describe the optical photometric data that we use and our analysis.

The Zwicky Transient Facility survey \citep[ZTF;][]{Bellm2019} is a optical time-domain survey that uses the Palomar 48 inch Schmidt telescope, in $gri$ bands. ZTF achieved first light in 2017 October. The Northern Sky Survey in ZTF is a three-day cadence survey of all fields with centers north of $\delta = -31^\circ$. Median 5$\sigma$ limiting magnitudes are 20.8, 20.6, and 19.9 mag in $g$, $r$, and $i$ bands, respectively. 
We used ZTF Data Release 20, which includes ZTF data up to 31 October 2023.

To reduce the impact of outliers and to highlight underlying trends in the ZTF photometry, we start with the 
the individual ZTF magnitudes and errors\footnote{https://irsa.ipac.caltech.edu/data/ZTF/docs/ztf\_explanatory\_supplement.pdf},
and use a smoothing window of 30 days around each point. 
We use the median value as the smoothed magnitude and half of the range between the 16th and 84th percentiles as the magnitude uncertainty. Then we calculate the intrinsic variability and require a significant variability in the ZTF $g$ band as SN$_g \geq 4$.

Our final CLQ candidate sample includes 653 sources, for which we sought to obtain follow-up spectroscopy to detect strong broad emission line variability. We summarize the details of the 653 candidates in Table \ref{tab:Candidates}. Throughout the manuscript, we use the short names instead of the full coordinates for the targets. 
Among the 653 candidates, there are 35 objects known as CLQs in literature, including J1011+5442 in \citet{Runnoe2016}, J1324+4802 \citep{Ruan2016, MacLeod2016}, iPTF16bco \citep{Gezari2017, Frederick2019}, 12 CL AGN in \citet[][; see Table \ref{tab:Candidates}]{Yang2018}, all (six) discovered CL low-ionization narrow emission-line regions (LINERs) in \citet{Frederick2019}, three CL AGN in \citet{Lopez-Navas2022}, and 12 CLQs in \citet{Wang2024}. Besides these, one candidate J0158-0052 was identified as a TDE in \citet{Blanchard2017}.  

\section{Spectroscopic Observations, Reduction and Analysis} \label{sec:data}

We performed spectroscopic follow-up using multiple telescopes with mirror diameters spanning from 2 to 10 meters, depending on their brightness. Table \ref{tab:telescope} summarizes the number of observations from each telescope. In total, there are 127 spectroscopic observations. Some objects were observed at more than one epoch. We next describe details for the telescopes and instruments used.

\begin{deluxetable}{lr}[htbp]
\tablecaption{Spectroscopic Follow-up \label{tab:telescope}}
\tablewidth{500pt}
\tabletypesize{\scriptsize}
\tablehead{
\colhead{Telescope} &
\colhead{Number}\\
\colhead{ } &
\colhead{of Spectra}
}
\startdata
    MMT & 59 \\
    LAMOST & 29 \\
    DBSP & 12 \\
    DESI & 8 \\
    SDSS-V & 5 \\
    SDSS & 2 \\
    HET & 4 \\
    LJT & 3\\
    XLT & 5 \\
\enddata
\end{deluxetable}

\subsection{MMT}
The majority of spectroscopic follow-up observations were obtained with the Binospec Spectrograph on the MMT\,6.5m Telescope situated on Mount Hopkins, Arizona \citep{Binospec}. Observations were carried out over several observational runs from 2021B to 2023B (see Table\,\ref{tab:Observations} for exact dates). 
The 270 $l$ mm$^{-1}$ grating was used with a LP3800 filter and a longslit width of 1$\arcsec$, providing a resolution of $R = 1340$. The central wavelength is 6500\,\AA, thus the wavelength coverage is 3900--9240\,\AA. In total, 59 candidates were observed by MMT.

\subsection{Hobby–Eberly Telescope}
Spectroscopic observations using the Hobby–Eberly Telescope (HET) were taken in April--July, 2023 \citep{Ramsey1998, Hill2021}.
The HET is a 10-meter aperture telescope located at the McDonald Observatory in Davis Mountains, Texas. We used the Low-Resolution Spectrograph 2 (LRS2) LRS2-R spectrograph. Objects with $z < 0.4$ was observed with the LRS2-B spectrograph, simultaneously covering from 370--470 nm (orange arm) and 460--700 nm (red arm) at a resolving power of 1900 and 1100 in each channel, respectively \citep{Chonis2016}. Objects with $z \geq 0.4$ was observed with the LRS2-R spectrograph, simultaneously covering from 650--842 nm (red arm) and 818--1050 nm (far red arm) at a resolving power of 1800 in each channel, respectively. We used the HET for four of our CLQ targets.

\subsection{Palomar Hale Telescope}
Observations using the Hale telescope was taken in May and Nov 2019.
The Palomar Hale telescope is a 200-inch (5.1 m) telescope at the Palomar Observatory, California. 
The Double Spectrograph (DBSP) was used with a 600 $l$ mm$^{-1}$ grating (blazed at 3780\,\AA) at blue side and 316 $l$ mm$^{-1}$ (blazed at 7150\,\AA) at red side. As the seeing was 1.2--1.5$\arcsec$, a slit width of 1.5$\arcsec$ was used, leading to $R\sim1100$ at 4500\,\AA\ and $R\sim950$ at 7500\,\AA. We used Palomar DBSP for 12 candidates.

\subsection{Lijiang 2.4m Telescope}
Spectroscopic follow-up of some bright candidates were taken by the Lijiang telescope (LJT) from Dec 2017 to Feb 2020.
The LJT is a 2.4 m telescope is located at Lijiang Observatory, China \citep{LJT}. It is equipped with the Yunnan Faint Object Spectrograph and Camera (YFOSC).
We used Grism 3 (G3) with a dispersion of 172\,\AA\ mm$^{-1}$ and wavelength coverage from 340 to 910 nm. We used a 1.8$\arcsec$ slit or 2.5$\arcsec$ slit when the seeing is smaller or larger than 2\arcsec, yielding a resolution of $R\sim670$ and $R\sim 250$, respectively. We observed 3 candidates using LJT.

\subsection{Xinglong 2.16m Telescope}
Several bright candidates were observed by the  Xinglong telescope (XLT) from 2018 to 2019.
XLT is a 2.16 m telescope located at the NAOC observatory, China \citep{Xinglong}.
It is equipped with the Beijing Faint Object Spectrograph and Camera (BFOSC).
using the BFOSC and Grism 4 (G4) with a dispersion of 198\,\AA\ mm$^{-1}$ and a wavelength coverage from 3850 to 8300\,\AA. 
We used a 1.8$\arcsec$ slit or 2.3$\arcsec$ slit when the seeing is smaller or larger than 2$\arcsec$, yielding a resolution of $R\sim340$ and $R\sim 265$, respectively. We observed 5 candidates using XLT.

\subsection{DESI Spectroscopy}

The DESI instrument is a robotically-actuated, fiber-fed spectrograph capable of capturing up to 5,000 simultaneous spectra across a wavelength range of 360 nm to 980 nm. The fibers feed an array of ten three-arm spectrographs with resolution $R = \lambda/\Delta \lambda$ between 2000 and 5500, depending on wavelength. This instrument covers an 8 square degree focal plane on the 4-m Mayall telescope in Kitt Peak, Arizona. We used spectra from the DESI Early Data Release (EDR). The EDR includes spectra and redshifts for 1.2 million galaxies and quasars, as well as nearly half a million stars \citep{DESI2024a, DESI2024b}. We obtained followup (bright state) spectroscopy for 8 of our CLQ candidates from the DESI EDR.

\subsection{LAMOST}

LAMOST is a 4-meter reflecting Schmidt telescope equipped with 4000 fibers and a 5-degree field of view \citep{Cui2012, Zhao2012}. It covers a wavelength range from 3700 -- 9000\,\AA\ with a blue arm (3700--5900\,\AA) and a red arm (5700--9000\,\AA; \citealt{Du2016}). The overall spectral resolution of LAMOST is approximately 1800. The data are processed using the LAMOST pipelines \citep{Luo2012}. In this paper, we utilize LAMOST spectra from data release eight (DR8; \citealt{Luo2015, He2016}) for the bright state spectroscopic epoch. The DR8 dataset includes a total of 17.23 million spectra. We obtained followup (bright state) spectroscopy for 29 CLQ candidates from LAMOST.   

\subsection{SDSS}
The final data release of SDSS-IV was DR17, in 2021 December. The SDSS eighteenth data release \citep[DR18;][]{SDSS-DR18} is the first release for the fifth generation of the SDSS survey (SDSS-V). SDSS-V uses the Sloan 2.5 m telescope at Apache Point Observatory (APO) in New Mexico, USA (Gunn et al. 2006) and the du Pont 2.5 m telescope at LCO in Chile. We found 5 candidates were observed within SDSS DR18. Also, two candidates were observed at more than one epoch and captured in both states within SDSS DR17.

\subsection{Data Reduction}
For DESI and SDSS, we used flux-calibrated spectra produced by their pipelines and provided in their data releases. 
For LAMOST, we followed \citet{Yang2018} to align the spectra in the two arms if there is a break around 5700--5900\,\AA.
The MMT spectra were reduced by the binospec pipeline.\footnote{https://bitbucket.org/chil\_sai/binospec/wiki/Home} 
The HET spectra were reduced with the Panacea software package\footnote{The
Panacea software package was written by Gregg Zeimann and running on
the Texas Advanced Computing center, available at https://github.com/grzeimann/Panacea/blob/master/README.md\#Code\-Description.}.
The other spectra were reduced using standard IRAF\footnote{IRAF is distributed by the National Optical Astronomy Observatory, which is operated by the Association of Universities for Research in Astronomy (AURA) under cooperative agreement with the National Science Foundation.} routines \citep{Tody1986, Tody1993}. 
The spectra were initially flux calibrated using standard stars observed on the same night. 
We discuss further flux calibration using imaging observations in \textsection \ref{sec:calibration}.

\subsection{Flux Calibration \label{sec:calibration}}

Even when using standard star observations during our follow-up, we were not always able to achieve reliable spectroscopic flux calibrations, critical for comparison of different states of CLQs, so we have analyzed other alternatives.

The narrow [O III] emission line has been widely used for flux calibration between mu;ti-epoch spectroscopy of AGN, by assuming that narrow-line flux is invariable on decade timescales both for reverberation mapping programs \citep[e.g.,][]{Shen2019} and CL AGN \citep[e.g.,][]{Yang2018, MacLeod2019}. However, some CL AGN have been observed to show narrow-line variability \citep{Yan2019, Li2022}. 
Measurement of narrow [OIII] lines (e.g., in J0936+2726) may be strongly affected by the iron lines appearing in the bright state, so the assumption of constant narrow [OIII] luminosity is not reliablefor flux calibration.
In addition, the use of narrow-emission line for flux calibration depends strongly on the spectral fitting code performing continuum and line emission decomposition, where results may be SNR dependent. 

To properly flux calibrate our bright state follow-up spectroscopy, we therefore chose to use imaging photometry from epochs close to the bright-state spectra. The SNR ratio for broad-band photometry is much higher than for narrow-emission lines, and  photometry in multiple bands also allows spectral shape corrections. As described in Section \ref{sec:photo}, ZTF is a three-day cadence optical image survey (covering 2017-2023 in ZTF DR20), and there are multiple images taken for the same position on the same night. We found ZTF images available for our targets close to our epochs of follow-up spectroscopy, mostly within 0.2--7 days (see $|\Delta t_{photo}|$ in Table \ref{tab:Observations} for each spectroscopic observation). Assuming the AGN flux does not dramatically change within such as a short period\footnote{Although our targets vary, the timescale is usually longer. For example, even for an extreme variable quasar that varies by 1 mag within a period of about a year, there is typically less than 0.003 (0.02) mag variability within 1 day (week).}, we therefore used the ZTF photometry to calibrate our bright state spectra

\begin{figure*}[htbp]
  \centering
  \hspace{0cm}
  \subfigure{
  \includegraphics[width=0.31\textwidth]{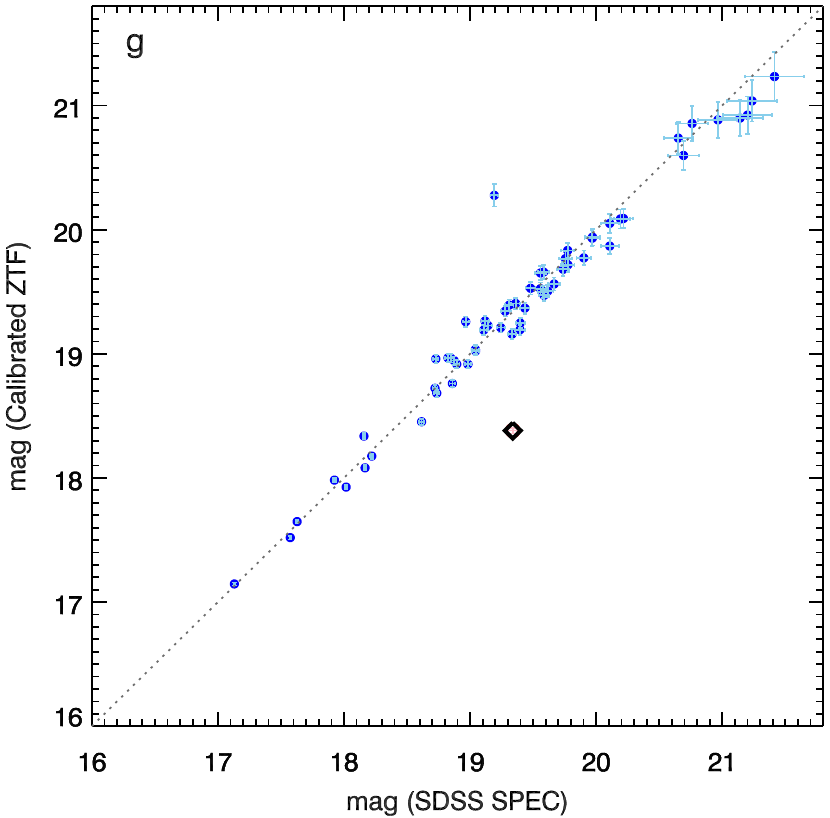}}
 \hspace{-0cm}
 \subfigure{
  \includegraphics[width=0.31\textwidth]{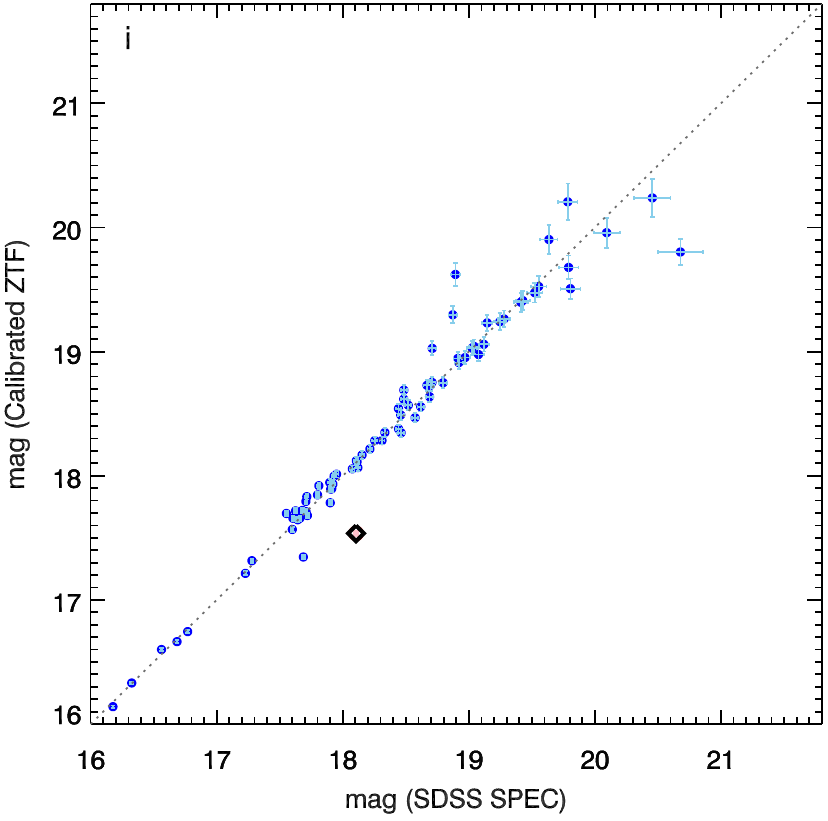}}
\hspace{-0cm}
 \subfigure{
  \includegraphics[width=0.31\textwidth]{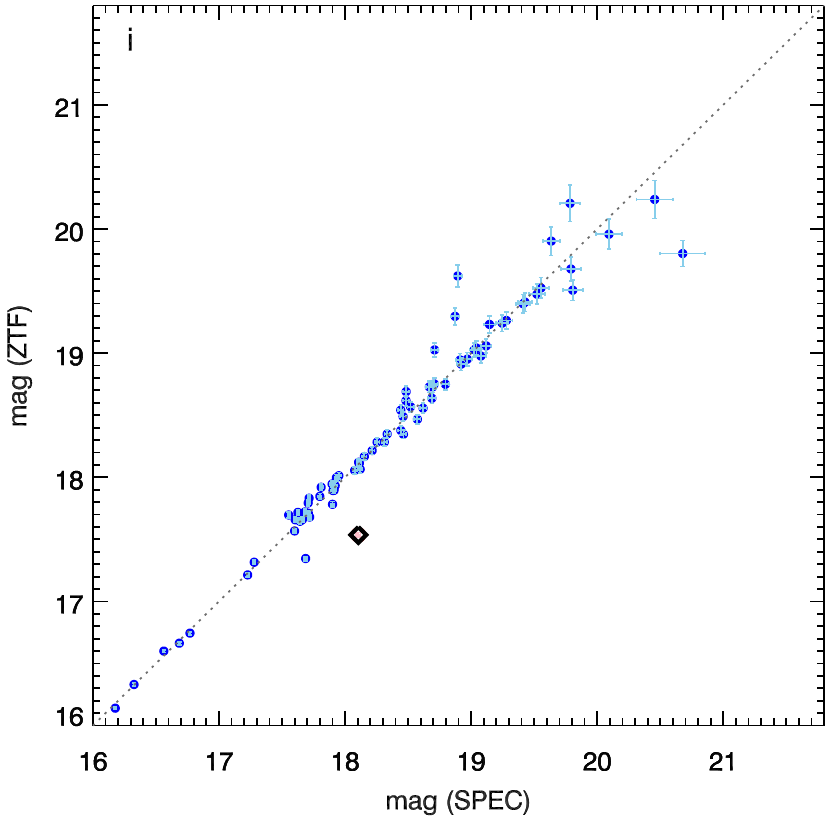}}
 \vspace{-0cm}
\caption{\label{fig:calibration} One example (J0936+2726) of spectroscopic flux calibration using ZTF and spec-photo magnitudes in $g$ (left panel), $r$ (middle panel), and $i$ (right panel) bands.  The x-axis shows SDSS spectral synthetic magnitudes, obtained at the quasar-faint state epoch. The y-axis shows ZTF magnitudes closest to the bright state epoch, calibrated to SDSS magnitudes, so that most (non-variable) objects fall near the 1-to-1 dashed line. Then, we measure the intrinsic variability between the bright and faint states for our target (pink point highlighted with black diamond frame) in each band (in SDSS image filters). The bright state spectra are renormalized using $g$, $r$ and $i$ offsets to obtain an absolute flux calibration consistent with their nearest-epoch ZTF observations.  
 }
\end{figure*}

For faint state photometry, we use the SDSS spectra themselves.  These are generally well flux-calibrated in both a relative and absolute sense \citep{Smee2013, Margala2016}. 
As the SDSS spectra were taken with 640 or 1000 fibers simultaneously in SDSS-I/II or BOSS multi-object plates, 
we can reliably calibrate the bright state spectra using non-variable objects in the field.
Specifically, for objects observed on the same SDSS plate on the same night, we obtain SDSS spectral synthetic  magnitudes using `SPECTROFLUX' from the SDSS DR17 catalog.\footnote{We obtain identical magnitudes when we convolve the published filter curves with the SDSS spectra.}
Most objects in the field are not variable, so we use their SDSS spectral synthetic magnitudes to recalibrate the ZTF $gri$ band photometry to the SDSS system, both to calculate variability in those bands, and for plotting purposes (e.g., Figure\,\ref{fig:example}). In Figure \ref{fig:calibration}, we illustrate the process in the field of J0936+2726, for which the narrow \OIII\ line is strongly blended with emerging iron emission at the bright state. The x-axis shows SDSS spectral synthetic magnitudes. 
 The y-axis shows ZTF magnitudes at the epoch closest to the bright-state spectroscopic observation. We use ZTF aperture photometry within a diameter of 2$\arcsec$ or 3$\arcsec$ if the faint-state SDSS observations were performed using a fiber diameter of 2$\arcsec$ or 3$\arcsec$, respectively.  The red point with black diamond frame is the target CLQ candidate J0936+2726, and the blue points are other objects with SDSS spectra in the field. 
 We calibrate the ZTF magnitudes to SDSS magnitudes,  
 weighted by the magnitude uncertainties for objects in the field. The majority of non-variable objects are along the 1-to-1 line, though our CLQ target strongly deviates, as expected. In this manner, we obtain the multi-band variability
 of the CLQ target at the time close to our spectroscopic follow-up (bright state), using SDSS-calibrated ZTF magnitudes. 
 
 In Table \ref{tab:Observations}, we summarize all the SDSS synthetic magnitudes at faint states, and the (SDSS-calibrated) ZTF magnitudes at bright states.
Then we apply a polynomial function to each bright state follow-up spectrum so that the spectral synthetic magnitudes (in SDSS filters) at the bright state are consistent with the SDSS-calibrated ZTF magnitudes in all bands. We used a 1st order polynomial (linear) function if three bands were available ($g$, $r$, and $i$ observations within ten days), otherwise we use a zero order (constant) polynomial function to normalize the spectroscopic flux.

For objects without near-epoch public ZTF photometric data to use for flux calibration of follow-up spectroscopy, we performed imaging observations in $g$, $r$, and $i$ bands using the 1.2 meter telescope at the Fred Lawrence Whipple Observatory (FLWO) on Mount Hopkins in Arizona. The FLWO images were observed within a short period after our MMT spectroscopic observations (triggered within 1 day and observed within 1-3 days after trigger).  This was used in about a dozen cases primarily to ensure that the nearest ZTF photometry still captured the object in the same state.

To aid the eye in evaluating variability, we add further high-cadence photometry to Figure\,\ref{fig:example} showing our optical light curves and multi-epoch spectra for CLQ candidates.
The Asteroid Terrestrial-impact Last Alert System \citep[ATLAS;][]{Tonry2018} is a high-cadence all-sky survey. It consists of four telescopes (two in Hawaii, one in Chile, and one in South Africa), which automatically scan the whole sky several times every night looking for moving objects. 
ATLAS obtains all sky coverage to $m\sim20~(5\sigma)$ with a 1 day multi-exposure cadence.
Specifically, the best 5$\sigma$ limiting magnitude ATLAS achieves in a 30 s exposure is 19.8, and the median over all lunations and sky conditions is 19.12. ATLAS observes in broad cyan ($c$, covering 420–650 nm) and orange ($o$, 560–820 nm) bandpasses.  
We applied constant offsets to ATLAS $c$- and $o$-band data to match the contemporaneous $r$- and $i$-band ZTF data (corrected to SDSS magnitudes as described above), respectively.

\subsection{Upper Limit of Broad-Emission Line Width \label{sec:FWHM}} 

In AGN, the technique of reverberation mapping \citep{Blandford1982, Peterson1993} has been used to measure the light-travel time delay of the broad-emission line flux responds to continuum luminosity and thus to derive (by multiplying the time delay by the speed
of light) the characteristic size of the BLR, $R_{\rm BLR}$, around the central, photoionizing source. A tight correlation between the $R_{\rm BLR}$ and continuum luminosity is found, as $R_{\rm BLR} \propto L^{\alpha}$, where $\alpha$ is the power-law slope. 
A theoretical expected slope is +0.5 for a virial relationship, and RM observations constrain the slope for the broad H$\beta$ emission in the range of 0.49 to 0.69, depending on the AGN sample, continuum luminosity wavelength, and fitting algorithm \citep[e.g.,][]{Peterson1999, Kaspi2005, Bentz2009}. 
By assuming that the emission lines are broadened primarily by the virial gas motions in the gravitational potential of the central SMBHs, the BLR size and the line width then give an estimate of the mass of the central SMBH \citep[e.g.,][]{Peterson1999, Kaspi2005, Bentz2009}.
The Keplerian velocity of the BLR, $V_{\rm BLR}$, at $R_{\rm BLR}$ is 
\begin{equation}
    V_{\rm BLR} = \sqrt{\frac{G M_{\rm BH}}{R_{\rm BLR}}}.
\end{equation}
We calculate the Eddington luminosity using $L_{\rm Edd} = 1.26 \times 10^{38}~M_{\rm BH}$ in units of erg s$^{-1}$.

Following \citet[][Eq. 7.18]{Netzer2013} for the emissivity-weighted radius of the H$\beta$ emitting region, 
\begin{equation}
    R_{\rm BLR} \simeq 0.12L_{46}^{0.6\pm0.1}~{\rm pc},
\end{equation}
in which $L_{46}$ is the bolometric luminosity, $L_{\rm bol}$, in units of $10^{46}$ erg s$^{-1}$ derived by using bolometric corrections from continuum luminosity at 5100\,\AA, BC$_{5100} = 9.26$. Thus the $V_{\rm BLR}$ can be calculated as \citet[][Eq. 7.22]{Netzer2013},
\begin{equation}
    V_{\rm BLR} \simeq 1700 M_8^{1/2} L_{46}^{-1/4}~{\rm km~s^{-1}},
\end{equation}
in which $M_8$ is the BH mass in units of $10^8~M_{\odot}$. For small Eddington ratio, $V_{\rm BLR}$ can be extremely large. For example, when $M_8=1$ and $R_{\rm Edd} = L_{\rm bol} / L_{\rm Edd}=10^{-4}$, typical of many low-Eddington ratio AGN found in low-ionization nuclear emission-line regions (LINERs), $ V_{\rm BLR} ({\rm H}\beta) \simeq 15,000~{\rm km~s^{-1}}$, larger than 99\% of type-I quasars \citep{Shen2011}. For low-luminosity AGN, the width of the broad emission lines can be extremely broad. Therefore, it is reasonable to assume a maximum limit to the broad line full width at half maximum, FWHM$_{\rm max}$, which may be related to cloud instability  close to the central BH, due e.g., to tidal forces.
Following \citet{Netzer2013}, we take this limit as ${\rm FWHM_{max}} = 20,000~{\rm km~s^{-1}}$.
We use the limit for fitting \Hb, but it is also a conservative limit for the \Ha\ and Mg\,II broad lines, as they are generated at larger radii according to reverberation mapping results (e.g., \citealt{Kaspi2000,Woo2018}).
Such a limit is important for practical purposes, since in the presence of very weak broad line emission, spectral fitting codes may fit extremely broad components that are unphysical, and which artificially reduce the estimated continuum luminosity.

\begin{figure*}[!ht]
\centering
\hspace{-0.4cm}
\includegraphics[width=0.53\textwidth]{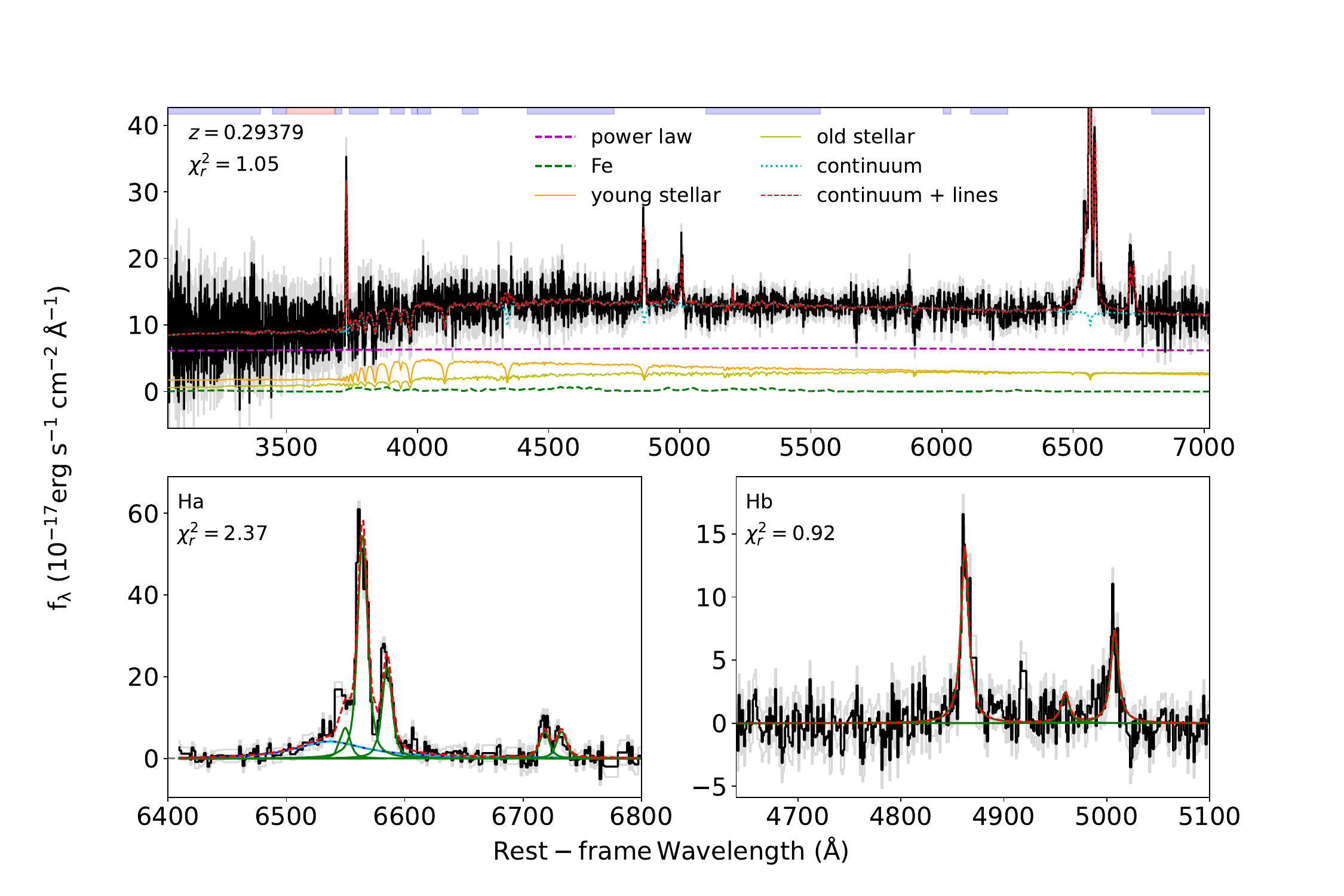}
\hspace{-1cm}
\includegraphics[width=0.53\textwidth]{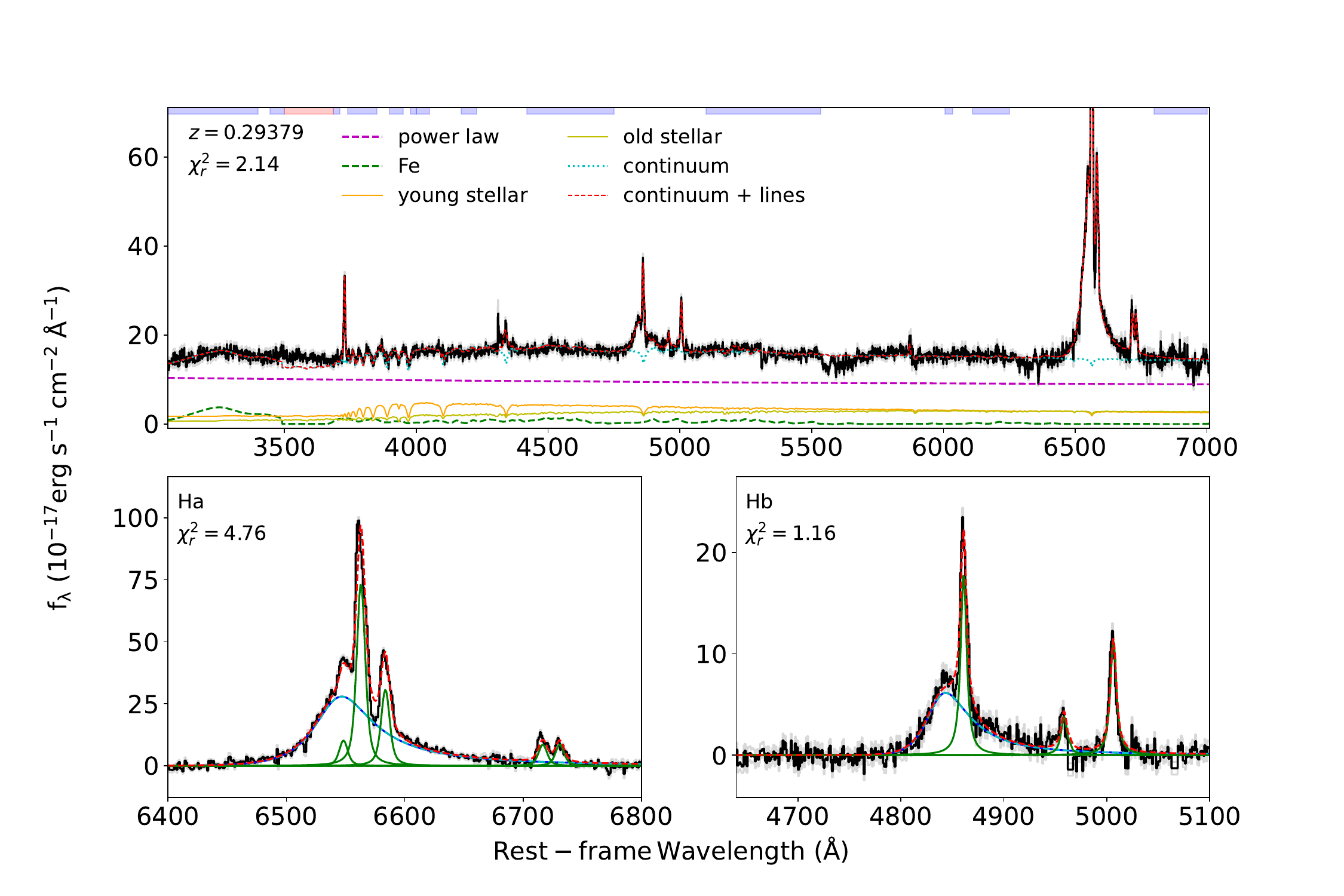}\\
\caption{Optical spectral fitting using \texttt{QGfit} to J0040+1609, as an example, in the faint (left) and bright (right) states. The top panels show the decomposition of continuum emission, including AGN power-law (purple), UV Fe II (green), and host galaxy stellar populations, both young (orange) and old (yellow). The continuum windows we used are shown as blue shaded bars on the top x-axis. The black and gray lines are the spectral flux densities (in units of $10^{−17}$ erg s$^{−1}$ cm$^{−2}$ \AA$^{-1}$) and their uncertainties. The cyan dotted lines are the total models from the continuum components described above, and the red dash lines are the total models including continuum and line emission. In the bottom panels, we show the emission-line fitting in the H$\alpha$ (left) and H$\beta$ (right) window, showing best-fit broad (cyan dash lines) and narrow (green solid lines) emission components. The red dash lines show the total model of broad and narrow line components. This object shows obviously skewed broad emission lines in both H$\alpha$ and H$\beta$ in the bright state.
}
\label{fig:fitting}
\end{figure*}

\subsection{Spectral Fitting} \label{sec:fitting}

To fit the spectra of CLQ both in the bright state (AGN dominated) and in the faint state (host galaxy dominated), we need a spectral fitting pipeline that can fit  both cases well. 
Previous spectral fitting codes were developed either for galaxies without accounting for AGN emission, e.g., pPXF \citep{Cappellari2004, Cappellari2017}, or for luminous quasars, e.g., QSOFIT \citep{Shen2019} without contribution from the host galaxies, or for quasars with host galaxies but only with a simple host galaxy template, e.g., QSfit \citep{Calderone2017}, or for quasars with host galaxy components using a PCA method that can not provide physical properties of the host galaxies (and sometimes even negative host component), such as PyQSOFit \citep{PyQSOFit2018}. 
Therefore, we develop a spectral fitting pipeline called ``\texttt{QGfit}", to fit spectra of diverse objects, ranging from luminous quasars to low-luminosity AGN and normal galaxies. The spectra are fit in the rest-frame of the quasar after correcting for Galactic reddening, again using the dust map of \citet{Schlegel1998} and the extinction curve from \citet{Cardelli1989}. 

To find the best-fit model parameters, \texttt{QGfit} can use the Python lmpfit or lmfit routines, 
with different optimization methods, such as Least-Squares minimization (leastsq), Nelder-Mead (nelder), Maximum likelihood via Monte-Carlo Markov Chain (emcee).  The code defaults to use leastsq to save time. To improve the fitting, \texttt{QGfit} will utilize other optimization methods when the reduced $\chi^2$ is larger than a user-defined criterion, for example, 1.5 we used in this work.  The code can automatically choose the optimization method yielding the smallest reduced $\chi^2$ value as the best fitting model. To quantify the measurement uncertainties, we adopted the spectral fitting uncertainties by using the emcee method,  specifically with 50 walkers, 1000 steps, and 50 burns. 

\texttt{QGfit} decomposes different components in the quasar spectrum, including power-law continuum, \FeII\ emission multiplets, and major broad and narrow emission lines. It uses empirical UV \FeII\ emission templates from the literature \citep{Vestergaard2001, Tsuzuki2006, Salviander2007} covering from 1000 to 3500\,\AA, and an optical \FeII\ template (3686--7484\,\AA) from \citet{Boroson1992}.  As an example, we show fits full to both faint and bright states for one object in Figure\,\ref{fig:fitting}.) from \citet{Boroson1992}.  As an example, we show fits full to both faint and bright states for one object in Figure\,\ref{fig:fitting}.

We take into account host galaxy stellar emission using the simple stellar population models \citep{Bruzual2003} with the \citep{Chabrier2003} initial mass function. We allow for two host components, one for young ($<$300 Myr)  and one for old ($>$300 Myr) stellar populations. We use 30 templates from \citet{Bruzual2003}, covering the metallicities of $Z$ = 0.004, 0.02 and 0.05, and the ages of young (0.005, 0.025, 0.10, and 0.29 Gyr) and old (0.64, 0.90, 1.4, 2.5, 5, and 11 Gyr) populations. 
We choose a few continuum windows and fit the continuum components described above together, following \citet{Shen2019}\footnote{The continuum windows can be easily modified by the user.}.  

After decomposing and subtracting the continuum emission, the algorithm proceeds to fit the emission lines, addressing both broad and narrow components. 
Three broad and one narrow Gaussian components are used for the H$\beta$ and H$\alpha$ emission. In the H$\beta$ (H$\alpha$) emission line window, the width and velocity shift of the narrow component of the H$\beta$ (H$\alpha$) emission is tied to the value of adjacent narrow emission line, such as \OIII\ ([NII]).
As the profile of broad emission line can be complicated, 
the code use a skewed Voigt function to fit emission lines 
as follows,
\begin{equation}
    f(x; A, \mu, \sigma, \gamma, \nu) = V(x; A, \mu, \sigma, \gamma) \Bigl \{1+{\rm erf}[\frac{\nu(x-\mu)}{\sigma \sqrt{2}}]\Bigl\}
\end{equation}
where erf() is the error function, $A$ is amplitude, $\mu$ is center, and $\nu$ is skew. 
$V(x; A, \mu, \sigma, \gamma)$ is the Voigt profile,
\begin{equation}
    V(x;\mu, \sigma,\gamma) = \int_{-\infty}^{\infty} G(x';\mu, \sigma)L(x-x';\mu, \gamma)dx'
\end{equation}
where $G(x;\mu, \sigma)$ is the Gaussian profile,
\begin{equation}
    G(x;\mu, \sigma) = \frac{e^{-(x-\mu)^2/2\sigma^2}}{\sigma \sqrt{2\pi}}
\end{equation}
and $L(x;\mu, \gamma)$ is the Lorentzian profile,
\begin{equation}
    L(x;\mu, \gamma) = \frac{\gamma}{\pi[\gamma^2 + (x-\mu)^2]}
\end{equation}
In the limiting cases of $\sigma=0$ and $\gamma=0$, $V(x;\mu, \sigma,\gamma)$ simplifies to $L(x;\mu, \gamma)$ and $G(x;\mu, \sigma)$, respectively. 

The FWHM of the Voigt profile can be found from the widths of the associated Gaussian and Lorentzian widths. An approximation of the FWHM of the Voigt profile with an accuracy of 0.02\% can be written as \citep{Kielkopf1973, OLIVERO1977}
\begin{equation}
    f_V = 0.5346f_L + \sqrt{0.2166f_L^2 + f_G^2}
\end{equation}
where $f_G$ is the FWHM of the Gaussian profile, $f_G = 2\sigma \sqrt{2 {\rm ln}2}$, and  $f_L$ is the FWHM of the Lorentzian profile, $f_L = 2\gamma$. We set a maximum limit on the broad-line emission width of $f_{V} \leq 20,000~{\rm km~s^{-1}}$, as discussed in \textsection \ref{sec:FWHM}. For broad-line emission, we  adopt a minimum of ${\rm FWHM_{min}} = 1200 ~{\rm km~s^{-1}}$, which is a widely used value for distinguishing between broad and narrow emission lines. 

In statistics, the Bayesian information criterion \citep[BIC][]{Schwarz1978} is often preferred for model selection. Models with lower BIC are generally preferred. \texttt{QGfit} uses BIC to automatically determine if a broad-emission line with ${\rm FWHM_{min}} \leq f_{V} \leq {\rm FWHM_{max}}$ exists. One skewed Voigt component is typically good enough to fit the majority broad-emission lines (see Figure \ref{fig:fitting} for examples), but \texttt{QGfit} also tests if two broad-emission lines is a better fit (with lower BIC). 
For narrow-emission lines, we use one symmetric Voigt function for simplicity, with $160~{\rm km~s^{-1}} < {\rm FWHM_{narrow}} < 1200~{\rm km~s^{-1}}$. The lower limit is set considering the instrumental dispersion of the SDSS spectrograph, which is $\sim 69~{\rm km~s^{-1}}$ per pixel (about $160~{\rm km~s^{-1}}$ FWHM). A notable feature of this pipeline is its capability to automatically detect broad-line emission, significantly reducing the need for labor-intensive visual inspections and also beneficial for quantitative comparisons.

% CLQ Candidates
\begin{deluxetable*}{cccrrrrrcc}
\tabletypesize{\tiny}
\tablecaption{CLQ Candidates \label{tab:Candidates}}
\tablewidth{1pt}
\tablehead{
\colhead{Name} &
\colhead{Coordinate} &
\colhead{Redshift} &
\colhead{SN$_{W1}$} &
\colhead{SN$_{W2}$} &
\colhead{$\Delta_{W1}$} &
\colhead{$\Delta_{W2}$} &
\colhead{SN$_g$} &
\colhead{Follow-up/Result} &
\colhead{Reference}
}
\startdata
J0003$+$0903 & 00:03:46.16$+$09:03:40.00 & 0.058 & 5.8 & 5.8 & 0.2 & 0.5 & 12.7 &  &  \\ 
J0004$+$0007 & 00:04:41.23$+$00:07:11.27 & 0.108 & 5.8 & 5.6 & 0.3 & 0.6 & 12.3 &  &  \\ 
J0009$+$3558 & 00:09:53.24$+$35:58:23.45 & 0.552 & 5.9 & 5.7 & 0.5 & 0.7 & 21.2 &  &  \\ 
J0010$+$0008 & 00:10:14.89$+$00:08:20.81 & 0.102 & 6.0 & 5.9 & 0.4 & 0.7 & 7.0 &  & \citet{Lopez-Navas2022} \\ 
J0016$+$0009 & 00:16:33.11$+$00:09:13.64 & 0.128 & 5.7 & 5.4 & 0.2 & 0.4 & 4.8 &  &  \\ 
J0020$+$0040 & 00:20:16.12$+$00:40:59.44 & 0.213 & 5.8 & 5.3 & 0.3 & 0.5 & 9.3 &  &  \\ 
J0022$+$1450 & 00:22:56.46$+$14:50:53.14 & 0.207 & 5.4 & 5.2 & 0.3 & 0.5 & 8.6 &  &  \\ 
J0037$+$0029 & 00:37:41.36$+$00:29:06.78 & 0.151 & 5.7 & 5.3 & 0.4 & 0.7 & 14.3 &  &  \\ 
J0039$-$0032 & 00:39:16.41$-$00:32:32.83 & 0.110 & 6.1 & 6.1 & 0.5 & 0.5 & 17.1 &  &  \\ 
J0040$+$1609 & 00:40:38.40$+$16:09:49.93 & 0.294 & 6.0 & 5.9 & 0.7 & 0.6 & 24.6 & CLQ &  \\ 
J0043$-$0938 & 00:43:11.60$-$09:38:16.01 & 0.054 & 6.1 & 6.1 & 0.4 & 0.6 & 10.3 &  &  \\ 
J0044$-$0803 & 00:44:50.01$-$08:03:32.79 & 0.457 & 6.0 & 5.4 & 0.8 & 0.8 & 16.9 &  &  \\ 
J0045$+$1555 & 00:45:51.47$+$15:55:47.72 & 0.115 & 6.0 & 5.9 & 1.1 & 1.0 & 17.7 &  &  \\ 
J0047$+$1541 & 00:47:30.34$+$15:41:49.41 & 0.031 & 5.9 & 5.9 & 0.2 & 0.4 & 16.2 & CLQ &  \\ 
J0052$+$0055 & 00:52:31.06$+$00:55:05.30 & 0.204 & 5.9 & 5.3 & 0.3 & 0.4 & 22.9 &  &  \\ 
J0052$+$1003 & 00:52:25.68$+$10:03:18.94 & 0.083 & 5.9 & 5.8 & 0.3 & 0.5 & 12.9 &  &  \\ 
J0053$+$3411 & 00:53:32.79$+$34:11:12.76 & 0.163 & 5.8 & 5.7 & 0.3 & 0.5 & 8.3 &  &  \\ 
J0053$-$1007 & 00:53:59.65$-$10:07:07.44 & 0.120 & 5.9 & 5.4 & 0.4 & 0.5 & 11.7 &  &  \\ 
J0100$+$0845 & 01:00:10.84$+$08:45:20.92 & 0.276 & 5.4 & 5.3 & 0.2 & 0.6 & 4.7 &  &  \\ 
J0100$-$0110 & 01:00:48.56$-$01:10:51.54 & 0.326 & 5.8 & 5.3 & 0.9 & 0.8 & 17.5 & Observed &  \\ 
J0103$+$1526 & 01:03:26.01$+$15:26:24.77 & 0.246 & 5.8 & 5.6 & 0.2 & 0.4 & 9.7 &  &  \\ 
J0104$+$1515 & 01:04:56.58$+$15:15:59.17 & 0.416 & 5.5 & 5.2 & 0.3 & 0.4 & 26.1 &  &  \\ 
J0107$+$2428 & 01:07:47.92$+$24:28:48.70 & 0.160 & 6.0 & 5.9 & 0.8 & 0.9 & 19.1 & CLQ &  \\ 
J0109$+$1816 & 01:09:08.17$+$18:16:07.52 & 0.444 & 6.0 & 6.0 & 1.1 & 1.1 & 11.7 &  &  \\ 
J0109$+$2400 & 01:09:14.67$+$24:00:34.41 & 0.493 & 5.8 & 5.4 & 0.4 & 0.6 & 4.8 &  &  \\ 
J0110$+$0026 & 01:10:59.31$+$00:26:01.14 & 0.019 & 6.0 & 5.9 & 0.5 & 0.8 & 16.6 & CLQ &  \\ 
\enddata
\tablecomments{The 653 CLQ candidates. For objects with `observed' or `CLQ' in the column Follow-up/Result, see observational details in Table\,\ref{tab:Observations}. The full table is available in the electronic version.
}
\end{deluxetable*}

% Spectroscopic Observations and Photometric data
\begin{deluxetable*}{cccccccccccc}
\tablecaption{Spectroscopic Observations and Photometric Data \label{tab:Observations}}
\tablewidth{1pt}
\tablehead{
\colhead{Name} &
\colhead{State} &
\colhead{Telescope} &
\colhead{MJD} &
\colhead{Date} & 
\colhead{$|\Delta t_{\rm photo}|$} &
\colhead{$g$} &
\colhead{$r$} &
\colhead{$i$} &
\colhead{$W1$} &
\colhead{$W2$} &
\colhead{$\Delta t_{\rm spec}$} \\
 & & & & & day & mag & mag & mag & mag & mag & year
}
\startdata
J0040+1609 & Faint & SDSS & 51884 & 2000-12-06 & 0.0 & 20.1 & 19.2 & 18.9 & 16.7 & 16.4 &   \\ 
J0040+1609 & Bright & MMT & 60181 & 2023-08-25 & 1.4 & 19.7 & 19.0 & 18.6 & 16.7 & 16.5 & 22.7 \\ 
\hline
J0047+1541 & Faint & SDSS & 51879 & 2000-12-01 & 0.0 & 17.3 & 16.4 & 16.0 & 14.8 & 15.5 &   \\ 
J0047+1541 & Bright & MMT & 59856 & 2022-10-04 & 0.3 & 17.3 & 16.3 & 15.9 & 14.9 & 15.4 & 21.9 \\ 
\hline
J0100-0110 & Faint & SDSS & 52254 & 2001-12-11 & 0.0 & 21.0 & 19.8 & 19.4 & 18.0 & 17.9 &   \\ 
J0100-0110 & Bright & DBSP & 58794 & 2019-11-07 & 2.7 & 20.3 & 19.7 & $\cdots$ & 17.4 & 17.2 & 17.9 \\ 
\hline
J0107+2428 & Faint & SDSS & 57367 & 2015-12-11 & 0.0 & 19.0 & 18.2 & 17.8 & 16.6 & 16.8 &   \\ 
J0107+2428 & Bright & MMT & 59970 & 2023-01-26 & 0.2 & 18.4 & 17.4 & 17.4 & 15.9 & 15.9 & 7.1 \\ 
\hline
J0110+0026 & Faint & SDSS & 51794 & 2000-09-07 & 0.0 & 17.6 & 16.8 & 16.5 & 15.8 & 16.3 &   \\ 
J0110+0026 & Bright1 & LAMOST & 59201 & 2020-12-18 & 1.2 & 17.3 & 16.7 & 16.3 & 15.4 & 15.7 & 20.3 \\ 
J0110+0026 & Bright2 & MMT & 59880 & 2022-10-28 & 1.4 & 17.0 & 16.4 & 16.1 & 15.5 & 15.8 & 22.2 \\ 
\hline
J0127+1530 & Faint & SDSS & 51884 & 2000-12-06 & 0.0 & 19.3 & 18.5 & 18.2 & 17.2 & 17.6 &   \\ 
J0127+1530 & Bright & MMT & 59822 & 2022-08-31 & 0.6 & 19.4 & 18.8 & 18.5 & 17.1 & 17.3 & 21.7 \\ 
\hline
J0132+1501 & Faint & SDSS & 51884 & 2000-12-06 & 0.0 & 19.9 & 18.9 & 18.4 & 16.9 & 17.0 &   \\ 
J0132+1501 & Bright & MMT & 60240 & 2023-10-23 & 1.7 & 19.7 & 18.8 & 18.4 & 16.9 & 17.0 & 22.9 \\ 
\hline
J0141+0105 & Faint & SDSS & 51788 & 2000-09-01 & 0.0 & 18.4 & 17.5 & 17.0 & 15.5 & 15.6 &   \\ 
J0141+0105 & Bright4 & DBSP & 58794 & 2019-11-07 & 5.8 & 17.8 & 17.1 & 16.7 & 15.0 & 15.0 & 19.2 \\ 
\hline
\enddata
\tablecomments{Spectroscopic epochs used for faint and bright states. Photometric magnitudes are
derived directly from SDSS spectra for the faint state.  The (observed frame) time difference  $|\Delta t_{\rm photo}|$ is the separation between our bright state spectral epoch and the nearest ZTF photometric epoch.
The full table is available in the electronic version. The WISE magnitudes are in AB magnitude.
The typical (median) $\Delta t_{\rm photo} = 0.18$ days in the bright state. The full table is available in the electronic version.
}
\end{deluxetable*}

\subsection{CLQ Criteria and Quantitative Definition} \label{sec:criteria}

A quantitative definition of CLQ is needed for clarity, reproducibility and comparisons between studies.  It is straightforward to define extreme variability by the dramatic continuum variability, for example, if the maximum $g$-band change is larger than 1 or 2 mag \citep{Rumbaugh2018, Graham2020, Shen2021}. However, the majority of the extremely variable quasars does not show the ``changing-look" phenomena. Also, the correlation between the continuum and broad-emission line variability is affected by the dominance of the host galaxy. 
For example, a host fraction of 0.3 in the bright state would limit the maximum dimming to be 1.3 magnitudes, and a host fraction of 0.1 would limit the dimming to 2.5 magnitudes \citep{Shen2021}. Especially in the ``turn-on" case, when the continuum is dominated by the host (almost entirely so in the faint state), a criterion of minimum continuum magnitude variability is not appropriate and not comparable to the ``turn-off" case and between different emission lines, with different fractions of host galaxy contribution. Physically, the AGN continuum emission and broad-line emission are from two different regions, from the accretion disk and the broad-line region, respectively. Therefore, we use a simple and straightforward definition -- we define the CL AGN by the changes in the broad-line emission, in which the flux of the broad-line emission (H$\beta$, H$\alpha$, or MgII) in the faint state is less than 30\% of that in the bright state. While the specific fraction is somewhat arbitrary, it is reasonable.
Based on global radiative MHD simulations, fluctuations of one magnitude variability (a factor of 2.5) in continuum can be reproduced on multiyear timescales in simulations of quasar accretion disk that are roughly in a steady state \citep{Jiang2019,Jiang2020}. Variations of two magnitudes (a factor of 6) or more in the continuum is too large to be consistent with steady-state accretion in the simulations. Therefore, for a criterion on broad-emission line flux change, we choose a reasonable minimum bright/faint ratio of 10/3.
The variability of emission lines are usually weaker than that in continuum emission. For example, \citet{Dexter2019_RM} found a linear correlation between the broad H$\beta$ variability and continuum variability at 2700\,\AA\ with factor of $0.35\pm0.05$, which is in agreement with the predictions of photoionization models \citep[e.g.,][]{Korista2004}. Therefore, we use a criterion of $f_{\rm line, faint} / f_{\rm line, bright} < 0.3$ or $\Delta_{\rm line} / f_{\rm line, bright} = (f_{\rm line, bright} - f_{\rm line, faint}) /f_{\rm line, bright} > 0.7$, a reasonable extreme emission line variability that is too large  in steady-state accretion.

\citet{Zeltyn2024} used a similar criterion for identifying CLQs based on the ratio of flux between the bright and faint states, as $C({\rm line}) = f_{\rm line, bright} / f_{\rm line, faint} - \Delta(f_{\rm line, bright} / f_{\rm line, faint}) > 2$, in which $\Delta(f_{\rm line, bright} / f_{\rm line, faint})$ is the uncertainty of ratio $f_{\rm line, bright} / f_{\rm line, faint}$. However, they require broad-line emission in the faint states.  By contrast, since we select from spectroscopic galaxies, our CLQ sample does not preclude broad-line emission in the faint states. Our criteria remain valid even when no broad line emission is detectable in the faint state i.e., when $f_{\rm line, faint}$ = 0.

\begin{figure}[htbp]
\hspace*{-0.5cm}
\epsscale{1.1}
\plotone{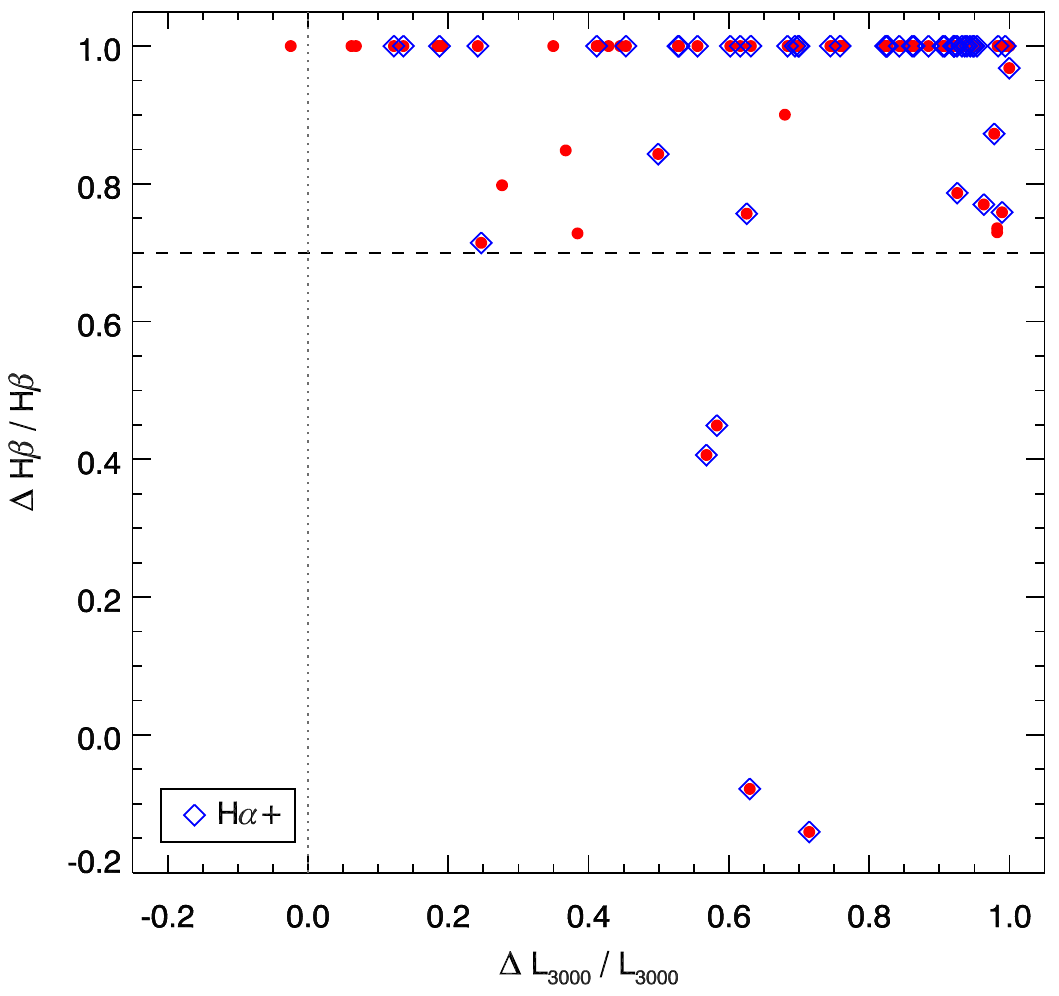}
\vspace{-0.0cm}
\caption{\label{fig:line_ratio} Fractional change of broad H$\beta$ luminosity plotted against the fractional change in the best-fit power law. Points show all spectral epochs contrasted against the designated dim epoch for each CLQ. The blue dashed lines indicate a 70\% fractional change in line flux, the minimum we require for a CLQ classification.}
\end{figure}

\section{Results} \label{sec:results}
\subsection{New Turn-on CLQs} \label{sec:new_CLQs}

% CLQ Emission Line Results
\normalsize
\begin{deluxetable*}{c|l|rrrr|rrrlr|c|l}
\tablecaption{CLQ Emission Line Results \label{tab:Individual1}}
\tablewidth{1pt}
\tablehead{
\colhead{Name} &
\colhead{Date} & % MJD or delta_MJD
\colhead{$f_{{\rm H}\beta,{\rm faint}}$} &
\colhead{$f_{{\rm H}\beta,{\rm bright}}$} &
\colhead{$R_{{\rm H}\beta}$} &
\colhead{$N_{\sigma}({\rm H}\beta)$} &
\colhead{H$\alpha$/Mg\,{\sc ii}} &
\colhead{$f_{{\rm line},{\rm faint}}$} &
\colhead{$f_{{\rm line},{\rm bright}}$} &
\colhead{$R_{{\rm line}}$} &
\colhead{$N_{\sigma}({\rm line})$} &
\colhead{Lines} &
\colhead{CLQ?} \\
\colhead{(1)} & \colhead{(2)} & \colhead{(3)} & \colhead{(4)} & \colhead{(5)} & \colhead{(6)} & \colhead{(7)} & \colhead{(8)} & \colhead{(9)} & \colhead{(10)} & \colhead{(11)} & \colhead{(12)} & \colhead{(13)}
}
\startdata
J0040$+$1609 & 2023-08-25 & 0.0 & 452.8 & 0.00 & 3.8 & H$\alpha$ & 448.4 & 2501.5 & 0.18 & 18.3 & H$\beta+$H$\alpha+$ & CLQ \\ 
\hline
J0047$+$1541 & 2022-10-04 & 0.0 & 673.3 & 0.00 & 4.9 & H$\alpha$ & 0.0 & 6768.3 & 0.00 & 13.9 & H$\beta+$H$\alpha+$ & CLQ \\ 
\hline
J0100$-$0110 & 2019-11-07 & 0.0 & 0.0 & $\cdots$ & 0.7 & H$\alpha$  & 901.0 & 1482.6 & 0.61 & 1.1 & H$\beta-$H$\alpha-$ &   \\ 
\hline
J0107$+$2428 & 2023-01-26 & 0.0 & 1095.7 & 0.00 & 3.5 & H$\alpha$ & 0.0 & 3634.3 & 0.00 & 11.4 & H$\beta+$H$\alpha+$ & CLQ \\ 
\hline
J0110$+$0026 & 2020-12-18 & 0.0 & 0.0 & $\cdots$ & 0.9 & H$\alpha$  & 1825.6 & 3369.4 & 0.54 & 10.2 & H$\beta-$H$\alpha-$ &   \\ 
  & 2022-10-28 & 0.0 & 4256.0 & 0.00 & 7.4 & H$\alpha$ & 1825.6 & 20545.2 & 0.09 & 13.9 & H$\beta+$H$\alpha+$ & CLQ \\ 
\hline
J0127$+$1530 & 2022-08-31 & 0.0 & 0.0 & $\cdots$ & 1.1 & H$\alpha$  & 0.0 & 0.0 & $\cdots$ & 2.2 & H$\beta-$H$\alpha-$ &   \\ 
\hline
J0132$+$1501 & 2023-10-23 & 0.0 & 623.3 & 0.00 & 1.9 & H$\alpha$  & 1153.9 & 2977.9 & 0.39 & 9.0 & H$\beta+$H$\alpha-$ & CLQ \\ 
\hline
J0141$+$0105 & 2019-11-07 & 0.0 & 1445.1 & 0.00 & 1.4 & H$\alpha$ & 0.0 & 7722.7 & 0.00 & 9.0 & H$\beta+$H$\alpha+$ & CLQ \\ 
\hline
J0146$+$1311 & 2023-09-09 & 0.0 & 872.8 & 0.00 & 4.8 & H$\alpha$ & 777.2 & 4624.9 & 0.17 & 18.8 & H$\beta+$H$\alpha+$ & CLQ \\ 
\hline
J0157$-$0036 & 2023-10-23 & 548.0 & 472.4 & 1.16 & 0.5 & H$\alpha$  & 916.4 & 1622.6 & 0.56 & 4.3 & H$\beta-$H$\alpha-$ &   \\ 
\hline
J0216$-$0337 & 2022-08-31 & 185.2 & 338.7 & 0.55 & 2.6 & H$\alpha$  & 572.5 & 980.4 & 0.58 & 6.5 & H$\beta-$H$\alpha-$ &   \\ 
\hline
J0225$+$0005 & 2019-11-06 & 0.0 & 0.0 & $\cdots$ & 0.9 & H$\alpha$  & 539.2 & 1286.3 & 0.42 & 5.1 & H$\beta-$H$\alpha-$ &   \\ 
\hline
J0254$-$0247 & 2020-09-19 & 0.0 & 0.0 & $\cdots$ & 0.9 & H$\alpha$  & 2604.8 & 2399.2 & 1.09 & 2.2 & H$\beta-$H$\alpha-$ &   \\ 
\hline
J0334$+$0051 & 2023-09-06 & 0.0 & 690.2 & 0.00 & 1.4 & Mg\,{\sc ii}  & 1280.1 & 1192.1 & 1.07 & 2.5 & H$\beta+$Mg\,{\sc ii}$-$ & CLQ \\ 
\hline
J0803$+$2207 & 2019-11-07 & 0.0 & 0.0 & $\cdots$ & 1.6 & H$\alpha$  & 1279.4 & 2491.5 & 0.51 & 10.2 & H$\beta-$H$\alpha-$ &   \\ 
  & 2024-01-13 & 0.0 & 846.5 & 0.00 & 7.3 & H$\alpha$  & 1279.4 & 3260.8 & 0.39 & 27.2 & H$\beta+$H$\alpha-$ & CLQ \\ 
\hline
J0813$+$4608 & 2021-02-05 & 878.9 & 2635.8 & 0.33 & 1.2 & H$\alpha$  & 3745.3 & 1717.5 & 2.18 & 7.3 & H$\beta-$H$\alpha-$ &   \\ 
  & 2024-01-13 & 878.9 & 3250.3 & 0.27 & 9.9 & H$\alpha$  & 3745.3 & 10421.5 & 0.36 & 21.6 & H$\beta+$H$\alpha-$ & CLQ \\ 
\hline
J0819$+$3019 & 2023-10-22 & 0.0 & 2493.3 & 0.00 & 7.4 & H$\alpha$ & 0.0 & 8735.4 & 0.00 & 31.1 & H$\beta+$H$\alpha+$ & CLQ \\ 
\hline
J0823$+$4202 & 2023-03-13 & 0.0 & 0.0 & $\cdots$ & 2.0 & H$\alpha$ & 0.0 & 557.8 & 0.00 & 5.6 & H$\beta-$H$\alpha+$ & CLQ \\ 
\hline
J0828$+$2202 & 2024-01-14 & 571.0 & 369.5 & 1.55 & 2.0 & H$\alpha$  & 1339.6 & 2103.0 & 0.64 & 3.5 & H$\beta-$H$\alpha-$ &   \\ 
\hline
J0829$+$2319 & 2024-01-14 & 324.2 & 588.7 & 0.55 & 1.2 & H$\alpha$ & 800.1 & 2974.1 & 0.27 & 7.9 & H$\beta-$H$\alpha+$ & CLQ \\ 
\hline
J0837$+$0356 & 2021-03-15 & 0.0 & 449.8 & 0.00 & 2.3 & H$\alpha$  & 797.3 & 1558.1 & 0.51 & 7.2 & H$\beta+$H$\alpha-$ & CLQ \\ 
\hline
J0851$+$0441 & 2021-03-19 & 1268.2 & 0.0 & $\cdots$ & 1.6 & H$\alpha$  & 1434.2 & 2167.0 & 0.66 & 5.3 & H$\beta-$H$\alpha-$ &   \\ 
\hline
J0854$+$1113 & 2019-11-07 & 0.0 & 718.2 & 0.00 & 2.3 & H$\alpha$ & 0.0 & 3977.0 & 0.00 & 9.4 & H$\beta+$H$\alpha+$ & CLQ \\ 
\hline
J0859$+$0922 & 2019-02-03 & 0.0 & 0.0 & $\cdots$ & 0.5 & H$\alpha$  & 0.0 & 0.0 & $\cdots$ & 4.3 & H$\beta-$H$\alpha-$ &   \\ 
  & 2019-11-07 & 0.0 & 0.0 & $\cdots$ & 0.5 & H$\alpha$  & 0.0 & 0.0 & $\cdots$ & 2.8 & H$\beta-$H$\alpha-$ &   \\ 
\hline
J0901$+$2907 & 2023-10-22 & 481.1 & 446.1 & 1.08 & 2.1 & H$\alpha$ & 851.8 & 3025.6 & 0.28 & 8.3 & H$\beta-$H$\alpha+$ & CLQ \\ 
\hline
J0906$+$4046 & 2024-01-14 & 0.0 & 303.1 & 0.00 & 1.7 & H$\alpha$  & 870.1 & 1110.5 & 0.78 & 4.6 & H$\beta+$H$\alpha-$ & CLQ \\ 
\hline
J0908$+$0755 & 2023-10-23 & 595.3 & 2083.8 & 0.29 & 0.9 & H$\alpha$ & 786.2 & 3956.9 & 0.20 & 6.5 & H$\beta+$H$\alpha+$ & CLQ \\ 
\hline
J0910$+$1907 & 2023-03-26 & 0.0 & 753.8 & 0.00 & 4.5 & H$\alpha$  & 1156.7 & 3838.5 & 0.30 & 8.9 & H$\beta+$H$\alpha-$ & CLQ \\ 
\hline
J0914$+$0126 & 2020-03-20 & 317.7 & 536.4 & 0.59 & 4.0 & H$\alpha$  & 1327.4 & 2868.5 & 0.46 & 13.5 & H$\beta-$H$\alpha-$ &   \\ 
\hline
J0914$+$0502 & 2020-01-31 & 616.7 & 2267.8 & 0.27 & 1.9 & H$\alpha$  & 1159.2 & 1702.5 & 0.68 & 7.2 & H$\beta+$H$\alpha-$ & CLQ \\ 
  & 2021-04-05 & 616.7 & 812.1 & 0.76 & 1.0 & H$\alpha$  & 1159.2 & 2414.8 & 0.48 & 7.7 & H$\beta-$H$\alpha-$ &   \\ 
  & 2021-04-11 & 616.7 & 862.1 & 0.72 & 1.6 & H$\alpha$  & 1159.2 & 2403.0 & 0.48 & 8.5 & H$\beta-$H$\alpha-$ &   \\ 
\hline
J0915$+$4814 & 2021-10-14 & 738.7 & 1244.2 & 0.59 & 2.2 & H$\alpha$ & 593.4 & 3575.4 & 0.17 & 7.4 & H$\beta-$H$\alpha+$ & CLQ \\ 
\hline
J0926$-$0006 & 2021-01-13 & 0.0 & 260.1 & 0.00 & 2.2 & H$\alpha$ & 0.0 & 1014.4 & 0.00 & 10.5 & H$\beta+$H$\alpha+$ & CLQ \\ 
\hline
J0936$+$2726 & 2021-11-09 & 0.0 & 1339.0 & 0.00 & 7.2 & H$\alpha$ & 0.0 & 5354.9 & 0.00 & 9.5 & H$\beta+$H$\alpha+$ & CLQ \\ 
\hline
J0947$+$5449 & 2019-11-07 & 0.0 & 1170.1 & 0.00 & 2.2 & Mg\,{\sc ii} & 0.0 & 779.8 & 0.00 & 1.4 & H$\beta+$Mg\,{\sc ii}$+$ & CLQ \\ 
  & 2023-12-13 & 0.0 & 724.7 & 0.00 & 5.3 & Mg\,{\sc ii} & 0.0 & 933.0 & 0.00 & 5.7 & H$\beta+$Mg\,{\sc ii}$+$ & CLQ \\ 
\hline
\enddata
\tablecomments{Cols. (3)-(4): Broad H$\beta$ emission line flux in the faint and bright states, respectively.
Col. (5): The H$\beta$ line ratio between the faint and bright states.
Col. (6): The $N_{\sigma}$ of the H$\beta$ emission used in the literature \citep{MacLeod2019, Green2022}.
Col. (7): Indicates whether the line is H$\alpha$ or Mg\,II. If it is H$\alpha$, then Cols. (8)-(11) correspond to H$\alpha$; if Mg\,II, these cols. correspond to Mg\,II.
Cols. (8)-(11): Similar to Cols. (3)-(6), but for the emission line described in Col. (7).
Col. (12): The emission line results. Based on the line flux ratio criteria defined in Section \ref{sec:criteria}, H$\beta+$, H$\alpha+$, or Mg\,II+ indicates that the source is a CLQ based on the H$\beta$, H$\alpha$, or Mg\,II emission, respectively.  H$\beta-$, H$\alpha-$, or Mg\,II- indicates the line failed to meet our CLQ criteria.
Col. (13): If any of the H$\beta$, H$\alpha$, or Mg\,II emission lines satisfy our CLQ criteria, the source is flagged as a CLQ. If none of the three lines satisfy our CLQ criteria, the source is not a CLQ. The full table is available in the electronic version.
}
\end{deluxetable*}

\begin{figure*}[!ht]
\centering
% \vspace{-2cm}
\hspace{-0.4cm}
\includegraphics[width=0.52\textwidth]{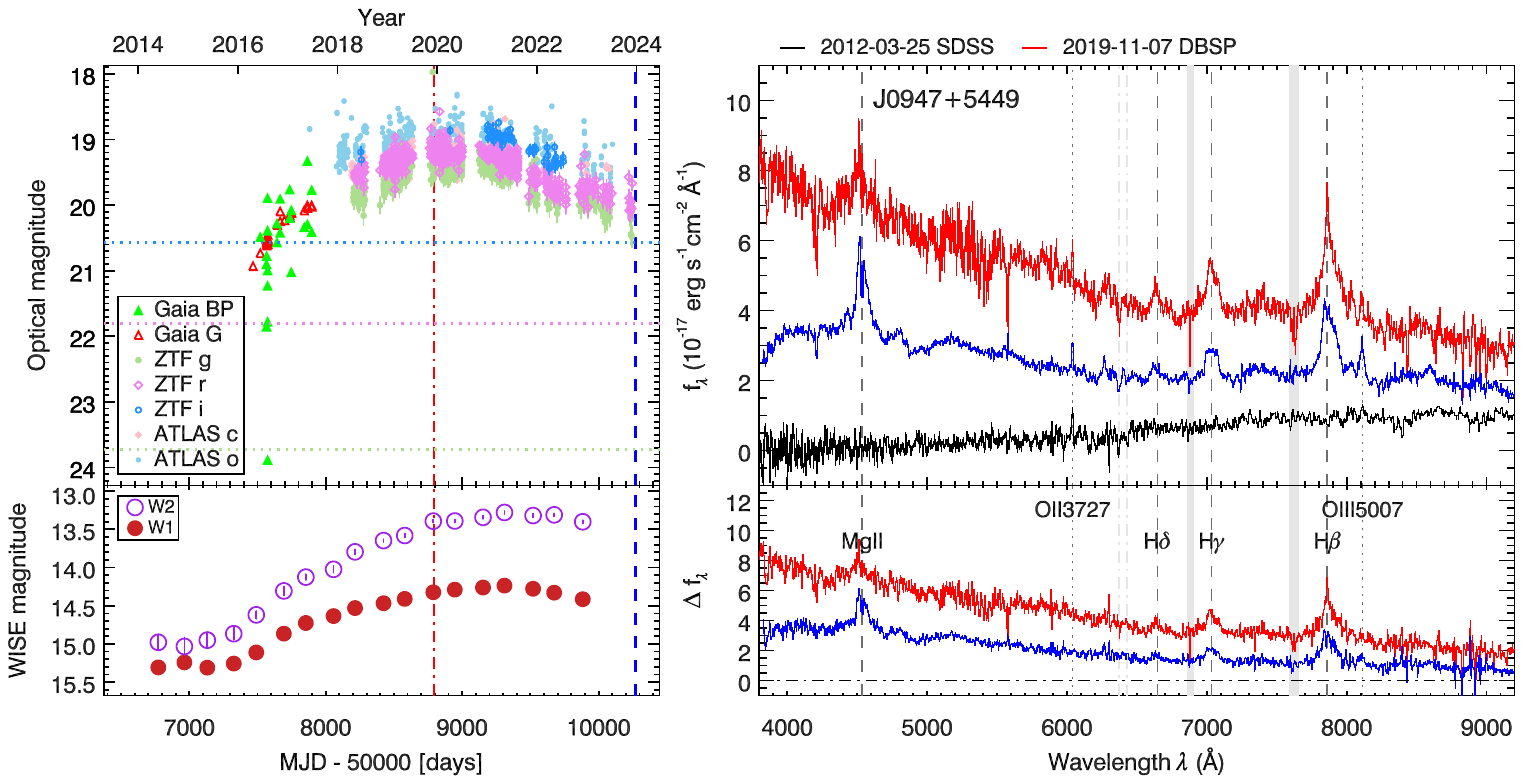}
\hspace{-0.7cm}
\includegraphics[width=0.52\textwidth]{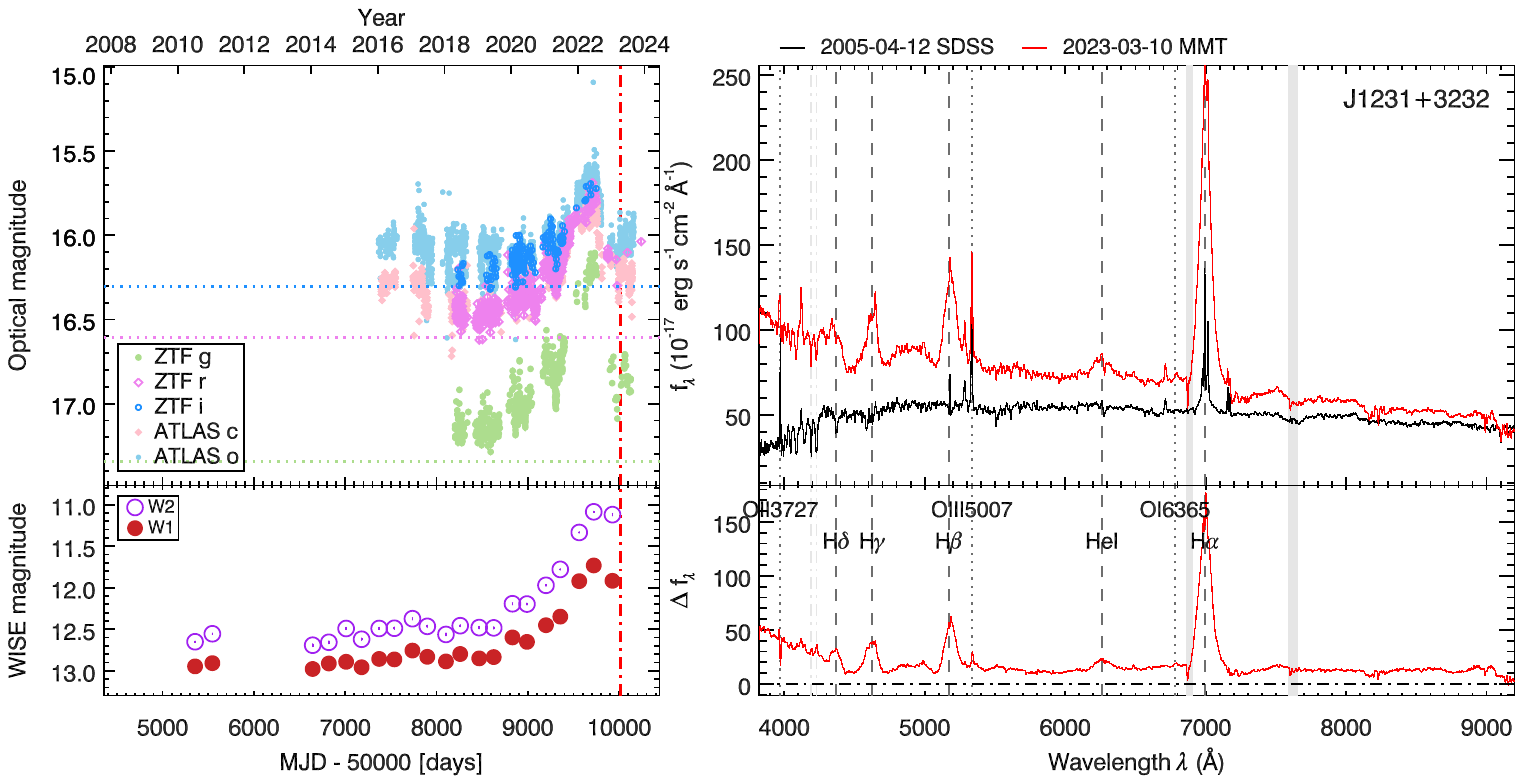}\\
\vspace{-1cm}
\caption{{\bf Left}: an example of a `fresh quasar', J0947+5449, likely awakening for the first time after a long quiescent phase, based on the lack of any optical spectroscopic AGN signatures in the faint state. It significantly brightened by $>5$ mag in the optical (upper left panel) and $\sim2$ mag in the MIR (bottom left panel). 
The horizontal dotted lines indicate the faint-state magnitudes in $g$ (green), $r$ (red), and $i$ (blue) bands, respectively. The bright state spectral epochs are marked by vertical dashed lines.
The right upper panel shows the spectra in the faint state (black) and bright state (red and blue), and the bottom panel shows the spectral difference between the two states, showing strong changes in continuum and broad-line luminosity. 
{\bf Right}: an example of a turn-on quasar with indications of AGN activity (based on narrow-line emission flux ratios) in the faint state. 
}
\label{fig:example}
\end{figure*}

\begin{figure*}[!ht]
\centering
\vspace{-1cm}
\hspace{-0.4cm}
\includegraphics[width=0.52\textwidth]{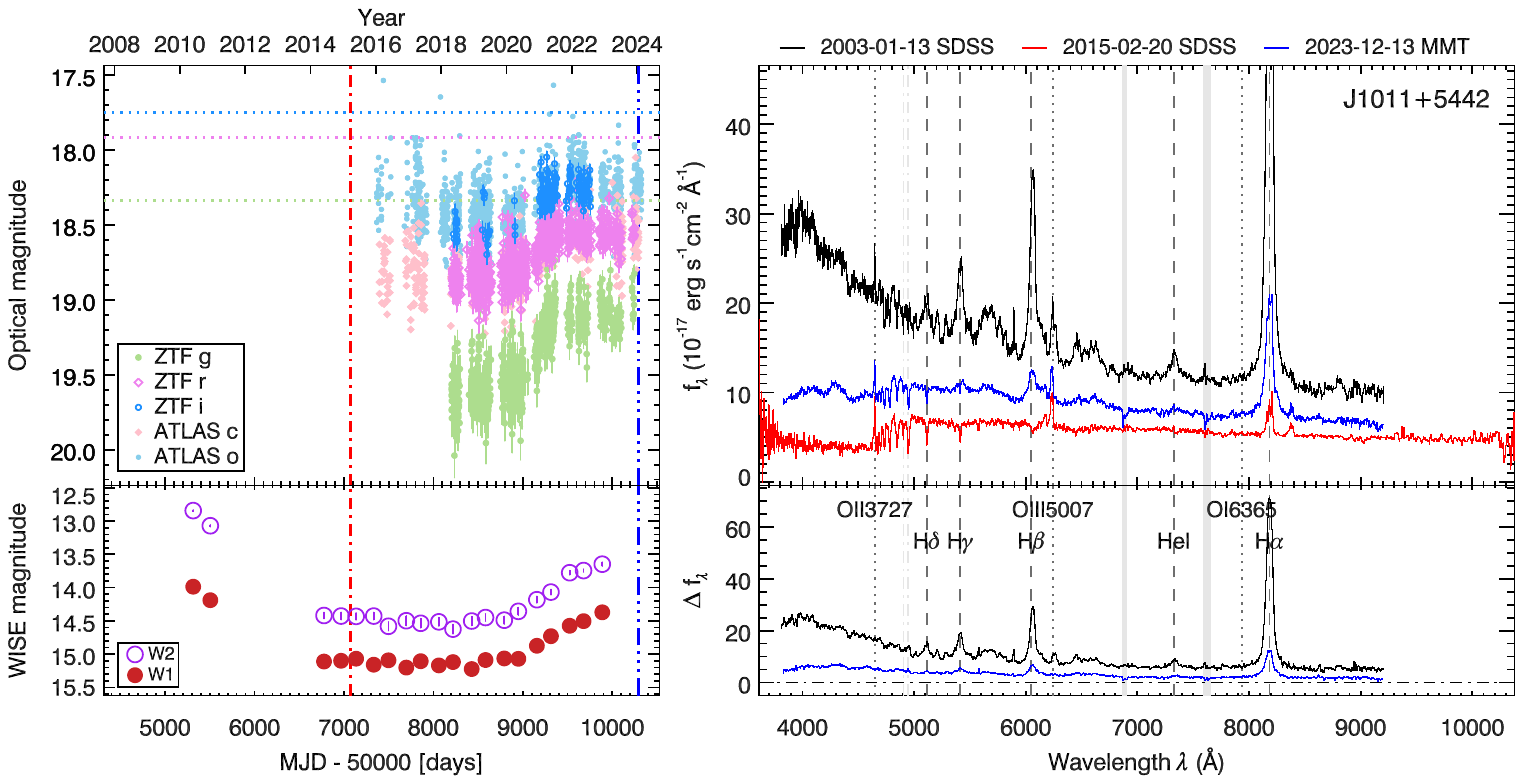}
\hspace{-0.7cm}
\includegraphics[width=0.52\textwidth]{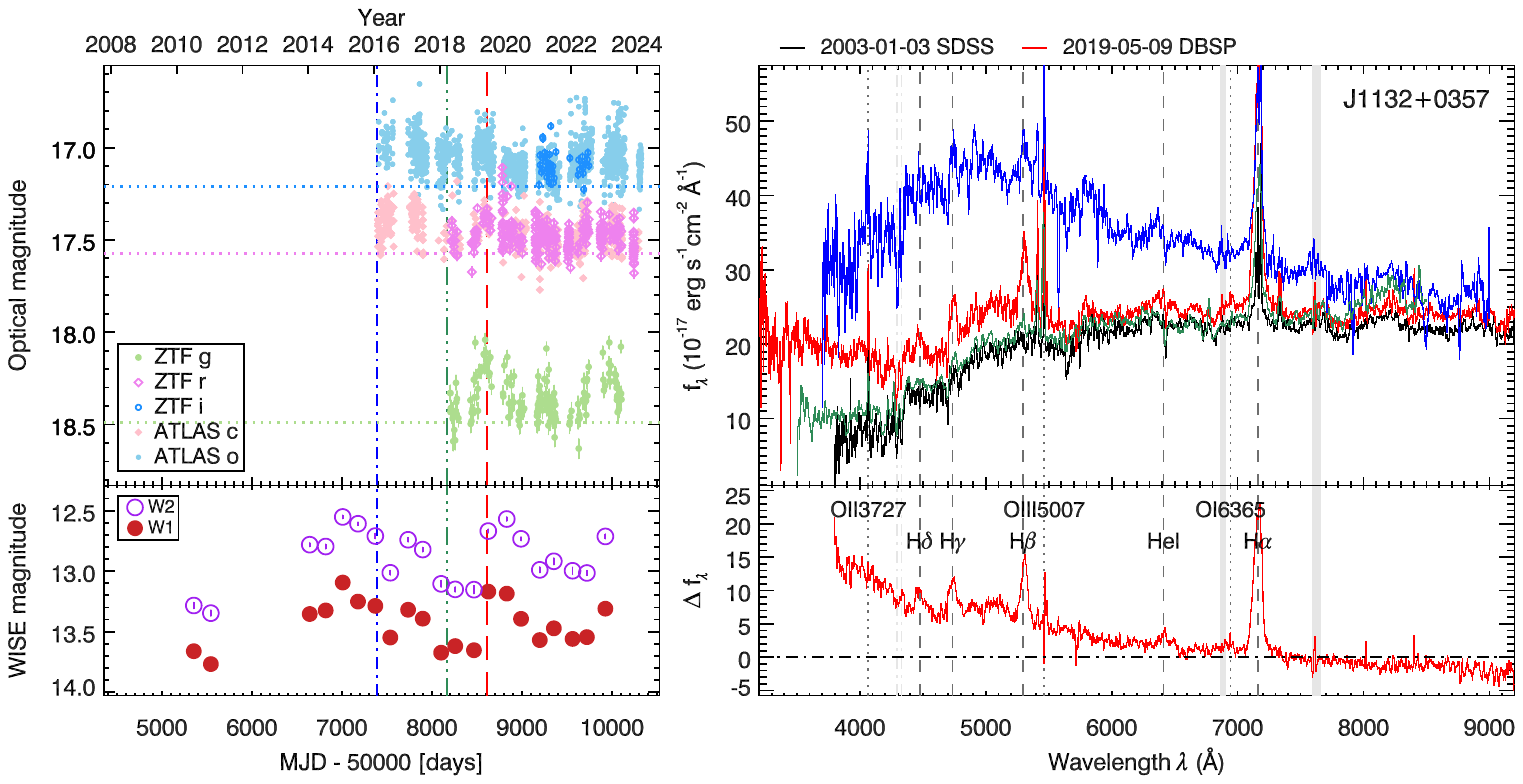}\\
\vspace{-1cm}
\caption{Two previously identified CLQs, J1011+5442 (left; \citealt{Runnoe2016}) and J1132+0357 (right; \citealt{Yang2018}) that have changed state again. Lines and symbols are explained in Figure \ref{fig:example}.
}
\label{fig:known}
\end{figure*}

In Section \ref{sec:data}, we described a total of 127 spectroscopic follow-up observations. Some objects were observed in more than one epoch, so these observations cover 115 candidates.\footnote{Note that we added two candidates that were not in the 653 CLQ candidate sample in Table\,\ref{tab:Candidates} J0141+0105 and J1442+5558. J1442+5558 is a radio detected source, but it turns out to be a turn-on CLQ. J0141+0105 is missed due to the WISE data quality (cc\_flags) limitation.} 
We fit the spectra of all the sources observed. As described in Section \ref{sec:fitting}, we fit the spectra of both states. As the host galaxy is not variable, and we have done absolute flux calibration, we fixed the non-variable host galaxy component to be the same for both bright and faint state spectral fitting, so that only the AGN components (AGN power-law and FeII emission) are variable in the continuum model. 

After fitting the continuum emission, we fit the emission lines for each epoch. Figure \ref{fig:line_ratio} shows the broad line ratio between the faint and bright state from the H$\beta$ line vs. H$\alpha$ line. As described in Section \ref{sec:criteria}, we use as a definition for CLQ classification a broad emission line ratio of $f_{\rm line, faint} / f_{\rm line, bright} < 0.3$, which is equivalent to $\Delta {\rm line}/ f_{\rm line, bright} > 0.7$. 

From the 115 observed objects, we found 82 sources with $f_{\rm faint} / f_{bright} < 0.3$ for at least one broad emission line. 
For the others, either there is no emerging broad emission lines, or the broad emission line is too weak. In Table \ref{tab:Individual1}, we summarize the broad-emission line flux of H$\beta$ and H$\alpha$ (or Mg II) in the faint and bright state.
 
We have identified the largest sample to date - 82 spectroscopically confirmed turn-on changing-look quasars (CLQ). Table \ref{tab:Candidates} 
summarizes the source information. 
The column `Follow-up/Result' marked as `Observed' denotes the targets we observed, while `CLQ' indicates that the target is confirmed as a CLQ by our criteria.
Our method efficiently detects CLQ candidates using optical and infrared time-domain surveys. We used the 82 turn-on CLQs as our turn-on CLQ sample for analysis of physical properties in following sections. 

Some of these turn-on CLQs were independently observed by other work. J0915$+$4814 is a turn-on LINER reported by \citet[ZTF18aaidlyq;][]{Frederick2019}. Two turn-on CLQs (J1003+3525 and J1115+0544) were published in \citet{Yang2018}. Seven turn-on CLQs were independently discovered in \citet{Wang2024} (J0819+3019, J0951+3416, J1020+2437, J1524+4327, J1538+4607, J1552+2102, and J2225+2019). Two CLQs, J1011+5442 \citep{Runnoe2016} and J1132+0357 \citep{Yang2018}, were discovered as turn-off CLQs, but have recently turned on again. Therefore, we newly discovered 70 new CLQs and two new turn-on CLQ events.

Figure \ref{fig:example} shows two example turn-on CLQs. For each object, we show the light curves in the optical (left) and MIR (bottom) on the left, and multi-epoch spectra on the right. In the optical light curves, ZTF magnitudes are calibrated to SDSS magnitudes as described in \S\,\ref{sec:calibration}, and the horizontal dashed lines show the SDSS spectral synthetic magnitudes at the faint state. In the right panel, the spectra taken at different epochs are indicated by different colors. The bottom right panel shows the difference spectra between the bright states and the SDSS faint-state spectra. The vertical lines in the light curves indicate the epochs at which the spectra were taken. Figures of all other CLQs are available in the appendix.

There is significant diversity among the turn-on CLQs. The left panel of Figure \ref{fig:example} shows an example of a `fresh quasar', J0947+5449.  It brightened by more than 5 mag in the optical and $\sim2$ mag in the MIR.  As there is no detectable \OIII\ emission in the faint-state spectrum, it is likely awakening after a long quiescent phase. In our sample, there are two cases, J0947+5449 and J2203+1124, without detectable \OIII\ emission in the faint state. The right panel shows an example of a turn-on CLQ with indications of AGN activity in the faint state, based on the narrow line emission line flux ratios. In some cases, strong Fe emission lines appear in the bright state, for example in J0936+2726. Within the 82 CLQs, 75 met our CLQ criteria in H$\beta$,  55 in H$\alpha$, and one in MgII (J0947+5449). Some CLQs met the criteria only for H$\alpha$, but not H$\beta$; these are highlighted by blue diamonds in Figure \ref{fig:line_ratio}.

For comparison to criteria used in the literature \citep{MacLeod2019, Green2022}, 38/75, 54/55, and 0/1 of them are designated CLQs using the $N_{\sigma} > 3$ criterion for H$\beta$, H$\alpha$, and MgII, respectively. As shown in Figure \ref{fig:example} (left), even though J0947+5449 is obviously a CLQ both in H$\beta$ and MgII, it does not pass the $N_{\sigma}$ criterion, which is more dependent on SNR\footnote{We note that the flux ratio criterion is also somewhat dependent on SNR when the broad-emission line is very weak.}. For comparison, we also list the $N_\sigma$ values in Table \ref{tab:Individual1}.

In Figure \ref{fig:known}, we show the two CLQs changed back and forth. J1011+5442 (left) is a turned off CLQ reported by \citet{Runnoe2016}. The recent optical and MIR data indicate this object brightened again. We obtained a new MMT spectrum and found it turned on with emerging broad emission lines. J1132+0357 (right) was found as a turn-on CLQ in \citet{Yang2018} from 2003 (SDSS) to 2016 (LAMOST). This object shows variability both in the optical and MIR back and forth. We found it turned off in 2018 (LJT), and then turned on again in 2019 (Palomar DBSP).

Our candidates are selected from a galaxy sample. A lot of them do not have detectable broad emission lines using our criteria described in Section \ref{sec:FWHM}. Among the 115 objects we observed, 85 objects do not have detectable H$\beta$ broad-line emission in the faint state, and 33 have no detectable broad \Ha\ emission. Among the 82 objects that we identified as turn-on CLQs, 69  objects do not have detectable broad H$\beta$, and 29 do not show broad \Ha\ emission in the faint state. Some objects, even without broad-line emission in the faint state, were nevertheless not identified as CLQs because the broad-line emission in the bright-epoch spectra were not strong enough.  In some cases this could be because the bright-epoch spectra were not taken close enough to the brightest epoch.  

\begin{figure}[htbp] \label{fig:variability}
\hspace*{-0.5cm}
\epsscale{1.1}
\plotone{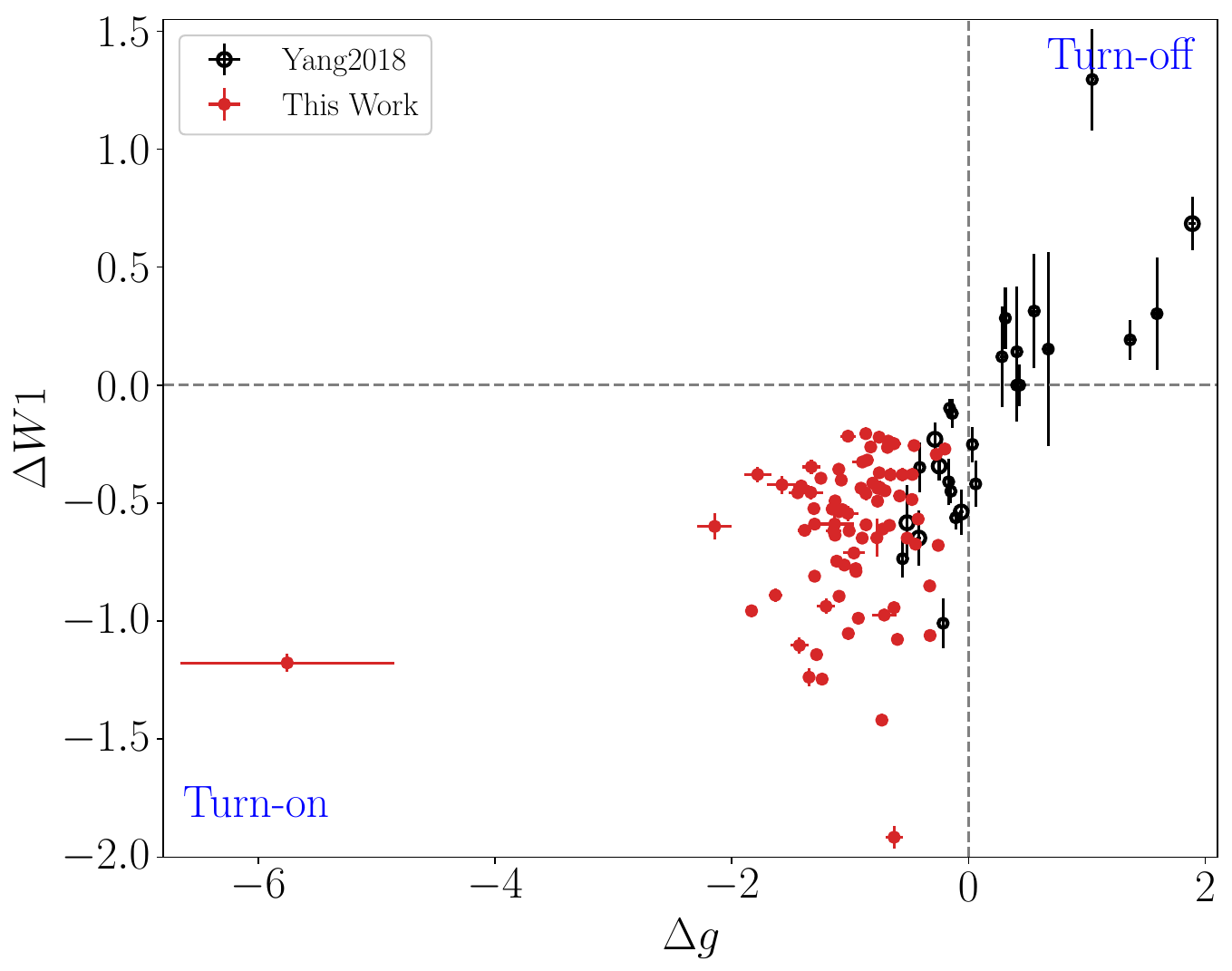}
\vspace{-0.0cm}
\caption{The variability of CL AGN in optical $g$-band and mid-infrared $W1$-band. The broadband flux in optical and mid-infrared have the same trend. The CL AGN marked as black dots are from \citet{Yang2018}.
The left bottom region, where both $g$ and $W1$ band brightened, is a region for turn-on CL AGN selection; the top right area, where both $g$ and $W1$ dimmed, is useful for turn-off CL AGN selection. }
\end{figure}

\begin{figure*}[htbp]
  \centering
  \hspace{0cm}
  \subfigure{
  \includegraphics[width=3.1in]{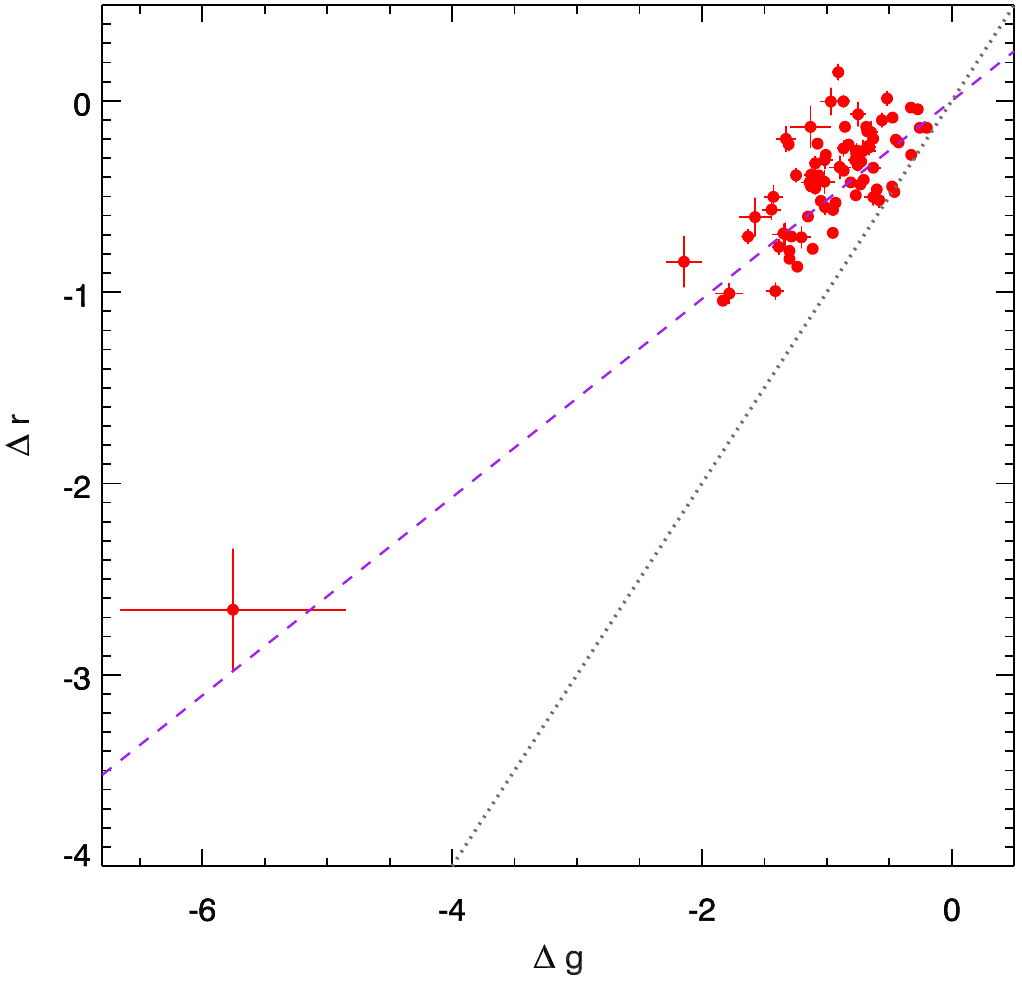}}
 \hspace{-0cm}
 \subfigure{
  \includegraphics[width=3.2in]{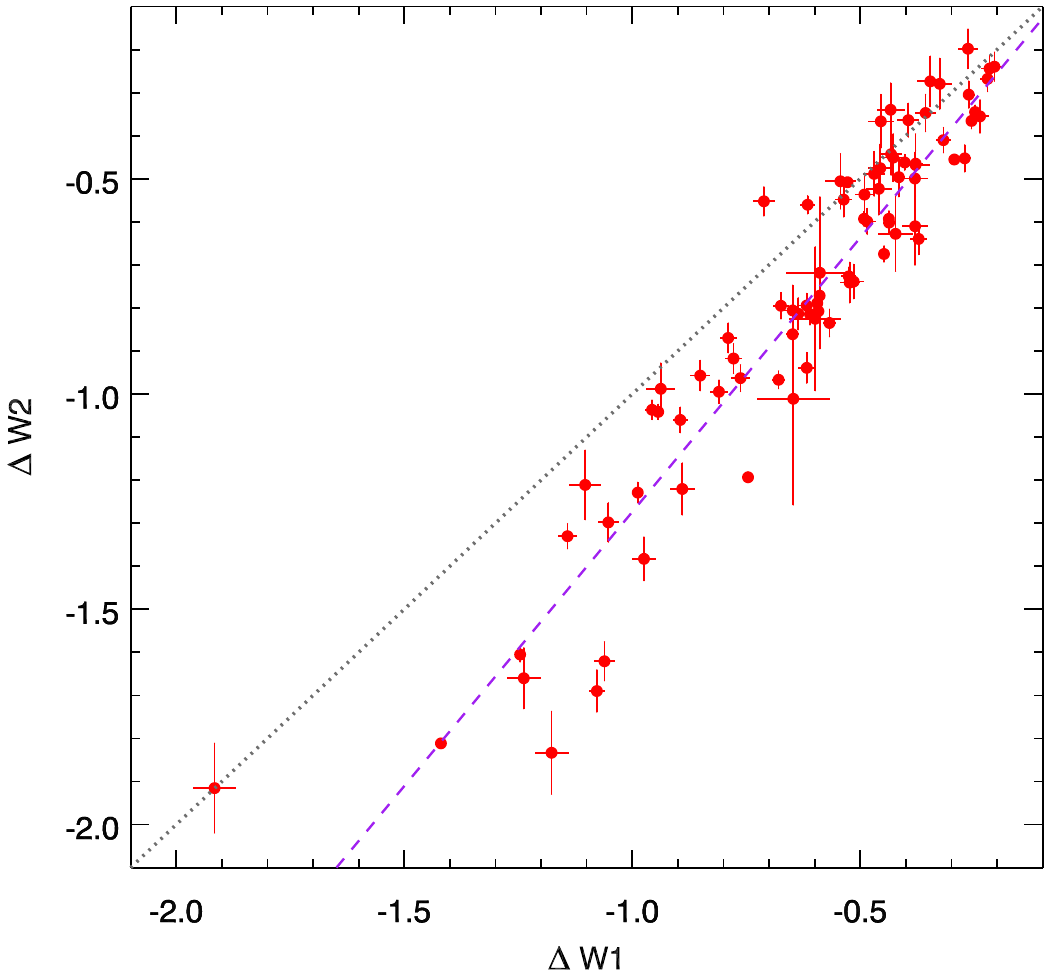}}\\
 \vspace{-0cm}
\caption{\label{fig:color} The multi-band magnitude variability of CL AGN in optical (left panel) and mid-infrared (right panel). Changes are calculated as later minus earlier epochs, so that negative values indicate brightening with time.
The gray dotted line is the 1-to-1 line, and the purple dashed line shows the fitted relation in Eq. \ref{eq:optical} (left) and Eq. \ref{eq:MIR} (right).
A bluer-when-brighter trend is confirmed in the optical. However, the mid-infrared $W1-W2$ color is redder when brighter. The opposite color change trend in the mid-infrared is possibly due to a stronger contribution from the AGN dust torus when the AGN turns on.
The variability amplitude in the $g$ band is larger than in the $r$ band.
The variability amplitude in the $W2$ band is larger than in the $W1$ band.
}
\end{figure*}

\subsection{Photometric Variability}

All the CLQs exhibit significant variability in both the MIR and optical from WISE and ZTF data. Figure \ref{fig:variability} shows the maximum MIR variability $\Delta W1$ vs. maximum optical variability $\Delta g$ of the turn-on CLQs. These objects became brighter at both wavelengths. The most extreme case is J0947+5449, which brightened by more than 5 mag in the optical. Figure \ref{fig:color} shows the color variability in optical (left panel) and MIR (right panel). We fitted the relation between the maximum optical ZTF $g$-band variability $\Delta g$\footnote{The variability compares the SDSS-calibrated ZTF magnitude as described in Section \ref{sec:calibration} to the SDSS spectral synthetic magnitude at the faint-state spectral epoch.} and the closest-epoch $r$-band variability $\Delta r$. We weighted the fitting by the uncertainties in both bands. We assumed $\Delta r = 0$ when $\Delta g = 0$, i.e., we fit only for a slope.
The best-fit is as follows,
\begin{equation} \label{eq:optical}
    \Delta r = 0.52 \times \Delta g.
\end{equation}
Therefore, there is stronger variability in the $g$ band than in the $r$ band, which is  consistent with the ``bluer-when-brighter" chromatism \citep[e.g.,][]{Wilhite2005, Schmidt2012}. The right panel of Figure \ref{fig:color} shows the maximum variability in the $W1$ band, $\Delta W1$, (compared to the epoch closest to the faint spectral epoch) and the simultaneous variability in the W2 band, $\Delta W2$. The best-fit is as follows,
\begin{equation} \label{eq:MIR}
    \Delta W2 = 1.27 \times \Delta W1
\end{equation}
So there is stronger variability at the longer wavelength in the MIR in the $W2$ band, compared to that in the $W1$ band. This is consistent with the ``redder when brighter" trend described in \citet{Yang2018}. It is possible that when an AGN turns on, the amount of hot dust increases. The UV/optical accretion-disk emission heats up the dust, which then thermally re-radiates this energy in the infrared spectrum at temperatures ranging from a few hundred K up to the dust sublimation of $\sim$1500 K. The thermal radiation from the torus dominates the total near- to mid-infrared emission of the AGN. The peak emission often falls in the wavelength range of approximately 5-30 $\mu$m \citep[e.g.,][]{Barvainis1987, Nenkova2008, Honig2010}. The larger variability in the $W2$ band than in $W1$ band is likely due to the hot dust emission peaking closer to the longer wavelength of the $W2$ band.

\begin{deluxetable*}{cclccclrr}
\tabletypesize{\tiny}
\tablecaption{Physical Properties of CLQs \label{tab:properties}}
\tablewidth{1pt}
\tablehead{
\colhead{Name} &
\colhead{log($M_{\rm BH}//M_{\odot}$)} &
\colhead{Line} &
\colhead{log($R_{{\rm Edd, faint}}$)} &
\colhead{log($R_{{\rm Edd, bright}}$)} &
\colhead{log($R_{{\rm Edd, brightest}}$)} &
\colhead{$\sigma_*$} &
\colhead{$M_{\rm s}/M_{\odot}$} &
\colhead{$M_{\rm stellar}/M_{\odot}$}\\
\colhead{(1)} & \colhead{(2)} & \colhead{(3)} & \colhead{(4)} & \colhead{(5)} & \colhead{(6)} & \colhead{(7)} & \colhead{(8)} & \colhead{(9)}
}
\startdata
% Name state line n_sigma ratio Result
J0040+1609 & 7.85 $\pm$ 0.04 & H$\beta$ & -1.71 & -1.49 & -1.28 & 198.4 $\pm$ 29.3 & 10.05 & 10.34 \\ 
J0047+1541 & 7.07 $\pm$ 0.21 & H$\beta$ & -2.79 & -2.22 & -1.98 & 134.3 $\pm$ 2.8 & 10.12 & 10.88 \\ 
J0107+2428 & 8.37 $\pm$ 0.05 & H$\beta$ & -2.80 & -2.07 & -1.93 & 183.1 $\pm$ 8.5 & 10.00 & 10.26 \\ 
J0110+0026 & 7.09 $\pm$ 0.03 & H$\beta$ & -2.97 & -2.12 & -1.91 & 71.8 $\pm$ 4.3 & 8.88 & 9.43 \\ 
J0132+1501 & 8.15 $\pm$ 0.02 & H$\beta$ & -2.95 & -2.21 & -1.95 & 161.4 $\pm$ 17.1 & 10.06 & 10.61 \\ 
J0141+0105 & 8.39 $\pm$ 0.08 & H$\beta$ & -3.81 & -2.07 & -1.95 & 173.1 $\pm$ 10.3 & 10.15 & 10.81 \\ 
J0146+1311 & 8.10 $\pm$ 0.02 & H$\beta$ & -2.81 & -2.07 & -1.79 & 137.5 $\pm$ 8.9 & 10.12 & 10.71 \\ 
J0334+0051 & 8.39 $\pm$ 0.05 & Mg\,II & -2.23 & -1.65 & -1.25 & 91.0 $\pm$ 31.0 & 10.23 & 10.59 \\ 
J0803+2207 & 7.59 $\pm$ 0.02 & H$\beta$ & -1.96 & -1.72 & -1.62 & 144.5 $\pm$ 11.5 & 9.95 & 10.24 \\ 
J0813+4608 & 7.37 $\pm$ 0.02 & H$\beta$ & -3.24 & -1.53 & -1.46 & 130.9 $\pm$ 3.9 & 9.83 & 10.35 \\ 
J0819+3019 & 7.98 $\pm$ 0.01 & H$\beta$ & -2.95 & -1.65 & -1.56 & 139.6 $\pm$ 7.3 & 10.14 & 10.45 \\ 
J0823+4202 & 7.56 $\pm$ 0.06 & H$\alpha$ & -2.56 & -1.65 & -1.50 & 116.4 $\pm$ 9.8 & 10.03 & 10.32 \\ 
J0829+2319 & 8.09 $\pm$ 0.01 & H$\alpha$ & -3.06 & -2.69 & -2.52 & 160.5 $\pm$ 11.0 & 10.16 & 10.58 \\ 
\enddata
\tablecomments{Col. (2): Black hole mass measured using the broad emission line indicated in col. (3). Col. (4)-(5): Eddington ratio measured in the faint and bright spectral epoch, respectively. Col. (6): The brightest Eddington ratio estimated used ZTF photometric data. Col. (7): stellar velocity dispersion $\sigma_*$ measured in the faint-state spectra, `VDISP' in SDSS catalogs. Col. (8): stellar mass measured from the faint-state spectra from \texttt{QGfit}. Col. (9): stellar mass corrected for fiber light loss. The full table is available in the electronic version.}
\end{deluxetable*}

\subsection{Physical Properties}
\label{sec:properties}

We use the broad-line emission in the bright state to derive the black hole mass. 
We prefer to use \hbeta\, and $L_{\rm 5100\AA}$ to calculate black hole mass, following \citet[][Eq. 5]{Vestergaard2006}. 
If \hbeta\, is not detected or is too weak, with FWHM larger than 10,000 \kms, we use \halpha, following \citet[][Eq. 1]{Greene2010}, which used the radius–luminosity relation calibrated using H$\beta$ \citep{Bentz2009b}. For J0334+0051, with weak H$\beta$ emission and a redshift at $z=0.429$, resulting in no coverage of H$\alpha$ emission, we used Mg II following \citet[][Eq. 2]{Shen2011}. 
We calculate the bolometric luminosity using 
log$L_{bol} = 0.975\,{\rm log}L_{\rm 3000} + 1.852$ \citep{Runnoe2012}, in which $L_{\rm 3000}$ is the luminosity at 3000\,\AA.   
In Table \ref{tab:properties}, we list the physical property measurements of the turn-on CLQs.

\begin{deluxetable}{llccc}
\tablecaption{Median Properties of Different Quasar Samples/States \label{tab:medians}}
\tablewidth{1pt}
\tabletypesize{\scriptsize}
\tablehead{
\colhead{Sample} & \colhead{State} & \colhead{log\,\Mbh} & \colhead{log\Lbol} & \colhead{log\REdd}
}
\startdata
\hline
SDSS Quasars &  & 8.46 & 45.1 & -1.4 \\
\hline
Turn-off CLQs & Bright & 8.48 & 45.0 & -1.4 \\
Turn-off CLQs & Faint & 8.48 & 44.5 & -2.0 \\
\hline
Turn-on CLQs & Brightest  & 7.87 & 44.2 & -1.8 \\
Turn-on CLQs & Bright & 7.87 & 44.0 & -2.0 \\
Turn-on CLQs & Faint  & 7.87 & 43.4 & -2.7 \\
\enddata
\end{deluxetable}

\begin{figure*}[!ht]
\centering
\includegraphics[width=1.0\textwidth]{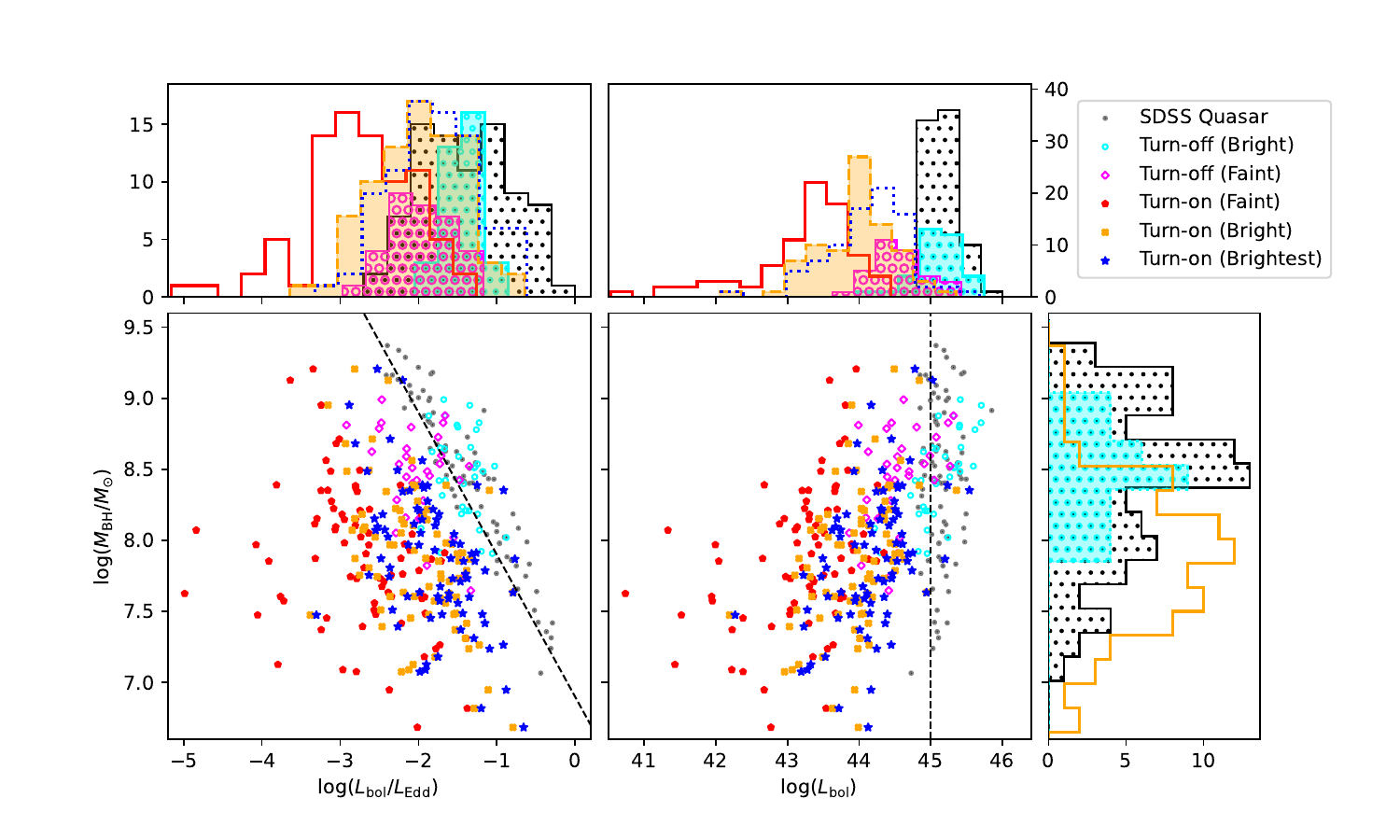}
\caption{Distribution of BH masses vs. $L_{\rm bol} / L_{\rm Edd}$ (lower left) and  $L_{\rm bol}$ (lower middle), all plotted in the logarithm. 
The gray dots are SDSS quasars \citep{Shen2011}, randomly sampled to match the redshift and magnitude distribution of our turn-on CLQ sample.  
The open cyan circles and open magenta diamonds are turn-off CLQs in the bright and faint states \citep{MacLeod2019, Green2022}.
The filled red pentagons are the turn-on CLQs in the faint state; the filled orange crosses are the turn-on CLQs in the bright state; and the filled blue stars are the turn-on CLQs in the brightest state. The dashed black lines indicate $L_{\rm bol}=10^{45}$ erg/s. Histograms show the distributions of $L_{\rm bol} / L_{\rm Edd}$ (upper left), $L_{\rm bol}$ (upper right) and $M_{BH}$. 
\label{fig:Ledd}}
\end{figure*}

Figure\,\ref{fig:Ledd} shows the black hole mass vs. Eddington ratio (bottom left panel) and bolometric luminosity (bottom middle panel) for three samples -- typical SDSS QSOs, turn-off CLQs and turn-on CLQs.  
We used SDSS quasars from \citet{Shen2011}. The redshift and apparent magnitude distribution of the SDSS quasar sample are different from our turn-on CLQ sample. To rule out redshift and magnitude effects, we randomly select the same sample size of 82 quasars from the SDSS quasars, matching both the redshift and $g$-band magnitude distribution of our CLQ sample (bright-state magnitude). We plot turn-off CLQs from \citet{MacLeod2019} and \citet{Green2022}, which have similar redshift and magnitude
distributions as our turn-on CLQ sample. For the turn-on CLQs, we plot epochs at the faint state (faint spectroscopic epoch), bright state (bright spectroscopic epoch), and the brightest state (brightest photometric epoch from ZTF data). 

Table\,\ref{tab:medians} shows the median values for the distributions of log\Mbh, log\LBol, and log\REdd\
for our comparison SDSS quasar sample, and for the bright and faint states of turn-off and turn-on CLQ samples.
The black hole mass of turn-on CLQs ranges widely (in the logarithm) from 6.4 to 9.2, with a median of 7.87.
This is significantly lower than the BH mass range for turn-off CLQs, with its median at about log\,$M_{\rm BH}$=8.48, 
whereas the overall distribution for SDSS quasars of similar redshift distribution is 8.46.
One possible interpretation might be that turn-on CLQs begin to accrete and continue active accretion in the quasar phase, gaining substantial BH mass, until material is no longer readily available and accretion stops and the AGN turns off.
However, the median masses listed would imply an increase in \Mbh\, via accretion by $\sim\,10^8$\Msun\, or more, a factor of $\sim 4$.  Such large gains in mass should take billions of years at reasonable Eddington ratios.  Observationally, we see CLQs turn on and off on decade timescales or shorter, which might occur millions of times to account for such large longterm mass increases.  Other AGN might cycle on thousand year timescales, which might occur thousands of times. Of course, accretion in a given quasar may also cycle on many different timescales over its lifetime, so that what we call the duty cycle of quasars may be represented simply as the fraction of time above a given \REdd, but may span a wide range of timespans above and below that value for any given quasar.  Observationally, we can only begin to probe the short end distribution of these timescales, but correlations of state change behavior with other quasar or host properties can inform theoretical expectations for all timescales, such as those derived from cosmological scale simulations (e.g., \citealt{AnglesAlcazar2021,Hopkins2024}.

In the bottom right panel of Figure \ref{fig:Ledd}, we show the bolometric luminosity, which is not strongly dependent on black hole mass.
The SDSS quasars are mostly at $L_{\rm bol} > 10^{45}~{\rm erg~s^{-1}}$, with a median of $10^{45.1}~{\rm erg~s^{-1}}$. 
Table\,\ref{tab:medians} contrasts the median $L_{\rm bol}$ values with those for the turn-off and turn-on CLQ samples.
We used ZTF photometric data to estimate the brightest epoch of each turn-on CLQ.  The median for their brightest $L_{\rm bol}$ values is $10^{44.2}~{\rm erg~s^{-1}}$.
\footnote{The bolometric luminosity of the vast majority of our turn-on CLQs are fainter than $10^{45}~{\rm erg~s^{-1}}$, which is a historical criterion demarcating quasars vs. Seyferts. However, for simplicity and continuity with the CLQ literature, we call them CLQs rather than CL AGN.}
The luminosity distribution for turn-on CLQs is significantly lower than for turn-off CLQs or normal quasars. One reason could be that due to our relatively short observational window, the AGN might need more time or multiple episodes of turbulent accretion to become as luminous. 
Longer-term tracking of turn-on CLQs could reveal that they become as bright as typical quasars. Better understanding of such timescales
and duty cycles will be helpful for understanding the relative luminosity functions and space densities of quasars, type 1 and type 2 AGN and even normal galaxies.

The top left panel of Figure \ref{fig:Ledd} shows histograms of log\,\REdd, and Table\,\ref{tab:medians} compares the median values of the three samples.
As described in literature, turn-off CLQs have lower Eddington ratio than typical quasars (e.g., \citealt{MacLeod2019,Green2022}).
Directly comparing the bright state of turn-on CLQs to general SDSS quasars, even when sampled to the same redshift distribution, is subject to selection effects, since we obtain follow-up spectroscopy to fainter magnitudes than the SDSS quasar surveys.
Matching both the redshift and apparent magnitude distributions, we find that the Eddington ratios in the bright state of turn-off CLQs are similar to those of SDSS quasars, with a median $R_{\rm Edd} = L_{\rm bol} / L_{\rm Edd} = 10^{-1.4}$. The median Eddington ratio of turn-off CLQs decreases  to $10^{-2}$ at faint states. However, we find even lower Eddington ratios for turn-on CLQs. In their faint state, the median Eddington ratio is as low as $10^{-2.7}$. It increases to $10^{-2}$, which is comparable to the faint state of turn-off CLQs. The estimated median brightest state Eddington ratio is only $10^{-1.8}$, still significantly lower than that of SDSS quasars.

For constant \Lbol, \REdd\ increases as \Mbh\, decreases.
However, our turn-on CLQ sample, with relatively lower black hole mass both than the normal quasars and turn-off CLQs, instead 
shows much lower Eddington ratios. Therefore, the low Eddington ratio is likely an intrinsic characteristic of turn-on CLQs. 

\subsection{Turn-on Narrow-line Seyfert 1s}
\label{sec:NLS1}

Narrow-line Seyfert 1 (NLS1) galaxies are defined to have FWHM(\Hb)$<2000$\kms \citep{Osterbrock1985}.
They typically show strong optical Fe\,II emission and have weak \OIII, with the ratio of \OIII\ to \Hb\ strength less than three.  NLS1s tend to show rapid variability, a soft X-ray excess \citep{Boller1996} and are thought to have high \REdd\ and relatively low \Mbh\ (below $\sim 10^{7.5}$; e.g., \citealt{Bian2004,ONeill2005}).  However, there are suggestions that inclination or radiation pressure effects may cause an underestimate of \Mbh\ \citep{Marconi2008,Rakshit2017}.

Given their strong variability, we might expect to detect a high fraction of NLS1 in our sample.  There are three sources, with FWHM of \Hb\ smaller than 2000\kms, and \Mbh\ smaller than $10^7$, including J2203+1124 ($M_{\rm BH} = 10^{6.8} M_{\odot}$), J0936+2726 ($M_{\rm BH} = 10^{6.7} M_{\odot}$), and J0952+2229 ($M_{\rm BH} = 10^{6.4} M_{\odot}$). 
There are two more sources with FWHM \Hb\ slightly larger than 2000\kms but with \Ha\ FWHM smaller than 2000\kms, including J0926-0006 ($M_{\rm BH} = 10^{6.9} M_{\odot}$) and J1000+0354 ($M_{\rm BH} = 10^{7.3} M_{\odot}$).
Another one with \Hb\ FWHM slightly larger than 2000, but with \Mbh\  smaller than $10^7$, J0926-0006 (\Ha\ FWHM  smaller than 2000). J1000+0354, with \Hb\ FWHM slightly larger than 2000, and \Mbh\  $10^{7.26}$, but \Ha\ FWHM smaller than 2000. 

The maximum probable fraction of NLS1 in our turn-on CLQ sample is about 7\%.  This is smaller than 
fractions found in other spectroscopic samples of AGN at similar redshift ranges (e.g., \citealt{Paliya2024}) and quite small compared to what one might expect for a variability-selected sample. We note that CLQ tend to have low \REdd\, whereas NLS1s may have higher values. Our variability criteria tend to select secular long-term rather than stochastic short-term variability, which may partly explain the different fractions. CLQs selected strictly from spectroscopic criteria may show a higher fraction of NLS1s.

\begin{figure}[!ht]
\centering
\includegraphics[width=0.5\textwidth]{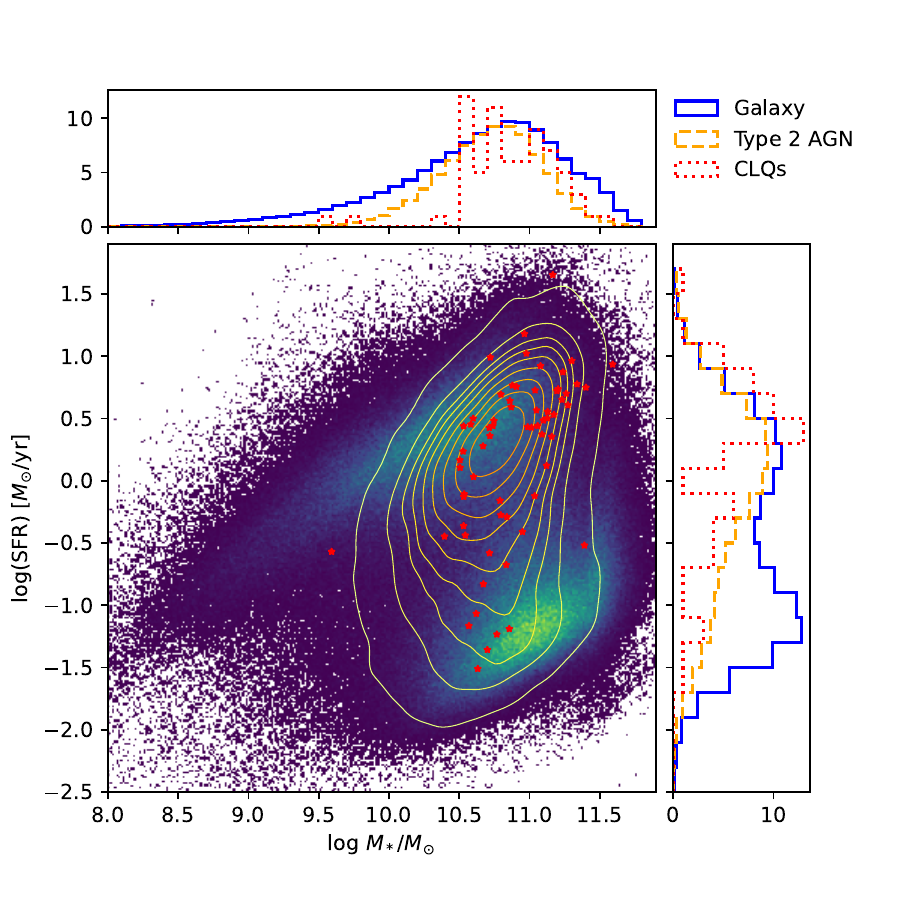}
\caption{The log star formation rate vs. stellar mass from our parent sample of 2.7 million SDSS DR8 galaxy spectra is shown by density with a color bar. Red dots represent measurements  
from the faint state SDSS spectra for the 70 turn-on CLQs hosts measured within that same DR8 sample. Orange contours show the distribution of 88,340 Type~2 AGN from \citet{Kauffmann2003}.  Relative to the parent galaxy sample, most turn-on CLQ hosts have high SFR with high stellar mass or low SFR with low stellar mass, or between. There are very few turn-on CLQs with low SFR and high stellar mass or high SFR and low stellar mass. Histograms show the distribution for these three samples of stellar mass (top panel) and star formation rate (right panel). 
The turn-on CLQ host distribution in stellar mass is high overall, and the host star formation rates are more similar to hosts of Type 2 AGN than of the overall galaxy population.}
\label{fig:host}
\end{figure}

\subsection{Host Properties} \label{sec:host}
The faint-epoch spectra of the CLQs provide a unique window to study their host galaxy properties. As there are few turn-on CLQs with black hole mass higher than $10^{8.5} M_{\odot}$, we checked the parent selection sample. We compare the host galaxy properties of our turn-on CLQs with normal galaxies. Figure \ref{fig:host} shows the star-formation rate (SFR) vs. stellar mass $M_*$ on a logarithmic scale. We used the parameters measured in the SDSS DR8 galaxy property catalog \citep{Kauffmann2003, Tremonti2004}. We cross-matched this sample with our 2.7 million parent galaxy sample and our turn-on CLQ sample. The color map represents the distribution the cross-matched galaxy sample. The red stars represent 70 turn-on CLQs cross matched.
The host galaxies of turn-on CLQs are found towards the high-mass end of the star-forming galaxies, and also have higher SFR. Several are at the low-mass end of passive galaxies, with a number scattered in between. Therefore, our turn-on CLQs are not biased towards low stellar mass galaxies. The CLQs have higher stellar masses compared to normal galaxies, but their black hole masses are lower than those of SDSS quasars with similar redshift and magnitude distributions.

We further compare the host galaxy properties of turn-on CLQs with type 2 AGN. From the SDSS DR8 galaxy sample, we use the BPT diagram (Eq. 1 in \citet{Kauffmann2003}) to identify 88,340 type 2 AGN. In Figure \ref{fig:host}, we overplot orange contours to show the host property distributions of type 2 AGN.
We run KS-tests between the turn-on CLQ sample and the type 2 AGN and normal galaxy samples, where a $p$-value below 0.05
indicates that two samples are not drawn from the same distribution. Turn-on CLQs differ more strongly from normal galaxies ($p_{M_*}=1.7\times 10^{-7},\ p_{{\rm SFR}}=8\times 10^{-10}$)
than from type 2 AGN
($p_{M_*}=0.001,\ p_{{\rm SFR}}=1.8\times 10^{-5}$).
As we might expect, the turn-on CLQ host galaxies are more similar to those of type 2 AGN.  Most significant, as can be easily seen from the histogram in the right panel of Figure\,\ref{fig:host}, is that turn-on CLQ hosts have higher star formation rate than normal galaxies.

\begin{figure}[!ht]
\centering
\includegraphics[width=0.5\textwidth]{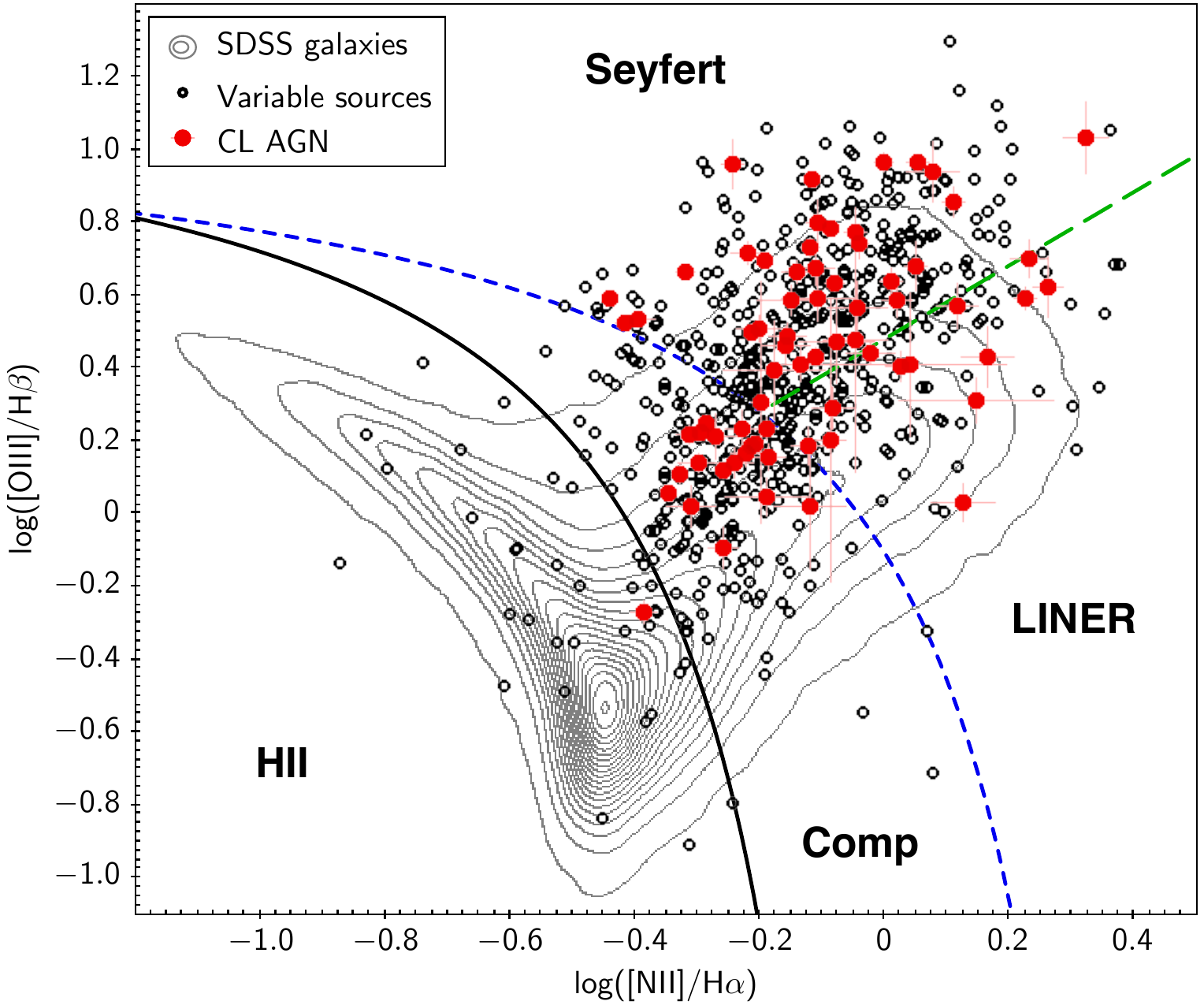}
\caption{BPT (Baldwin et al. 1981) diagram, i.e. the emission-line flux ratio [O III]/H$\beta$ versus the ratio [N II]/H$\alpha$. The gray contour are from the SDSS galaxies, showing the distribution of our parent sample \citep{Kauffmann2003}. The black open circles are the 653 galaxies significantly variable in the optical and MIR (see Table 1). The red dots show the line ratios measured from the faint state SDSS spectra for the hosts of CL AGN confirmed in this work. The blue dashed line shows the demarcation between HII and AGN regions in \citet{Kewley2001}, and the solid black curve shows the demarcation in \citet{Kauffmann2003}. The region between the black and blue lines is the composite region between HII and AGN regions. In the AGN region, the green long dash line is the separation between Seyferts and LINERs in \citet{BPT3}.} 
\label{fig:BPT}
\end{figure}

\subsection{The BPT diagram}
To avoid the contamination of bright continuum emission and broad-line emission in the bright-state epoch of CLQs, we can study their ionization properties in the faint state. We measured the narrow-line flux of \Hb, \OIII, \Ha, and [N II].

Among the 82 turn-on CLQs, we successfully measured the narrow line flux of 70 objects in their faint states. Narrow line fluxes are sometimes not detectable due to absence of e.g., \OIII\ emission (e.g., J0947+5449 and J2203+1124 described in Section \ref{sec:new_CLQs}), the absence of narrow H$\beta$ emission (e.g., J0854+1113), or no spectral coverage of the H$\alpha$ emission (e.g., J0334+0051 and J0947+5449), and etc.

For comparison, we show the host galaxy narrow-line ratios of galaxies from the SDSS DR8 sample as described in Section \ref{sec:host}. In Figure\,\ref{fig:BPT}, we show the narrow-line flux ratio of \OIII/H$\beta$ vs. [N II]/H$\alpha$. The contours are from galaxies, and the red solid dots are our turn-on CLQs. We also cross-match our parent 653 CLQ candidate sample with the SDSS DR8 catalog (open black circles in Figure \ref{fig:BPT}). The solid black and dash blue lines represent the criterion in \citet{Kauffmann2003} and \citet{Kewley2001}, respectively. Objects above the dash line are AGN (Seyferts/quasars or LINERS); objects under the solid line are star-forming (H\,II region) galaxies; and objects between the two lines are in the composite region with relatively weaker ionization contribution from the AGN than objects in the AGN region. We used \texttt{QGfit} to measure the narrow-line flux ratio of \OIII/H$\beta$ and [N II]/H$\alpha$ for 70 CLQs. We found that the vast majority (50/70) of our turn-on CLQs are in the AGN region. Some confirmed turn-on CLQs (19/70) are in the composite region. Only one confirmed turn-on CLQ (J0040+1609) is just inside the star-forming region, although we detected weak broad H$\alpha$ emission in its faint-epoch spectrum, indicating some AGN contribution. Therefore, we conclude that there are AGN contributions for all the 70 faint-state turn-on CLQs with narrow-line flux measurements of the \Hb, \OIII, H$\alpha$, and [N II] lines. However, for sources without such measurements, especially for sources lacking \OIII\ or \Hb\ narrow emission in the faint state, there may have been no AGN activity in the last few hundred or thousand years. The green long dash line represents the separation between Seyfert and LINERs in \citet{BPT3}. Among the 50 AGN, 37 sources are within the Seyfert region, and 13 sources are LINERs.

\subsection{Timescale Constraints}
Our observations provide a way to estimate the timescale of quasars turning on. The observation duration between the SDSS faint state and the bright state spectral epochs ranges from 4.8 to 23.1 years. In the rest frame, the duration ranges from 4.3 to 22.0 years. As the spectra bracket to change in states, this provides an upper limit of the time it takes for those quasars to turn on.

However, some objects were observed multiple times. For example, J0110+0026 at $z = 0.019$ was first observed by SDSS on 2000-09-07 in the faint state. It was re-observed by LAMOST on 2020-12-18, when there is stronger continuum emission, but no detectable H$\beta$ broad-line emission and very weak H$\alpha$ emission (less than a factor of 2 compared to its SDSS epoch). We observed this object on 2022-10-28 with MMT, and found the continuum emission had strengthened and strong broad H$\beta$ and H$\alpha$ emission emerged. This particular case shows that the broad H$\beta$ emission emerged in a time as short as 1.8 years in the rest frame.

Similar to J0110+0026, we have multiple observations for J0803+2207 (at $z = 0.125$) and J0813+4608 (at $z = 0.054$). These two cases constrain the emergence of broad H$\beta$ emission to 3.7 and 2.8 years in the rest frame, respecitvely.

We found two cases that changed back and forth. J1011+5442 turned off within 7.1 years, and then turned on within 2.6 years in the rest frame. J1132+0357 (at $z = 0.091$) was first observed by SDSS in 2003, and later observed to have turned on by LAMOST  \citep{Yang2018}. We re-observed it on 2018-02-16 with LJT, and found it turned off. On 2019-05-09, we re-observed it with Palomar DBSP, and found it turned on again. The turn-off and turn-on occur within 1.9 and 1.1 years. Its optical and MIR light curves shows it changed back and forth in both wavelengths, with consistent trends and a slight time delay between the optical and MIR light curves.

\section{Discussion} \label{sec:discussion}

\subsection{Coevolution of SMBH and Their Host Galaxies}

\begin{figure*}[!ht]
\centering
\includegraphics[width=0.48\textwidth]{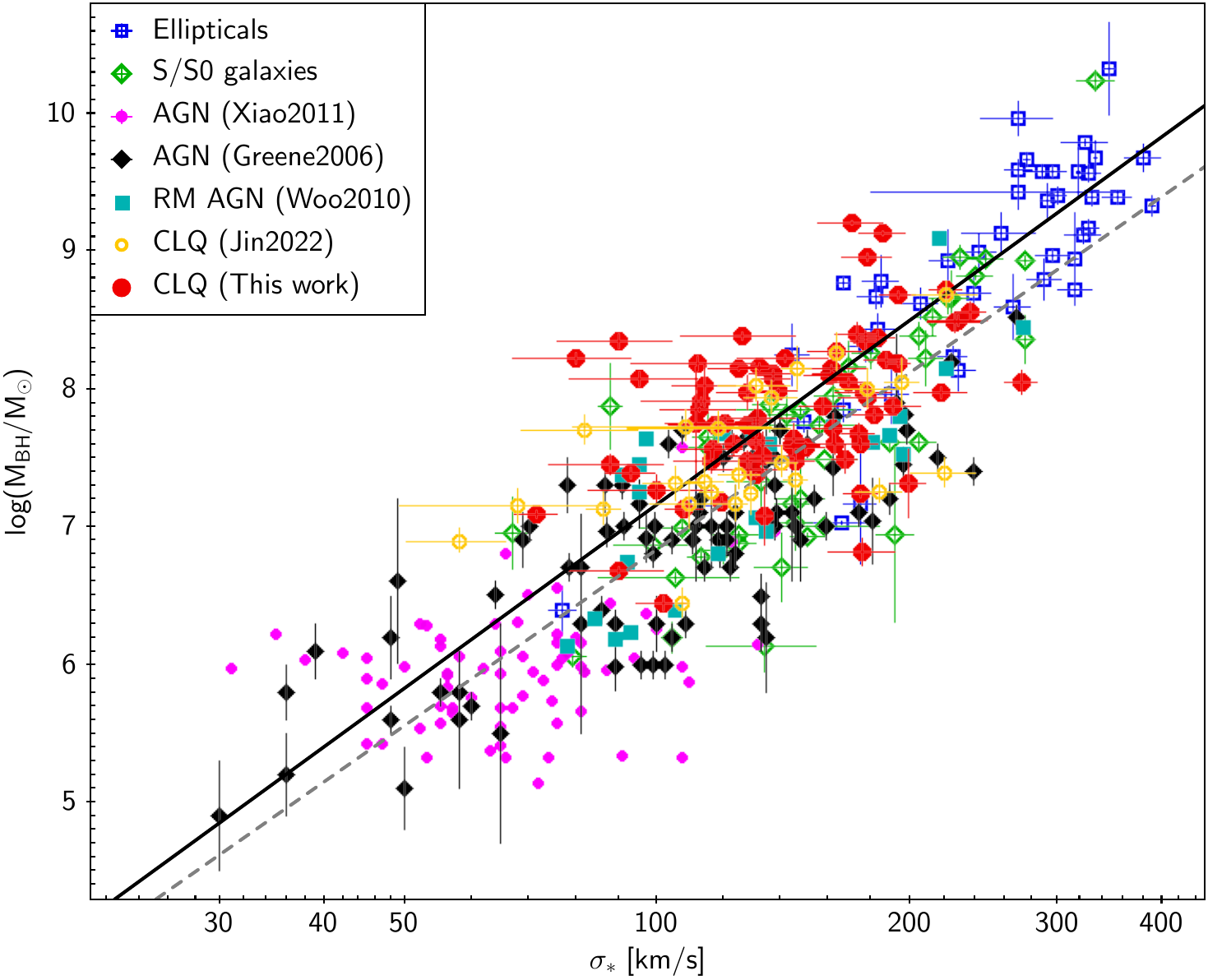}
\includegraphics[width=0.48\textwidth]{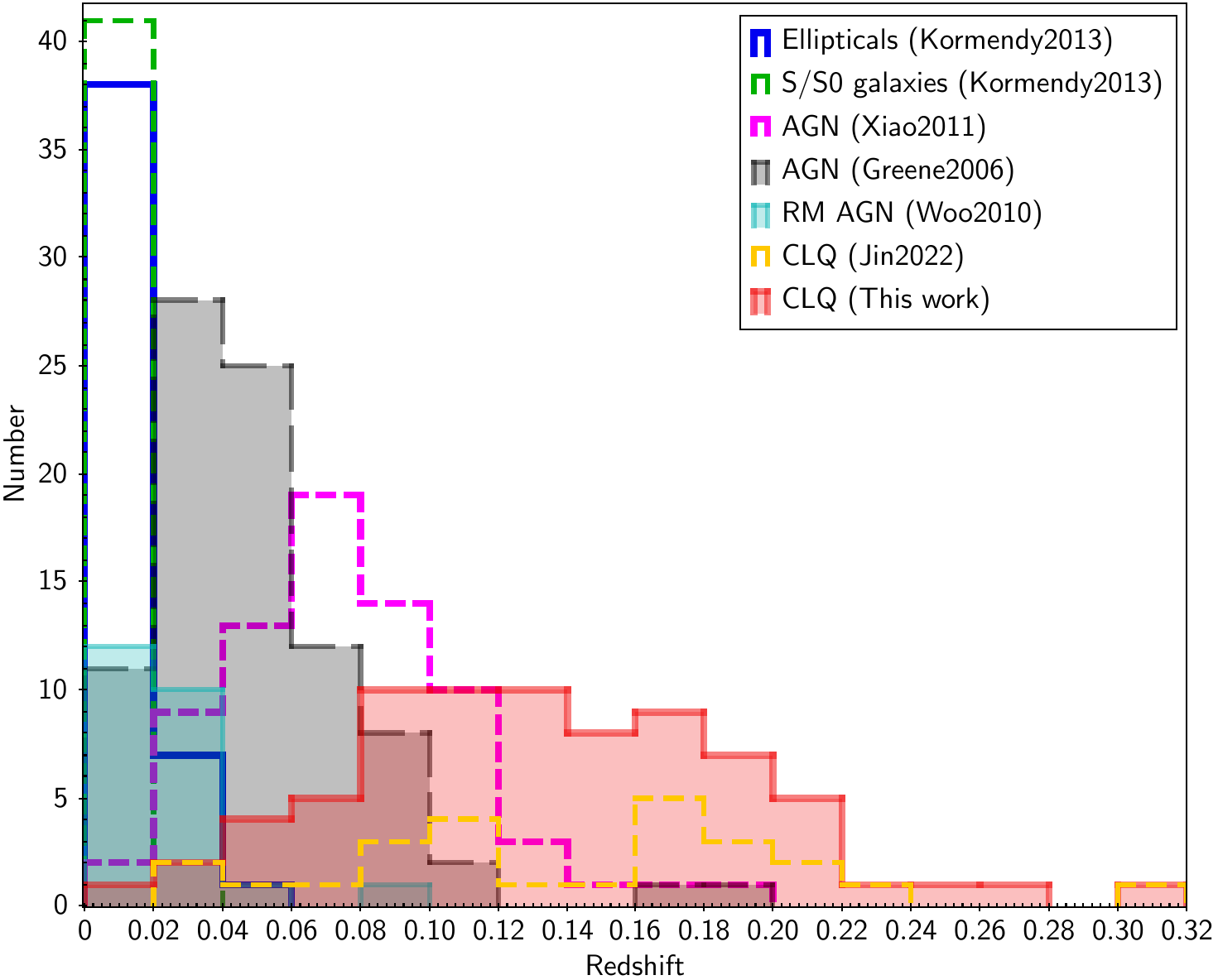}
\caption{Left: $M_{\rm BH}-\sigma_{*}$ relation. The black solid line is from \citet{Kormendy2013}, and the gray dashed line is from \citet{Gultekin2009}. Right: redshift distribution of the plotted samples. }
\label{fig:MBHsigma}
\end{figure*}

The observed scaling relations between SMBH mass and host galaxy properties {in the local universe} (e.g., the $M_{\rm BH}-\sigma_*$ or $M_{\rm BH}-M_*$ relations) form the bedrock of the prevailing theoretical framework of AGN feedback and the co-evolution of SMBHs and galaxies \citep[e.g.,][]{Gultekin2009, Ferrarese2000, McConnell2013, Kormendy2013}. Nonetheless, understanding the extent of this co-evolution over cosmic history remains challenging due to the difficulties in measuring both BH masses and host galaxy properties at higher redshifts \citep[e.g.,][]{Shen2015}. {The unique nature of CLQs provides unparalleled opportunities to investigate the BH-to-galaxy correlation beyond the nearby universe}, with enhanced sensitivity for measurements of the host galaxies in the dim state and of the central black hole masses in the bright state. As shown in Figure\,\ref{fig:MBHsigma}, the tight $M_{\rm BH}-\sigma_*$ relationship is evident for nearby galaxies at $z<0.05$ from \citet{Kormendy2013} and AGN at $z \lesssim 0.1$ \citep{Xiao2011, Greene2006, Woo2010}. Our CLQs, with median redshift 0.13, closely follow this relationship. The orange open circles are CLQs in literature \citep{Jin2022_Host}. The right panel shows the redshift distribution of samples with $M_{\rm BH}$ and $\sigma$ measurements. 

\begin{figure}[!ht]
\centering
\includegraphics[width=0.5\textwidth]{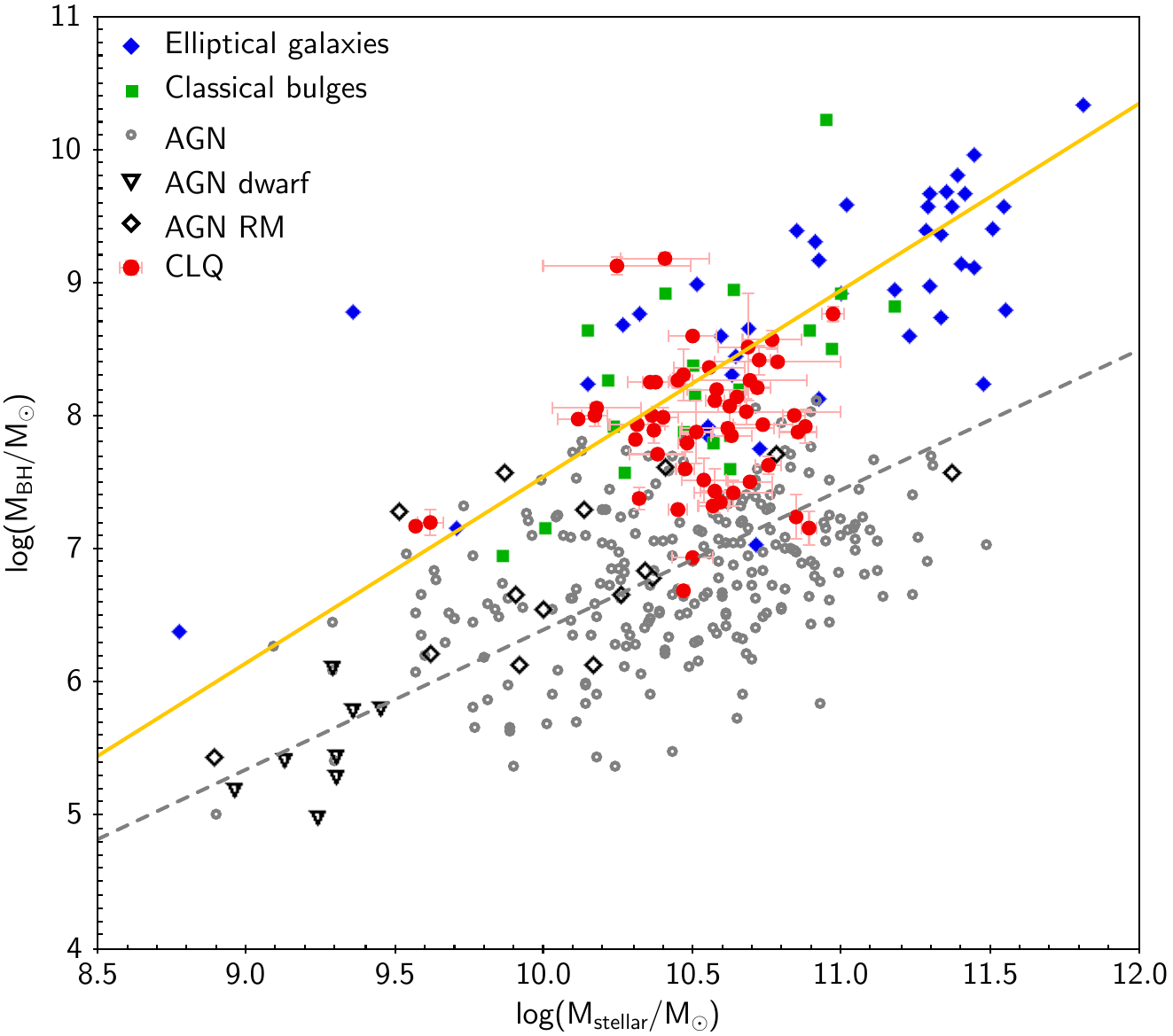}
\caption{The AGN black hole mass vs. total stellar mass. The blue diamonds and green squares are elliptical galaxies and S/S0 galaxies with classical bulges from \citet{Kormendy2013}, using the conversion from bulge mass to the total stellar mass by \citet{Reines2015}. The open gray circles are broad-line AGN from \citet{Reines2015}; the open black triangles are broad-line AGN and composite dwarf galaxies from \citet{Reines2013}. The black open diamonds are reverberation-mapped AGN with BH masses taken from \citet{Bentz2015}.  Solid red circles show values for the confirmed CL AGN in this paper, as measured from their initial faint state SDSS host spectra.  The solid orange and dashed gray lines are the BH-to-total stellar mass relation for elliptical galaxies and AGN, respectively \citet{Reines2015}. }
\label{fig:MBHstellar}
\end{figure}

Figure \ref{fig:MBHstellar} shows the relation between the black hole mass and host stellar mass. We measured the host stellar mass from the faint-state spectra, by decomposing the AGN and host contributions. We added the total stellar mass from the old and young stellar components, as described in Section \ref{sec:fitting}. Because there can be some fiber light loss in the SDSS spectra, we used the magnitude offset between the SDSS catalogued fibermag and modelmag to correct for fiber light loss in the $r$-band photometry data. We use the closest photometric epoch to the faint-state SDSS spectral observations. In Table \ref{tab:properties}, we list the corrected stellar masses of the host galaxies. In Figure \ref{fig:MBHstellar}, the solid red circles represent the confirmed CL AGN in this paper. The blue diamonds and green squares are elliptical galaxies and S/S0 galaxies with classical bulges \citep{Kormendy2013}, and has been converted from bulge mass to the total stellar mass by \citet{Reines2015}. The open gray circles are broad-line AGN \citep{Reines2015}; the open black triangles are broad-line AGN and composite dwarf galaxies from \citet{Reines2013}. The open black diamonds are reverberation-mapped AGN with BH masses taken from \citet{Bentz2015}. The solid orange and dashed gray lines are the BH-to-total stellar mass relation for the elliptical galaxies and AGN, respectively \citet{Reines2015}. The relation of AGN is about one order of magnitude lower than for ellipticals and classic bulges. \citet{Caplar2015} advocate a lower $M_{\rm BH}/M_{\rm stellar}$ ratio for AGN compared to inactive galaxies that have ``quenched" at earlier times. The CLQs are positioned between the two relations, possibly indicating that they are in a transitional phase between active and inactive galaxies.

\subsection{Turn-on CLQs without Recent AGN Activity}

We observed several turn-on CLQs lacking detectable \OIII\ narrow emission lines in the faint state. 
The left panel of Figure \ref{fig:example} shows one example of such a `fresh' quasar, J0947+5449. In the faint-state SDSS spectrum, there is no detectable Balmer emission, Mg II, or \OIII\ emission. However, there is a prominent [O II]$\lambda$3727 narrow line, along with noticeable K and H absorption lines. J0947+5449 significantly brightened by more than 5 mag in the optical and $\sim2$ mag in the MIR. We re-observed this object with Palomar DBSP in November 2019, and found emerging broad Mg II and Balmer emission. After 2020, J0947+5449 dimmed again, so we took another MMT spectrum in December 2023. The MMT spectrum confirms that the object has indeed faded, but it now shows more prominent \OIII\ narrow line emission (see Figure \ref{fig:example}). This confirms that the radius of the narrow-line region is larger than that of the broad-line region, leading to a delayed response to the brightening compared to the broad-line emission. The time interval between the two follow-ups is approximately four years (2.5 years in the rest frame). Another case is J2203+1124, which we also noted as a NLS1 in S\,\ref{sec:NLS1}. It was first observed by SDSS in 2001 in the faint state, without any detectable \OIII\ narrow emission.  The spectrum taken by LJT in 2017 shows emerging broad Balmer emission.  

Those turn-on CLQs in our sample lacking narrow emission lines could be showing their first AGN activity in a long time. In other words, such systems provide a {\em lower limit} on a quasar's quiescent phase. However, we note that there are rare but well-studied cases where the \OIII-emitting region can be small.  \citep{Peterson1993} find from variability that the NLR in the Seyfert 1 galaxy NGC 5548 has a radius of only 1-3 pc and is denser ($n_e\sim\ 10^5$ cm$^{-3}$) than previously supposed. \citep{Zheng1995} found similar results for 3C\,390.3, which has quasar-like luminosities, though results may be affected by jet-related variability.

Evidence suggests that the growth of SMBHs is not a continuous process throughout their lifetimes, instead, the growth of a SMBH is possibly self-regulated, so that significant growth only occurs during the active phases \citep{Mo2010}.  

Other observational methods of contraining quasar accretion cycling on long timescales exist.  Hanny's Voorwerp \citep{Lintott2009}, an \OIII\-emitting region about 27\,kpc from the inactive spiral galaxy IC\,2497 illustrates that IC\,2497 likely emitted quasar-level ionizing luminosities within the last $\sim 10^5$ years, e.g., the system provides an {\em upper limit} on a quasar quiescent phase.  

Some observations and simulations connecting AGN with merger activity suggest that active episodes of $10^8$ years duration may be typical \citep{DiMatteo2005}.  Evidence from comparing the local BH mass function to previous AGN activity and the cosmic X-ray background suggests similar typical durations \citep{Marconi2004}, even disregarding the role of mergers.  However, the size of regions of ionized intergalactic medium around quasars, proximity zones, 
provide interesting and complementary constraints on quasar lifetimes.  Using He\,II Ly$\alpha$ proximity zones at $3\lax z\lax 4$,  \citet{Khrykin2021} find active lifetimes from 1 to 30\,Myr, with a mean near the lower end.
Our study of CLQs is most sensitive to timescales under about a decade, but also in the case of `fresh' quasars can provide lower limits for the quiescent phases that are much longer.

\citet{Greenwell2024} identify a sample of (47) optically quiescent quasars (OQQs) with no detectable broad line or narrow \OIII\ emission, but quasar-like $W1-W2$ colors and 12$\mu m$ luminosities.  OQQs may be heavily obscured with a high nuclear covering fraction, where the broad line region is obscured and nuclear emission has not (yet) been able to ionize extended regions of the host galaxy.  Such objects would not be detected by our methods.  The space density of OQQs, determined both by the requisite physical conditions (SMBH, significant accretion and high obscuration) and the 
lifetime of those conditions, is not well-constrained.

\subsection{Mechanism and Models}

Several different scenarios have been proposed to explain CL phenomena. The changing-obscuration mechanism involves variations in the line of sight to the quasar due to changes of material obscuring the central engine (the supermassive black hole and its accretion disk). The optical and MIR changes are crucial for understanding the mechanisms driving their variability \citep{Sheng2017, Yang2018, Stern2018, Yang2023, Ricci2022_review}. All the CLQs in this work show strong variability both in optical and MIR. Significant MIR variability is not consistent with the scenario of dust obscuration. The mid-infrared flux is not significantly affected by dust extinction. In the scenario of variable obscuration, the variation in the W1 band due to dust extinction yields a factor of $\sim21$ times stronger variability in $g$-band magnitude, according to the extinction curve in the optical and mid-infrared \citep{Yang2018}. Our CLQs were selected by a variability amplitude larger than 0.2 mag both in WISE $W1$ and $W2$ bands. In an extinction scenario, a factor of 21 corresponds to more than 4.2 mag  variability in $g$ band. Only one object, J0947+5449, has more than 4.2 mag variability in the $g$ band, however, its WISE $W1$-band variability is as large as 1.2 mag.  
The optical changes do not exhibit the level of reddening expected if obscuration by dust were the primary cause, indicating that intrinsic changes in the central engine may play a more significant role. 
If the MIR light is reprocessed light from the dust torus, its echo of the continuum variations in UV/optical measures the average light-crossing time of the dust torus to accretion disk. The measured time delays between the MIR and optical of some CLQs  is consistent with the light-crossing time of the torus, further evidence that those CLQs' activity is consistent with intrinsic changes in the central engine \citep{Yang2023}.

TDEs are considered as one possible explanation for the dramatic changes observed in CLQs. TDEs occur when a star gets too close to a supermassive black hole and is torn apart by its tidal forces. The typical maximum SMBH mass for observable TDEs is around $10^8$ to $10^{8.5}$ solar masses \citep{Stone2016, vanVelzen2018}. Beyond this range, the event horizon is larger than the tidal radius for most stars. The SMBH masses for the vast majority of known TDEs are less than $10^8~M_{\odot}$ \citep{Yao2023}. Although our turn-on CLQs has slightly lower median black hole mass ($10^{7.78}~M_{\odot}$) than SDSS quasars at similar redshift ($10^{8.46}~M_{\odot}$), nearly half of our turn-on CLQs has black hole mass higher than $10^8~M_{\odot}$. The timescale of TDEs is typically months to years, with an initial brightening followed by a decline that can often be described by a $t^{-5/3}$ power-law decay \citep{Rees1988}. The timescale of TDEs are similar but slightly shorter than the observed timescale of CLQs.
However, the light curve trends of most CLQs in the optical are chaotic, inconsistent with the typical sharp rise and monotonic decay of most TDE light curves.  In the X-rays, turn-off CLQs observed to date show hardening of the X-ray spectrum, and changes in $f_X/f_{\rm opt}$ expected from a decrease in accretion rate \citep{Yang2023}; AGN are softer at higher \REdd. In contrast, TDEs show little evidence for X-ray spectral changes with time or luminosity \citep{Auchettl2018}.  Furthermore, repeating CL events in some objects is difficult to reconcile with the expectation that TDE rates of one per 10$^{4-5}$ years per galaxy (e.g., \citealt{Gezari2009,vanVelzen2018}).

It is more likely that CL phenomena are due to the changing accretion rate of the SMBHs \citep{LaMassa2015, MacLeod2016, Green2022, Yang2023}. This mechanism  explains well the consistent multiwavelength variability in optical, MIR, and X-ray, the emergence/disappearance of broad-line emission, and the time delay between optical and MIR emission \citep{Yang2023}. However, theoretical viscous timescales \sout{in accretion theory} \citep{Krolik1999, Frank2002} suggest that dramatic state changes in AGN should span $10^{4-7}$ years. The CL timescale is as short as 1-20 years, and it is 3--7 orders of magnitude shorter than quasar viscous timescale. The short timescale of CLQs poses challenges to the theoretical model of AGN accretion. Several possible mechanisms have been proposed to overcome this inconsistency.
For example, magnetically-driven disk winds, which carry away most of the gas's angular momentum, could considerably shorten the viscous timescales of the accretion disk \citep{Feng2021b}. 
\citet{Sniegowska2020} reproduce observed variability timescales in the case of radiation pressure instabilities between the standard gas-dominated outer disk and the hot optically thin inner advection-dominated accretion flow. Their model predicts regular outbursts, but accounting for realistic additional components such as winds or magnetic fields \citep{Hameury2020} easily creates more complex patterns. An accretion disk misaligned with the BH spin may cause tearing, which can lead to a rapid burst of accretion \citep{Liska2023}.

\section{Summary} \label{sec:summary}
Targeting galaxies with significant mid-IR and optical variability, we identify and present the largest sample of 82 turn-on CLQs. Among them, 70 CLQs were discovered here for the first time. Two previously detected turn-off CLQs recently re-awakened. Our main conclusions are as follows:

\begin{itemize}
\item Turn-off CLQs have lower Eddington ratio than typical quasars. 
Turn-on CLQs have lower Eddington ratios than either typical quasars or turn-off CLQs.

\item There is large amplitude variability both in the MIR and optical, inconsistent with the changing-obscuration mechanism. CLQ light curve shapes and timescales are inconsistent with TDEs. Therefore, the CLQs are most likely due to changing accretion rate from the central engine.

\item CLQs have stronger variability at shorter wavelength in the optical, but larger variability at longer wavelength in the $W2$ band than in $W1$ band. This is likely due to larger contribution from hot dust in the $W2$ band when the AGN turn on.

\item For CLQs in the faint state with narrow-line emission measurements of H$\beta$, H$\alpha$, \OIII, and [N II],  AGN contributions place the spectra in the AGN or composite region of the BPT diagram, and may produce with weak \Ha\ broad-line emission.

\item For two of the turn-on CLQs, there is no \OIII\ narrow-line emission detectable in the faint-state spectra. This evidence indicates they might be fresh quasars, likely awakening for the first time in decades or longer.

\item We measured the black hole mass from the bright-state spectra and the host properties in the faint-state spectra. The CLQs, beyond the local Universe, obey the local relations between the central SMBHs and host galaxy properties, such as stellar mass and velocity dispersion.

\item The host galaxy properties of the turn-on CLQs tend to higher mass than normal inactive galaxies, with star formation rates more similar to hosts of Type 2 AGN than to the overall galaxy population

\end{itemize}

\begin{acknowledgments} 
 Q.Y. is partially supported for this work by the National Aeronautics and Space Administration through Chandra Award Numbers GO0-21084X, GO1-22090X, GO3-24094X, and GO4-25059X issued by the Chandra X-ray Center, which is operated by the Smithsonian Astrophysical Observatory for and on behalf of the National Aeronautics Space Administration under contract NAS8-03060. P.G. acknowledges support from the Smithsonian Institution and the Chandra X-ray Center through NASA contract NAS8-03060.  We thank Allyson Bieryla for helping us obtain the FLWO 1.2m imaging observations.  This research has made extensive use of NASA's Astrophysics Data System (ADS).
We acknowledge the Texas Advanced Computing Center (TACC) at The University of Texas at Austin for providing high performance computing, visualization, and storage resources that have contributed to the results reported within this paper.

% % Telescopes
% MMT
We acknowledge the use of the MMT telescope. Observations reported here were obtained at the MMT Observatory, a joint facility of the Smithsonian Institution and the University of Arizona.
% HET
We acknowledge the use of the Hobby–Eberly telescope (HET).
Based on observations obtained with the HET, which is a joint project of the University of Texas at Austin, the Pennsylvania State University, Ludwig-Maximillians-Universitaet Muenchen, and Georg-August Universitaet Goettingen. The HET is named in honor of its principal benefactors, William P. Hobby and Robert E. Eberly.
We acknowledge the use of the Palomar Hale 5-meter telescope, the Lijiang 2.4-meter telescope, and the Xinglong 2.16-meter telescope. 

% % SDSS
We acknowledge the use of SDSS data. Funding for SDSS-III has been provided by the Alfred P. Sloan Foundation, the Participating Institutions, the National Science Foundation, and the U.S. Department of Energy Office of Science. The SDSS-III website is \url{http://www.sdss3.org/}. SDSS-III is managed by the Astrophysical Research Consortium for the Participating Institutions of the SDSS-III Collaboration including the University of Arizona, the Brazilian Participation Group, Brookhaven National Laboratory, Carnegie Mellon University, University of Florida, the French Participation Group, the German Participation Group, Harvard University, the Instituto de Astrofisica de Canarias, the Michigan State/Notre Dame/JINA Participation Group, Johns Hopkins University, Lawrence Berkeley National Laboratory, Max Planck Institute for Astrophysics, Max Planck Institute for Extraterrestrial Physics, New Mexico State University, New York University, Ohio State University, Pennsylvania State University, University of Portsmouth, Princeton University, the Spanish Participation Group, University of Tokyo, University of Utah, Vanderbilt University, University of Virginia, University of Washington, and Yale University.

% LAMOST
We acknowledge the use of LAMOST data. The Large Sky Area Multi-Object Fiber Spectroscopic Telescope (LAMOST, also named Guoshoujing Telescope) is a National Major Scientific Project built by the Chinese Academy of Sciences. Funding for the project has been provided by the National Development and Reform Commission. LAMOST is operated and managed by the National Astronomical Observatories, Chinese Academy of Sciences.

% DESI
This research used data obtained with the Dark Energy Spectroscopic Instrument (DESI). 
DESI construction and operations is managed by the Lawrence Berkeley National Laboratory. This research is supported by the U.S. Department of Energy, Office of Science, Office of High-Energy Physics, under Contract No. DE–AC02–05CH11231, and by the National Energy Research Scientific Computing Center, a DOE Office of Science User Facility under the same contract. Additional support for DESI is provided by the U.S. National Science Foundation, Division of Astronomical Sciences under Contract No. AST-0950945 to the NSF’s National Optical-Infrared Astronomy Research Laboratory; the Science and Technology Facilities Council of the United Kingdom; the Gordon and Betty Moore Foundation; the Heising-Simons Foundation; the French Alternative Energies and Atomic Energy Commission (CEA); the National Council of Science and Technology of Mexico (CONACYT); the Ministry of Science and Innovation of Spain, and by the DESI Member Institutions. The DESI collaboration is honored to be permitted to conduct astronomical research on Iolkam Du’ag (Kitt Peak), a mountain with particular significance to the Tohono O’odham Nation.

% % WISE
This publication makes use of data products from the \emph{Wide-field Infrared Survey Explorer}, which is a joint project of the University of California, Los Angeles, and the Jet Propulsion Laboratory/California Institute of Technology, funded by the National Aeronautics and Space Administration. 

% % ZTF
We acknowledge the use of ZTF data. Based on observations obtained with the Samuel Oschin Telescope 48-inch and the 60-inch Telescope at the Palomar Observatory as part of the Zwicky Transient Facility project. ZTF is supported by the National Science Foundation under Grant No. AST-2034437 and a collaboration including Caltech, IPAC, the Weizmann Institute for Science, the Oskar Klein Center at
Stockholm University, the University of Maryland, Deutsches Elektronen-Synchrotron and Humboldt University, the TANGO
Consortium of Taiwan, the University of Wisconsin at Milwaukee, Trinity College Dublin, Lawrence Livermore National
Laboratories, and IN2P3, France. Operations are conducted by COO, IPAC, and UW.

\end{acknowledgments} 

\appendix
The light curves and spectra of the other 78 CLQs, in addition to the CLQs shown in Figure \ref{fig:example} and \ref{fig:known}. The sources are ordered by their coordinates.

\facilities{MMT (Blue Channel spectrograph, BinoSpec), FLWO:1.2m, Sloan, PS1, PTF, CRTS, ZTF, WISE} 

\software{Astropy \citep{astropy2022}, IDL \citep{idl1992}, IRAF \citep{Tody1986}, Matplotlib \citep{matplotlib2023}, Numpy \citep{numpy2020}, Scipy \citep{scipy2022}, TOPCAT \citep{topcat2022}}

\bibliography{references.bib}

\clearpage
\begin{figure*}[!ht]
\centering
\vspace{-1cm}
\hspace{-0.4cm}
\includegraphics[width=0.52\textwidth]{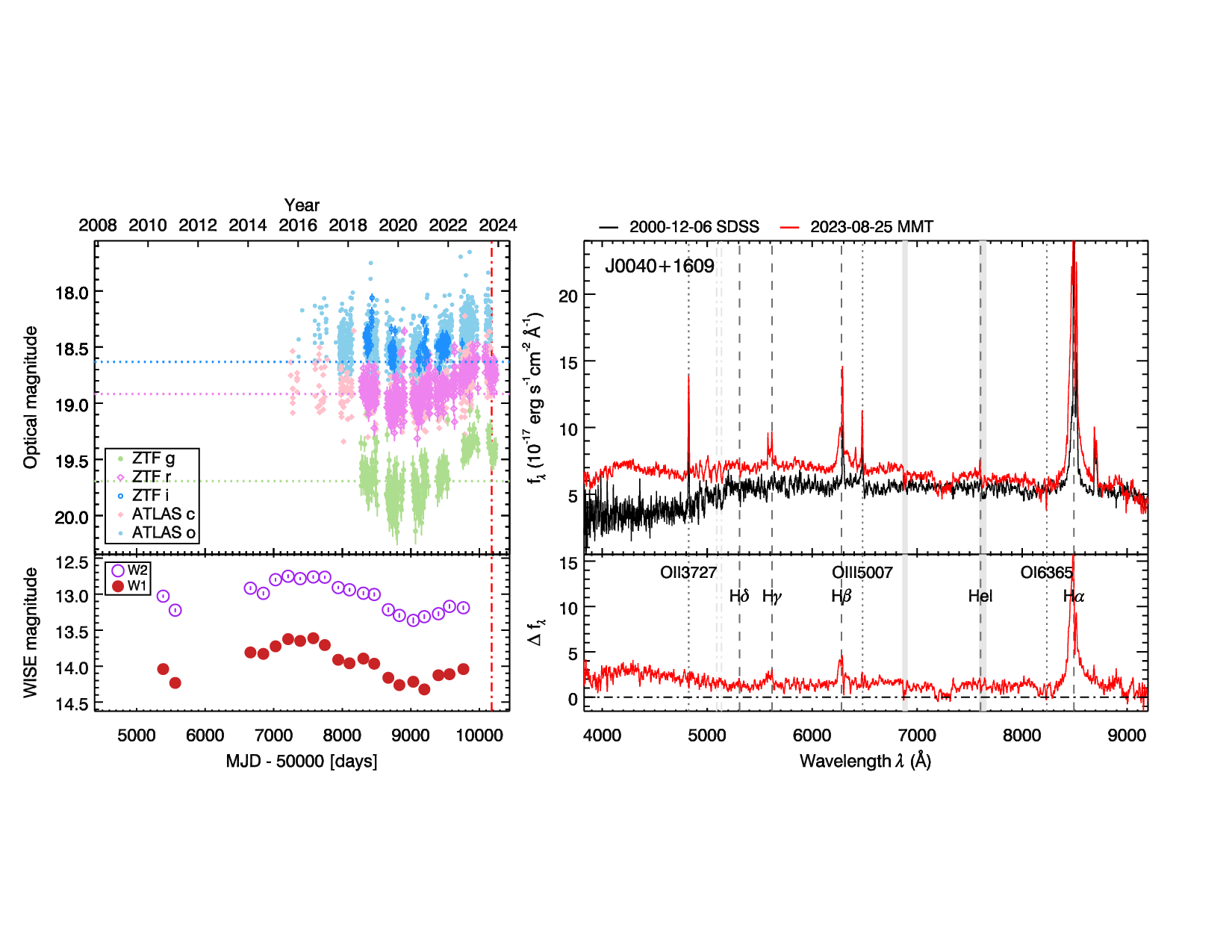}
\hspace{-0.7cm}
\includegraphics[width=0.52\textwidth]{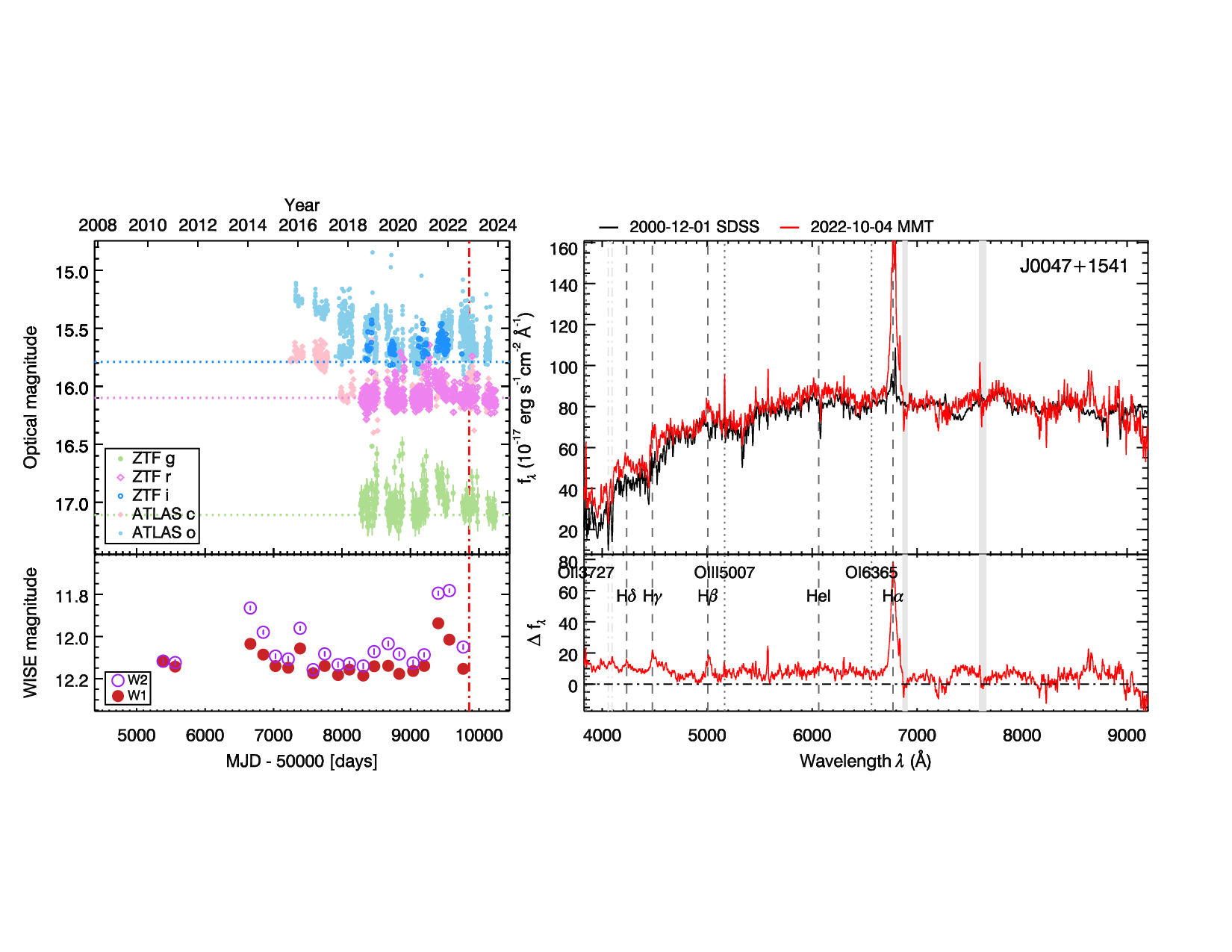}\\
\vspace{-2.7cm}
\hspace{-0.4cm}
\includegraphics[width=0.52\textwidth]{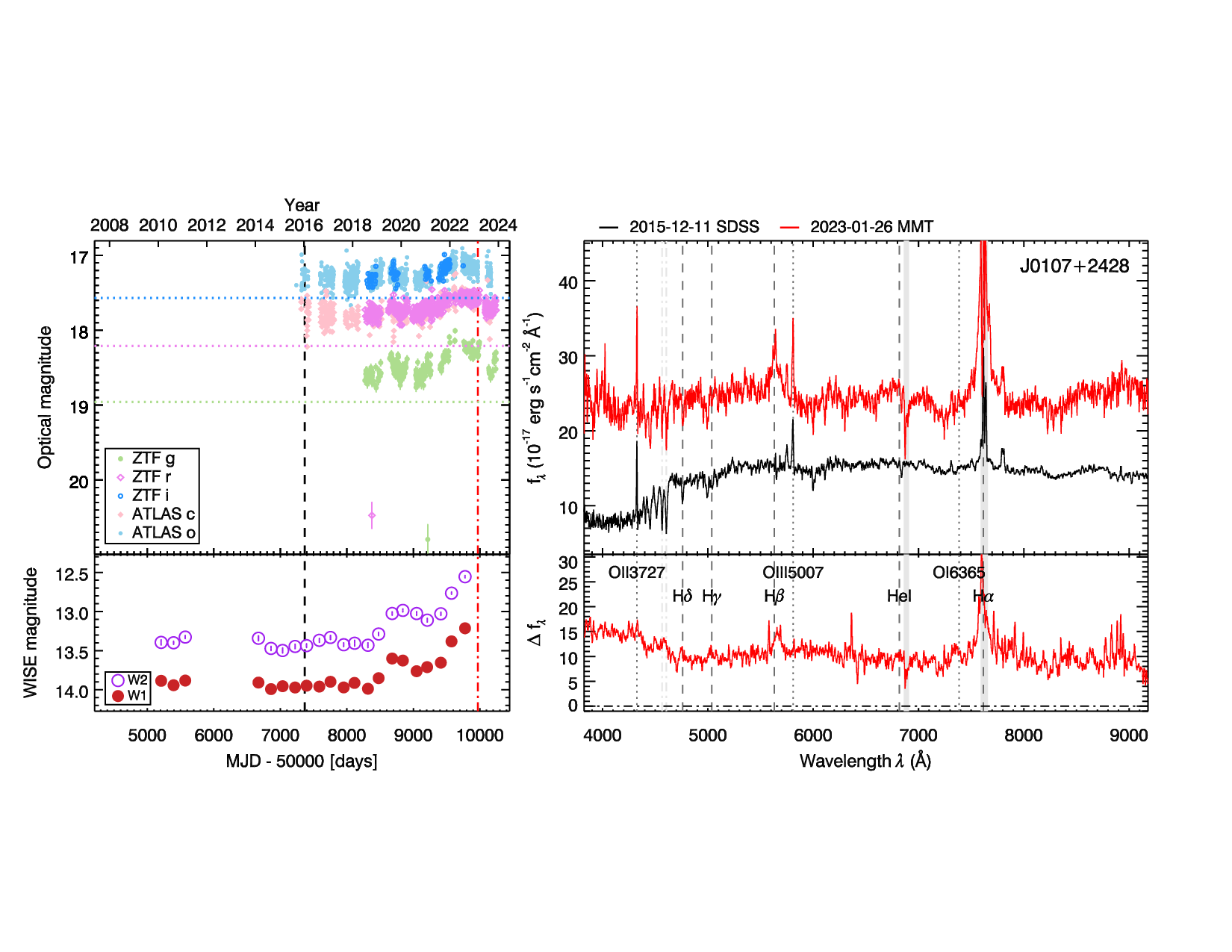}
\hspace{-0.7cm}
\includegraphics[width=0.52\textwidth]{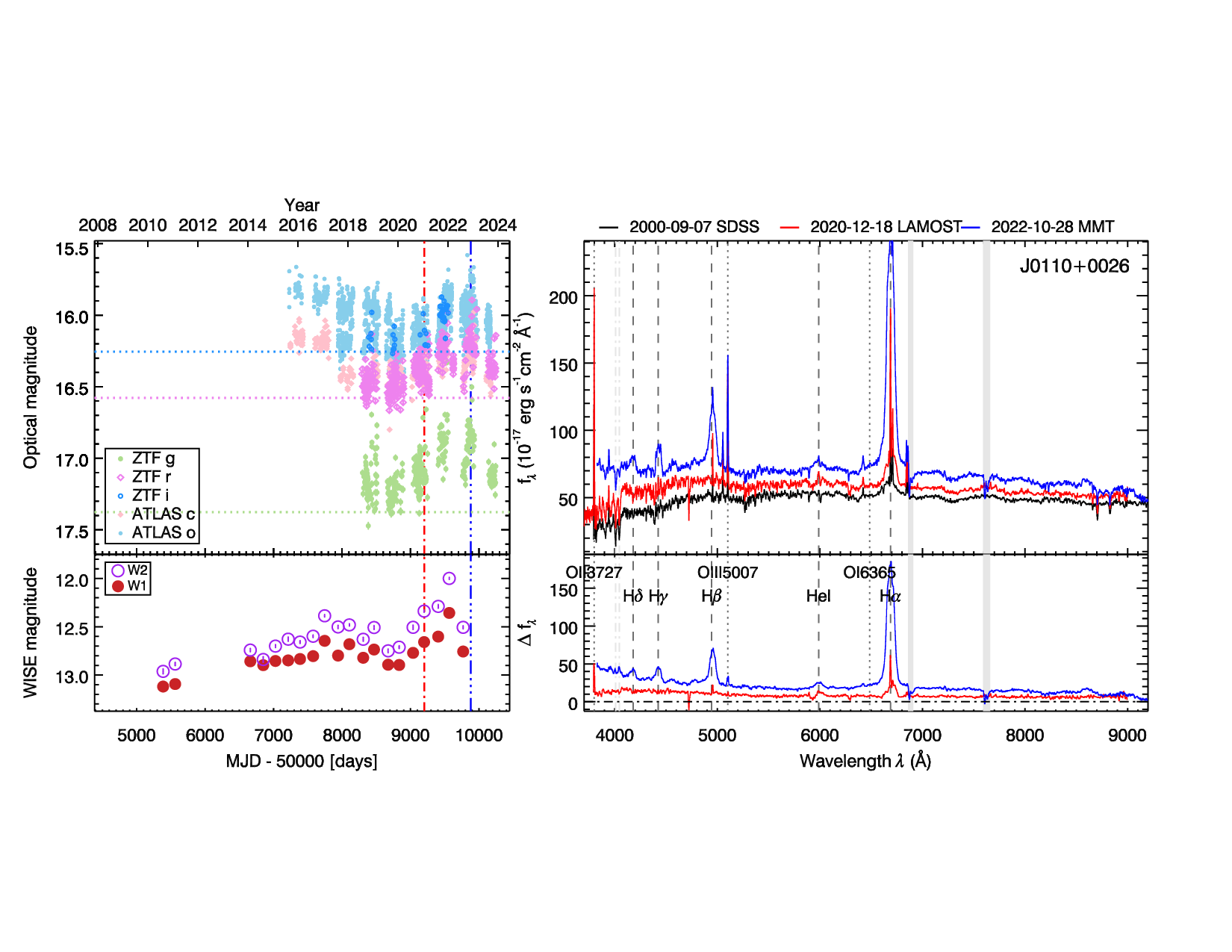}\\
\vspace{-2.7cm}
\hspace{-0.4cm}
\includegraphics[width=0.52\textwidth]{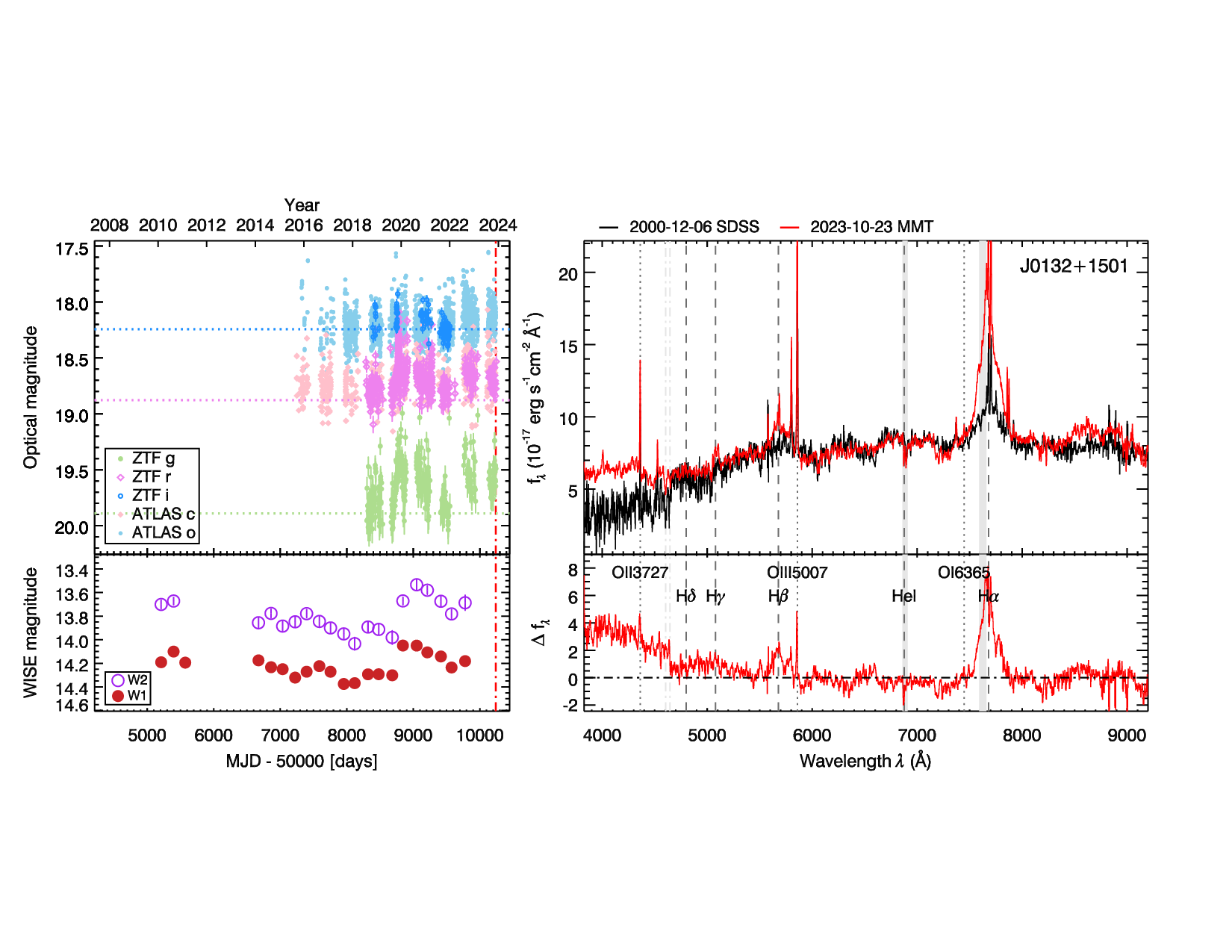}
\hspace{-0.7cm}
\includegraphics[width=0.52\textwidth]{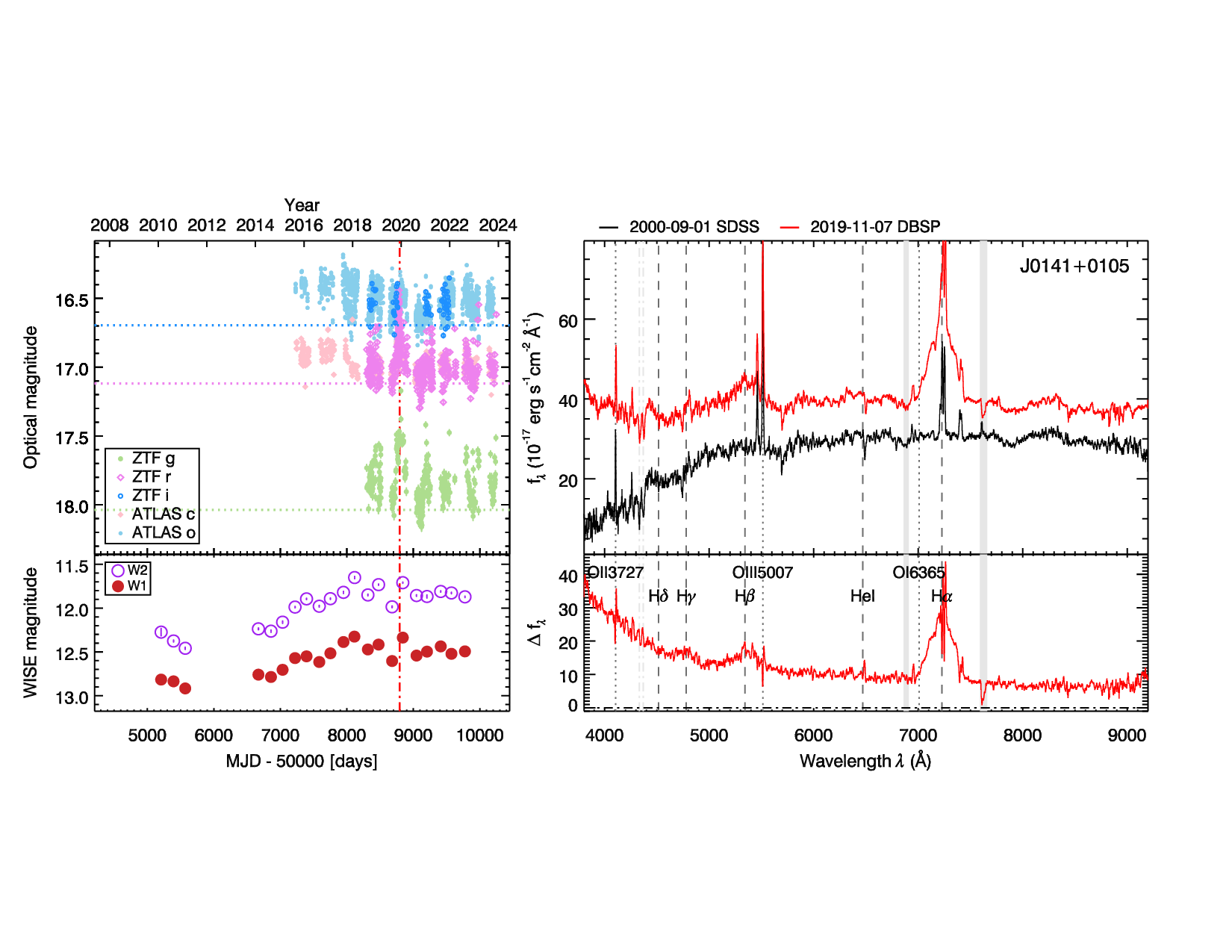}\\
\vspace{-2.7cm}
\hspace{-0.4cm}
\includegraphics[width=0.52\textwidth]{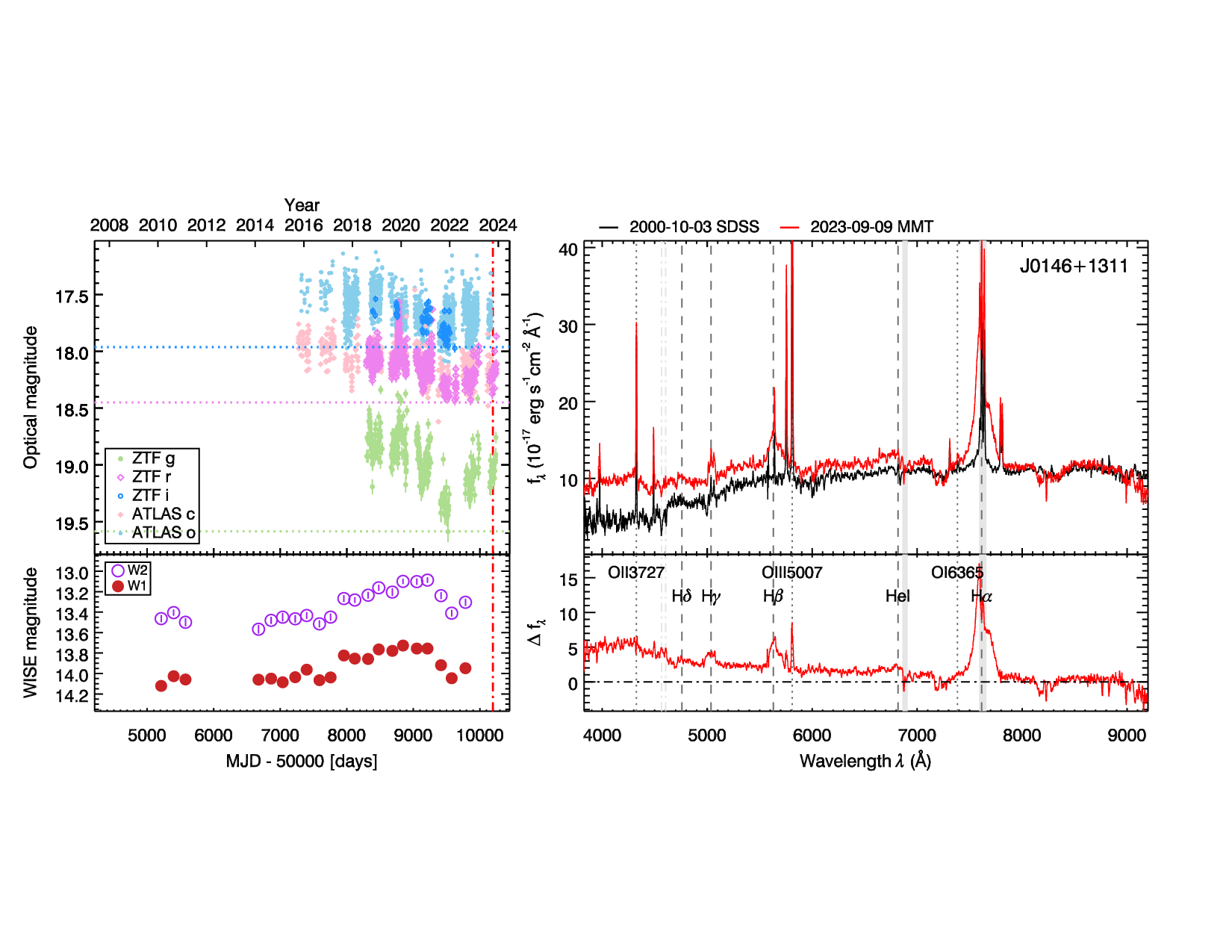}
\hspace{-0.7cm}
\includegraphics[width=0.52\textwidth]{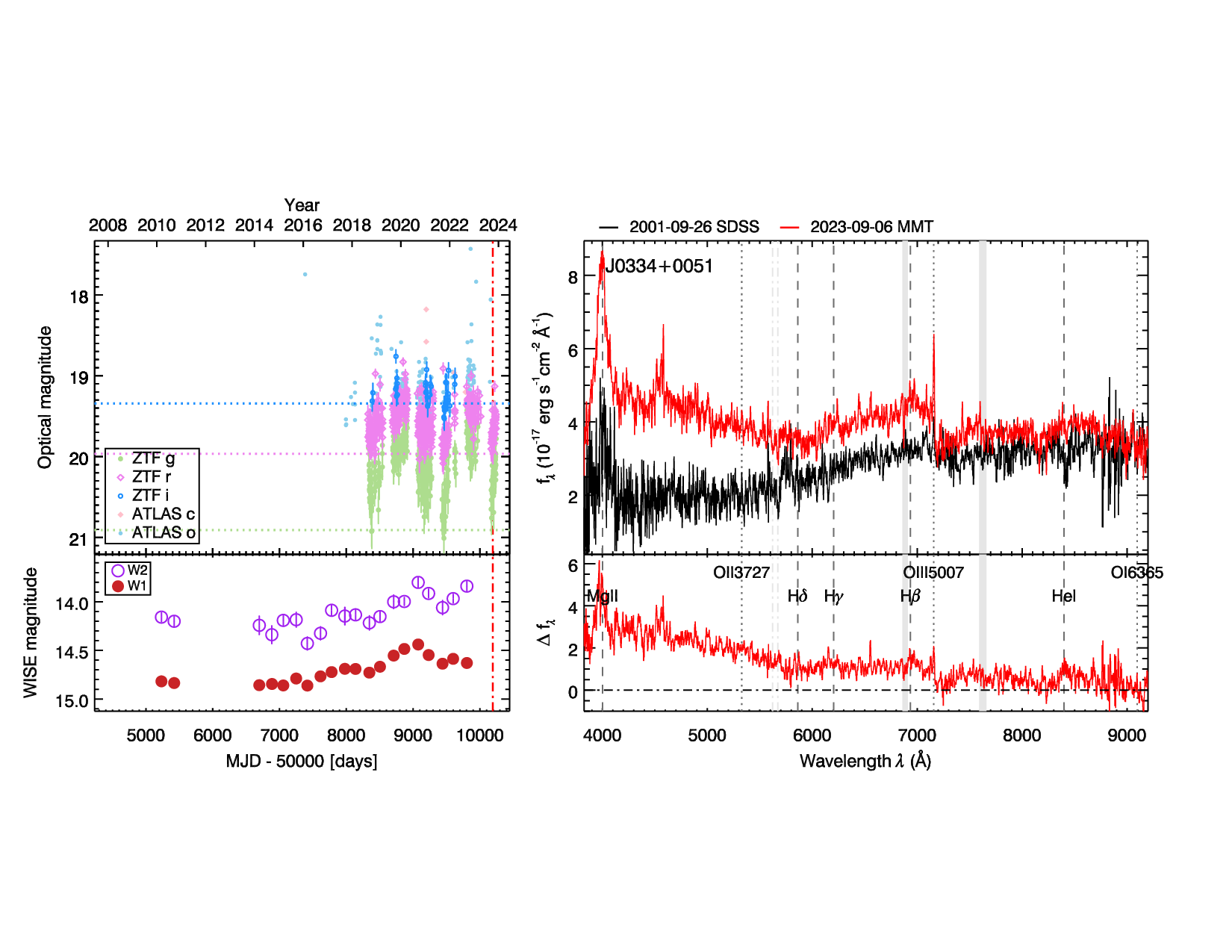}\\
\vspace{-2.7cm}
\hspace{-0.4cm}
\includegraphics[width=0.52\textwidth]{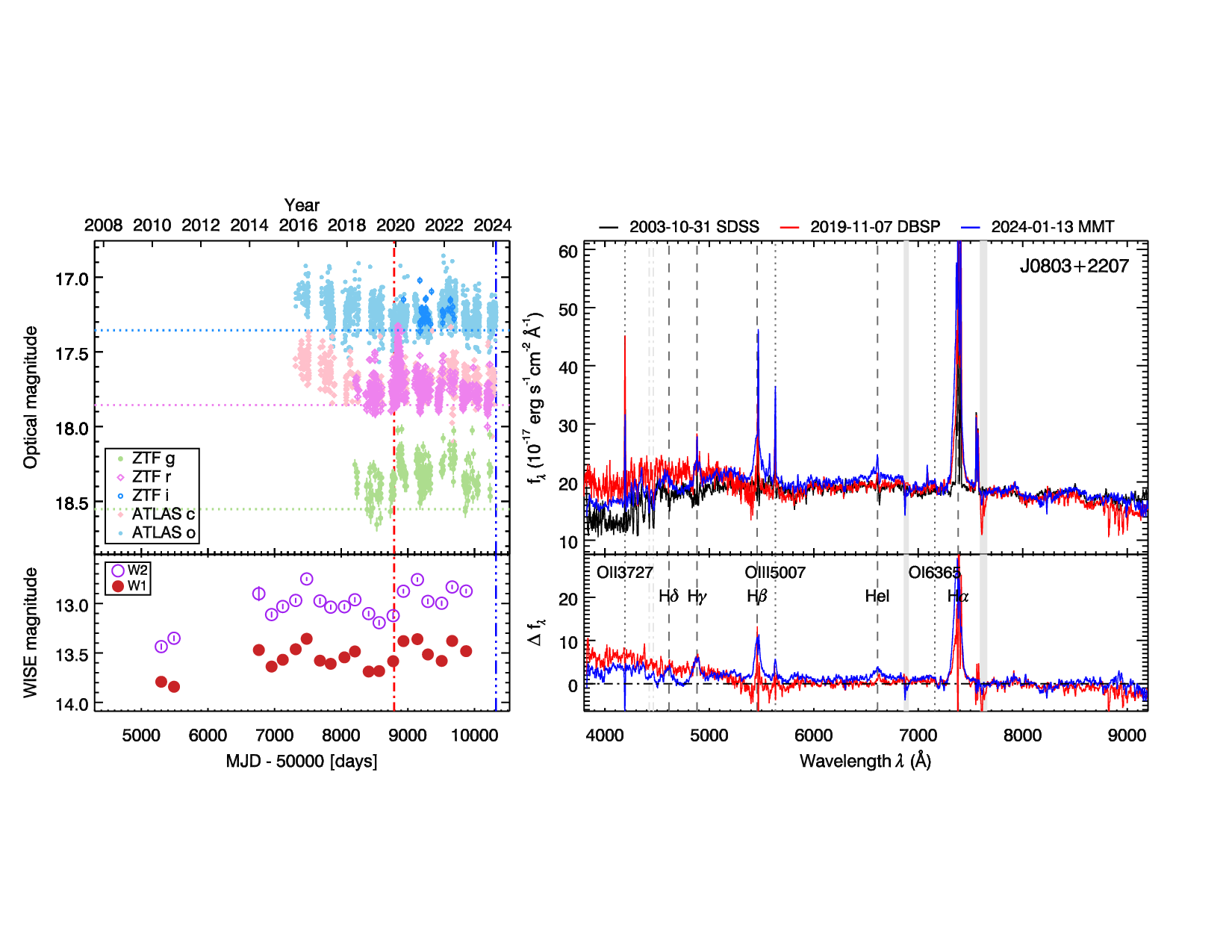}
\hspace{-0.7cm}
\includegraphics[width=0.52\textwidth]{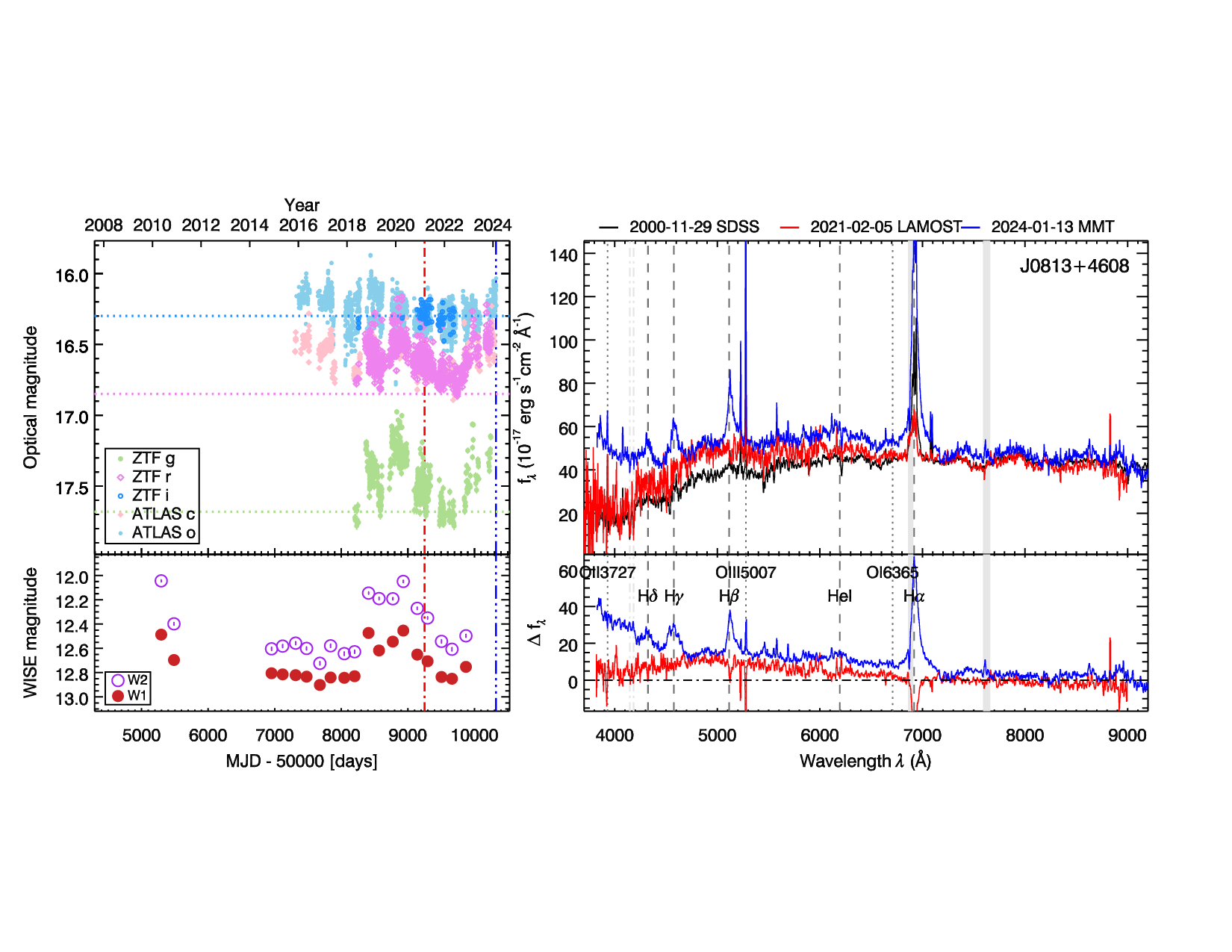}\\
\caption{Confirmed CLQs. Same as Figure \ref{fig:example}.}
\label{fig:optical_spec}
\end{figure*}

% \clearpage
\begin{figure*}[!ht]
\centering
\vspace{-1cm}
\hspace{-0.4cm}
\includegraphics[width=0.52\textwidth]{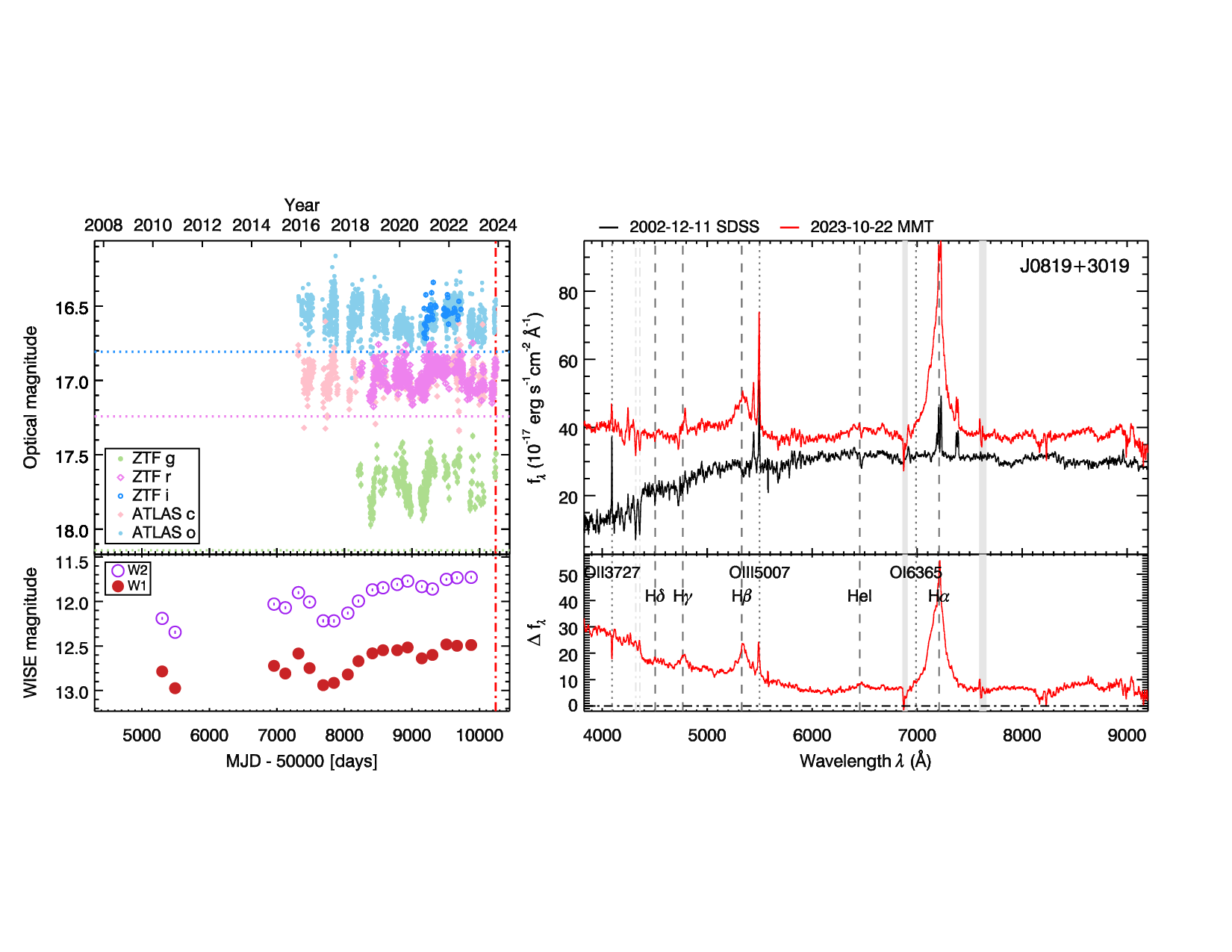}
\hspace{-0.7cm}
\includegraphics[width=0.52\textwidth]{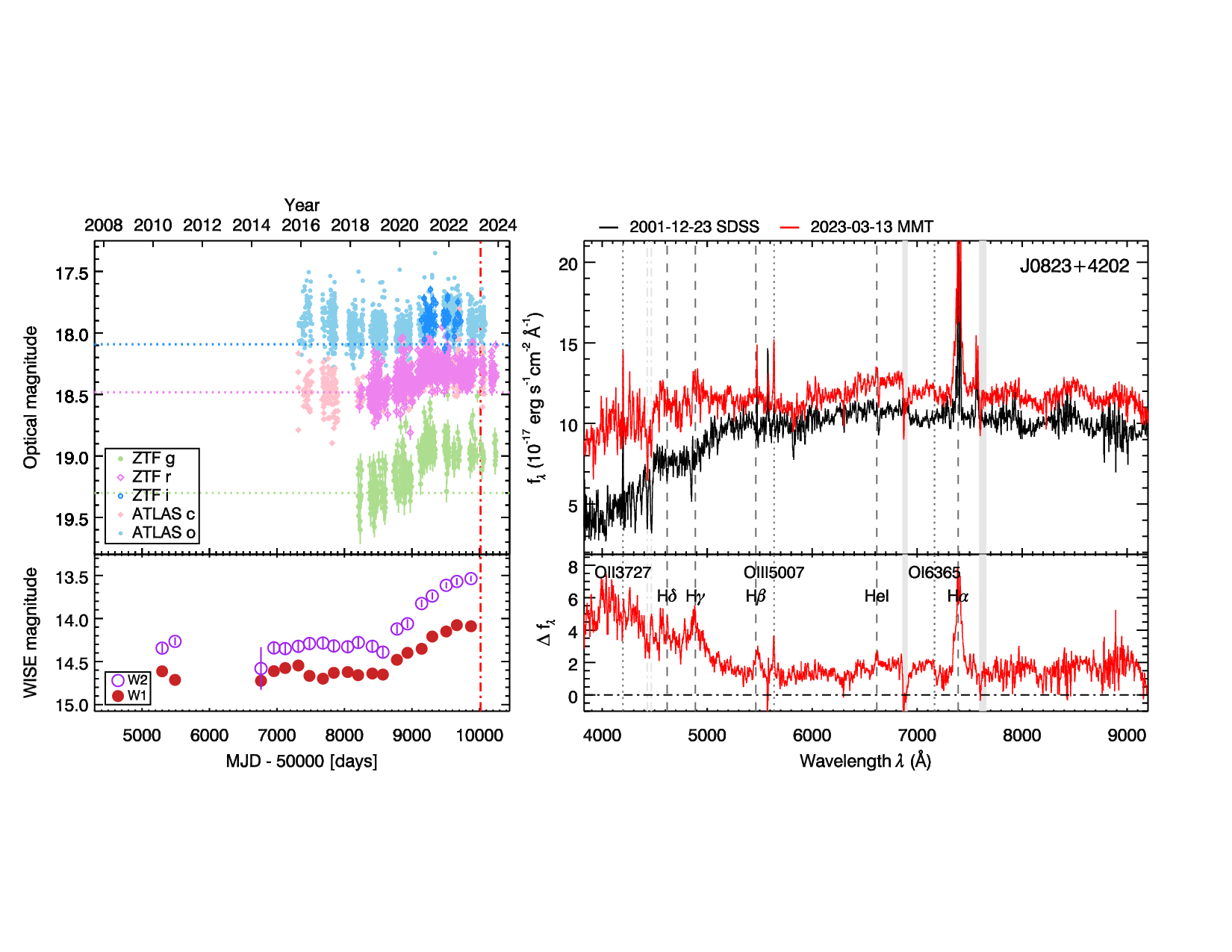}\\
\vspace{-2.7cm}
\hspace{-0.4cm}
\includegraphics[width=0.52\textwidth]{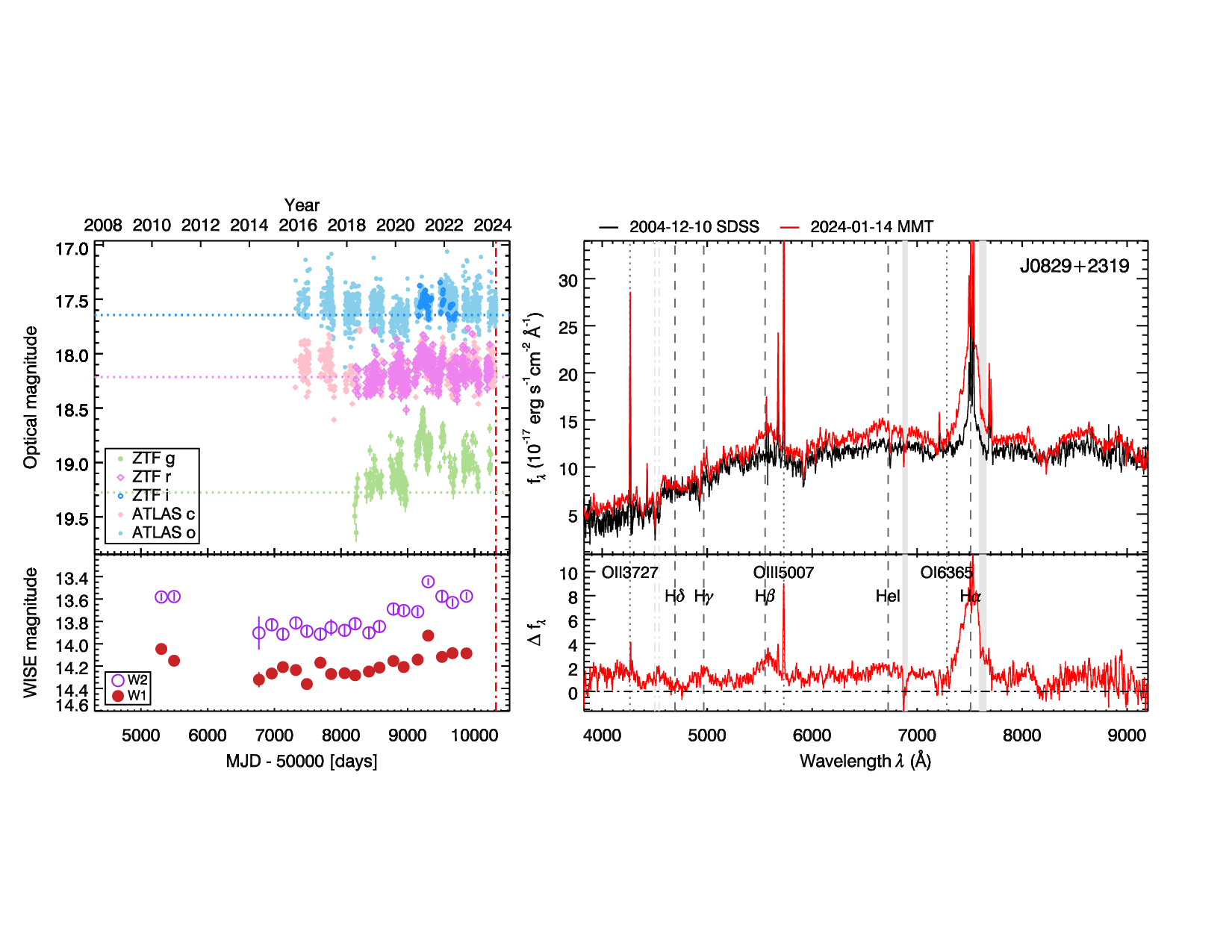}
\hspace{-0.7cm}
\includegraphics[width=0.52\textwidth]{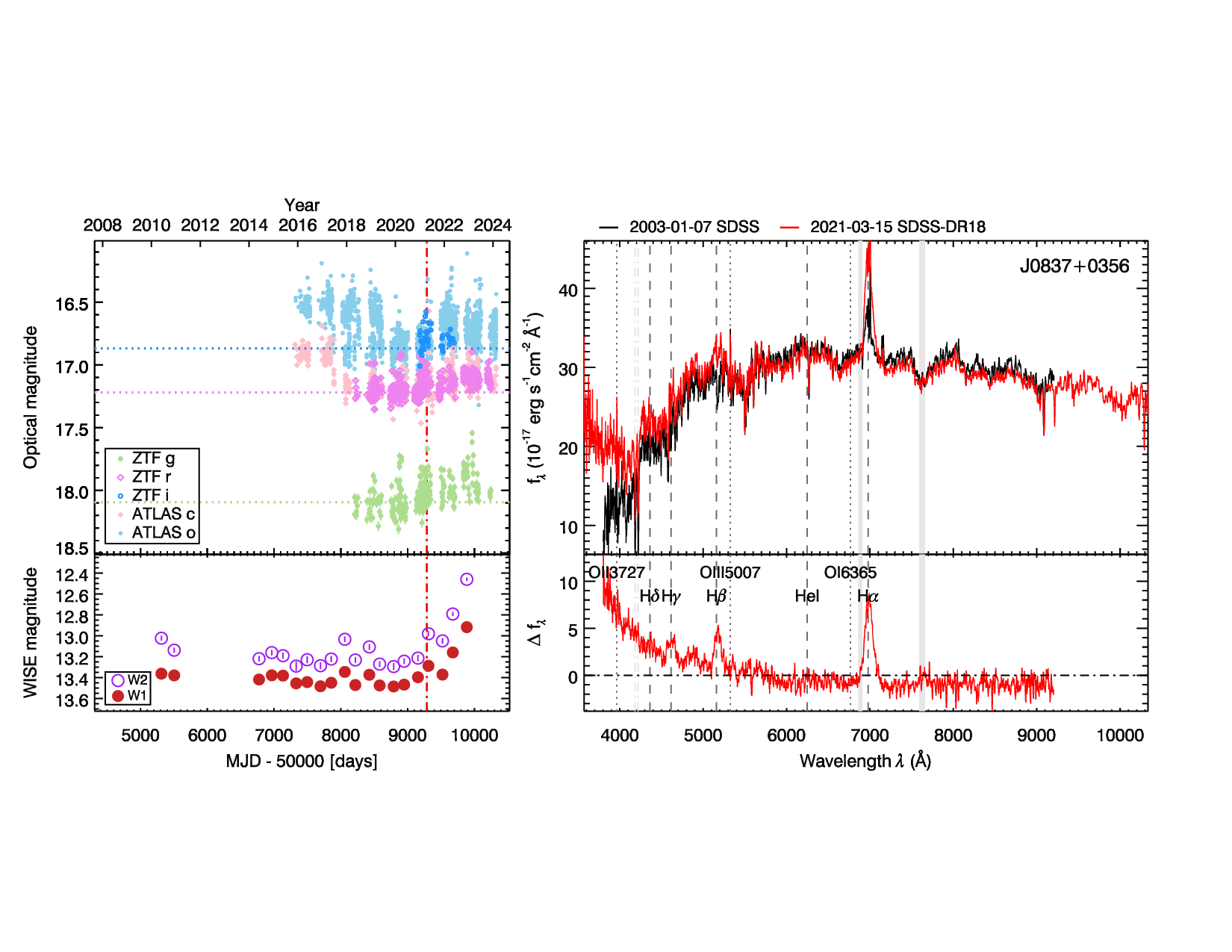}\\
\vspace{-2.7cm}
\hspace{-0.4cm}
\includegraphics[width=0.52\textwidth]{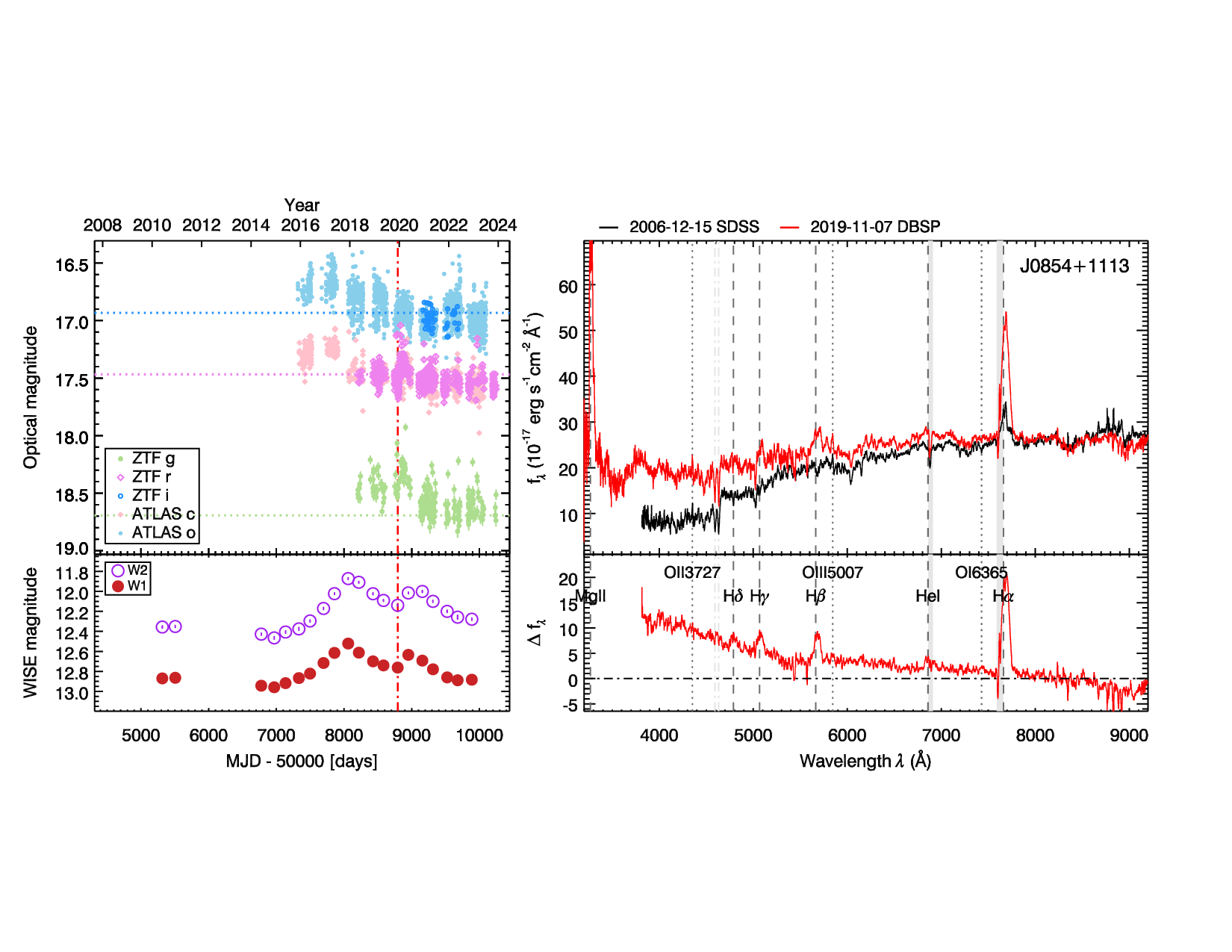}
\hspace{-0.7cm}
\includegraphics[width=0.52\textwidth]{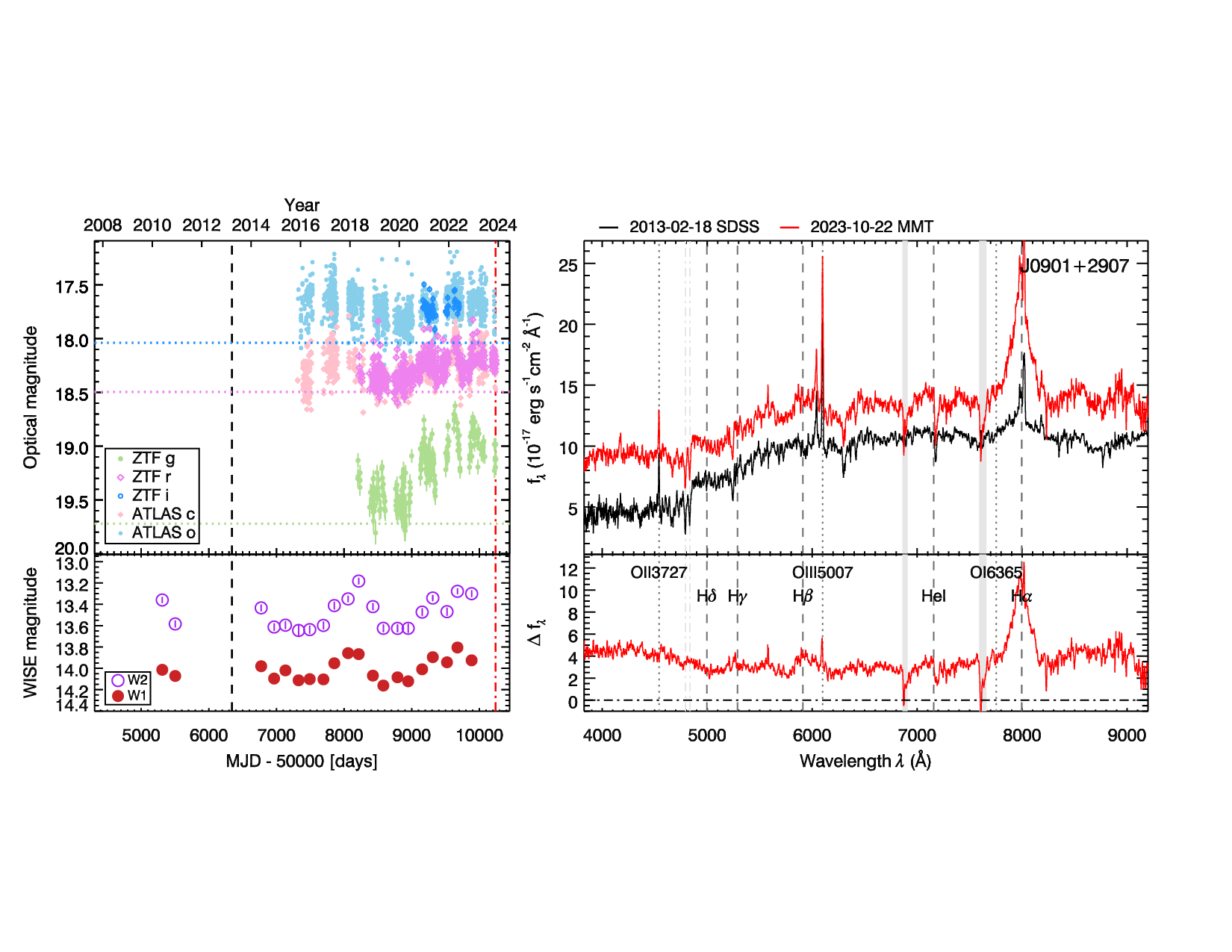}\\
\vspace{-2.7cm}
\hspace{-0.4cm}
\includegraphics[width=0.52\textwidth]{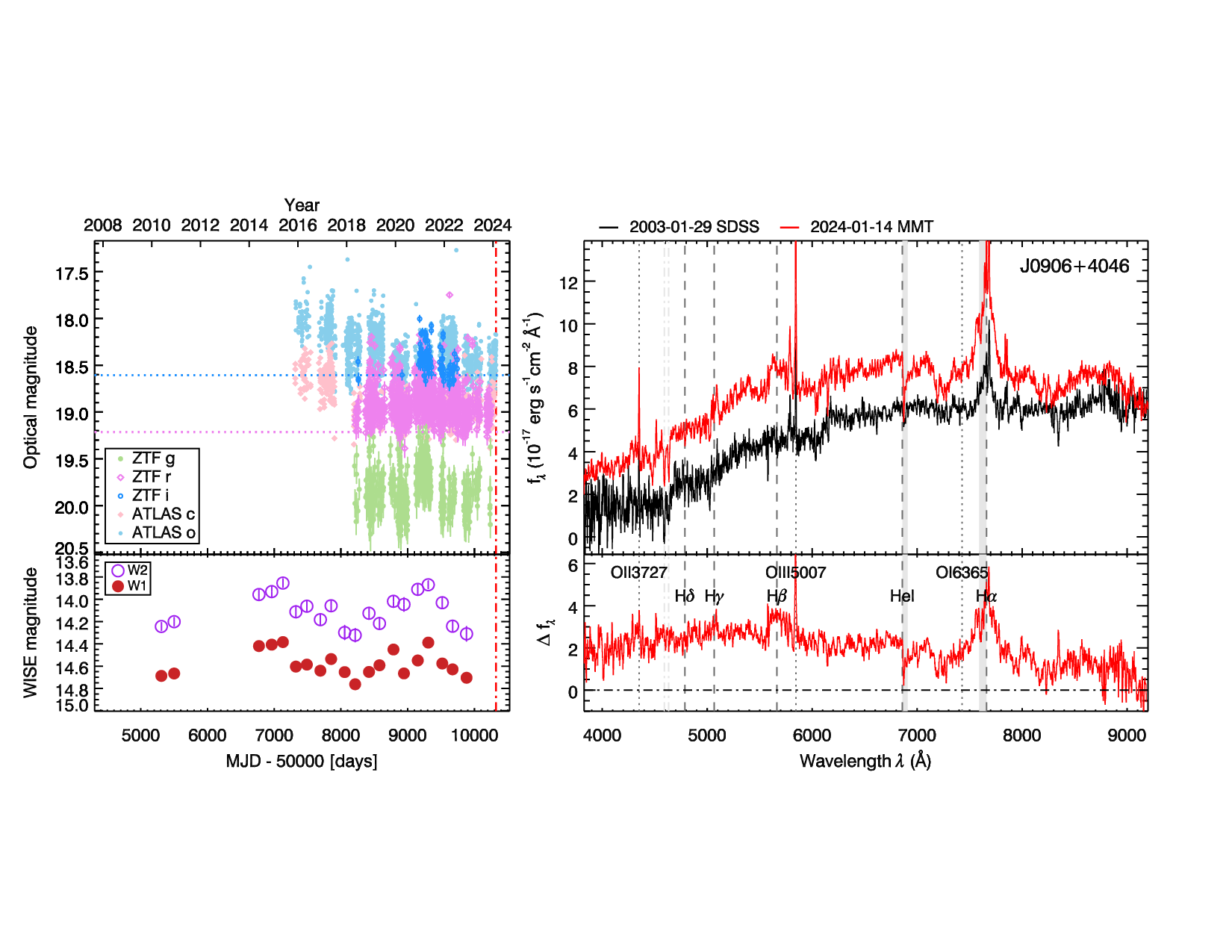}
\hspace{-0.7cm}
\includegraphics[width=0.52\textwidth]{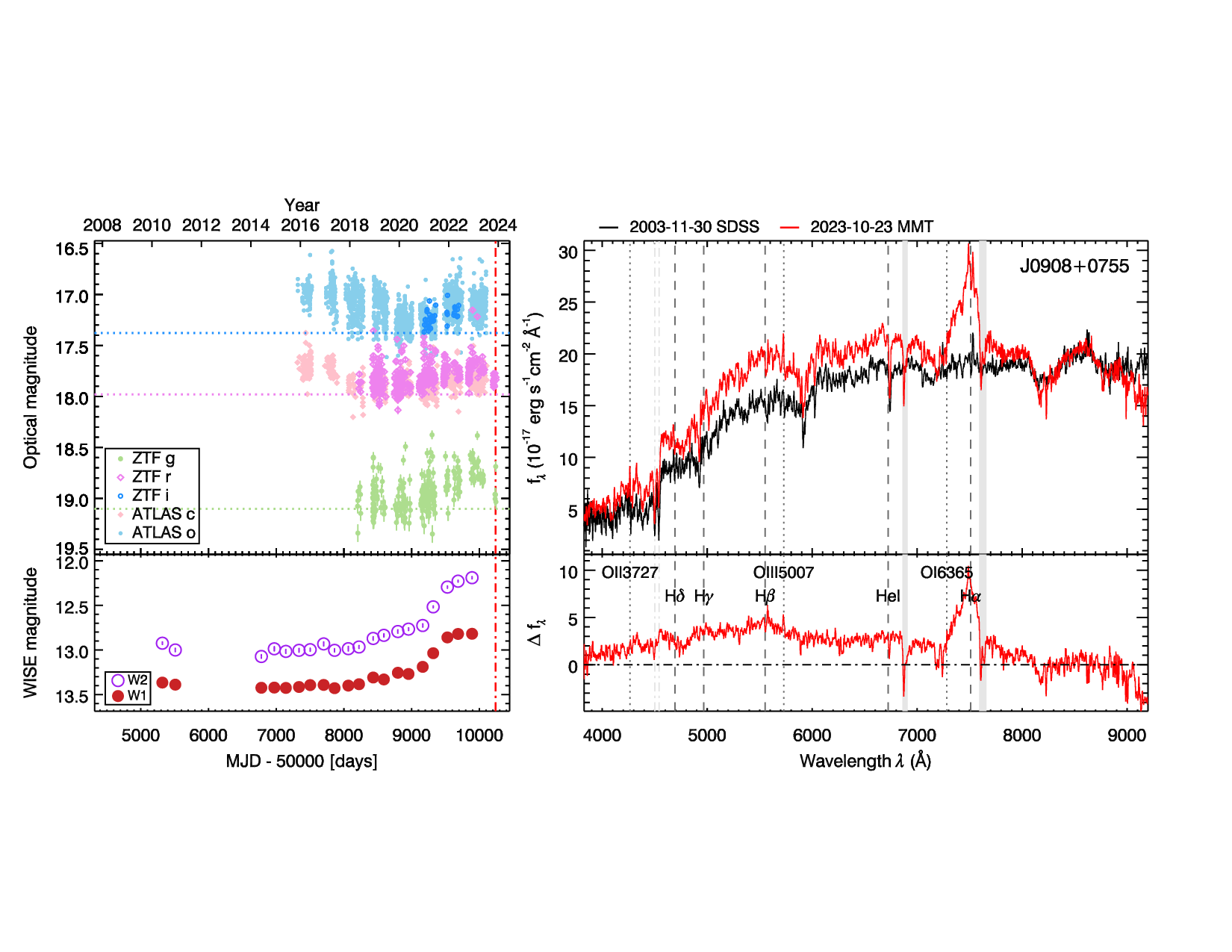}\\
\vspace{-2.7cm}
\hspace{-0.4cm}
\includegraphics[width=0.52\textwidth]{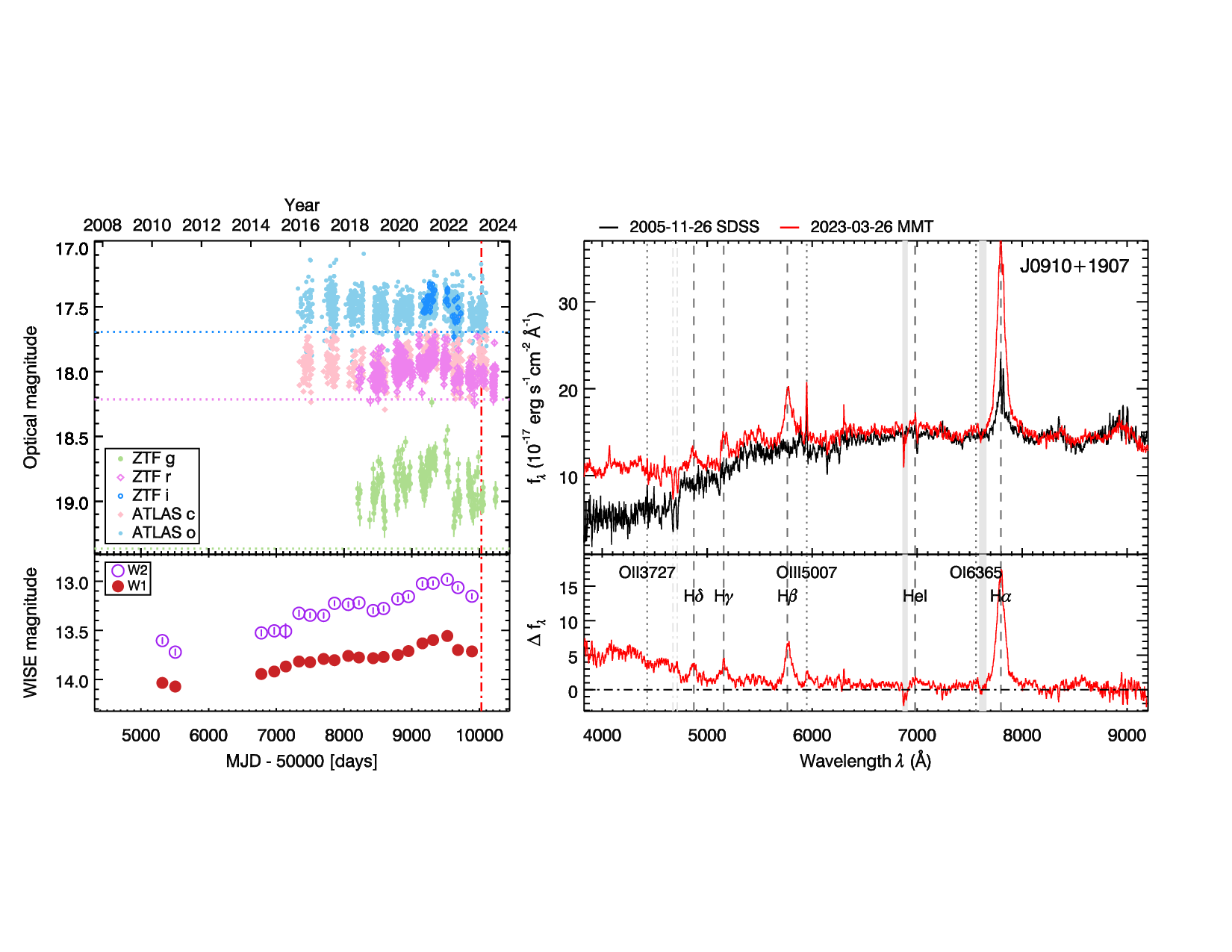}
\hspace{-0.7cm}
\includegraphics[width=0.52\textwidth]{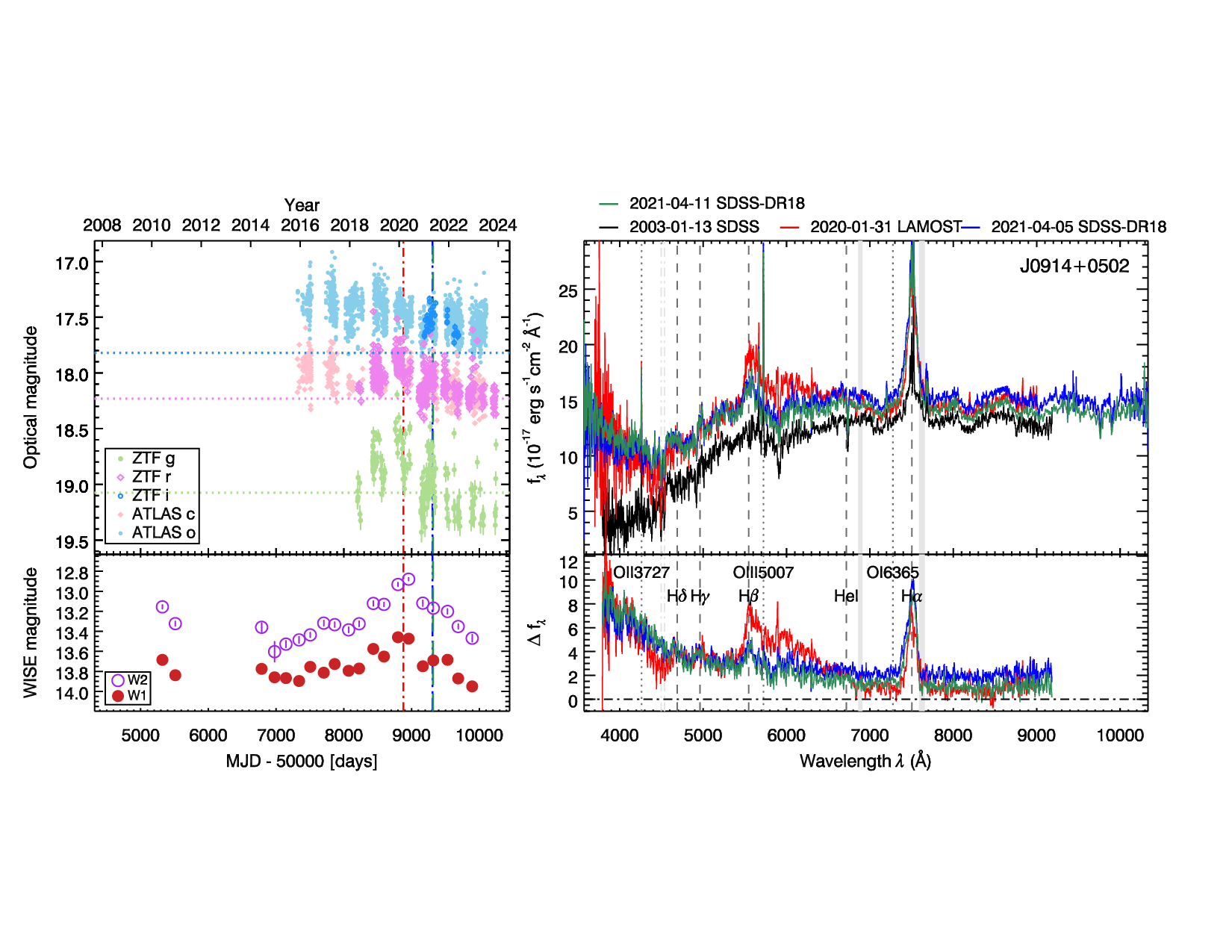}\\
\caption{Confirmed CLQs. Same as Figure \ref{fig:example}.}
\label{fig:optical_spec}
\end{figure*}

% \clearpage
\begin{figure*}[!ht]
\centering
\vspace{-1cm}
\hspace{-0.4cm}
\includegraphics[width=0.52\textwidth]{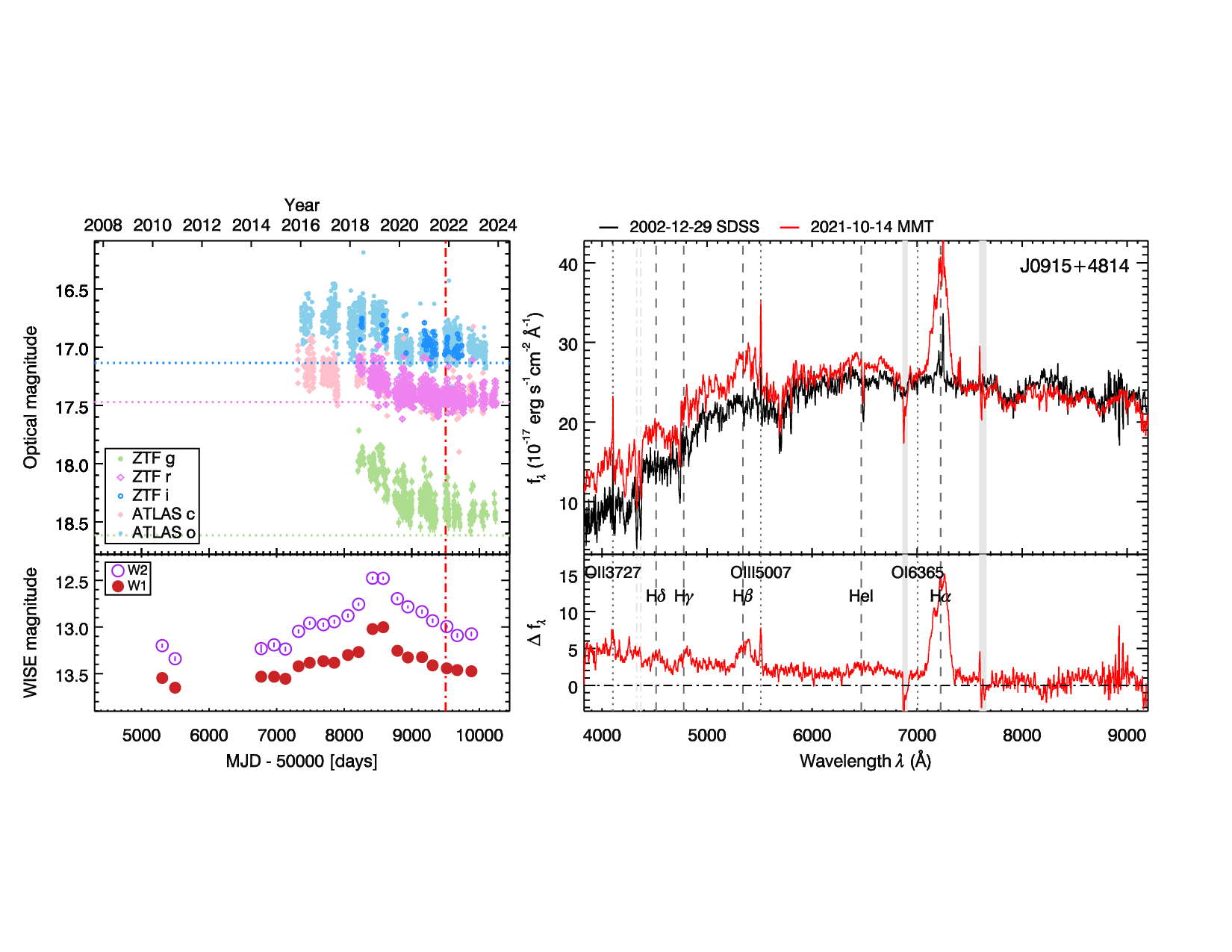}
\hspace{-0.7cm}
\includegraphics[width=0.52\textwidth]{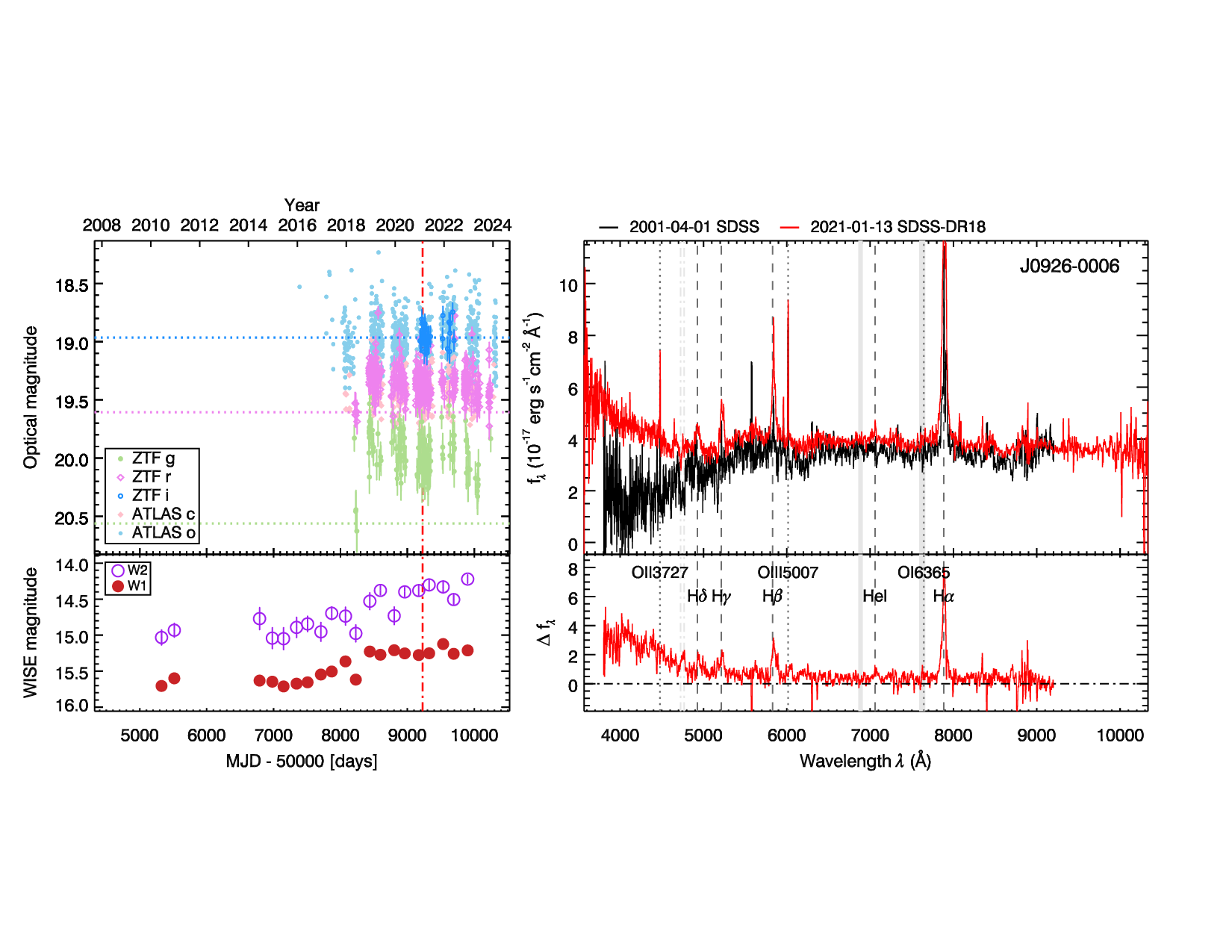}\\
\vspace{-2.7cm}
\hspace{-0.4cm}
\includegraphics[width=0.52\textwidth]{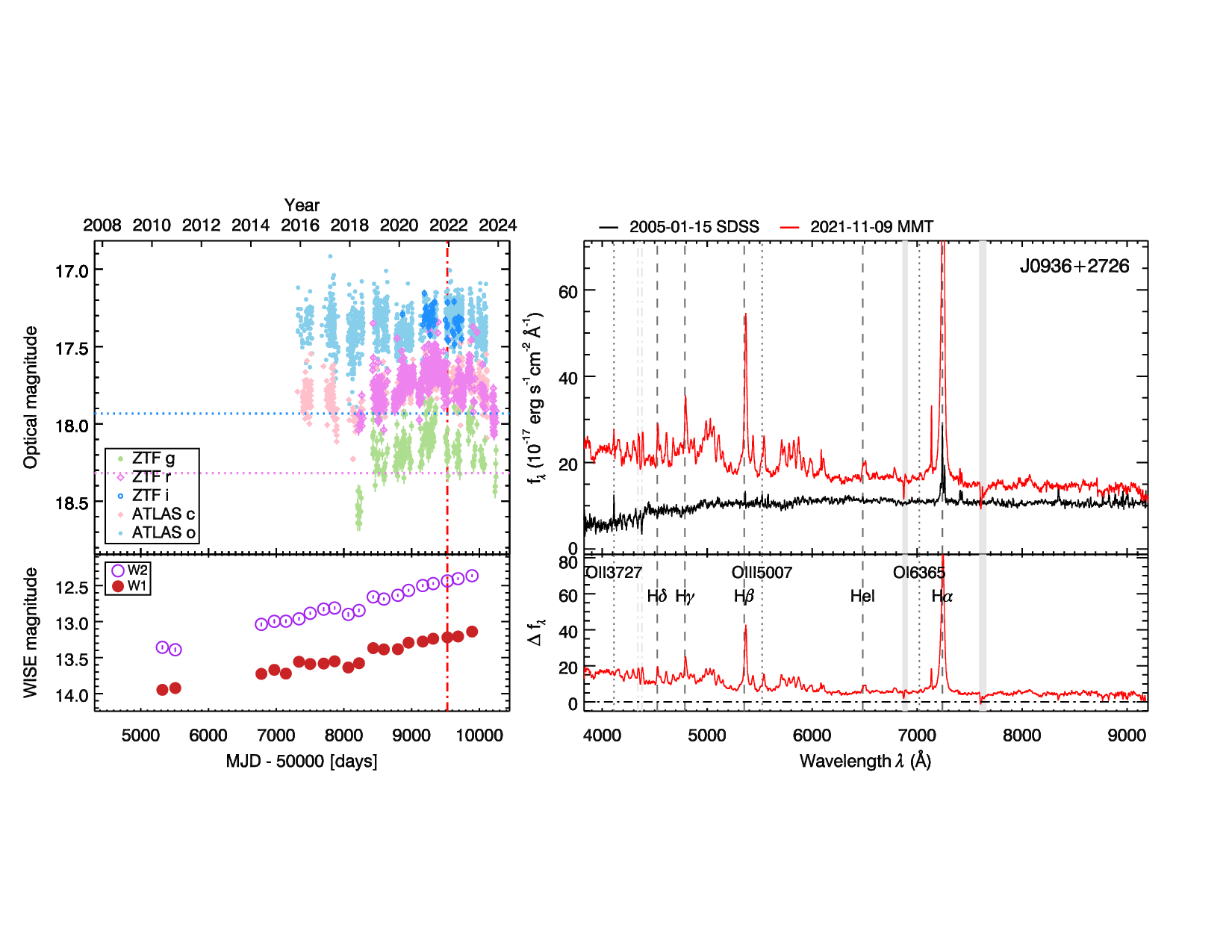}
\hspace{-0.7cm}
\includegraphics[width=0.52\textwidth]{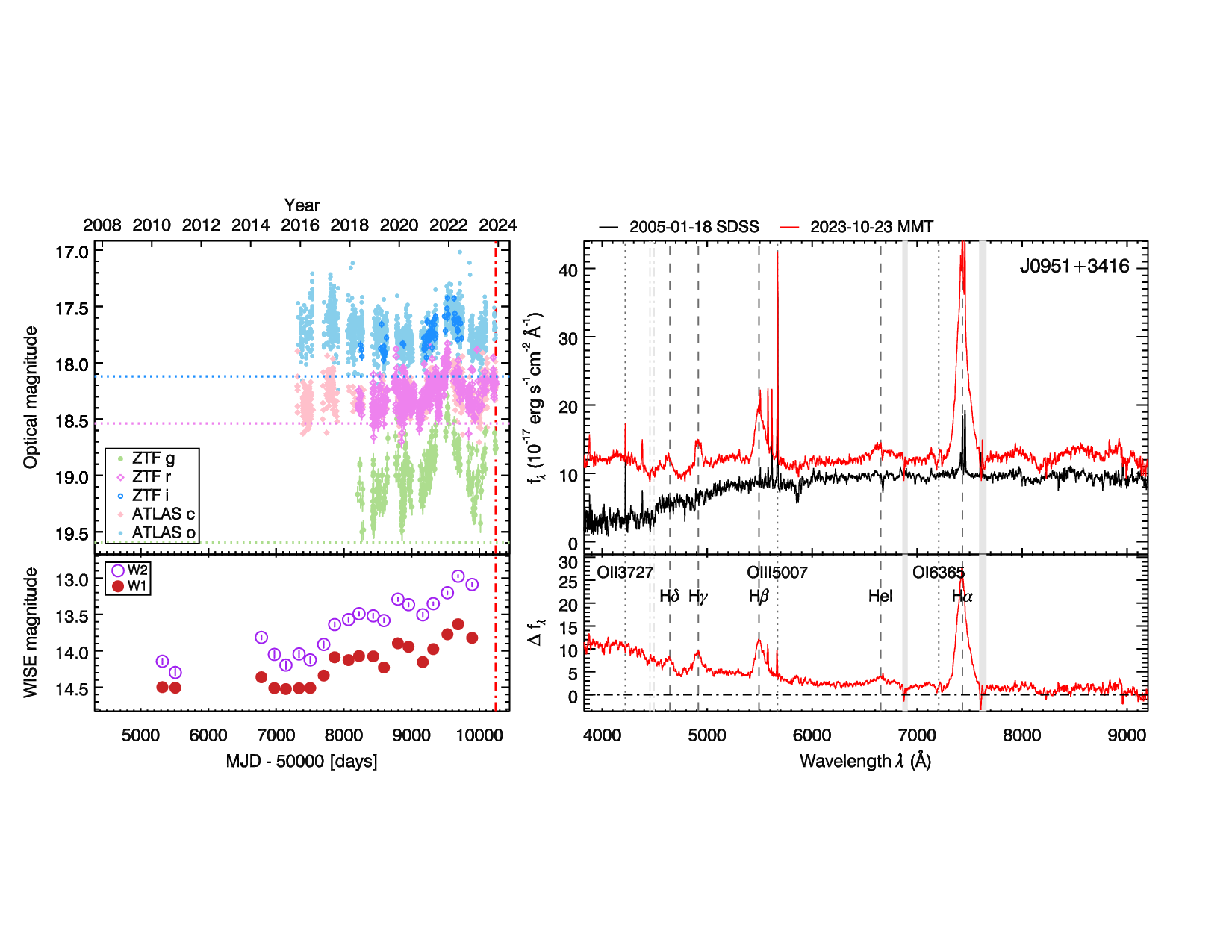}\\
\vspace{-2.7cm}
\hspace{-0.4cm}
\includegraphics[width=0.52\textwidth]{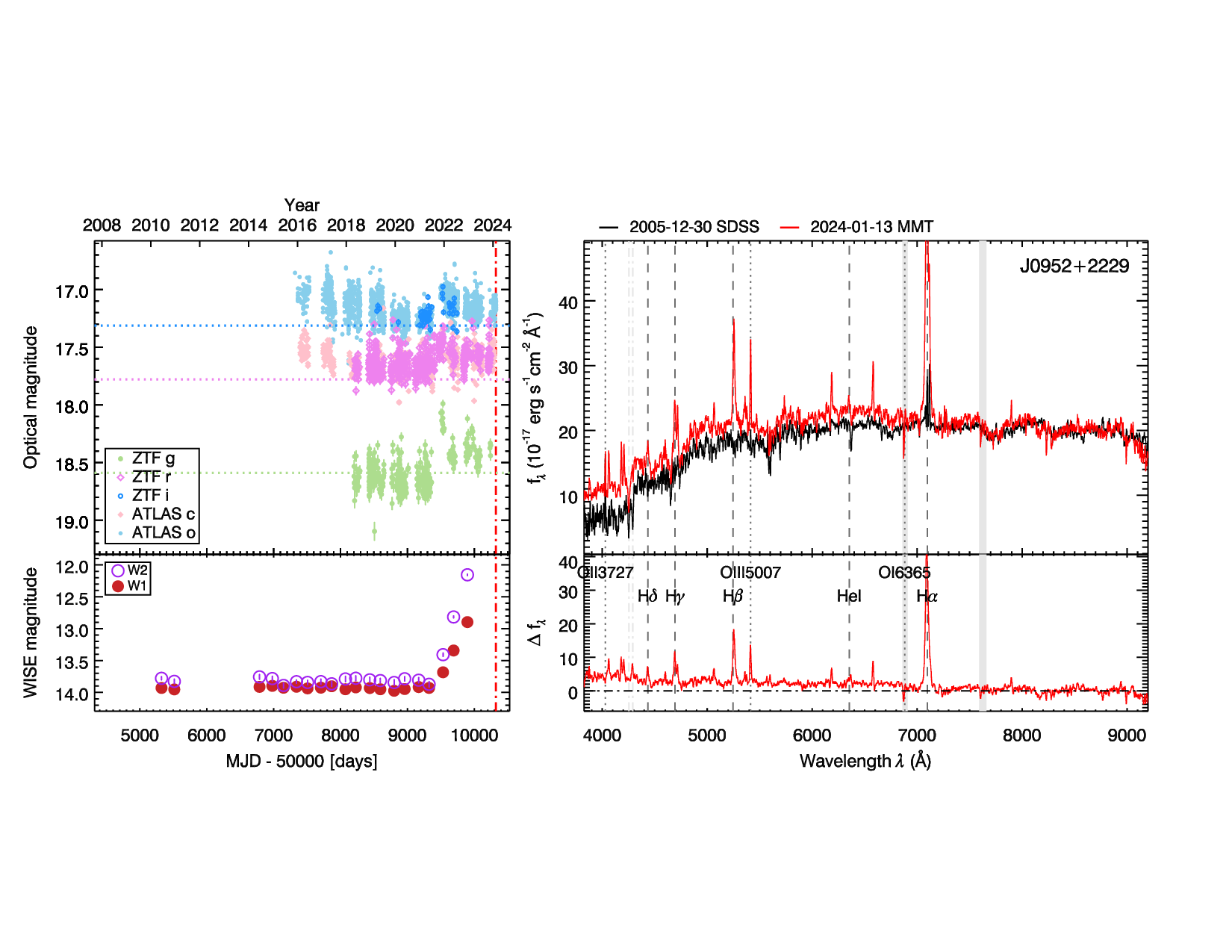}
\hspace{-0.7cm}
\includegraphics[width=0.52\textwidth]{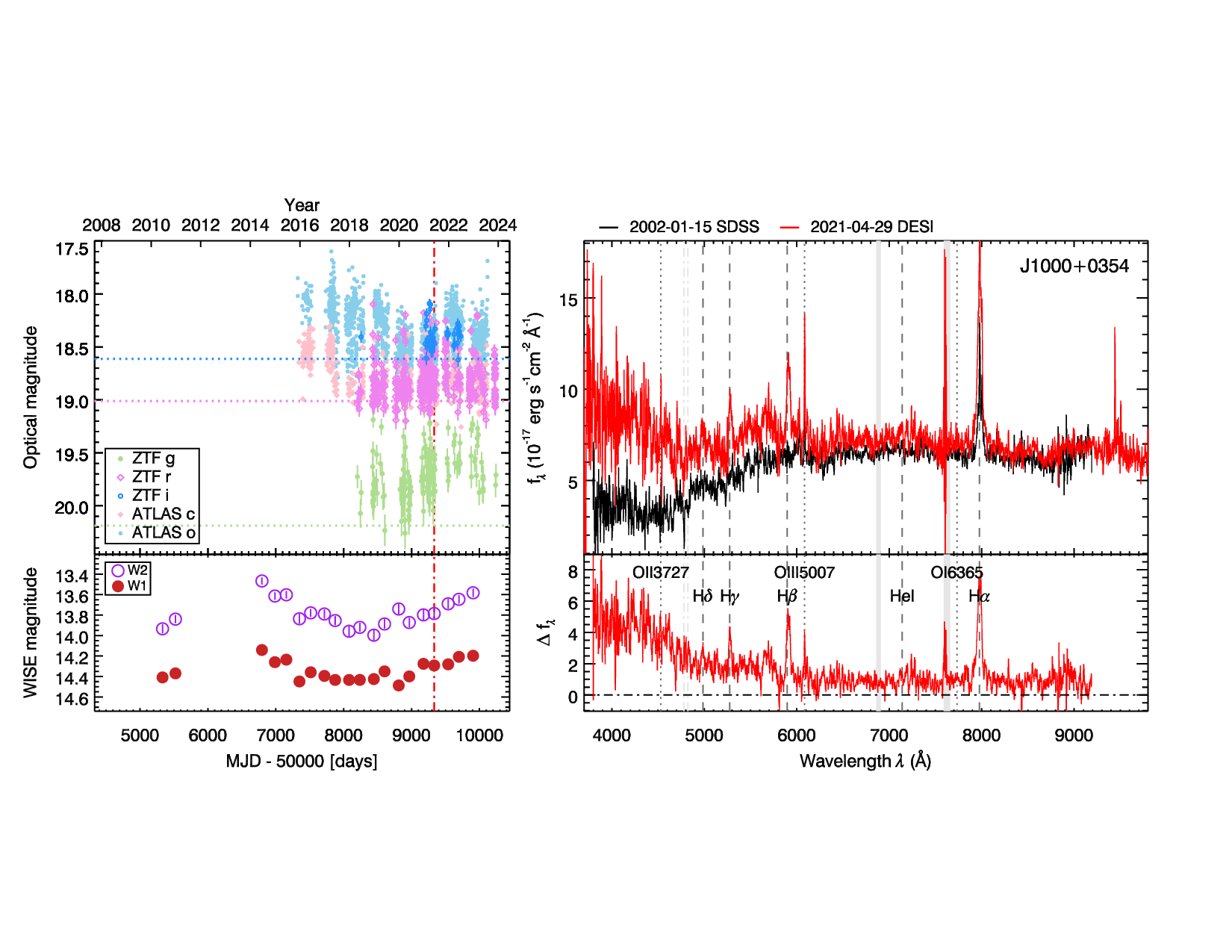}\\
\vspace{-2.7cm}
\hspace{-0.4cm}
\includegraphics[width=0.52\textwidth]{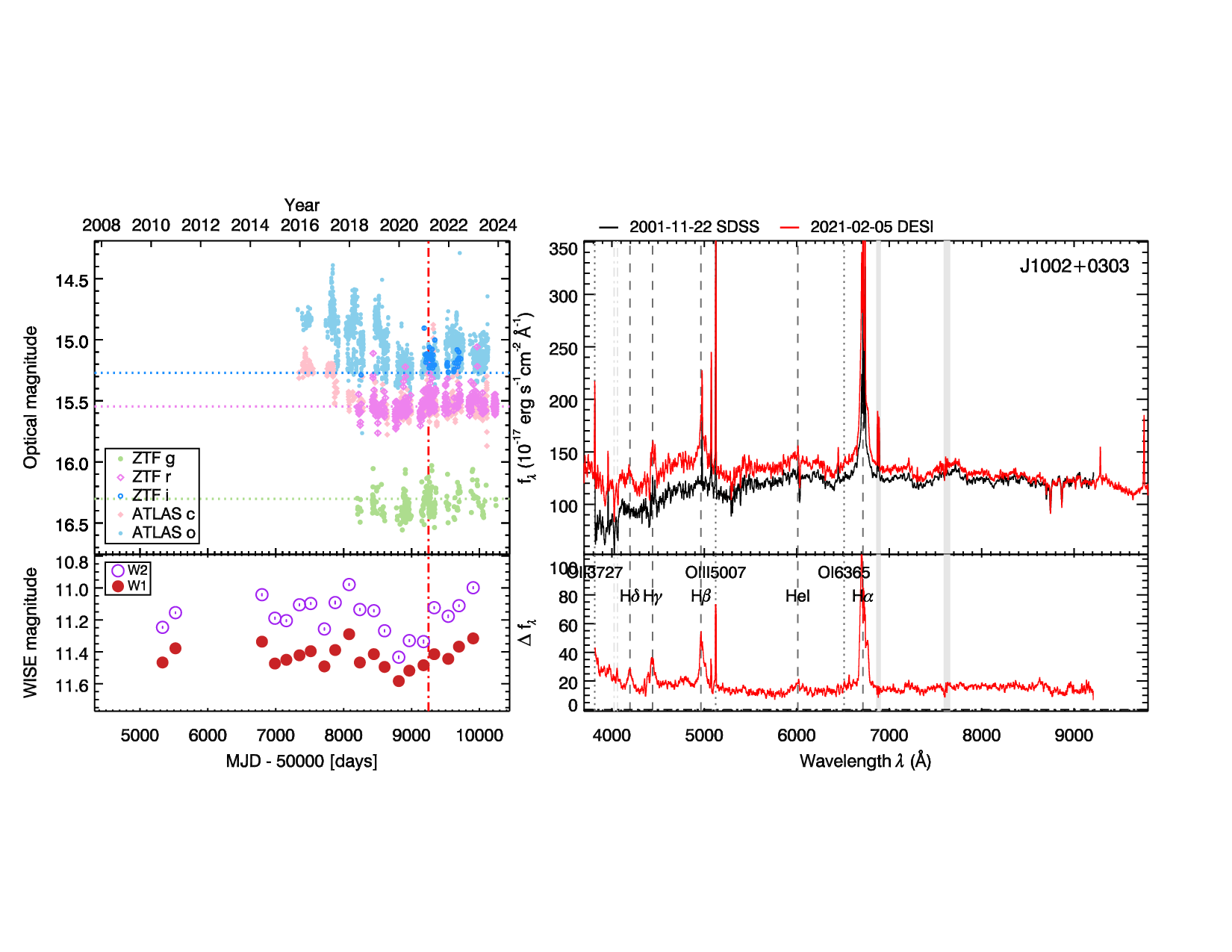}
\hspace{-0.7cm}
\includegraphics[width=0.52\textwidth]{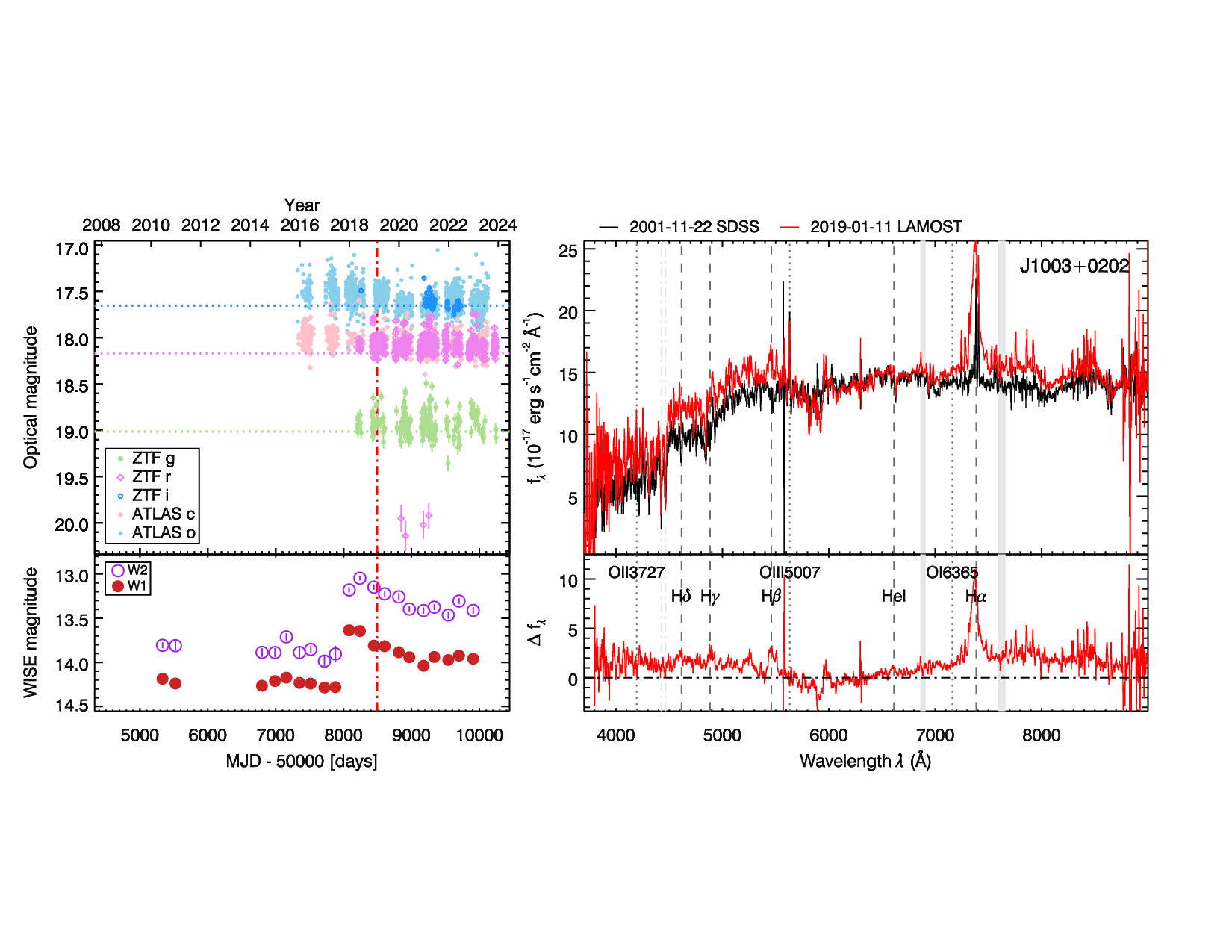}\\
\vspace{-2.7cm}
\hspace{-0.4cm}
\includegraphics[width=0.52\textwidth]{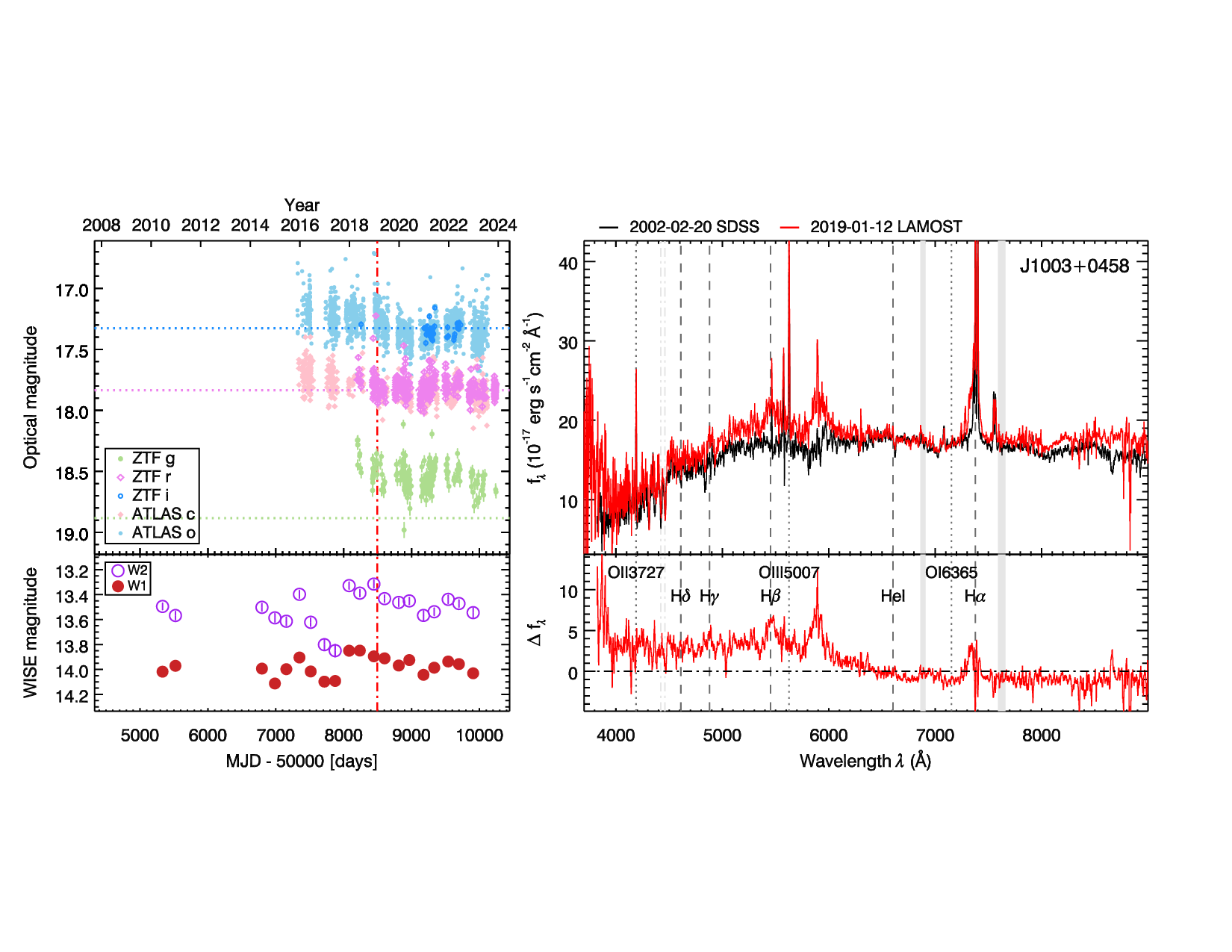}
\hspace{-0.7cm}
\includegraphics[width=0.52\textwidth]{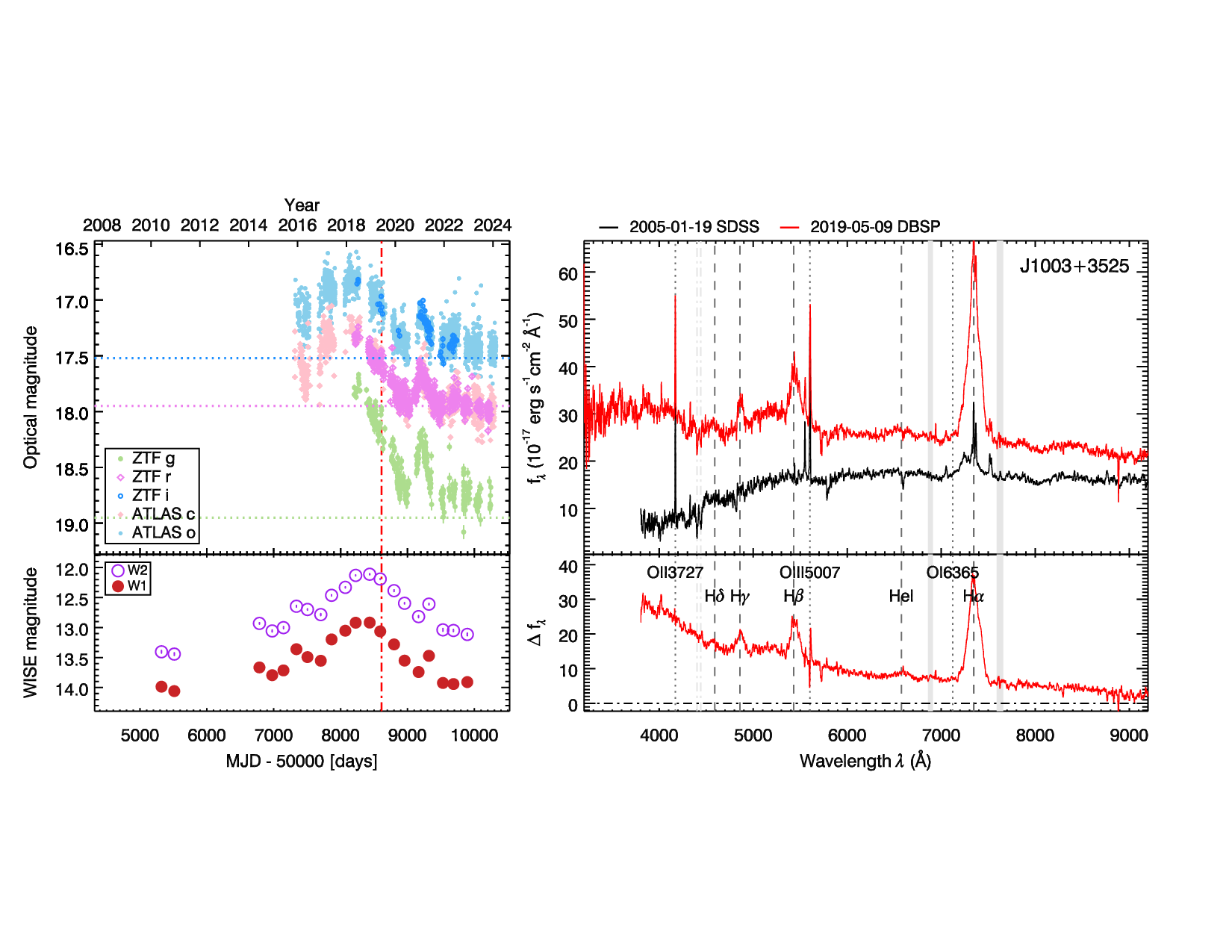}\\
\caption{Confirmed CLQs. Same as Figure \ref{fig:example}.}
\label{fig:optical_spec}
\end{figure*}

% \clearpage
\begin{figure*}[!ht]
\centering
\vspace{-1cm}
\hspace{-0.4cm}
\includegraphics[width=0.52\textwidth]{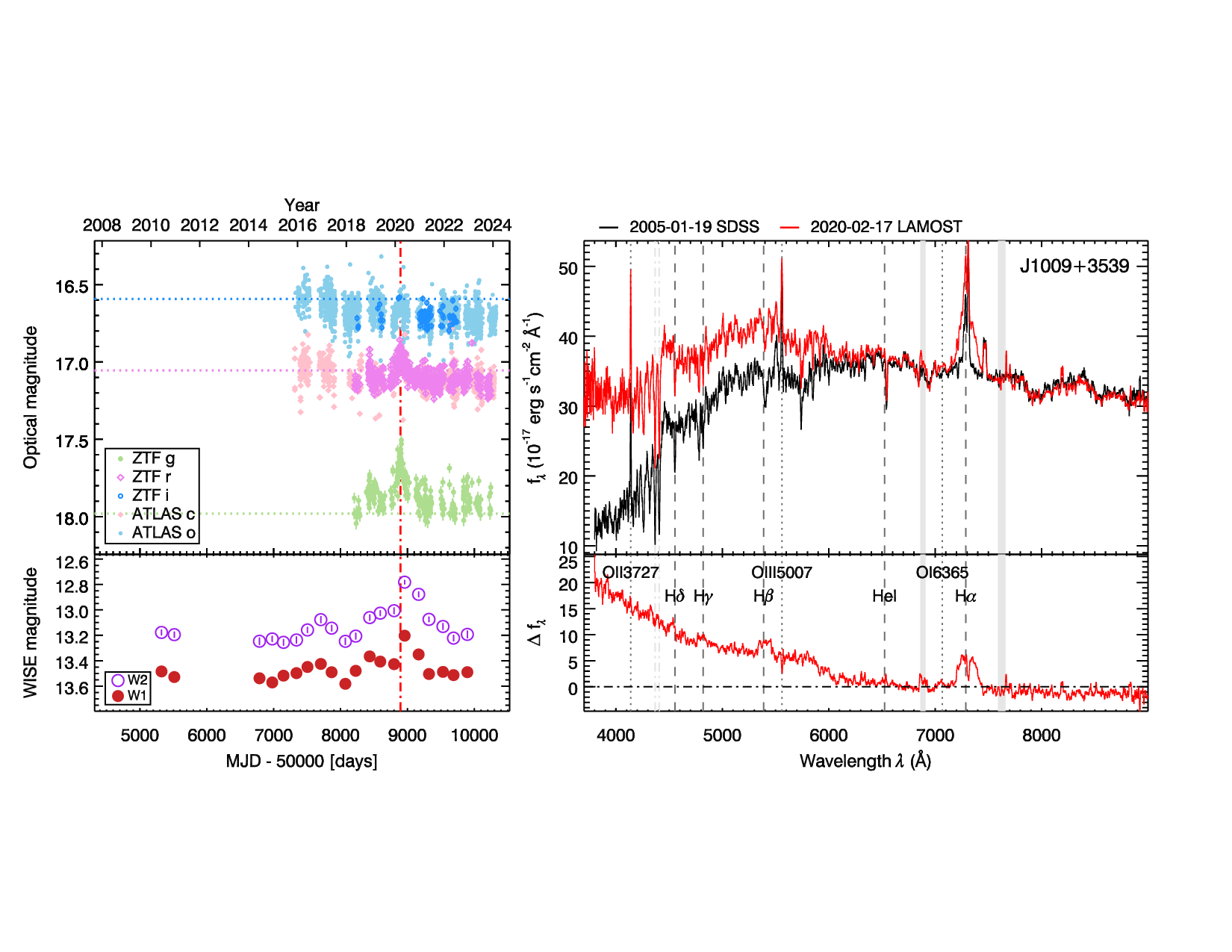}
\hspace{-0.7cm}
\includegraphics[width=0.52\textwidth]{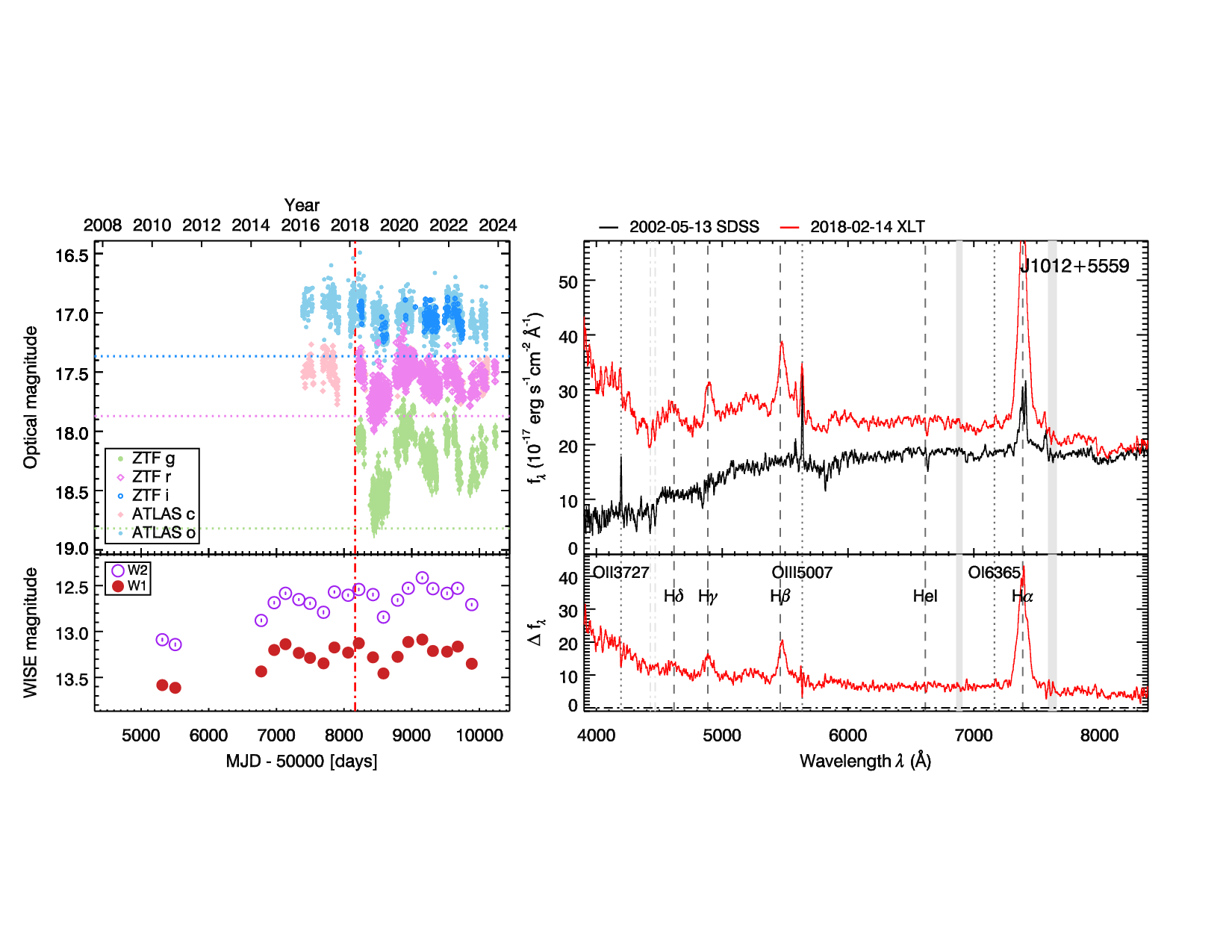}\\
\vspace{-2.7cm}
\hspace{-0.4cm}
\includegraphics[width=0.52\textwidth]{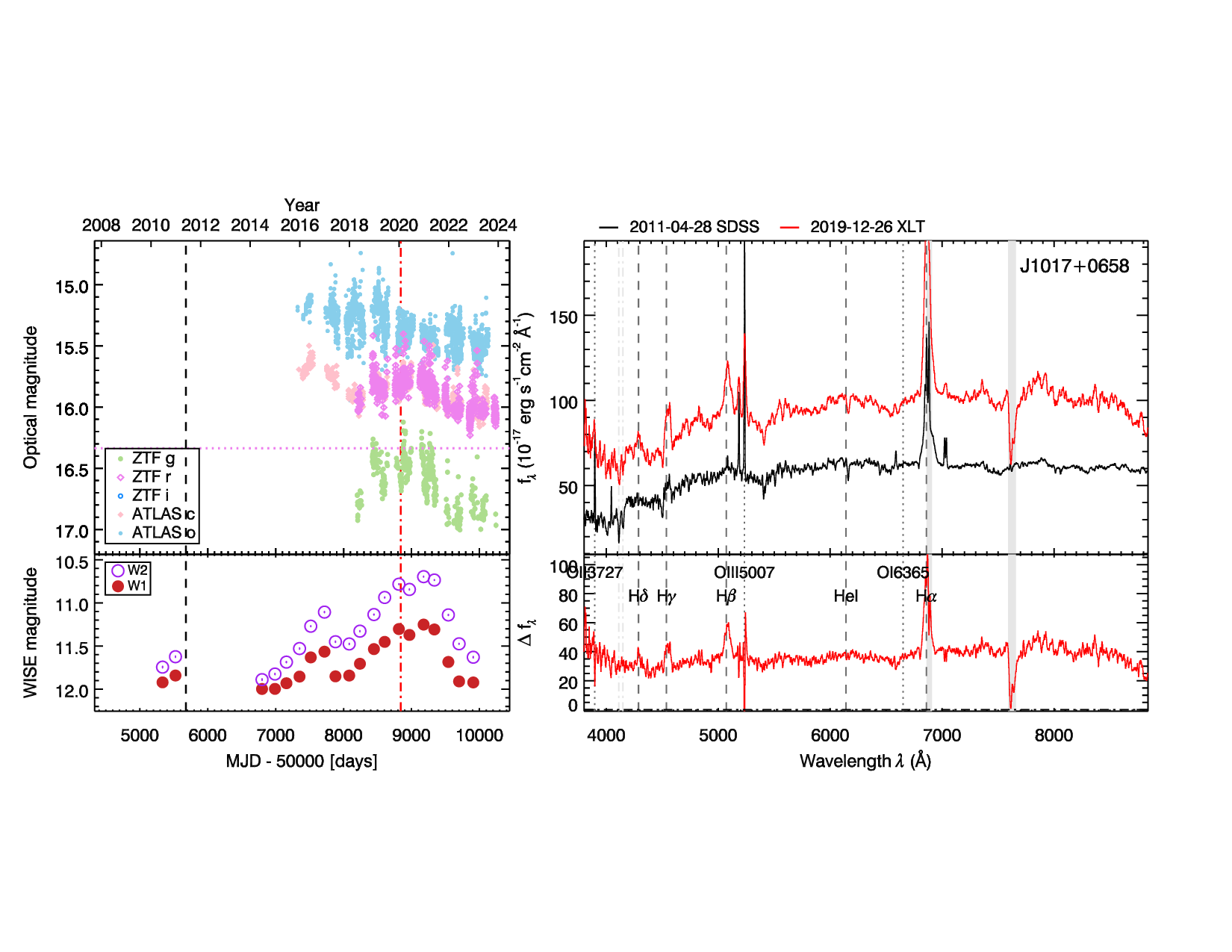}
\hspace{-0.7cm}
\includegraphics[width=0.52\textwidth]{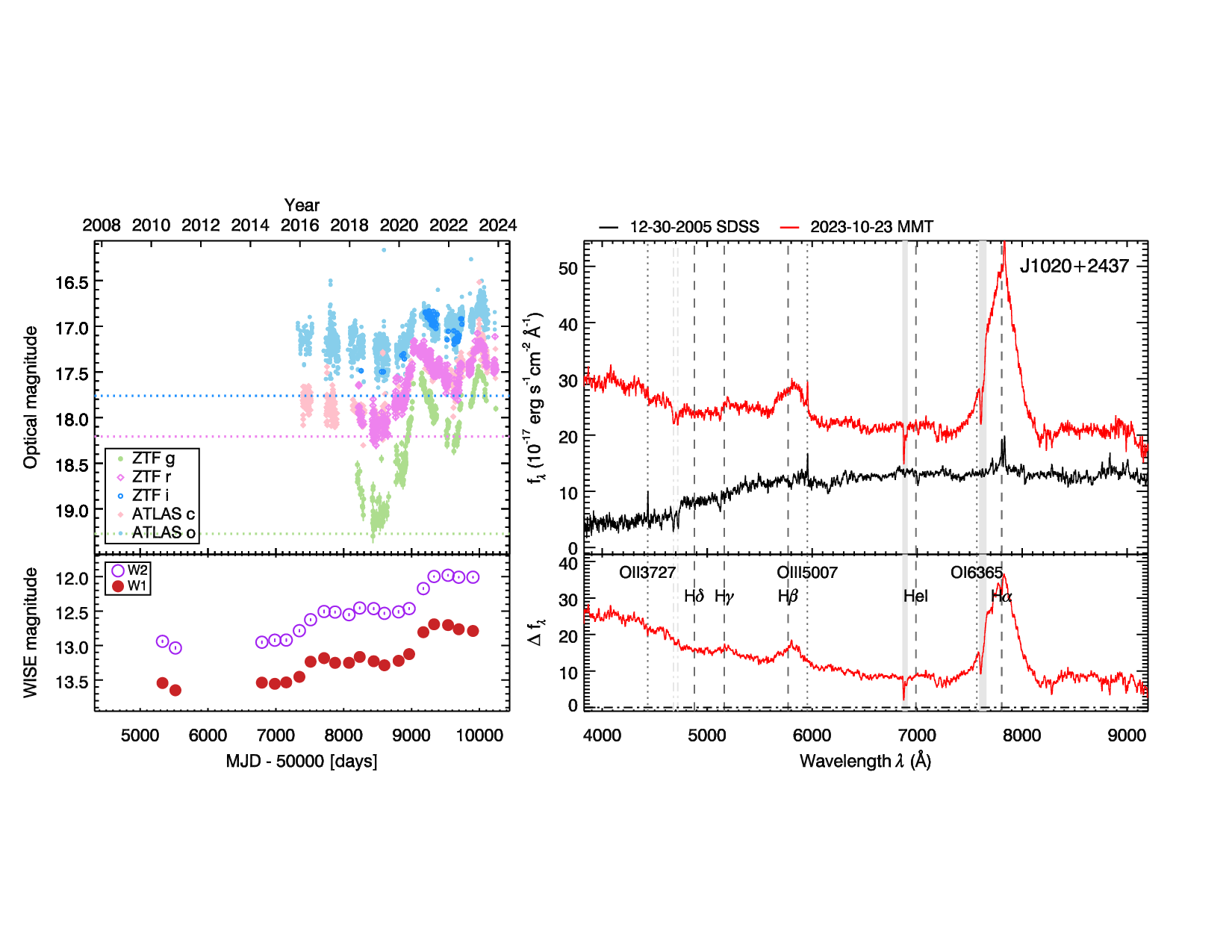}\\
\vspace{-2.7cm}
\hspace{-0.4cm}
\includegraphics[width=0.52\textwidth]{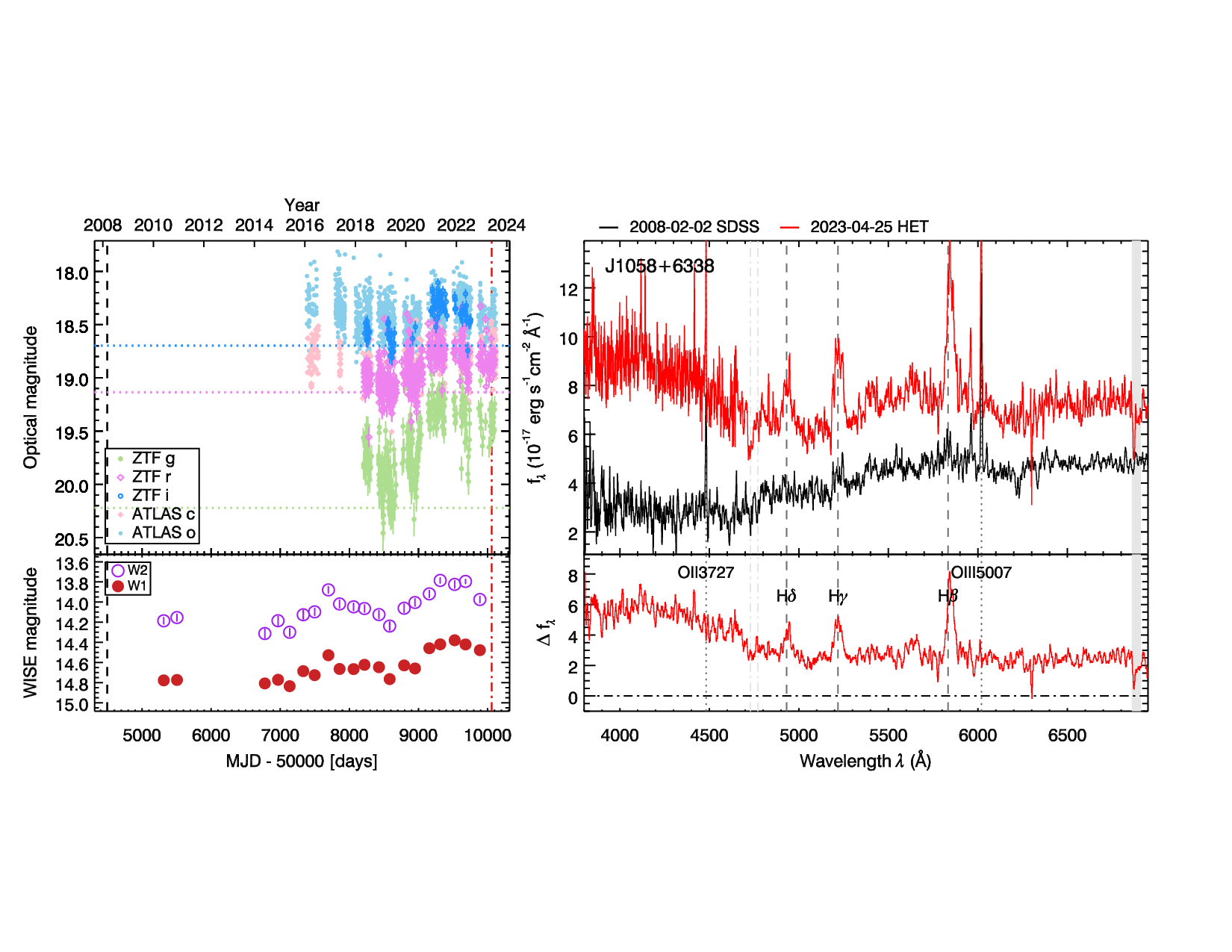}
\hspace{-0.7cm}
\includegraphics[width=0.52\textwidth]{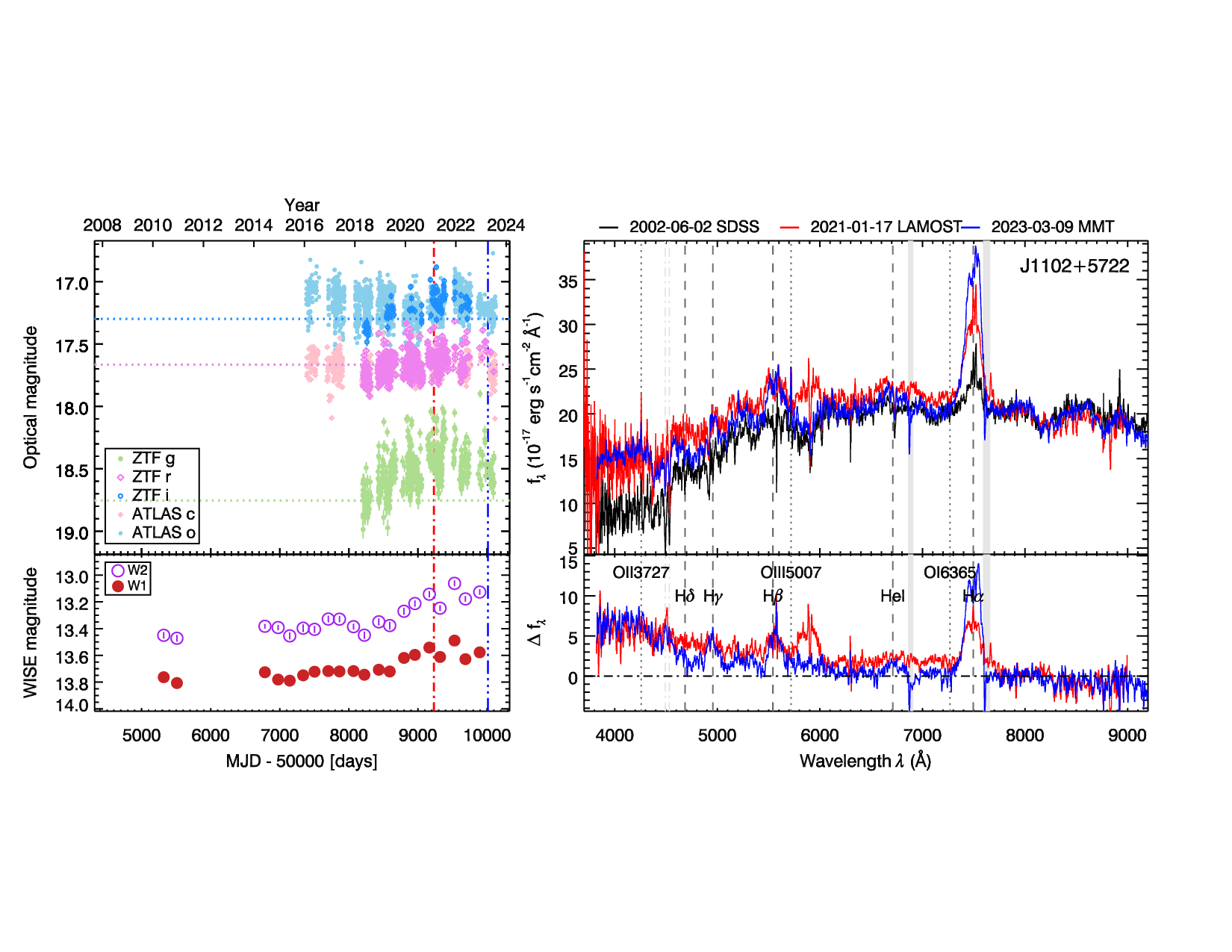}\\
\vspace{-2.7cm}
\hspace{-0.4cm}
\includegraphics[width=0.52\textwidth]{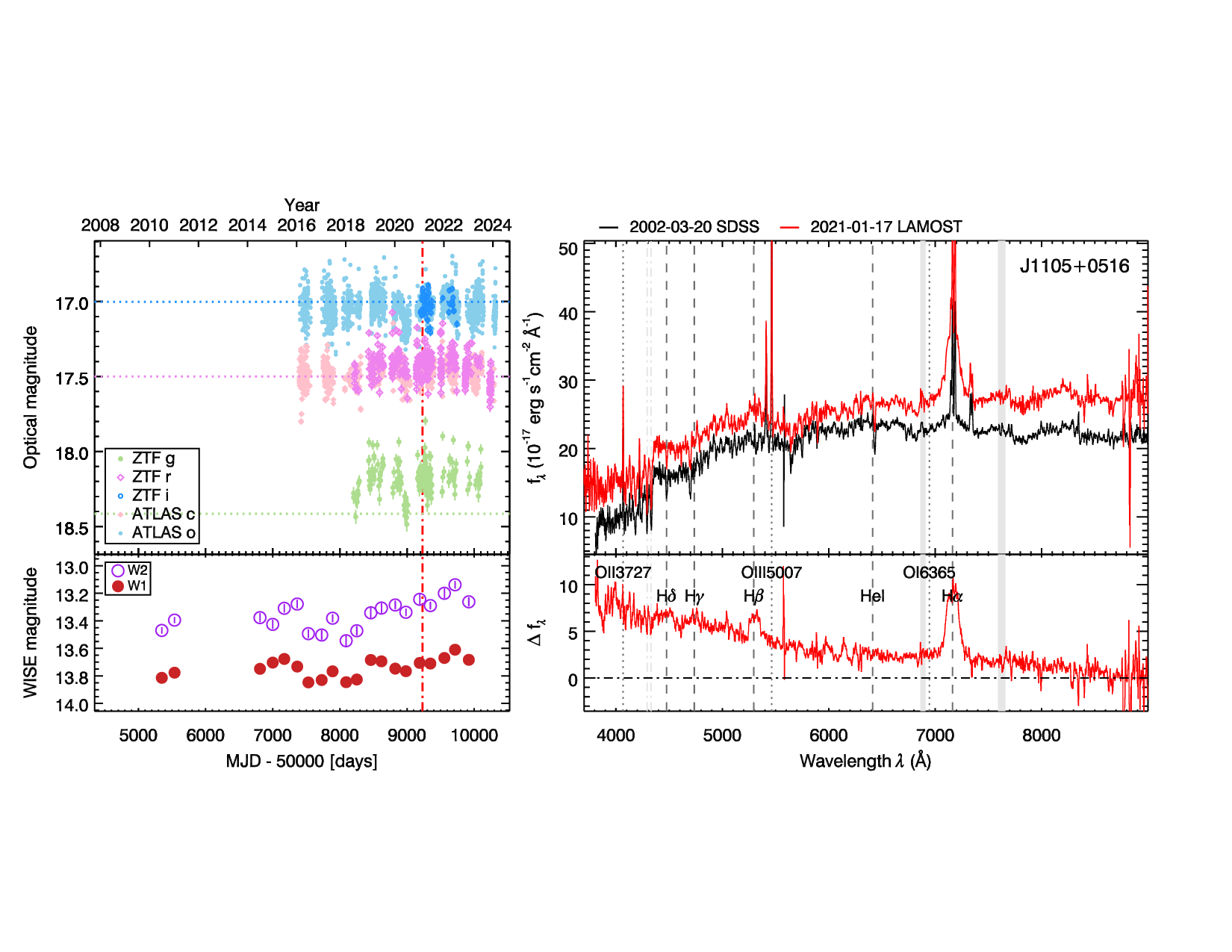}
\hspace{-0.7cm}
\includegraphics[width=0.52\textwidth]{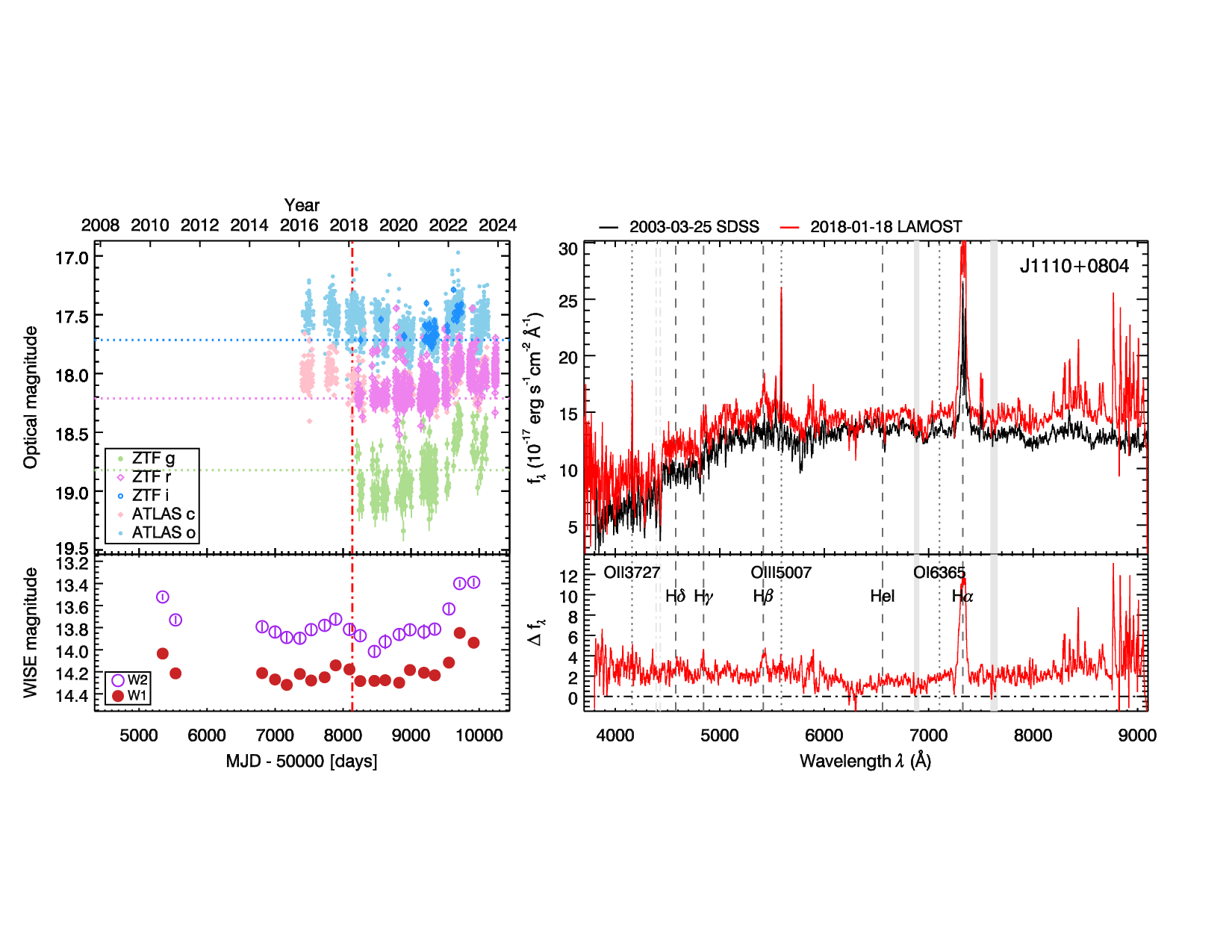}\\
\vspace{-2.7cm}
\hspace{-0.4cm}
\includegraphics[width=0.52\textwidth]{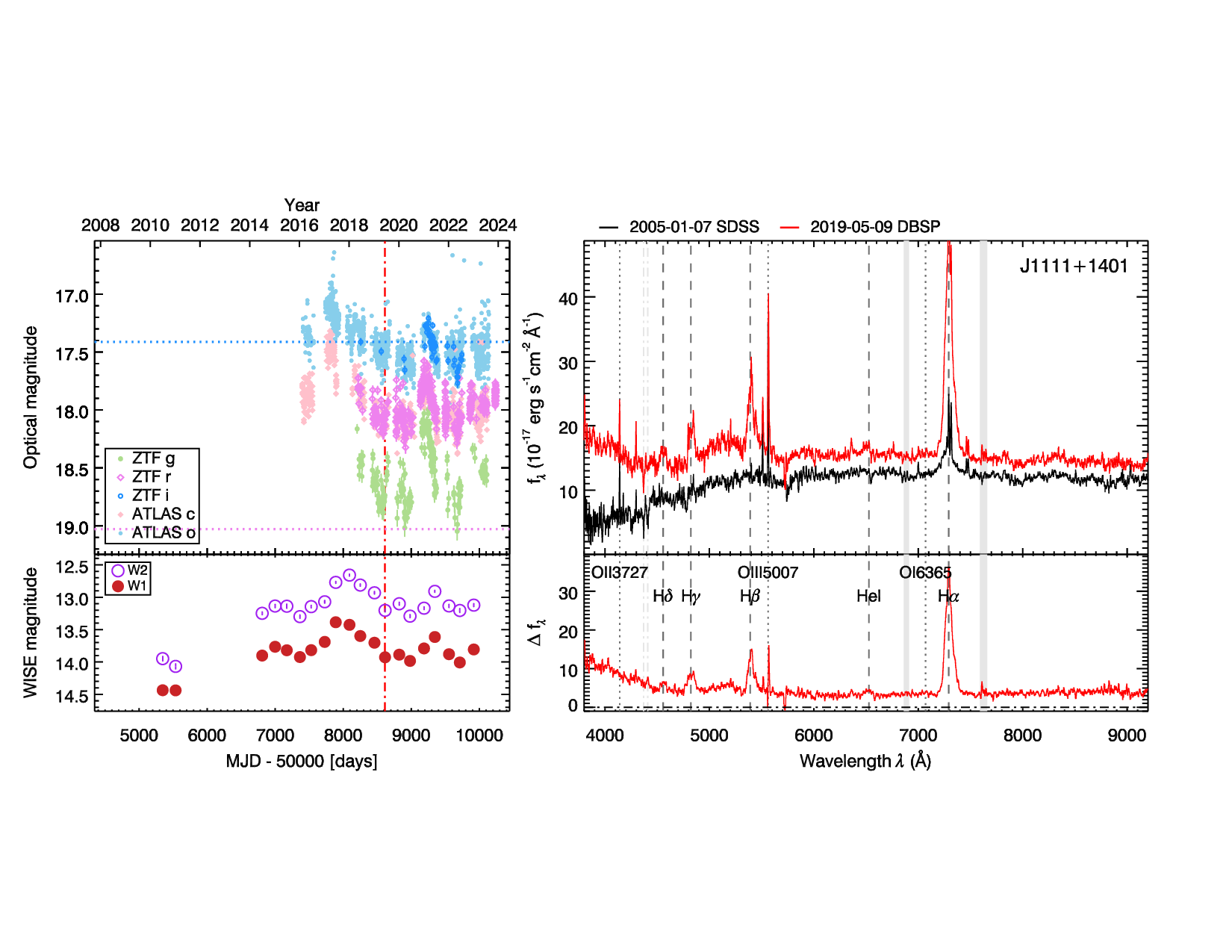}
\hspace{-0.7cm}
\includegraphics[width=0.52\textwidth]{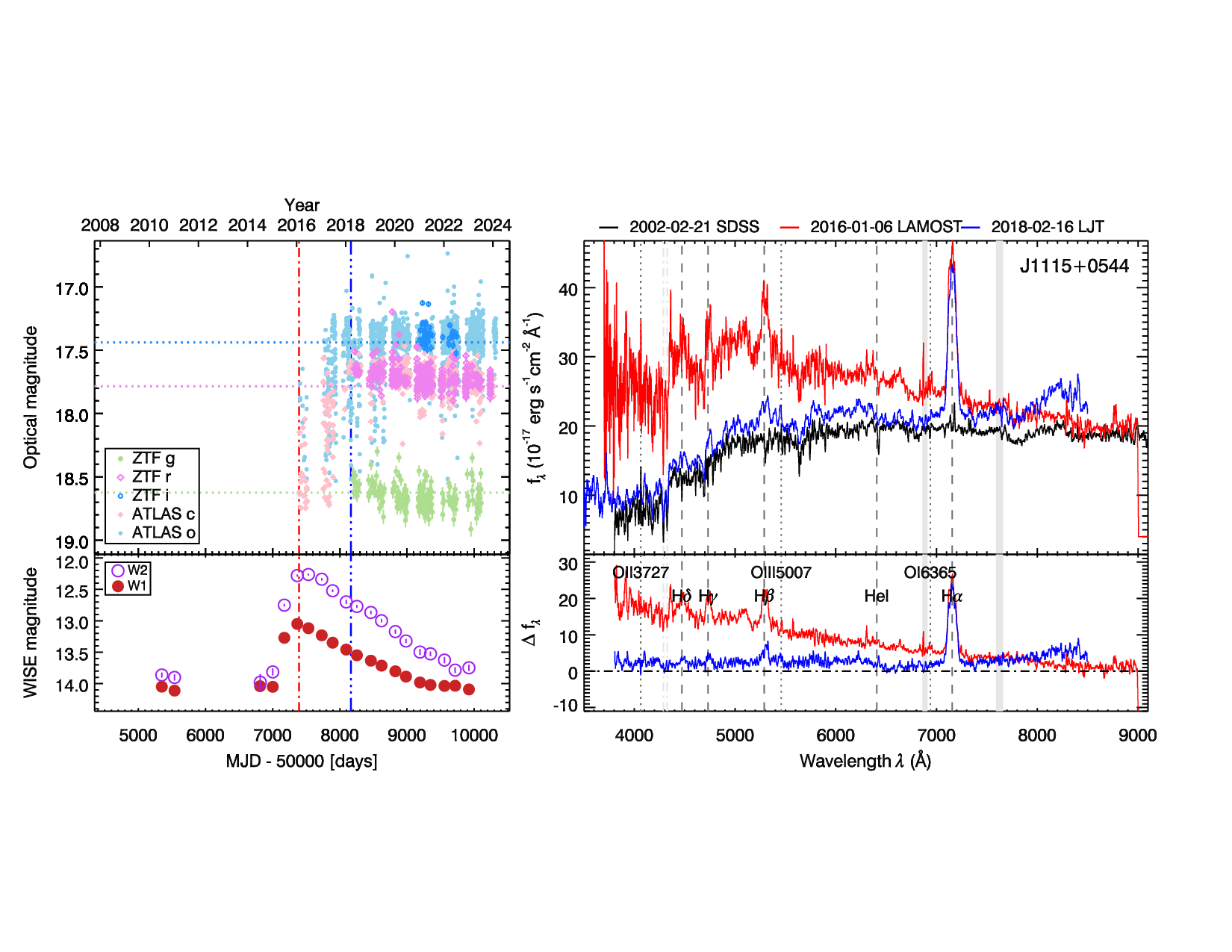}\\
\caption{Confirmed CLQs. Same as Figure \ref{fig:example}.}
\label{fig:optical_spec}
\end{figure*}

% \clearpage
\begin{figure*}[!ht]
\centering
\vspace{-1cm}
\hspace{-0.4cm}
\includegraphics[width=0.52\textwidth]{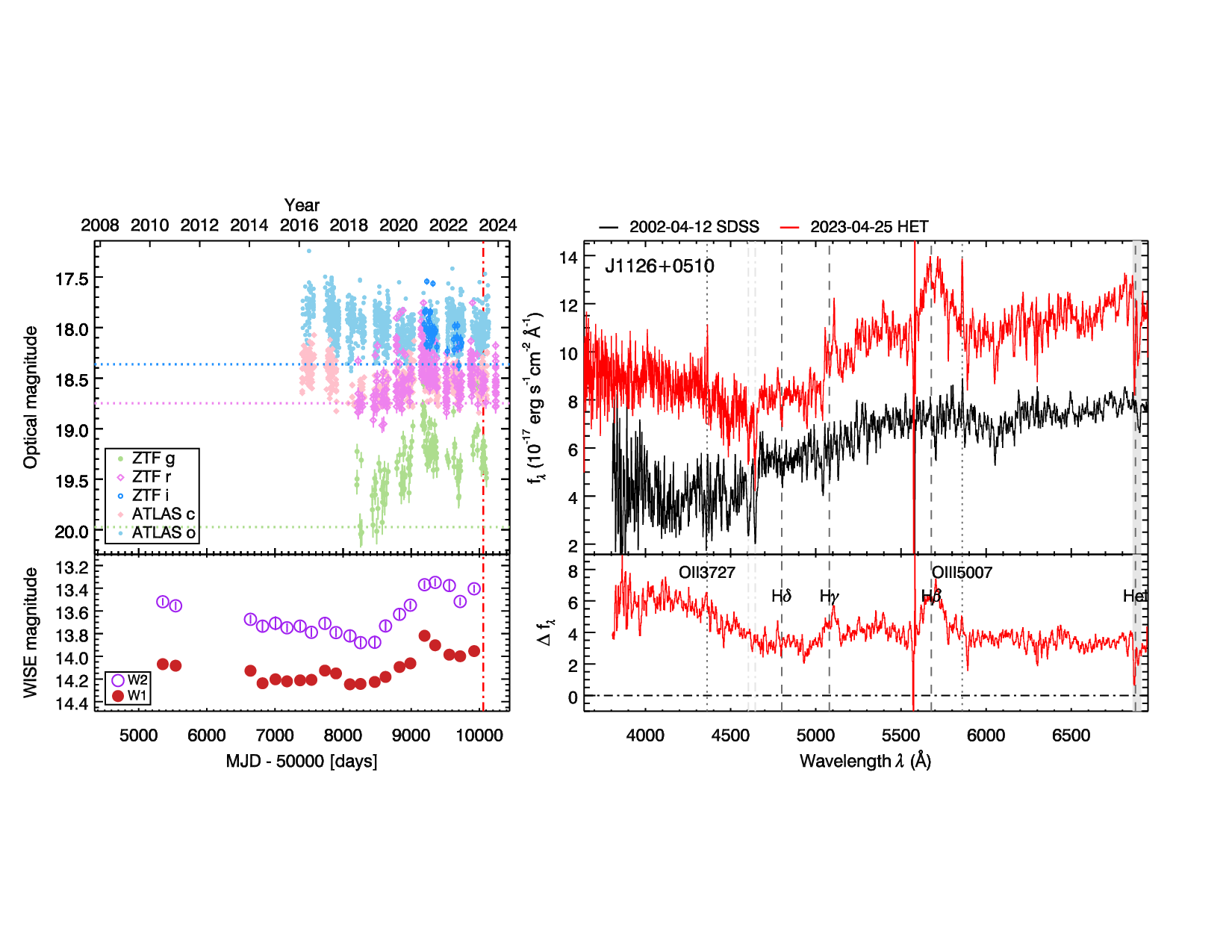}
\hspace{-0.7cm}
\includegraphics[width=0.52\textwidth]{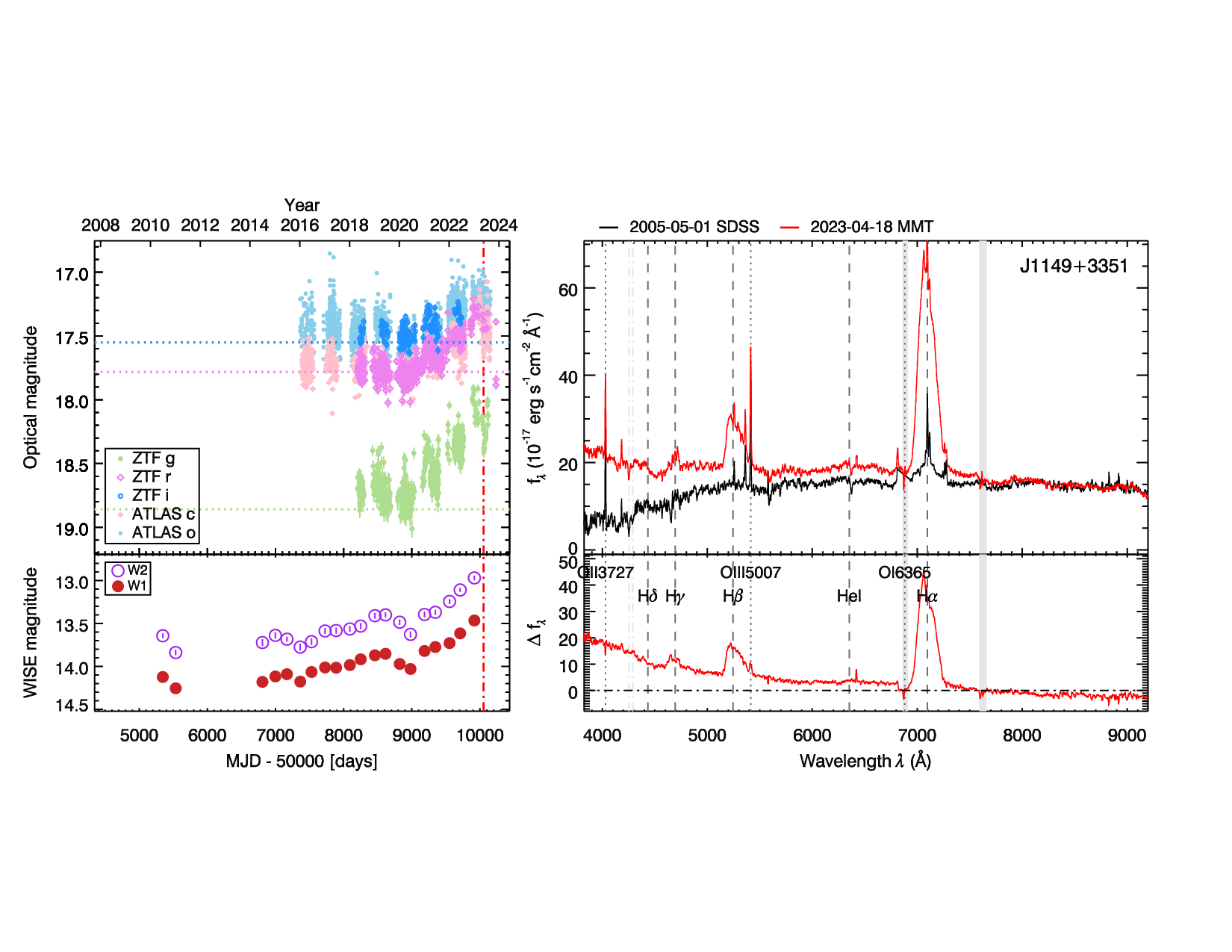}\\
\vspace{-2.7cm}
\hspace{-0.4cm}
\includegraphics[width=0.52\textwidth]{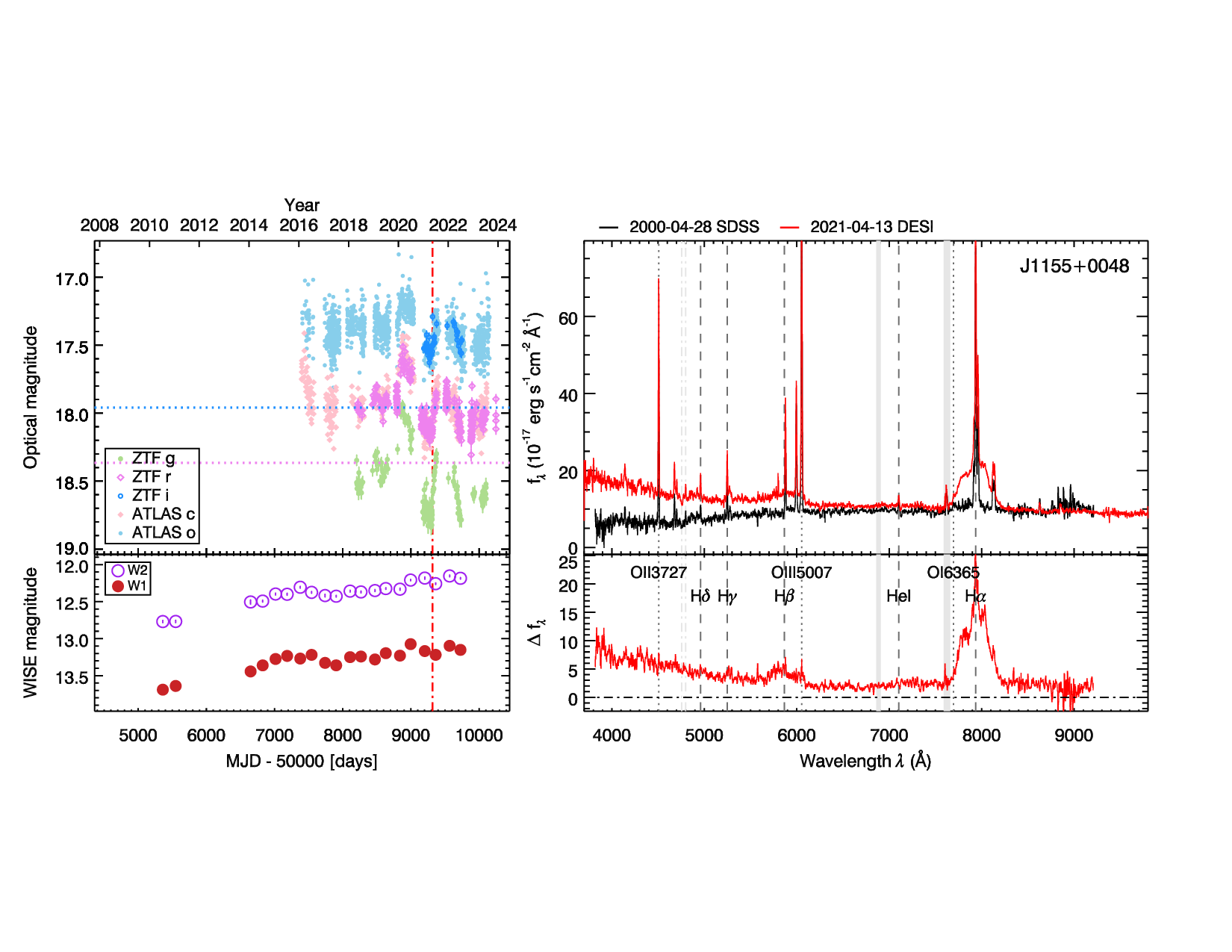}
\hspace{-0.7cm}
\includegraphics[width=0.52\textwidth]{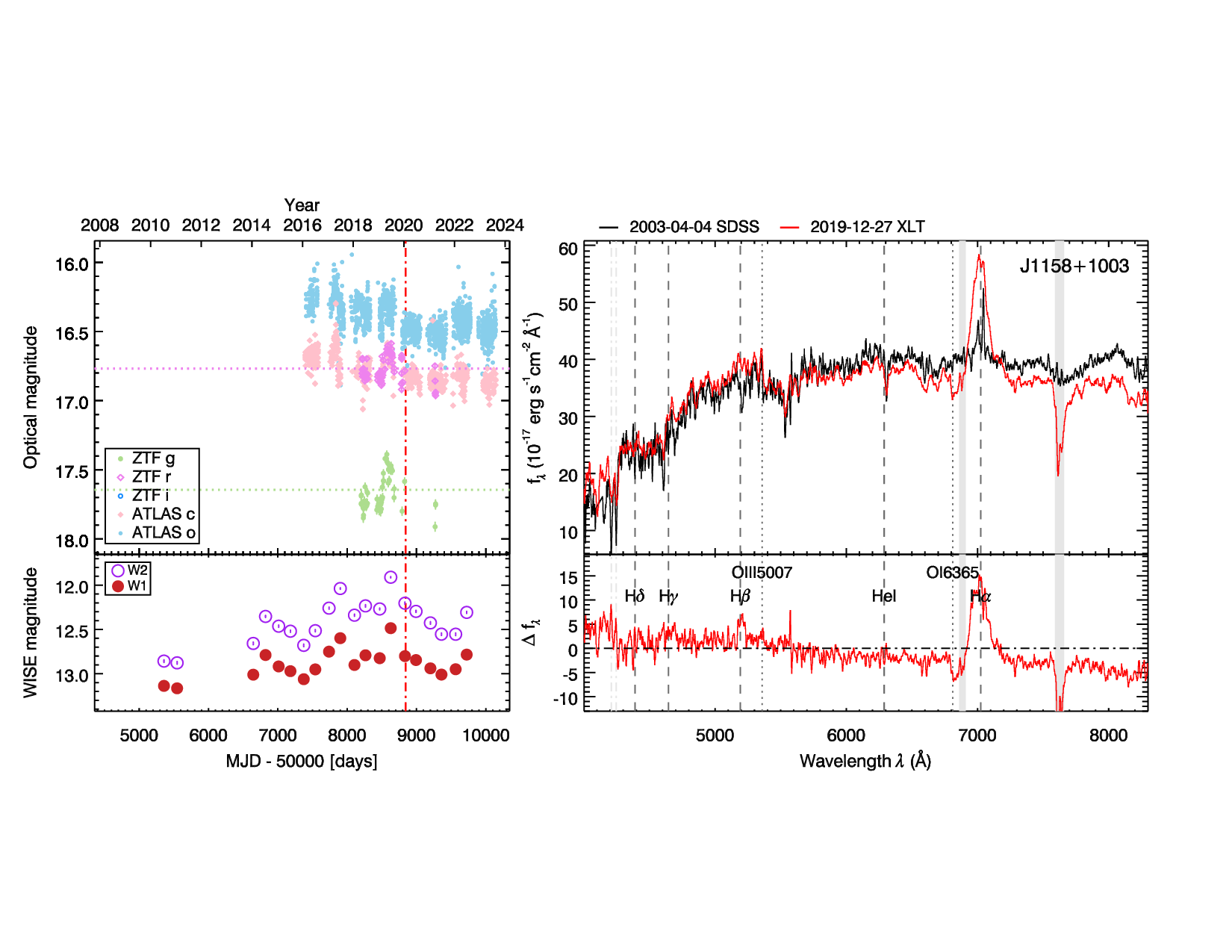}\\
\vspace{-2.7cm}
\hspace{-0.4cm}
\includegraphics[width=0.52\textwidth]{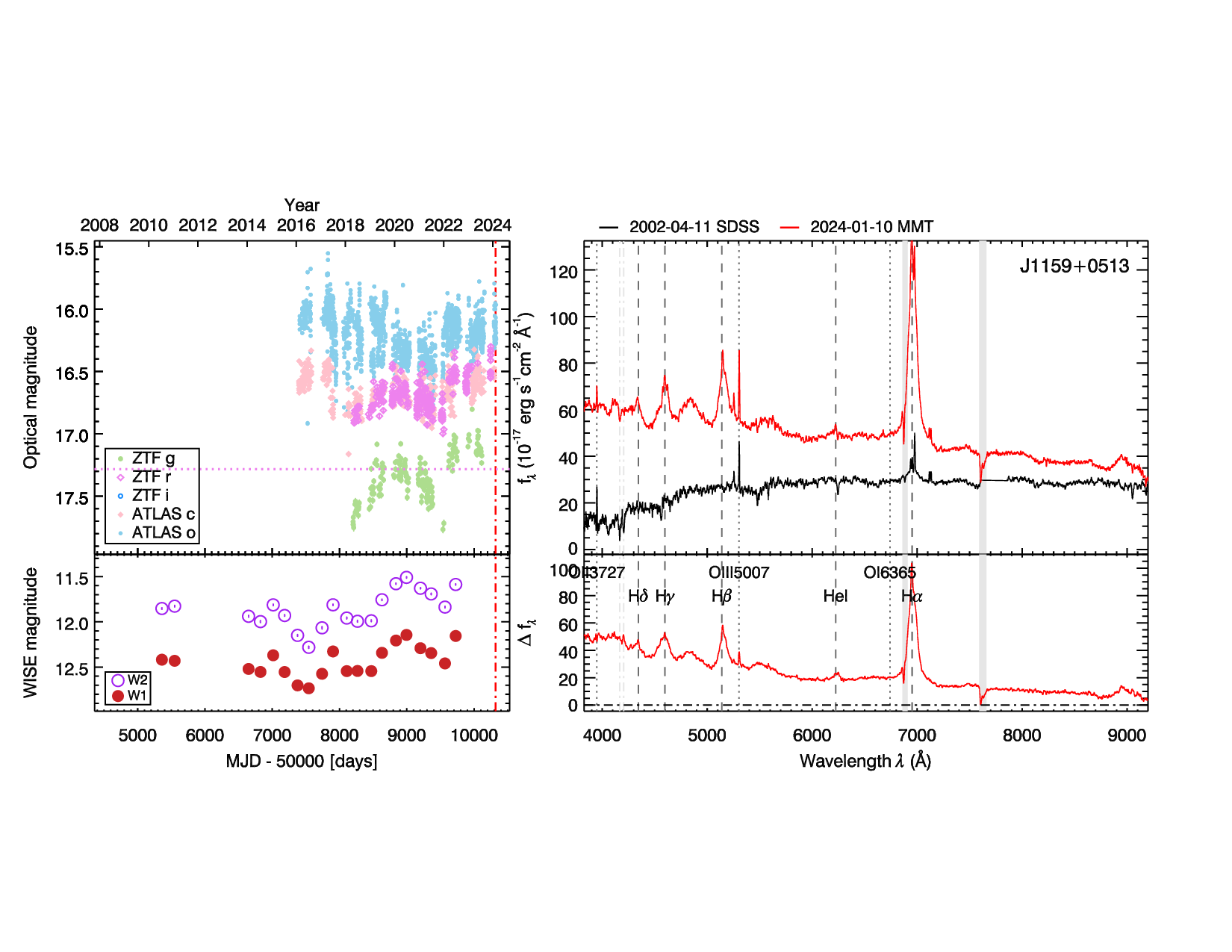}
\hspace{-0.7cm}
\includegraphics[width=0.52\textwidth]{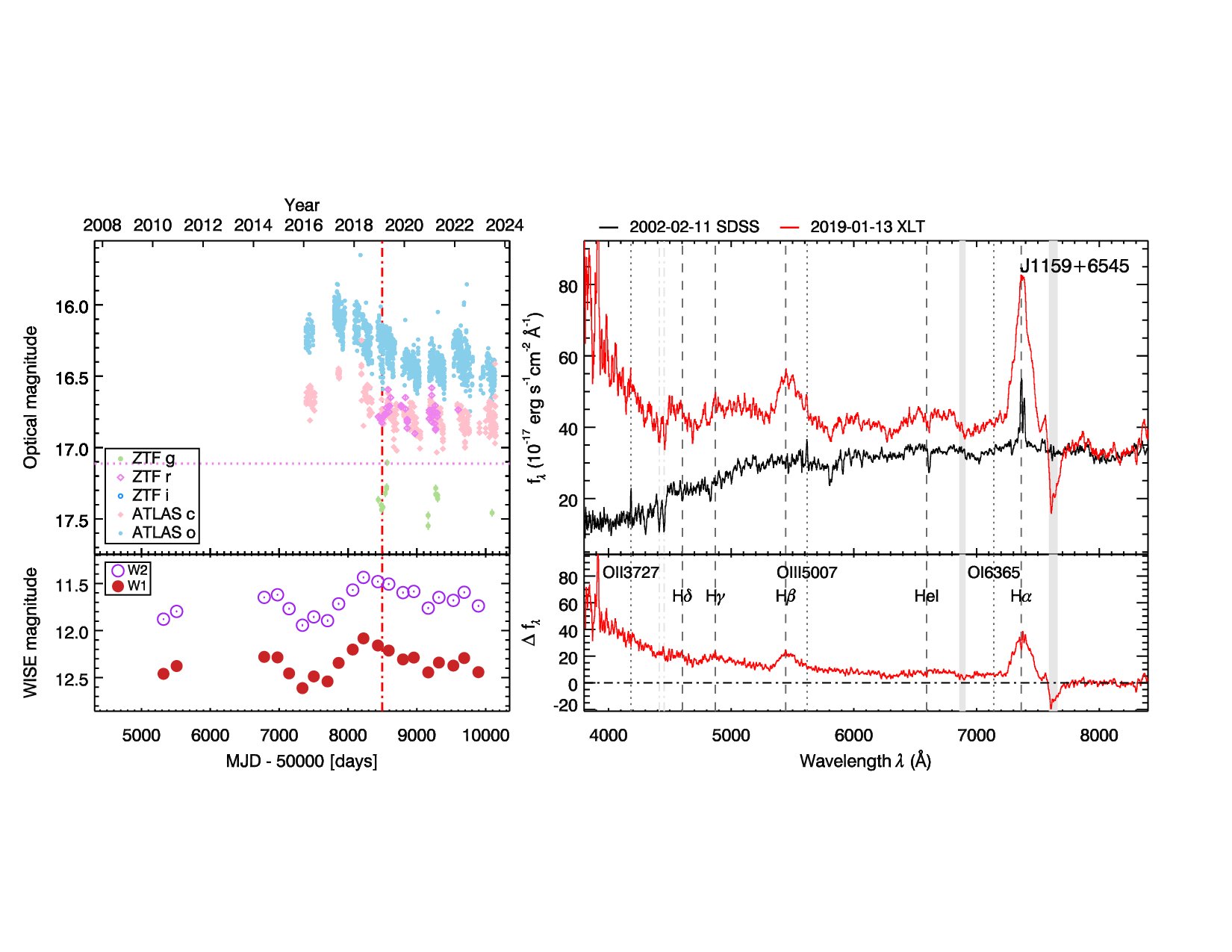}\\
\vspace{-2.7cm}
\hspace{-0.4cm}
\includegraphics[width=0.52\textwidth]{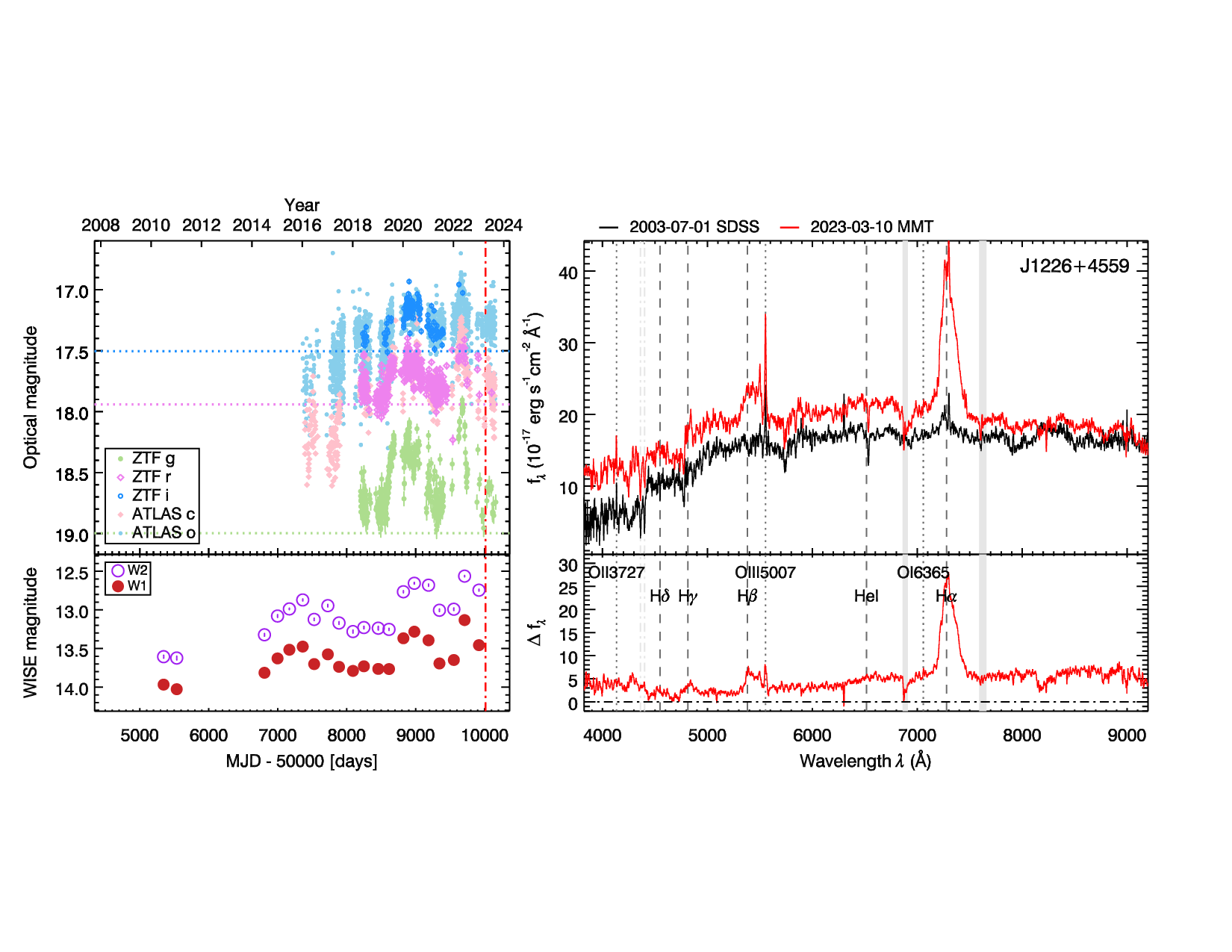}
\hspace{-0.7cm}
\includegraphics[width=0.52\textwidth]{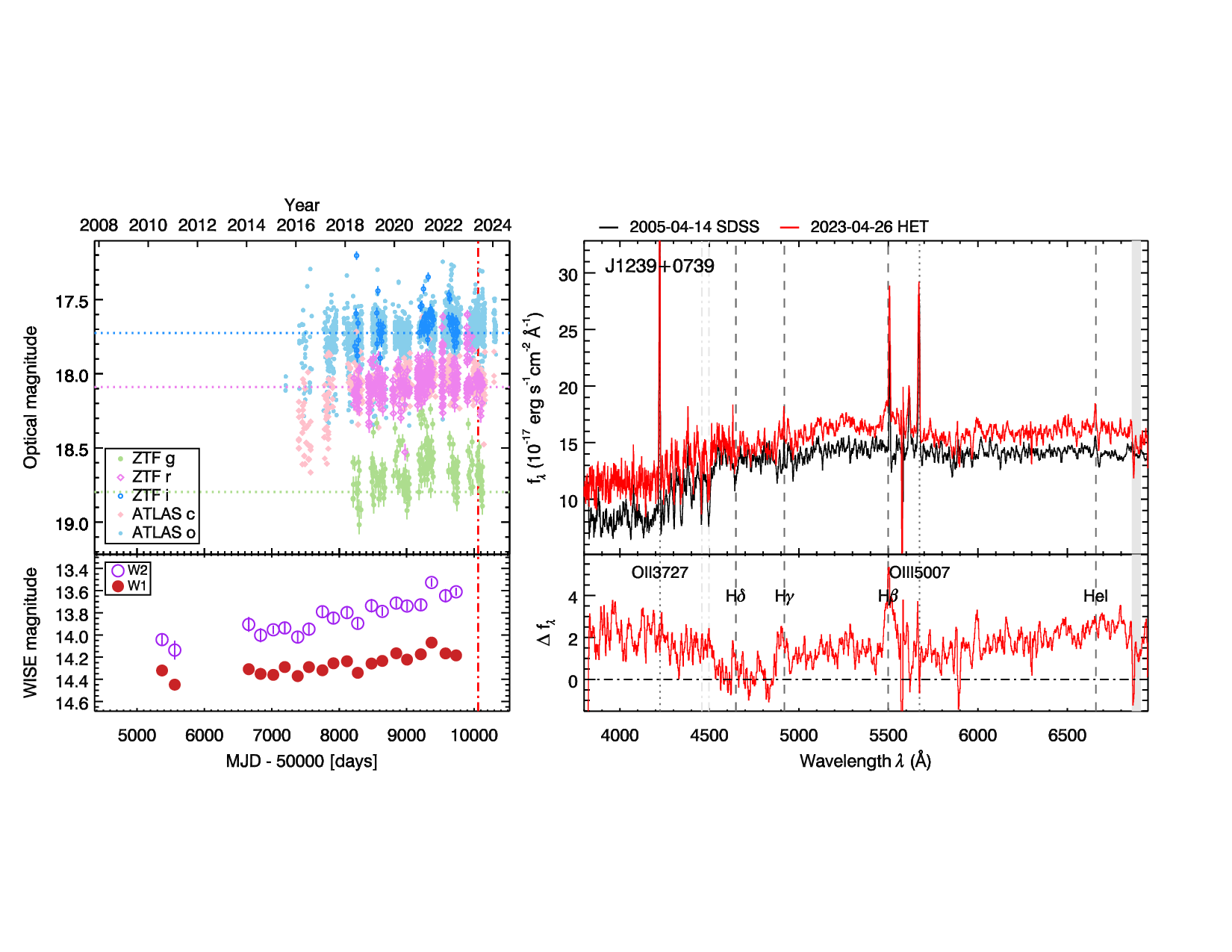}\\
\vspace{-2.7cm}
\hspace{-0.4cm}
\includegraphics[width=0.52\textwidth]{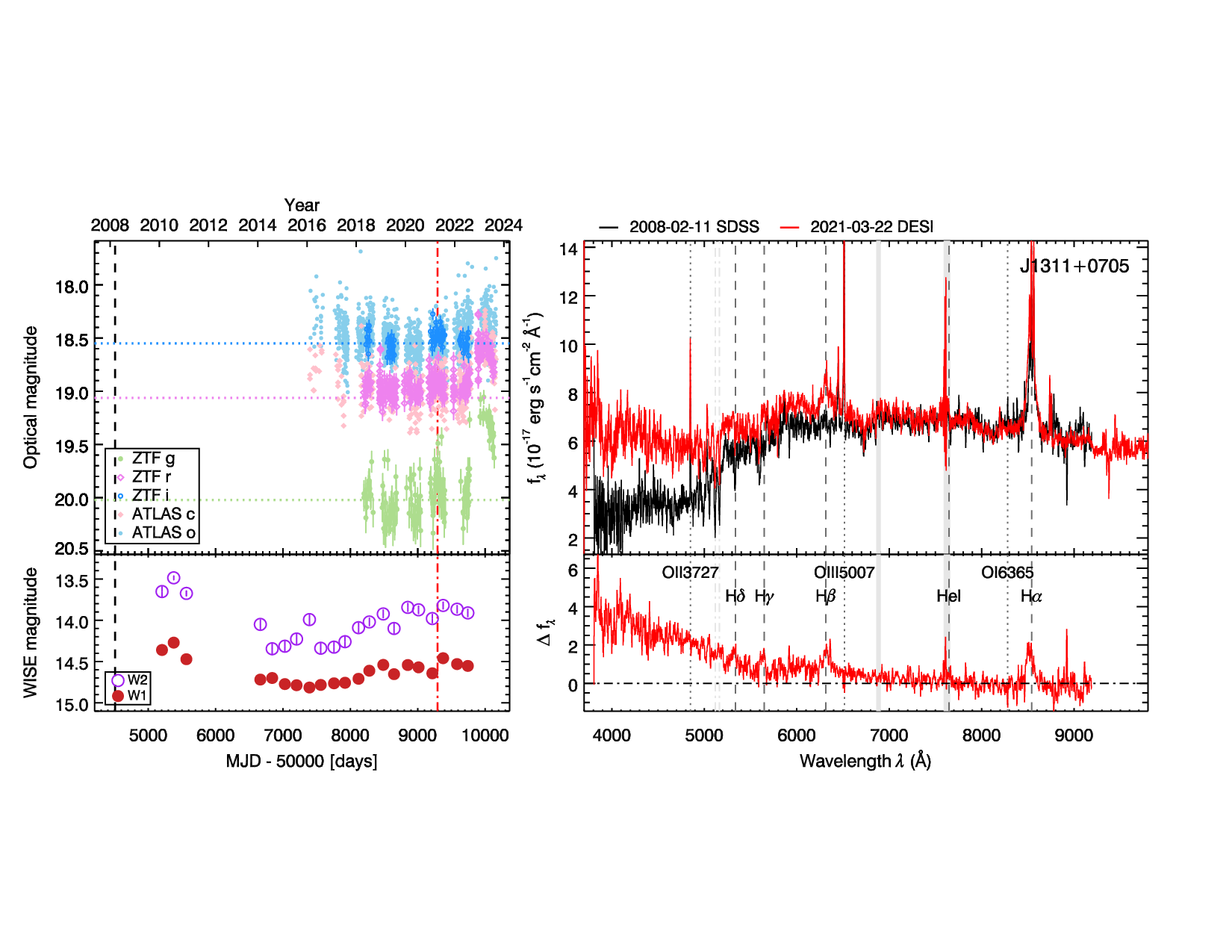}
\hspace{-0.7cm}
\includegraphics[width=0.52\textwidth]{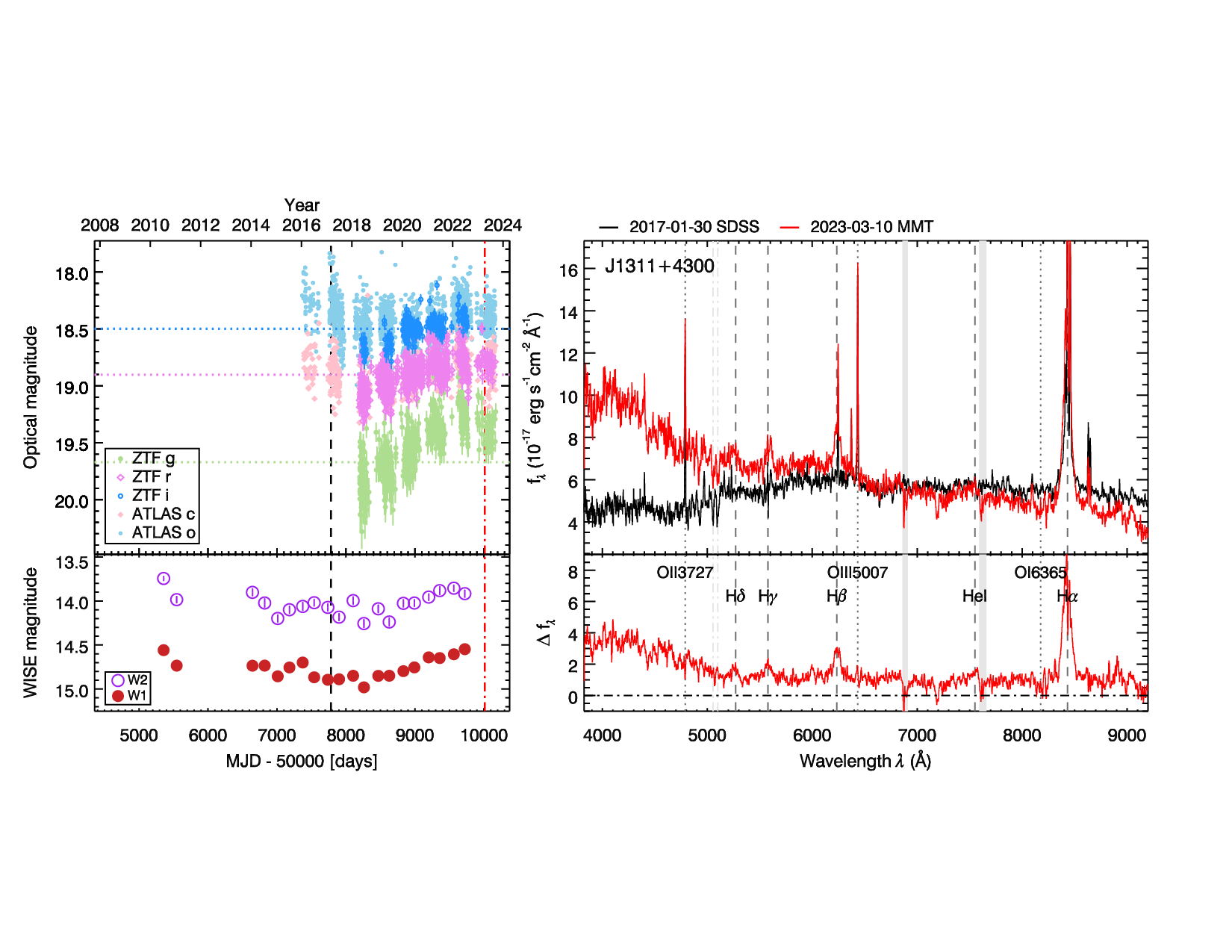}\\
\caption{Confirmed CLQs. Same as Figure \ref{fig:example}.}
\label{fig:optical_spec}
\end{figure*}

% \clearpage
\begin{figure*}[!ht]
\centering
\vspace{-1cm}
\hspace{-0.4cm}
\includegraphics[width=0.52\textwidth]{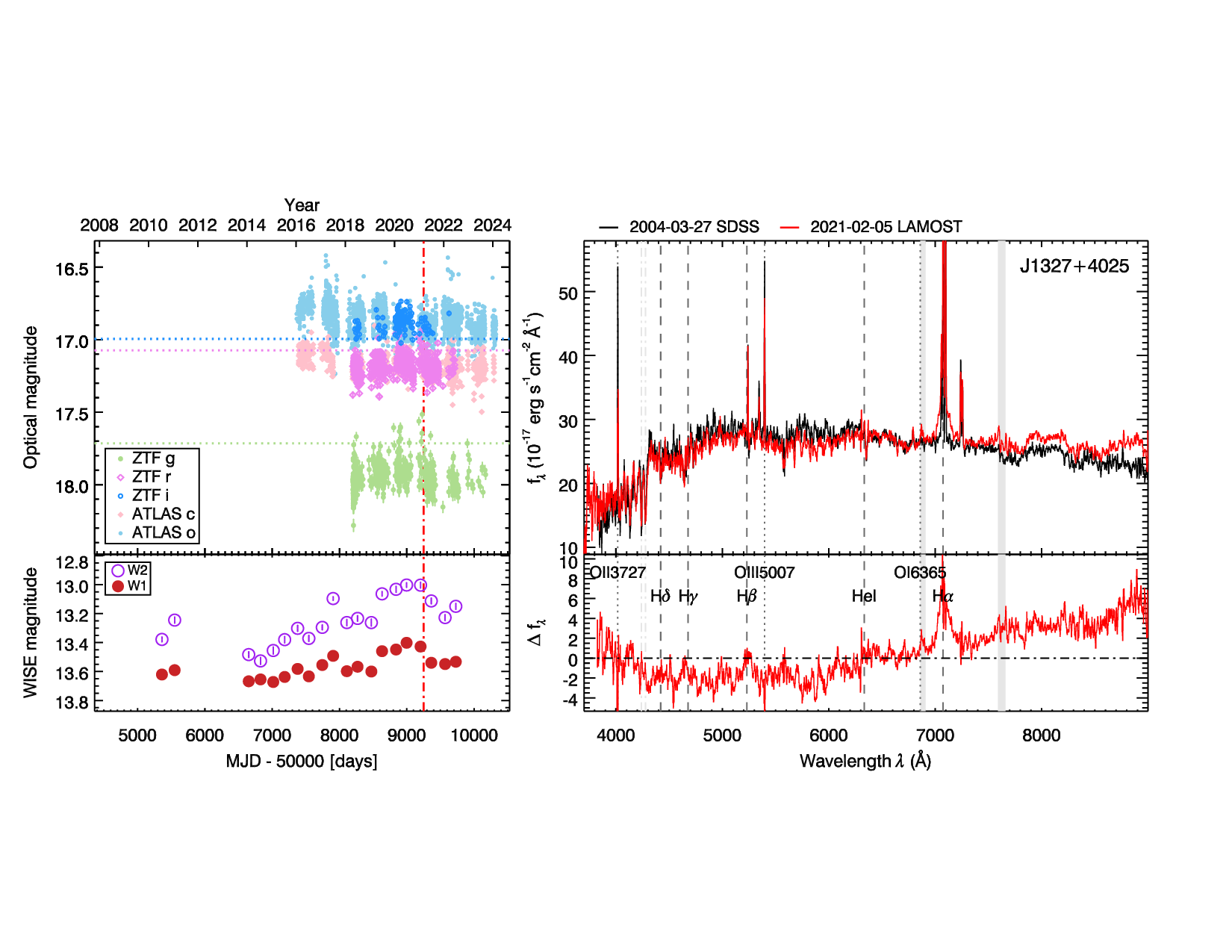}
\hspace{-0.7cm}
\includegraphics[width=0.52\textwidth]{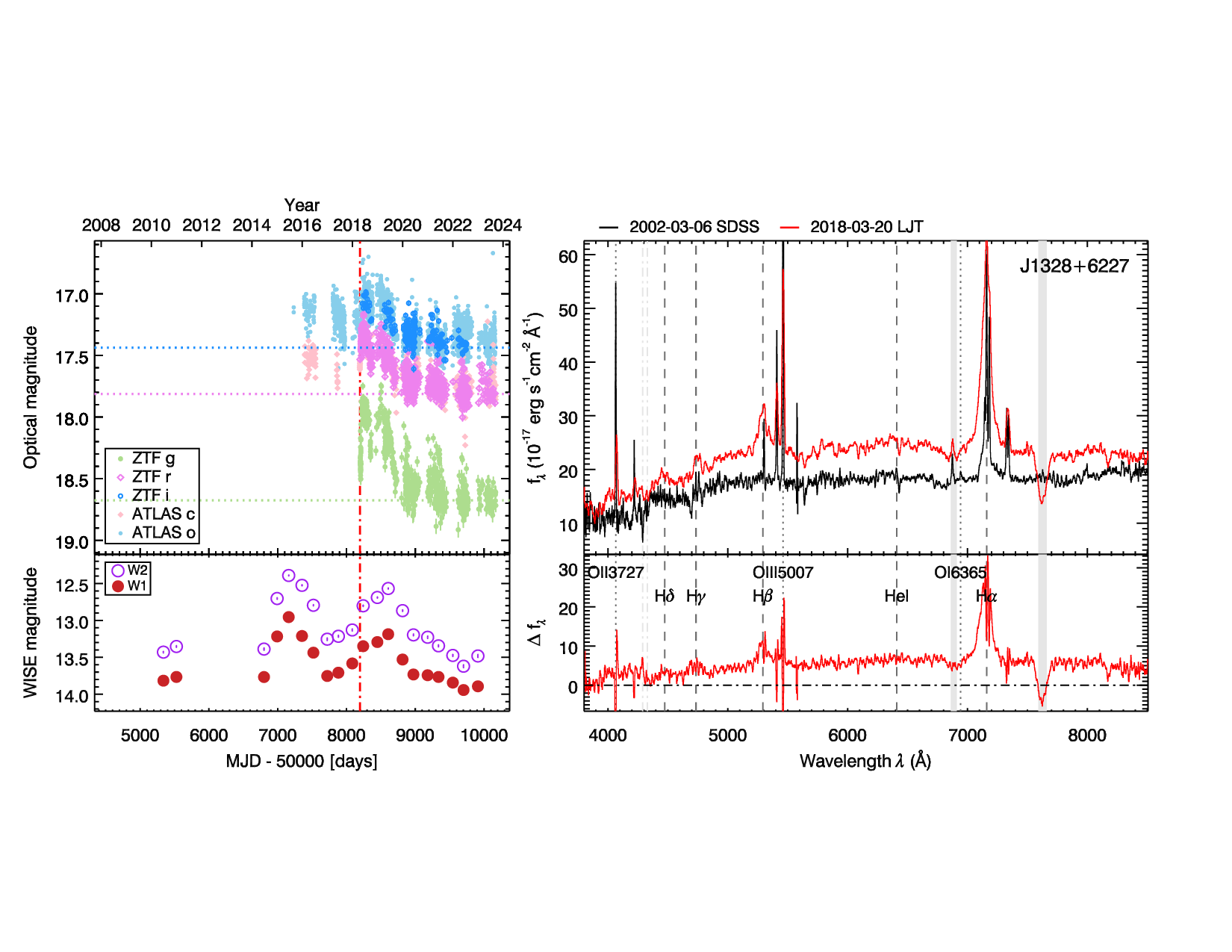}\\
\vspace{-2.7cm}
\hspace{-0.4cm}
\includegraphics[width=0.52\textwidth]{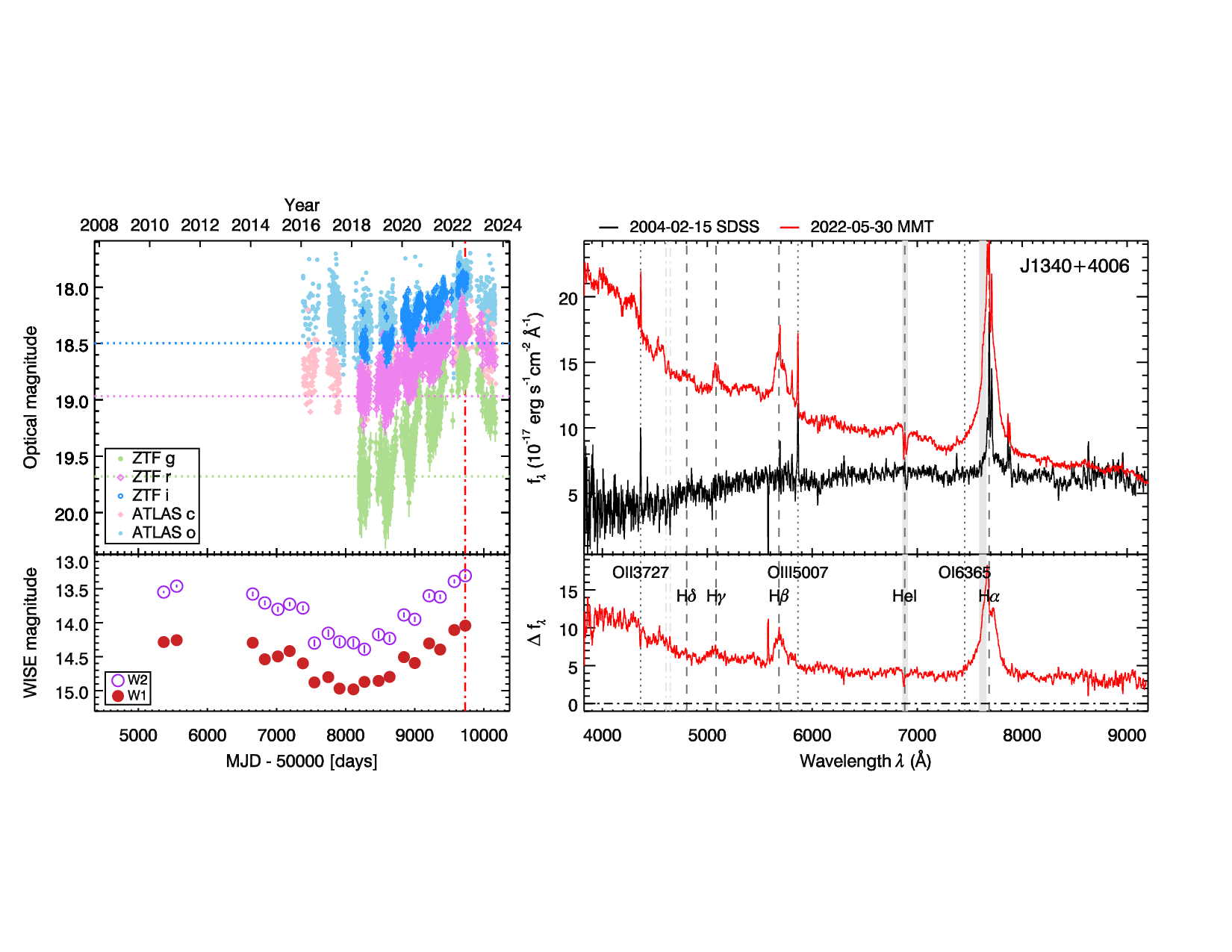}
\hspace{-0.7cm}
\includegraphics[width=0.52\textwidth]{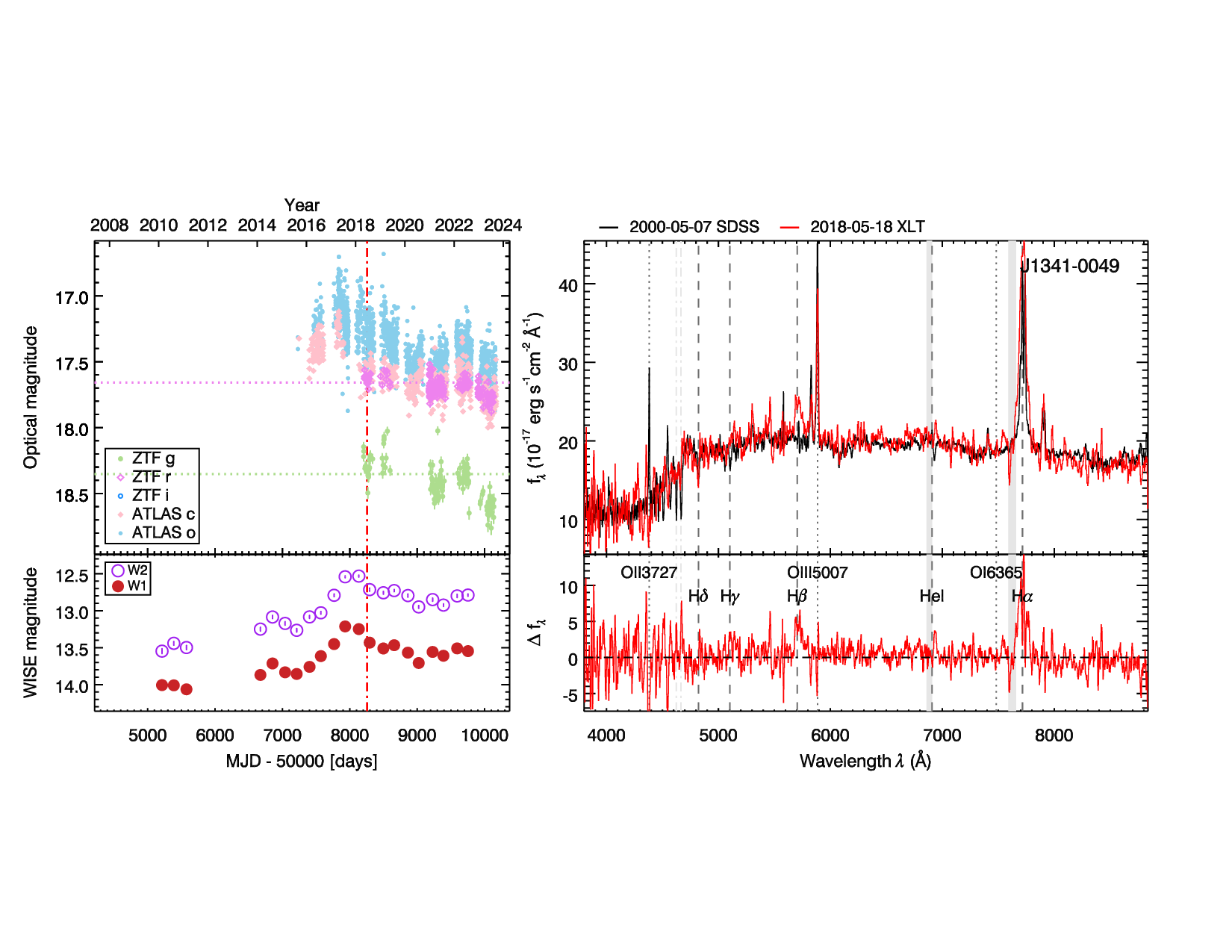}\\
\vspace{-2.7cm}
\hspace{-0.4cm}
\includegraphics[width=0.52\textwidth]{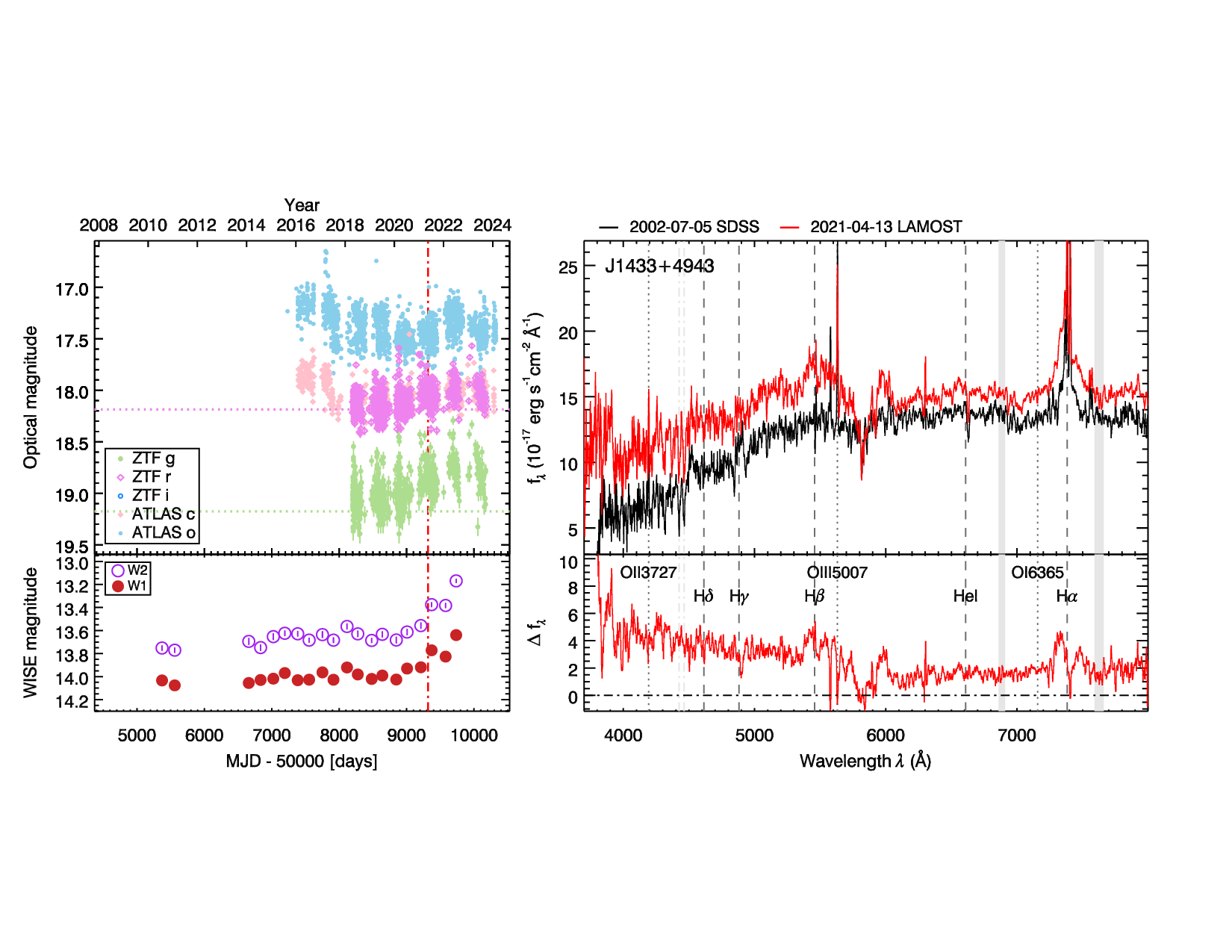}
\hspace{-0.7cm}
\includegraphics[width=0.52\textwidth]{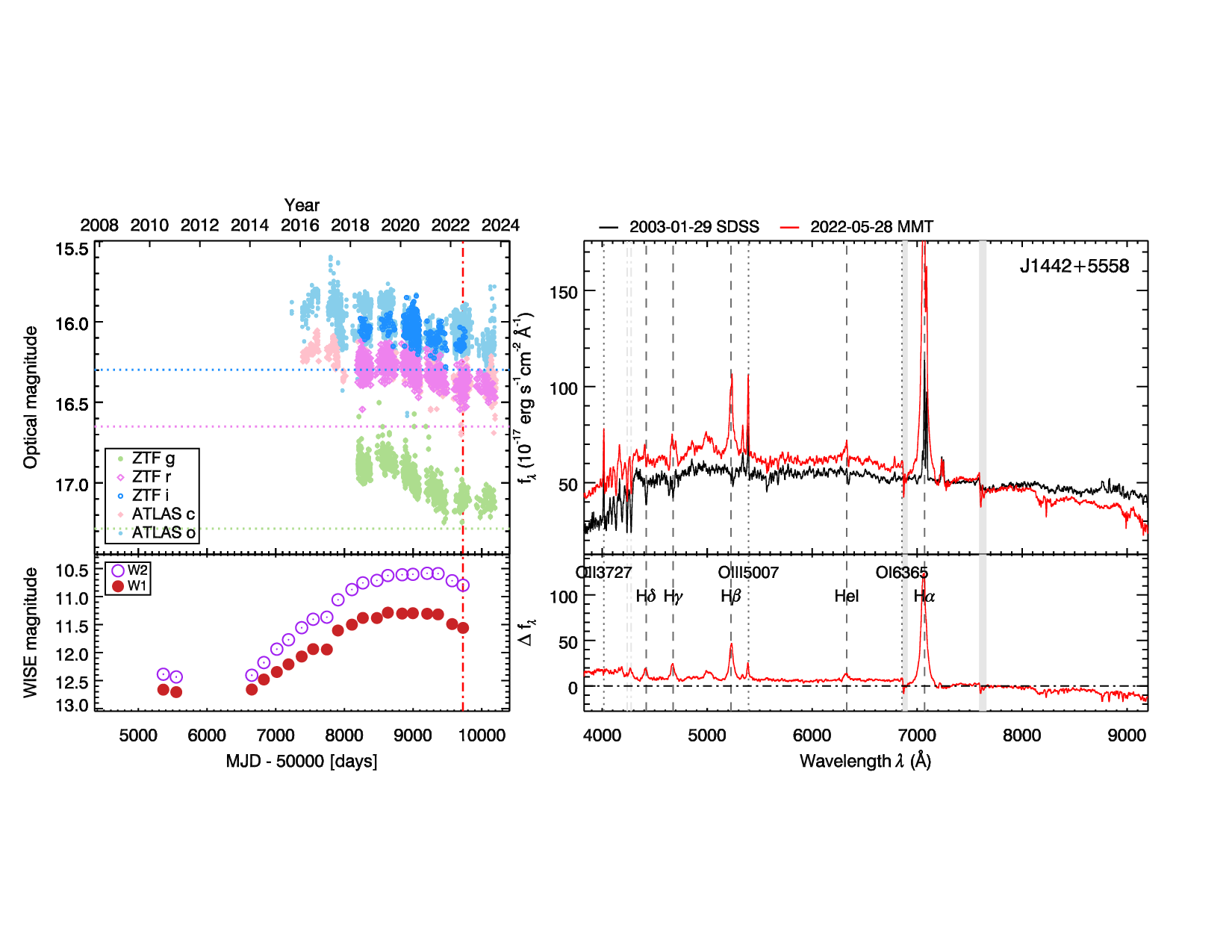}\\
\vspace{-2.7cm}
\hspace{-0.4cm}
\includegraphics[width=0.52\textwidth]{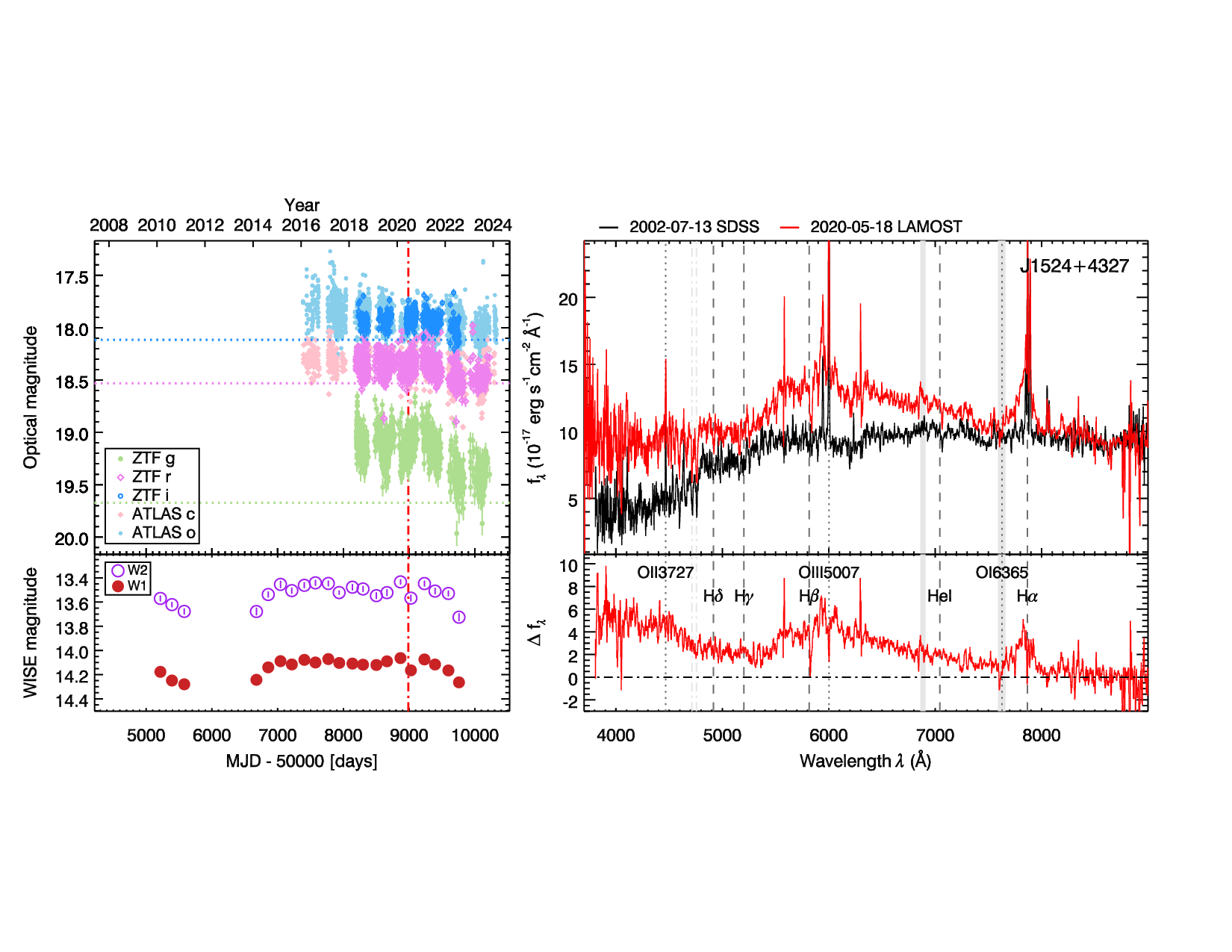}
\hspace{-0.7cm}
\includegraphics[width=0.52\textwidth]{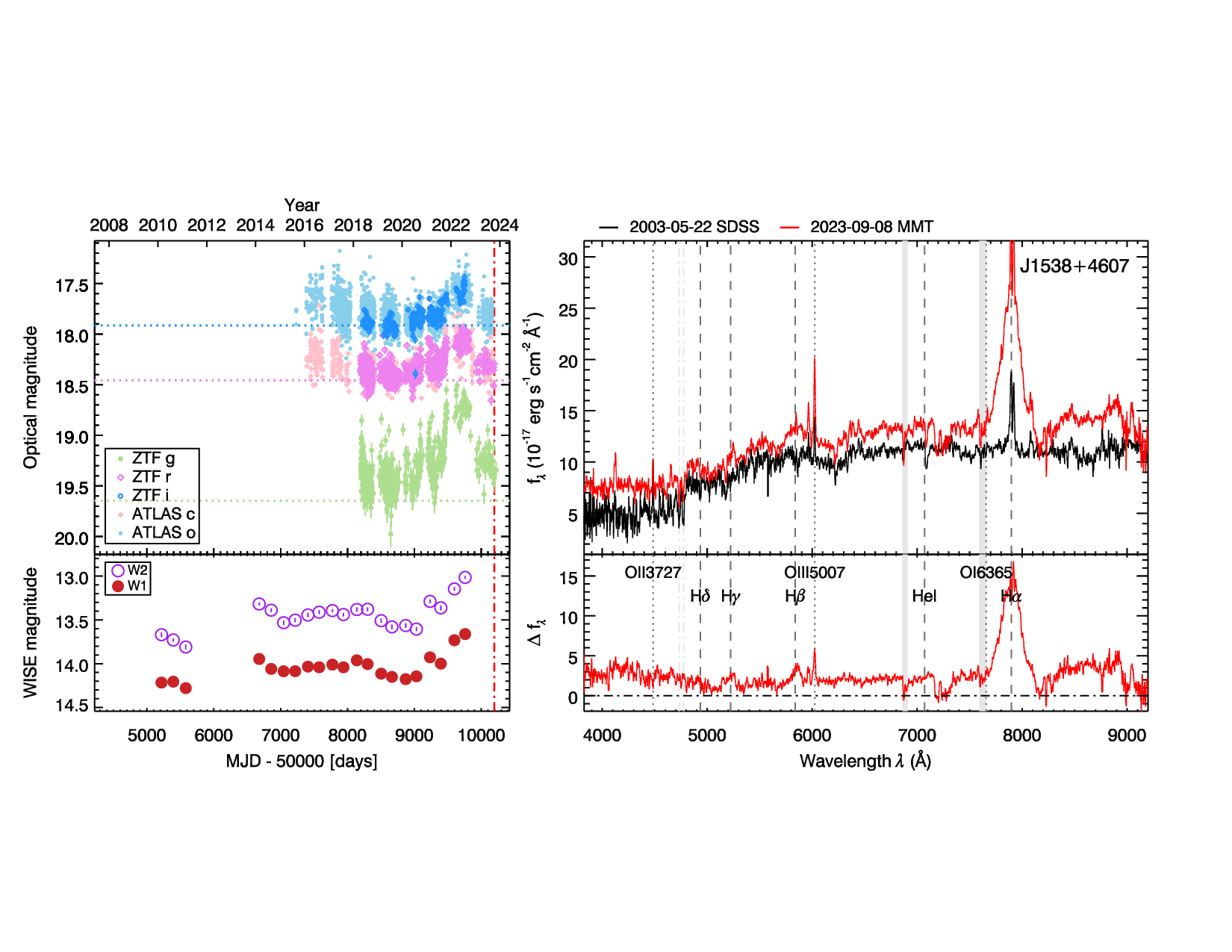}\\
\vspace{-2.7cm}
\hspace{-0.4cm}
\includegraphics[width=0.52\textwidth]{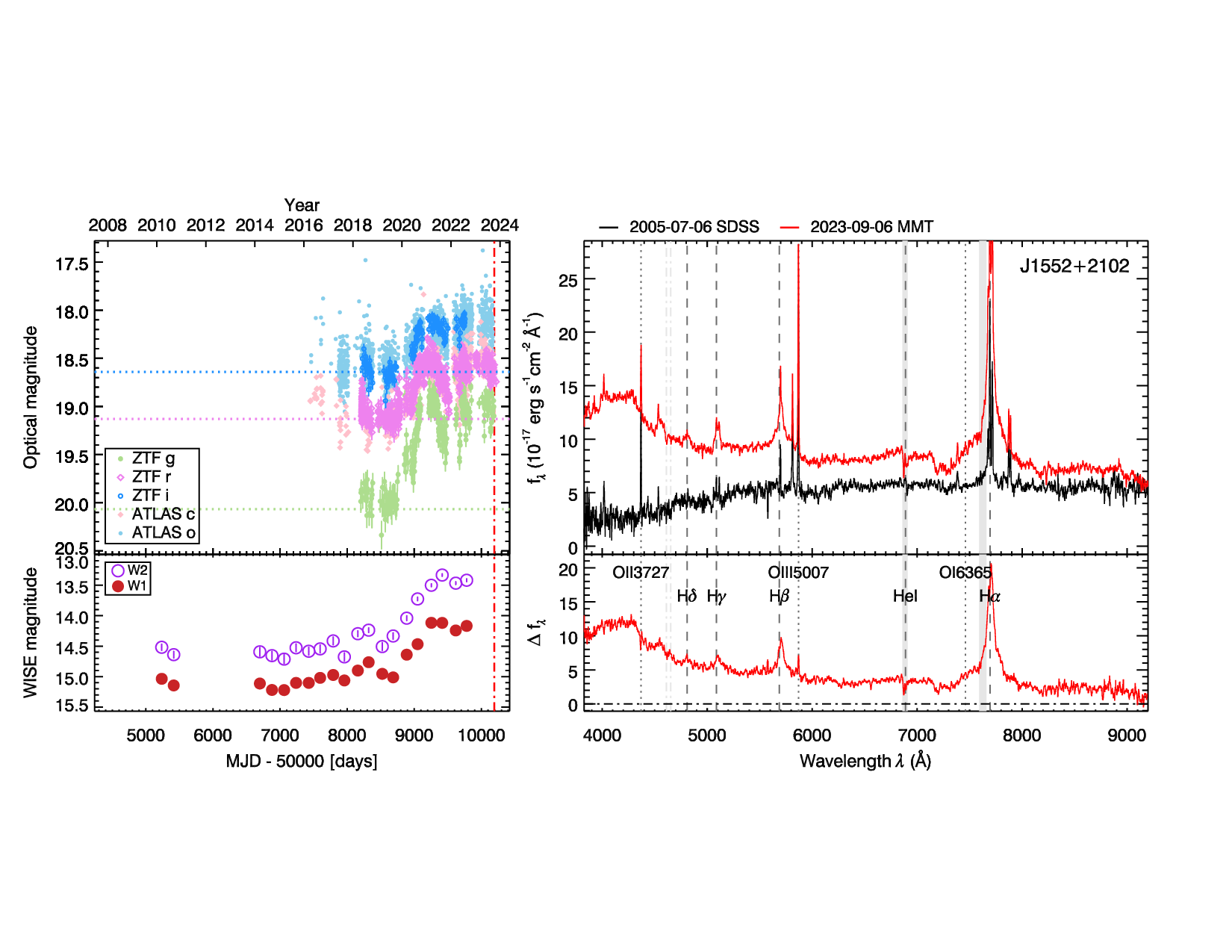}
\hspace{-0.7cm}
\includegraphics[width=0.52\textwidth]{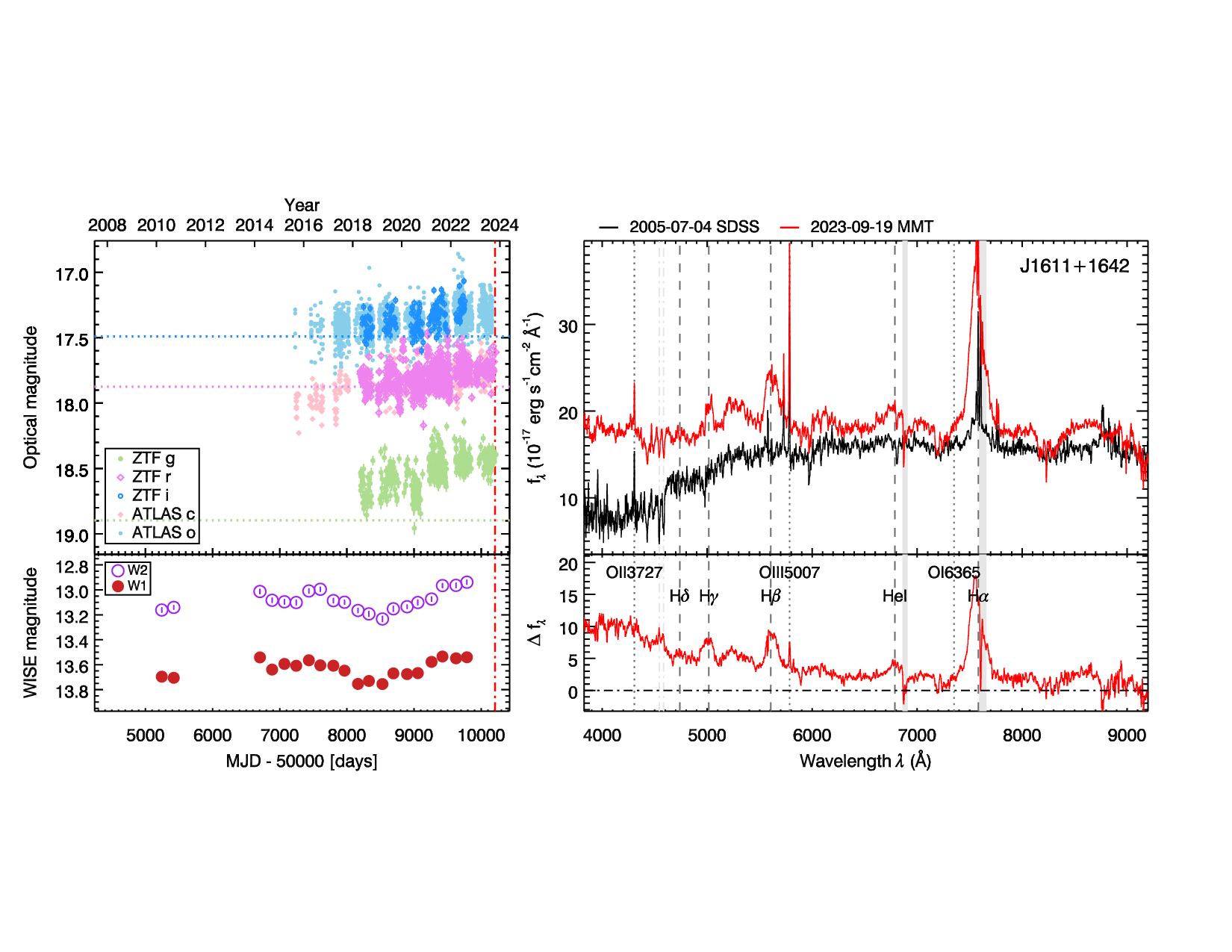}\\
\caption{Confirmed CLQs. Same as Figure \ref{fig:example}.}
\label{fig:optical_spec}
\end{figure*}

% \clearpage
\begin{figure*}[!ht]
\centering
\vspace{-1cm}
\hspace{-0.4cm}
\includegraphics[width=0.52\textwidth]{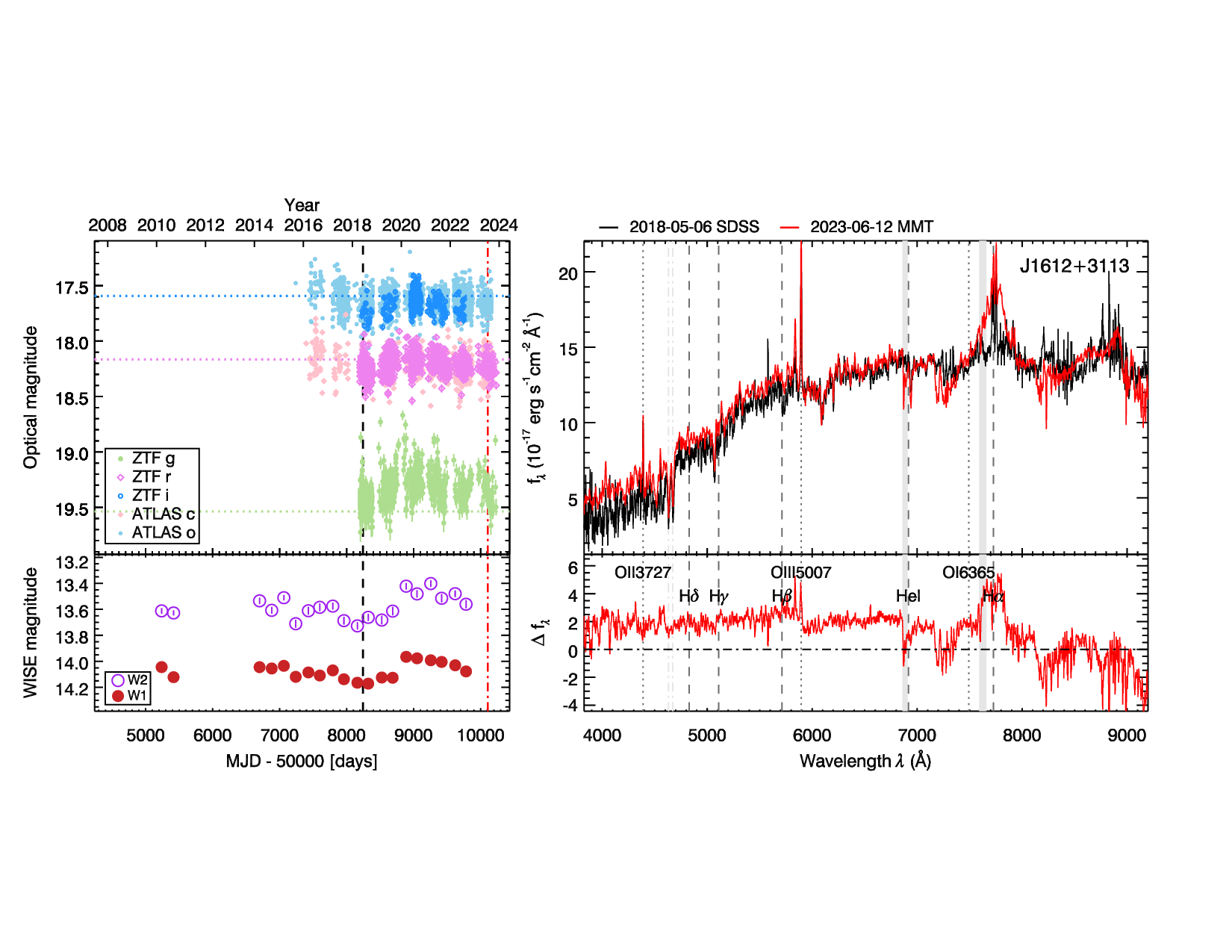}
\hspace{-0.7cm}
\includegraphics[width=0.52\textwidth]{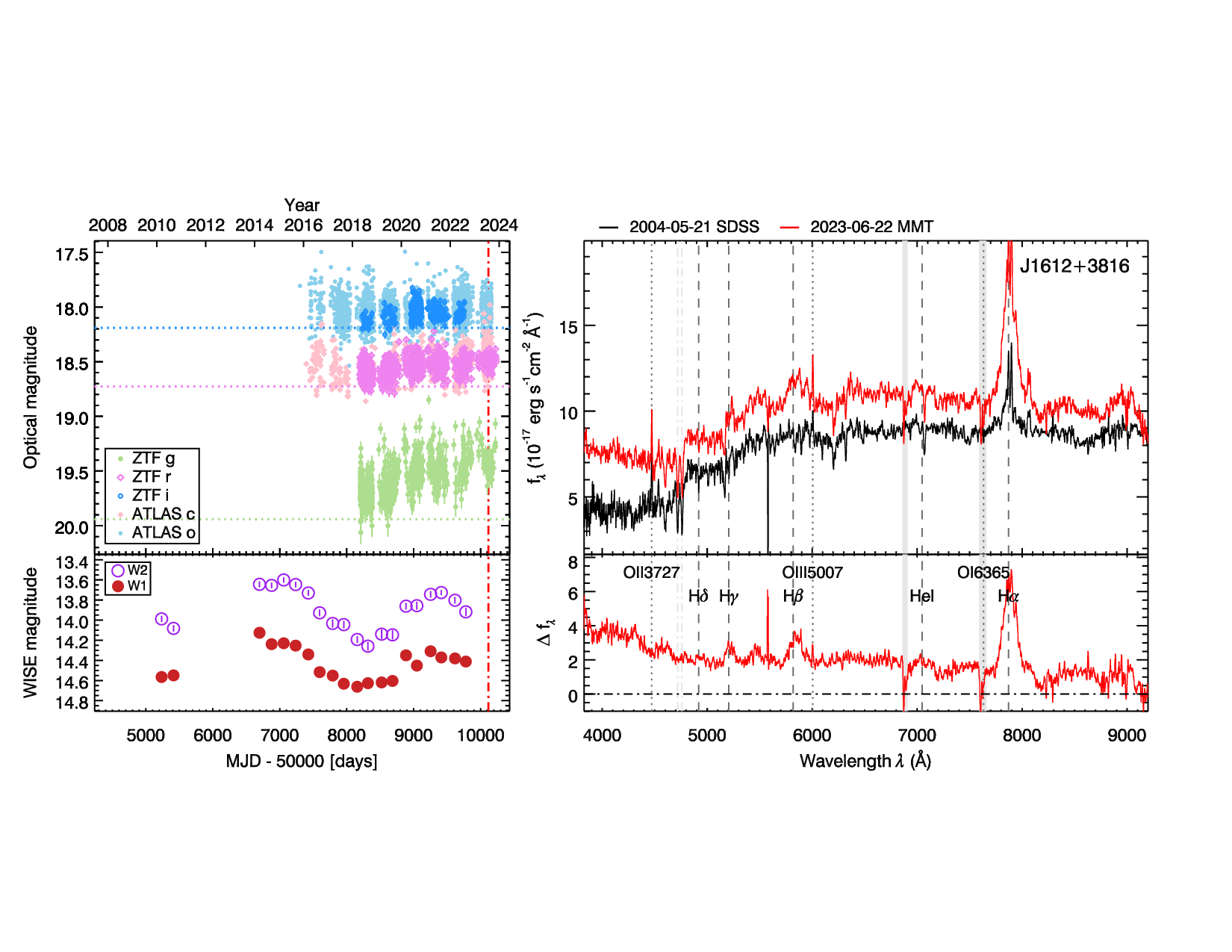}\\
\vspace{-2.7cm}
\hspace{-0.4cm}
\includegraphics[width=0.52\textwidth]{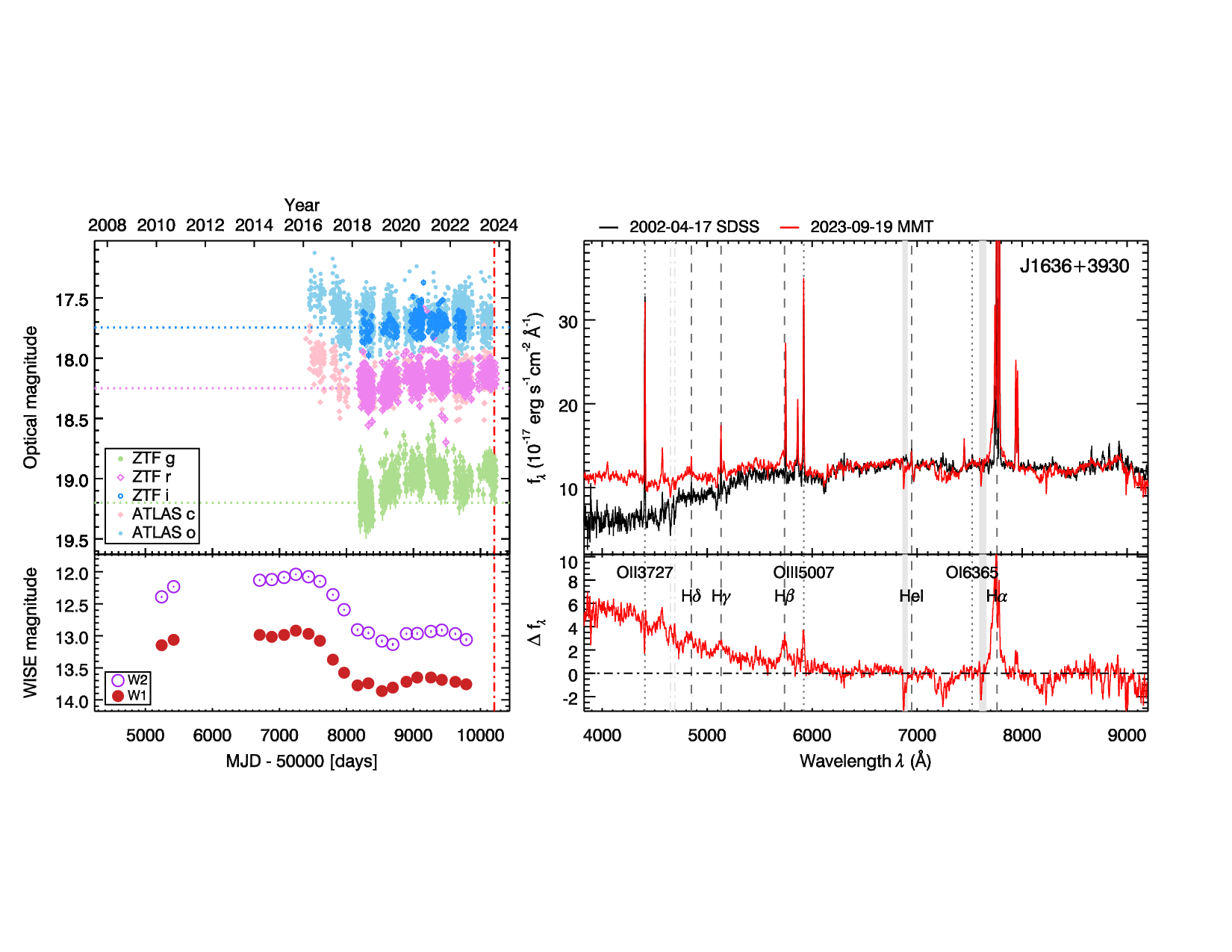}
\hspace{-0.7cm}
\includegraphics[width=0.52\textwidth]{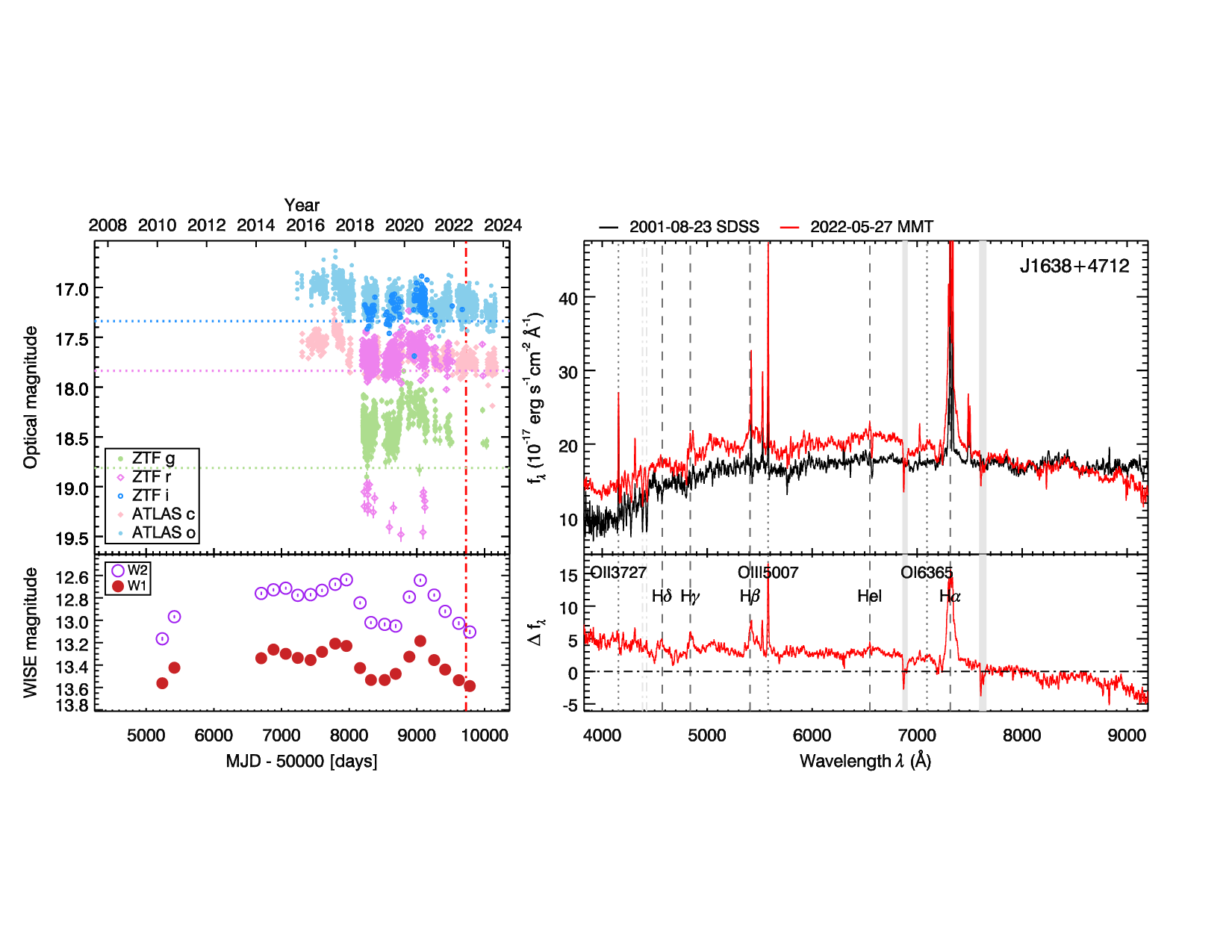}\\
\vspace{-2.7cm}
\hspace{-0.4cm}
\includegraphics[width=0.52\textwidth]{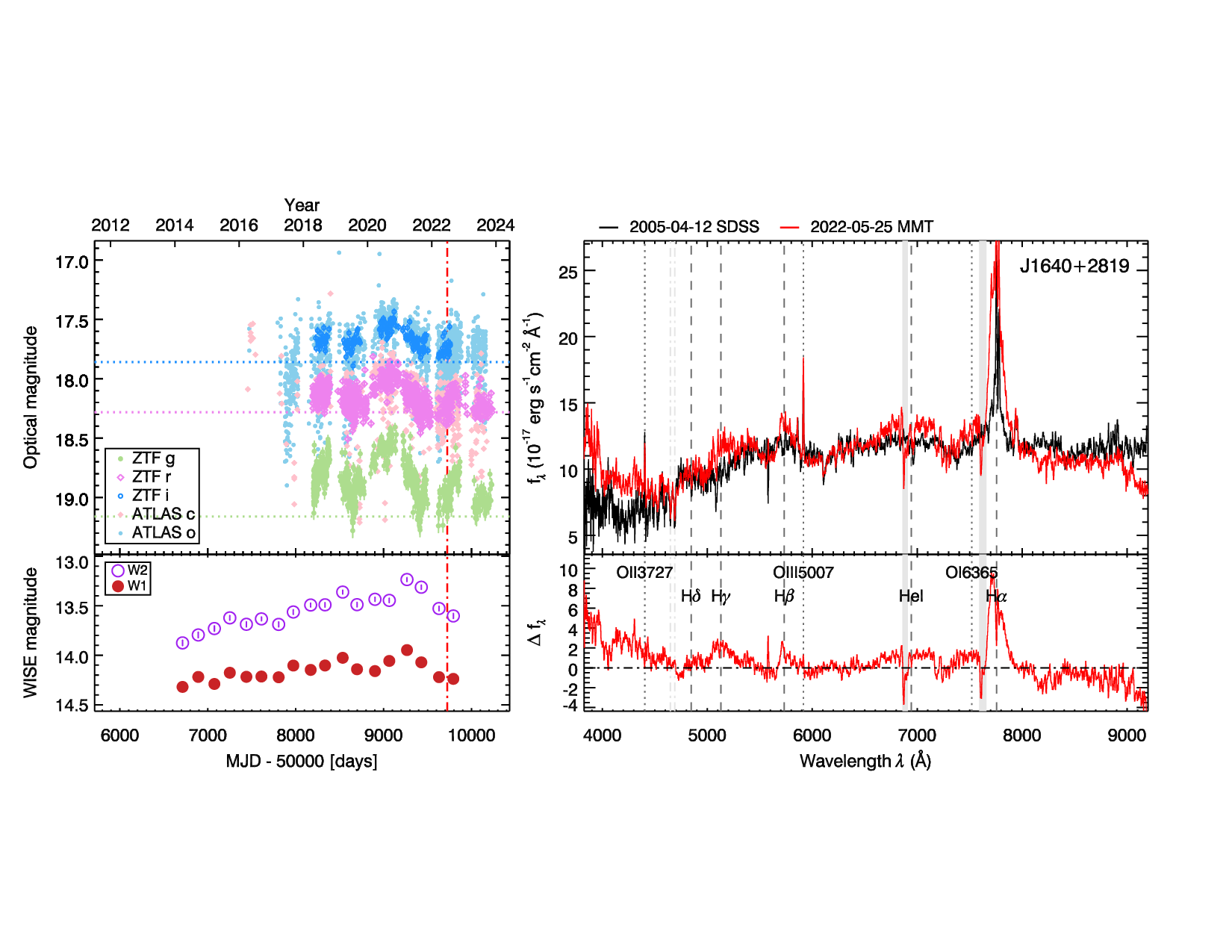}
\hspace{-0.7cm}
\includegraphics[width=0.52\textwidth]{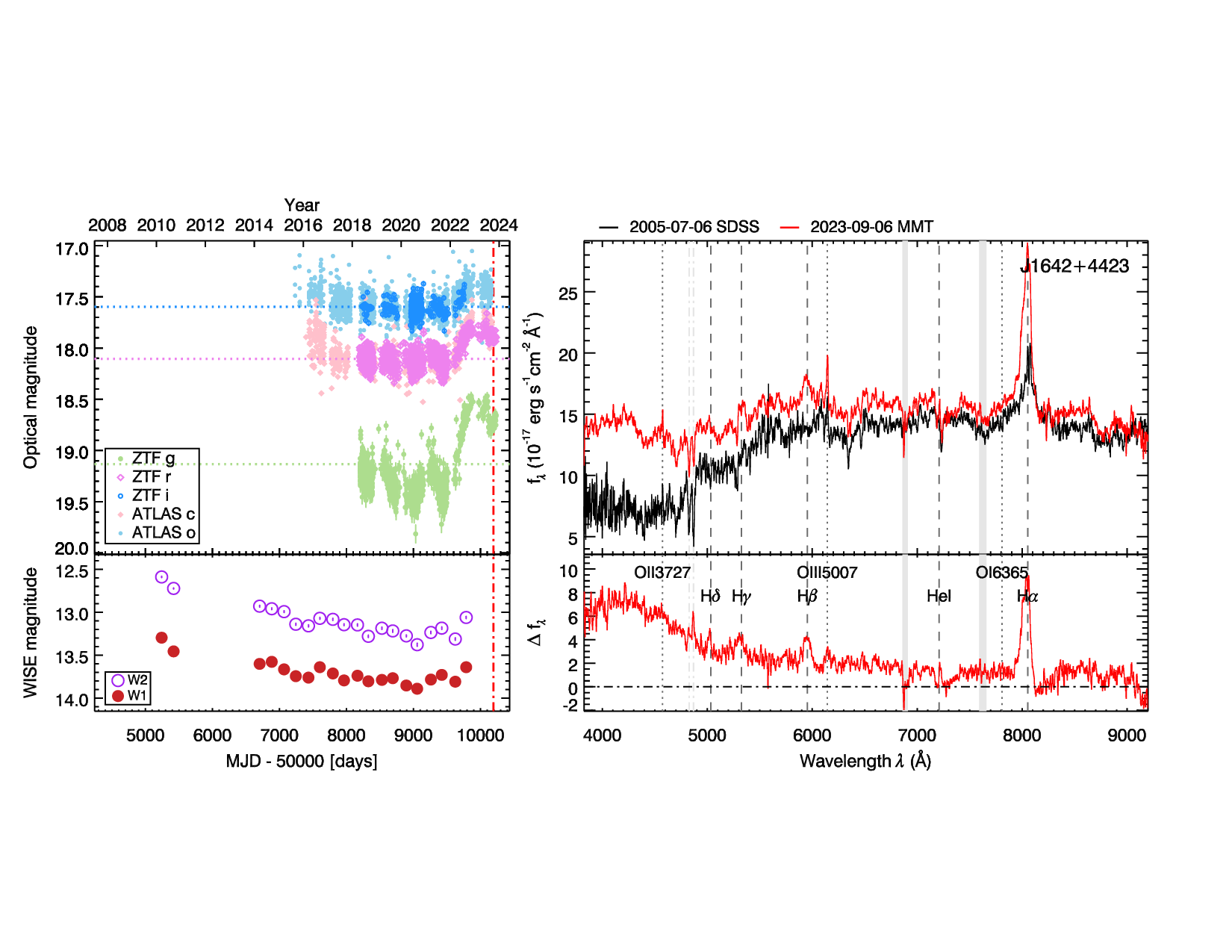}\\
\vspace{-2.7cm}
\hspace{-0.4cm}
\includegraphics[width=0.52\textwidth]{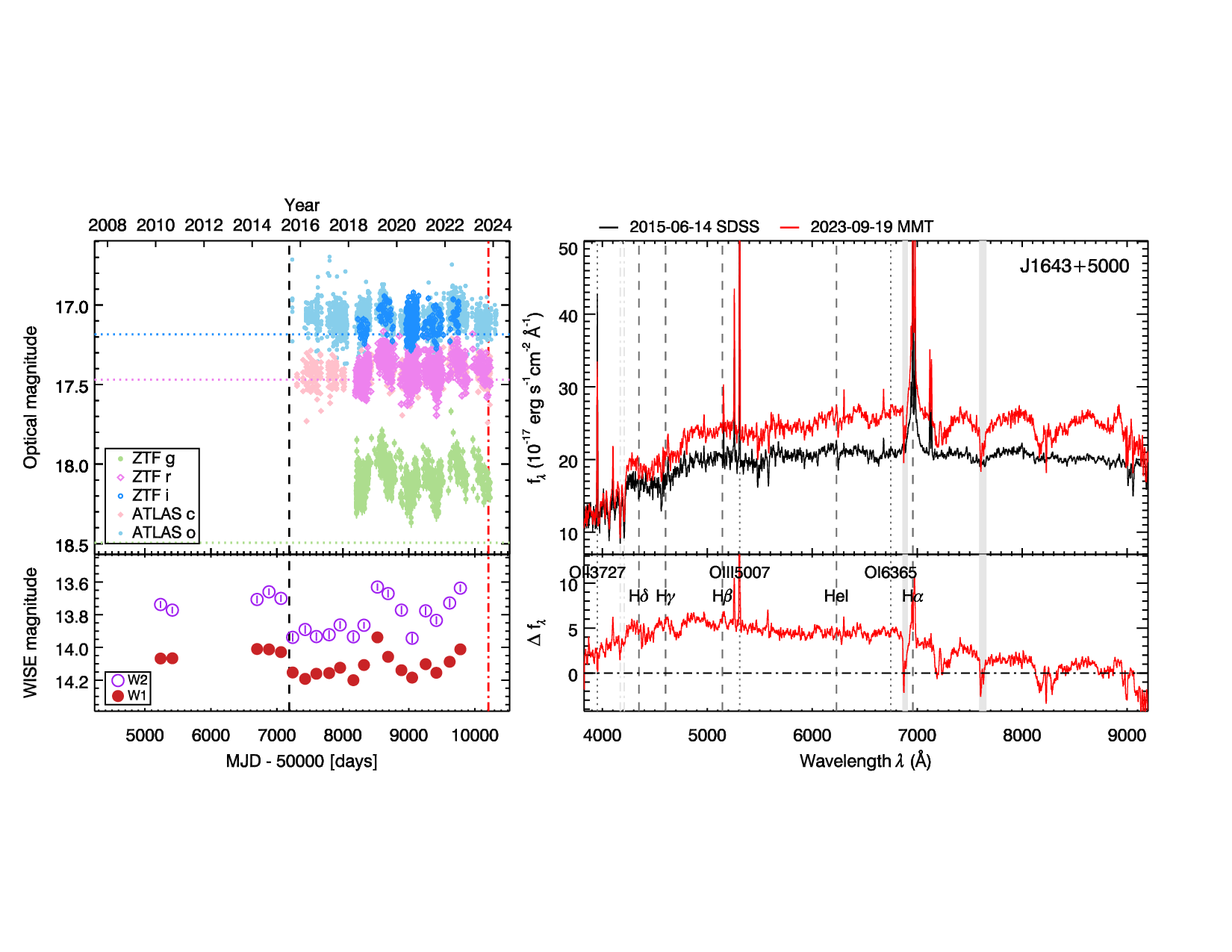}
\hspace{-0.7cm}
\includegraphics[width=0.52\textwidth]{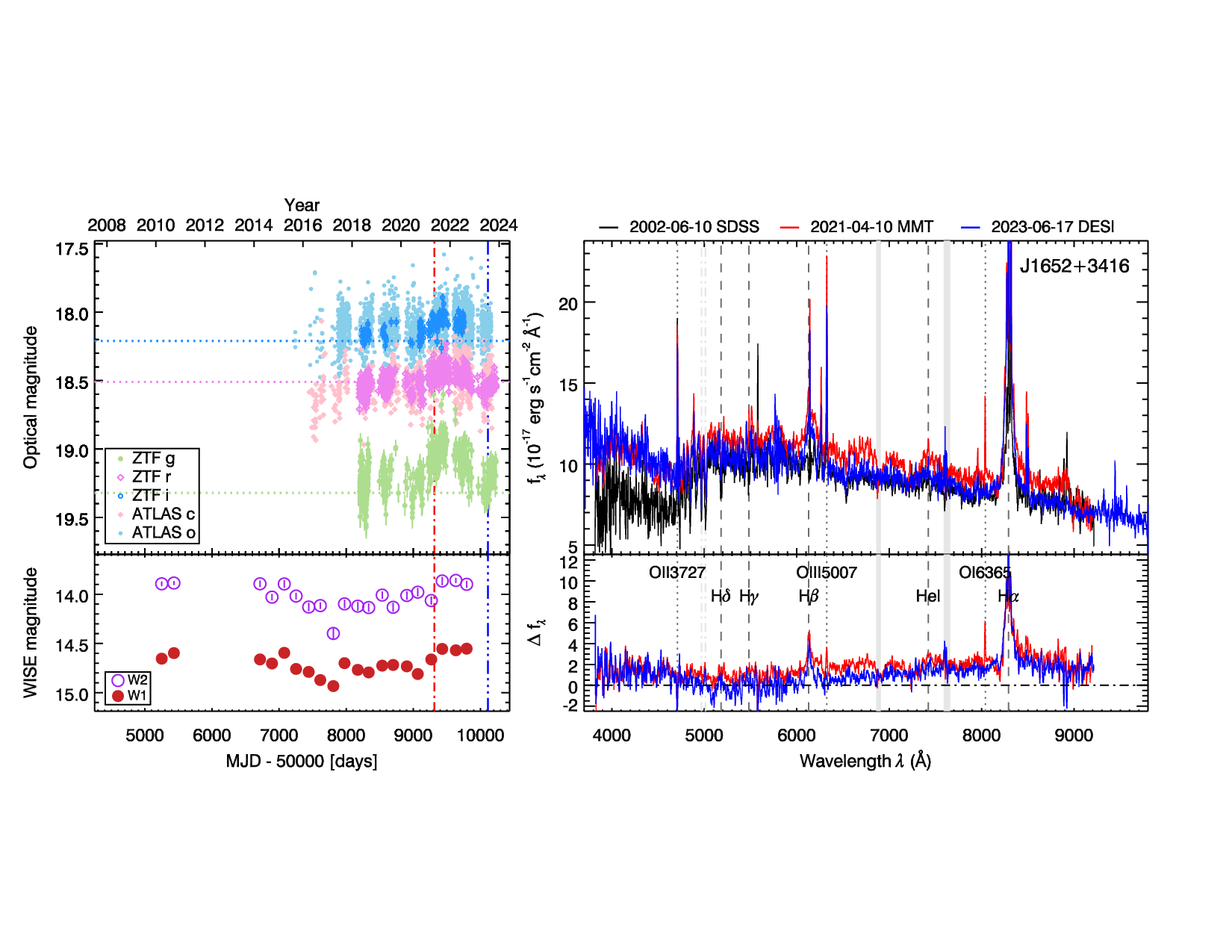}\\
\vspace{-2.7cm}
\hspace{-0.4cm}
\includegraphics[width=0.52\textwidth]{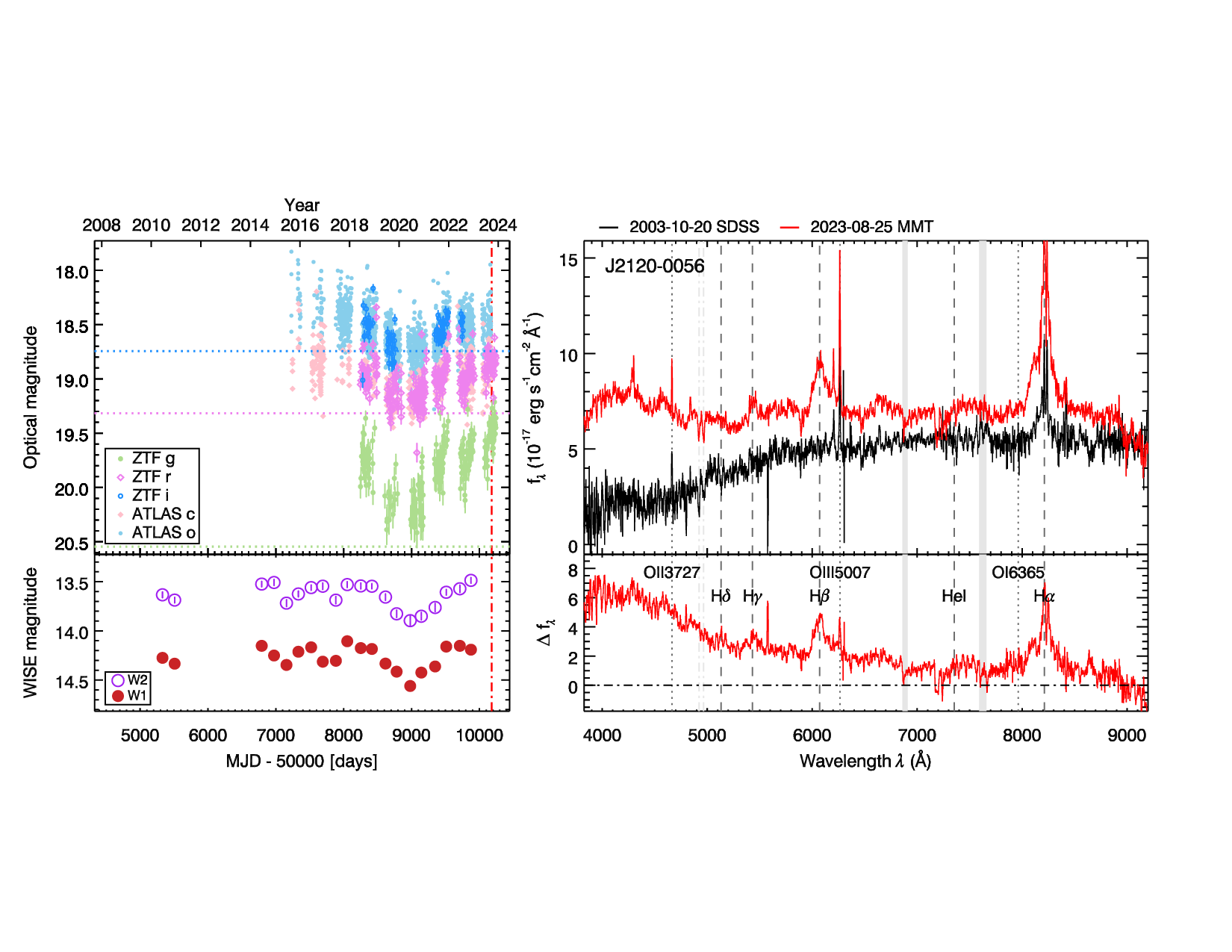}
\hspace{-0.7cm}
\includegraphics[width=0.52\textwidth]{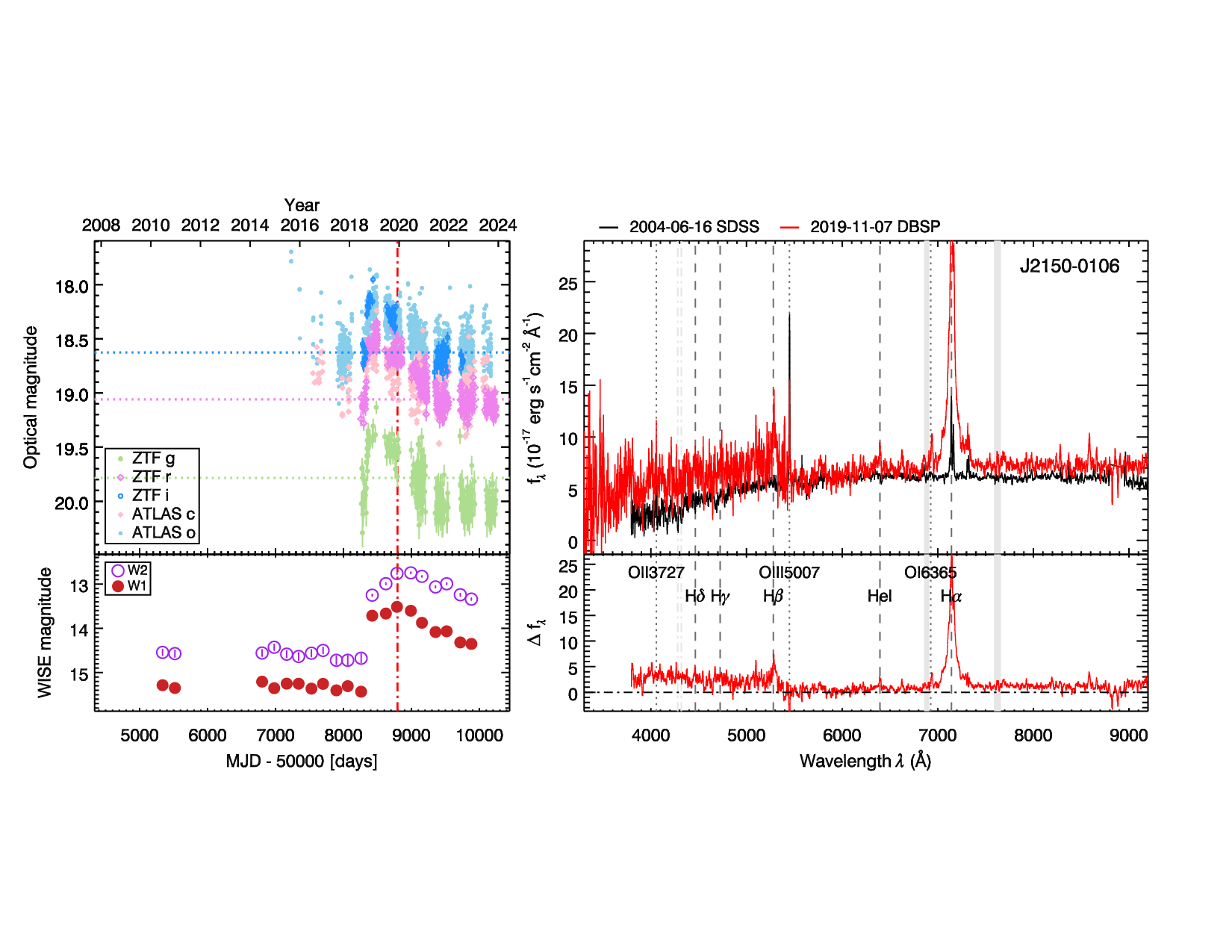}\\
\caption{Confirmed CLQs. Same as Figure \ref{fig:example}.}
\label{fig:optical_spec}
\end{figure*}

% \clearpage
\begin{figure*}[!ht]
\centering
\vspace{-1cm}
\hspace{-0.4cm}
\includegraphics[width=0.52\textwidth]{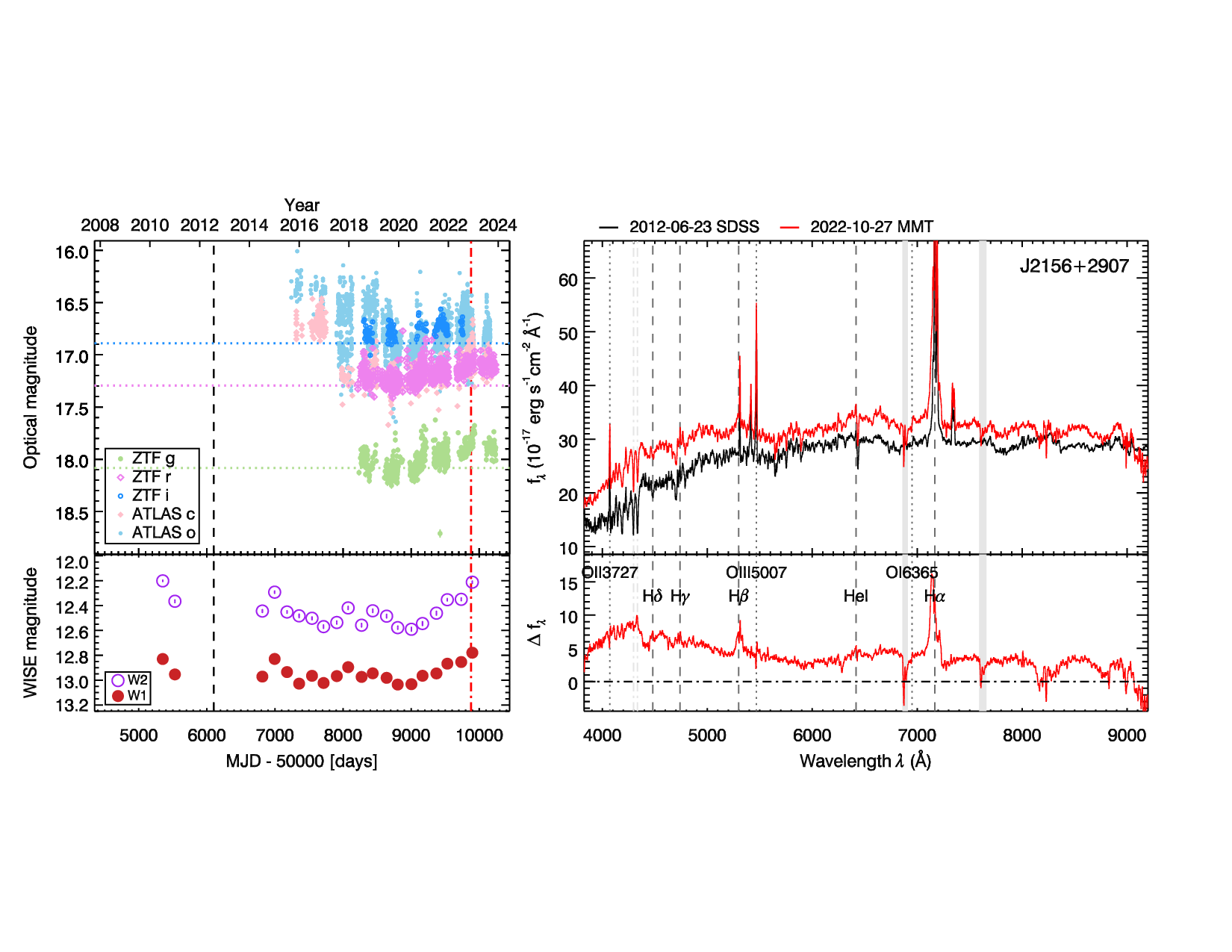}
\hspace{-0.7cm}
\includegraphics[width=0.52\textwidth]{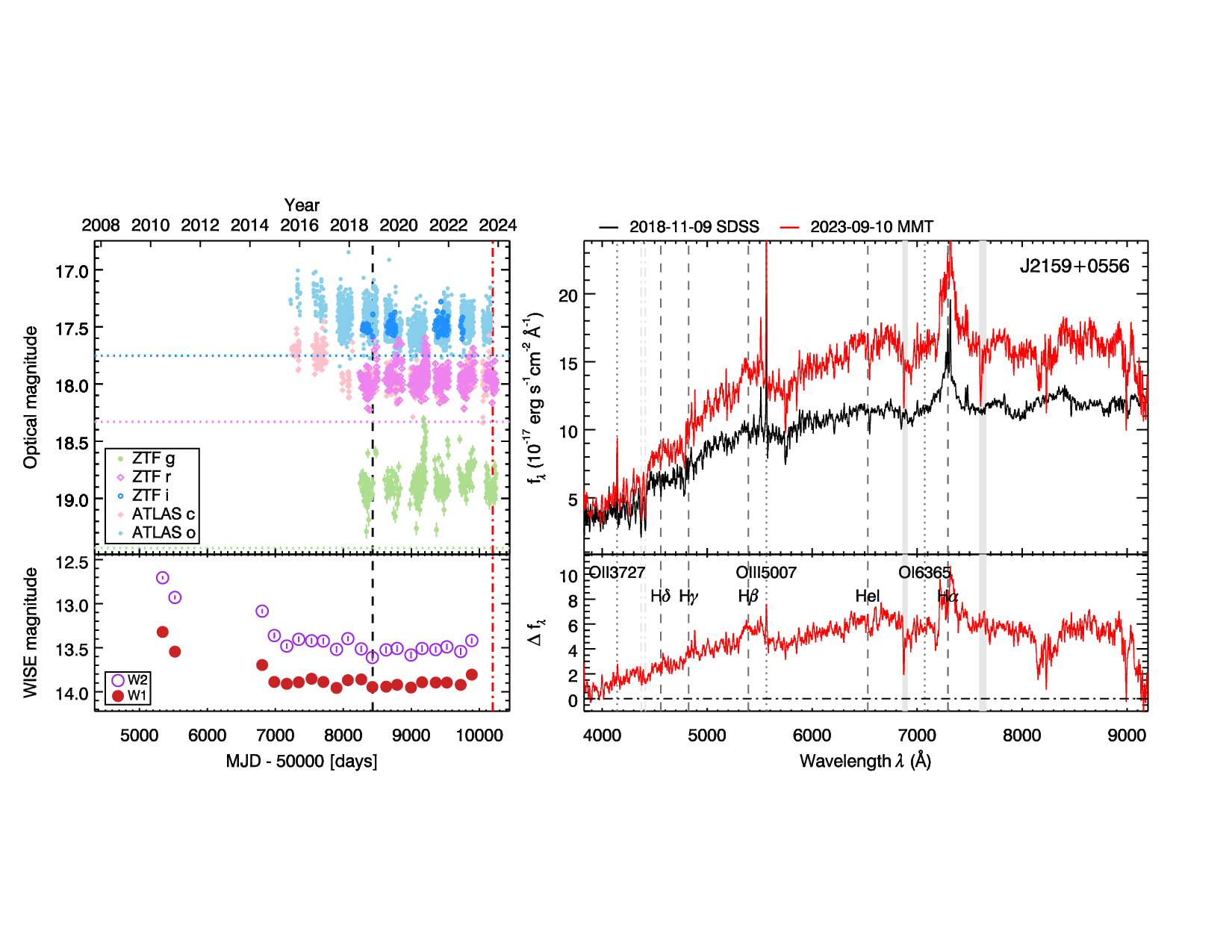}\\
\vspace{-2.7cm}
\hspace{-0.4cm}
\includegraphics[width=0.52\textwidth]{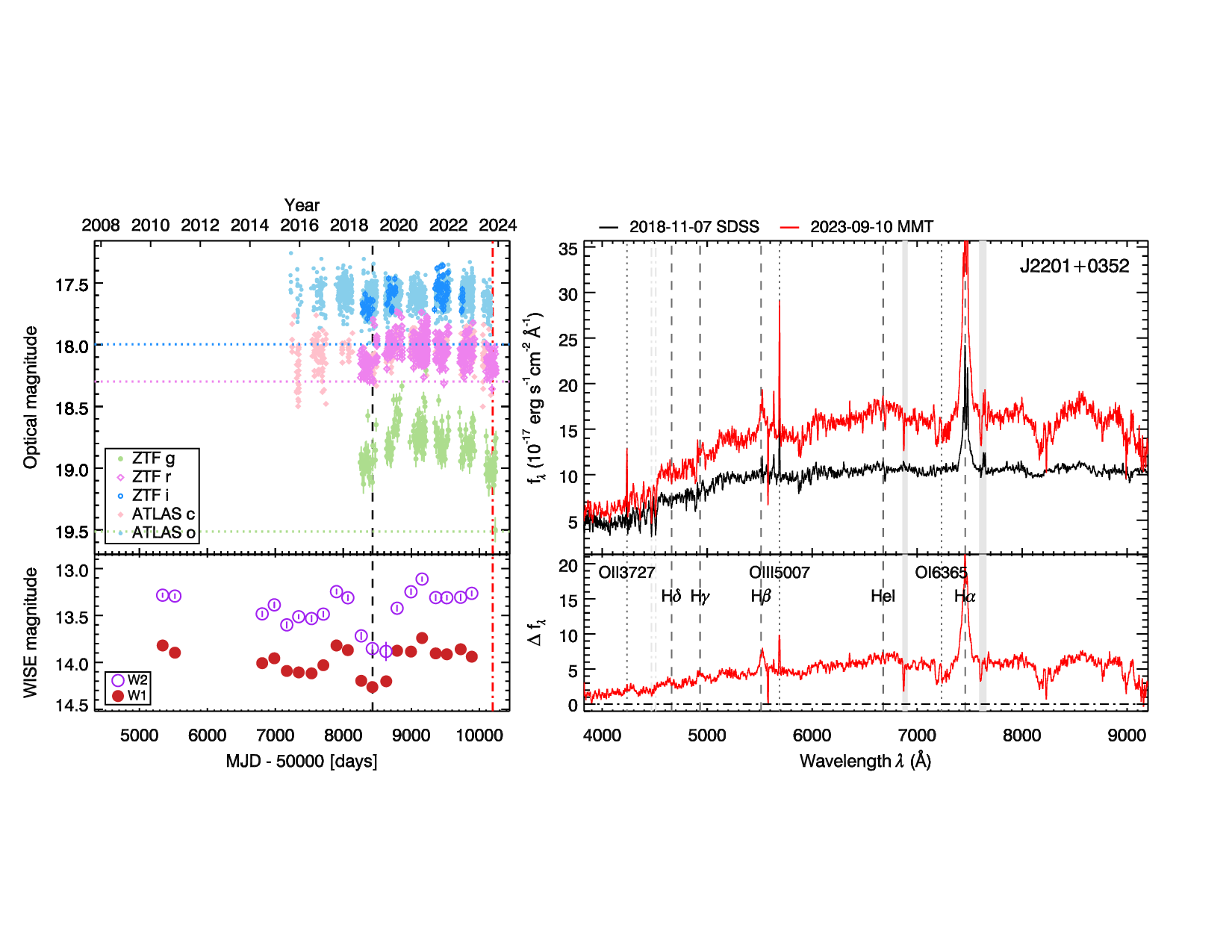}
\hspace{-0.7cm}
\includegraphics[width=0.52\textwidth]{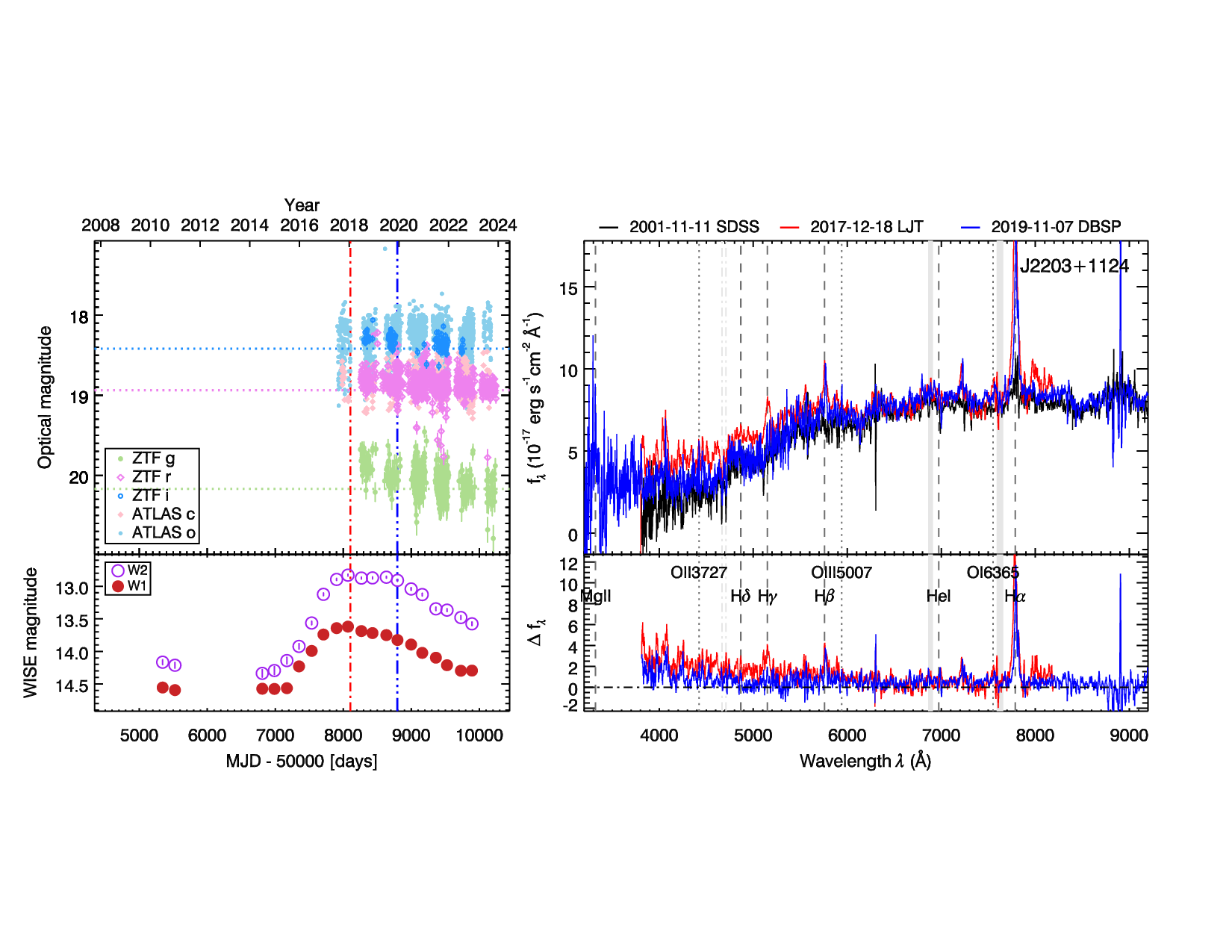}\\
\vspace{-2.7cm}
\hspace{-0.4cm}
\includegraphics[width=0.52\textwidth]{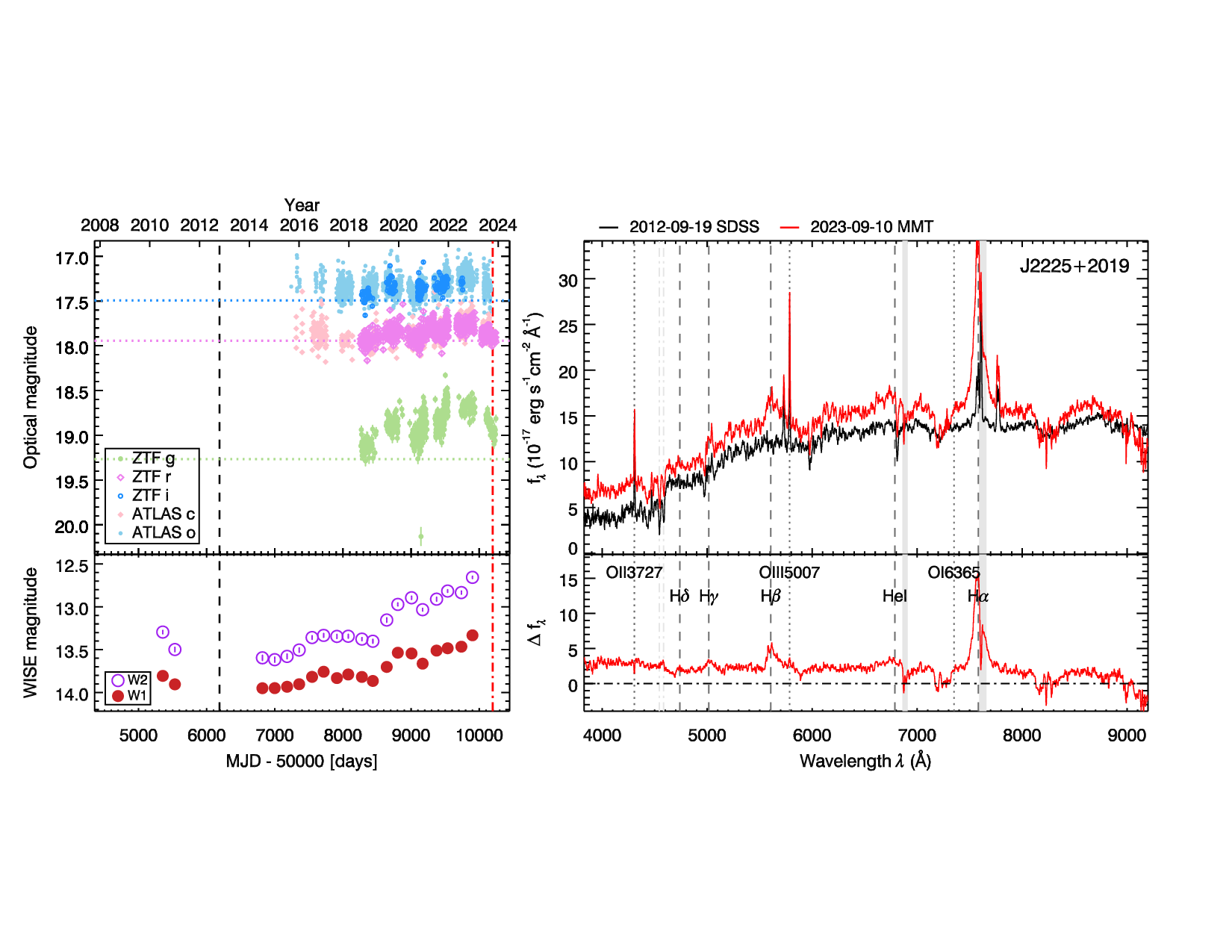}
\hspace{-0.7cm}
\includegraphics[width=0.52\textwidth]{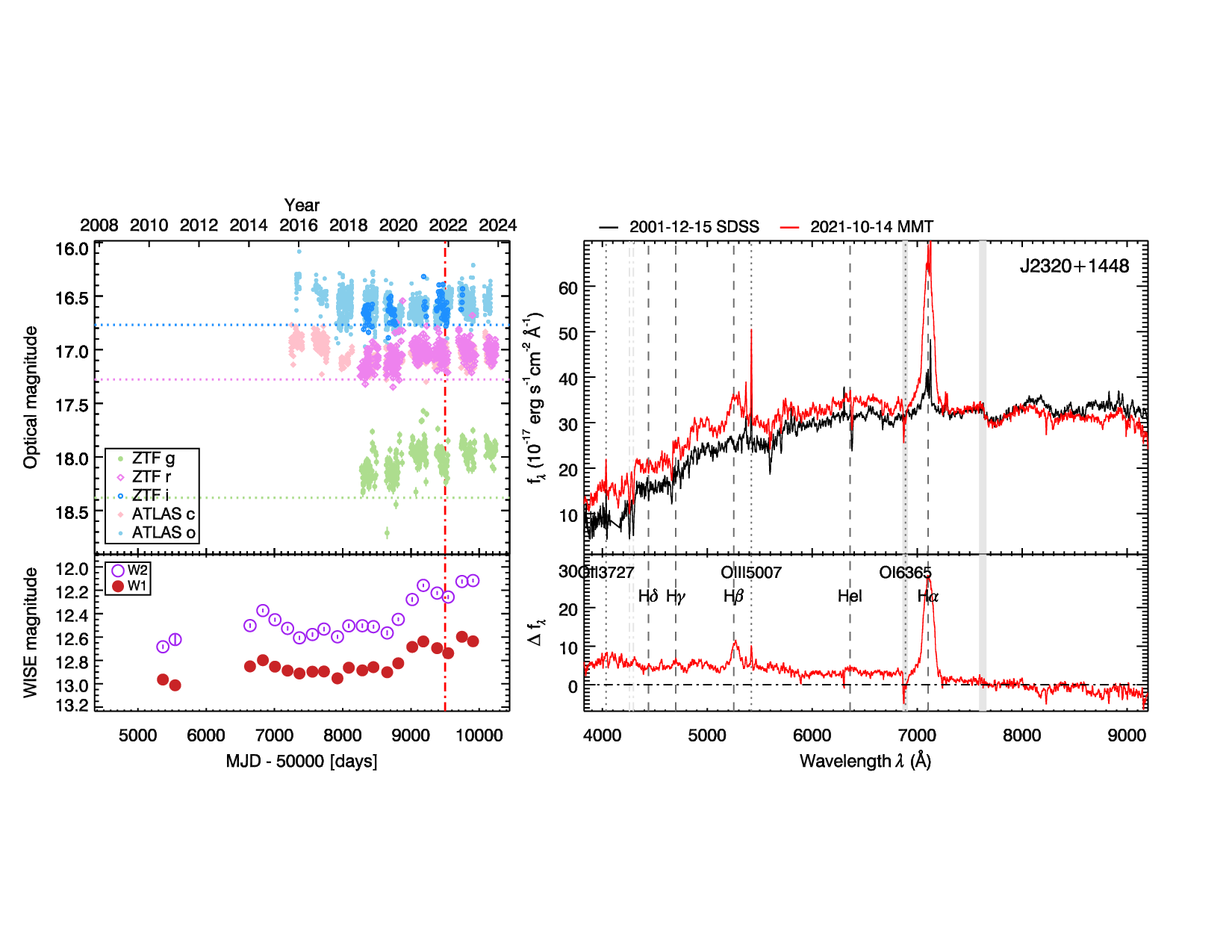}\\
\vspace{-2.7cm}
\hspace{-0.4cm}
\includegraphics[width=0.52\textwidth]{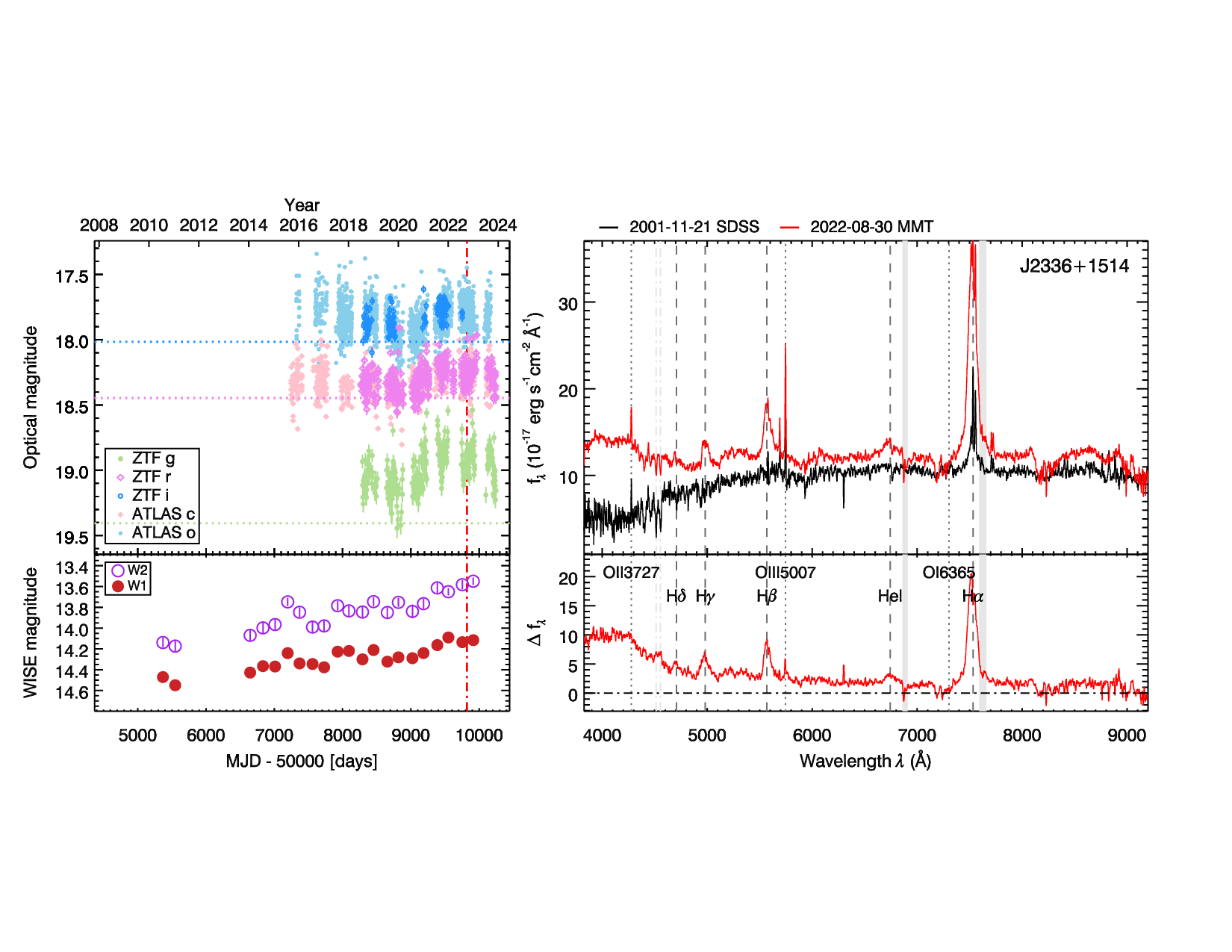}
\hspace{-0.7cm}
\includegraphics[width=0.52\textwidth]{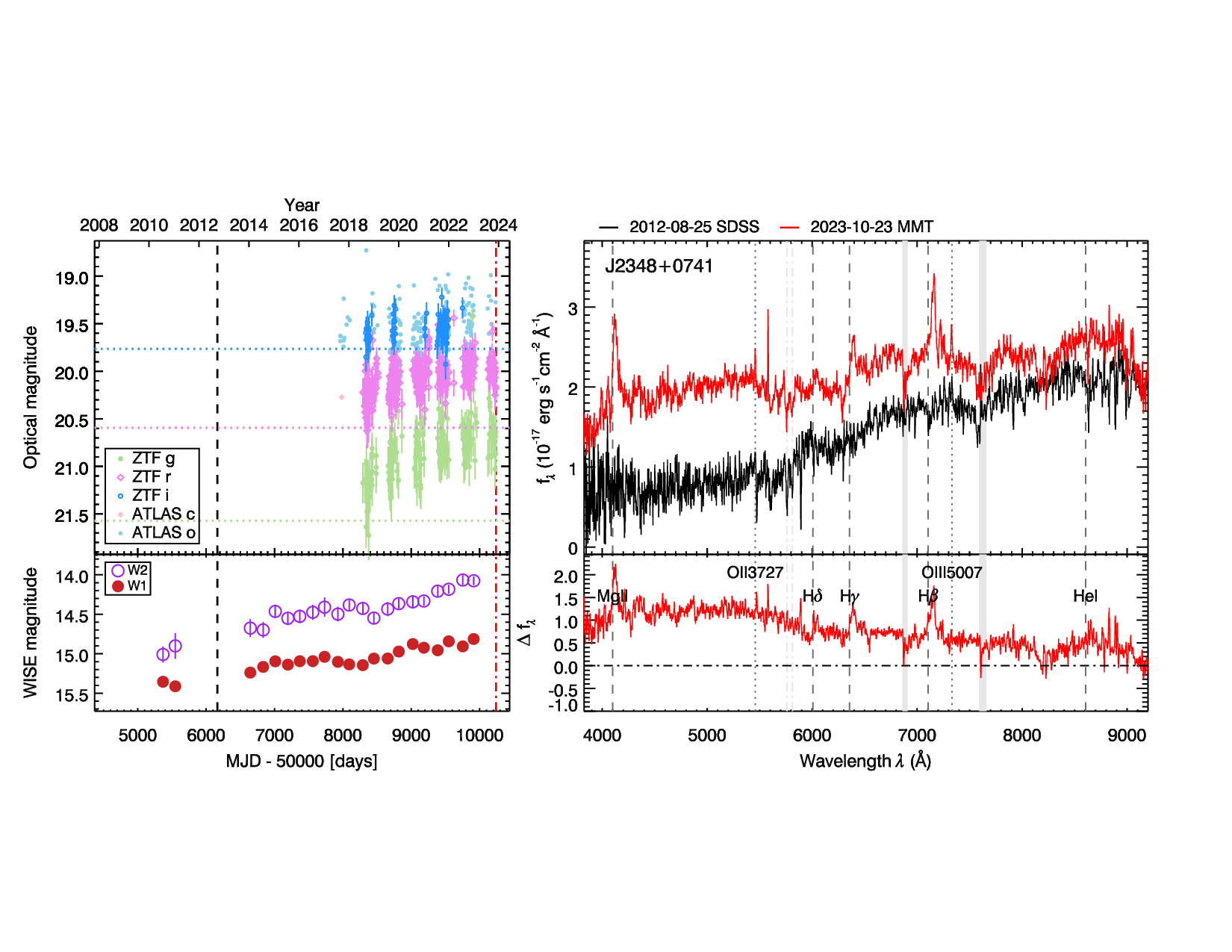}\\
\caption{Confirmed CLQs. Same as Figure \ref{fig:example}.}
\label{fig:optical_spec}
\end{figure*}

\end{document}